\documentclass[aps,superscriptaddress,floatfix,twocolumn,longbibliography]{revtex4-1}
\usepackage[latin9]{inputenc}
\usepackage{amsmath}
\usepackage{amssymb}
\usepackage{graphicx}
\usepackage{esint}
\usepackage{epstopdf}
\usepackage{braket}
\usepackage{color}
\usepackage{hyperref}
\usepackage{graphicx}
\usepackage{stackengine}
\makeatletter
\@ifundefined{textcolor}{}
{
 \definecolor{BLACK}{gray}{0}
 \definecolor{WHITE}{gray}{1}
 \definecolor{RED}{rgb}{1,0,0}
 \definecolor{GREEN}{rgb}{0,1,0}
 \definecolor{BLUE}{rgb}{0,0,1}
 \definecolor{CYAN}{cmyk}{1,0,0,0}
 \definecolor{MAGENTA}{cmyk}{0,1,0,0}
 \definecolor{YELLOW}{cmyk}{0,0,1,0}
}

\makeatother

\begin{document}

\title{The BHL-BCL crossover: from nonlinear to linear quantum amplification}

\author{Juan Ram\'on Mu\~noz de Nova}
\email{jrmnova@fis.ucm.es}
\affiliation{Departamento de F\'isica de Materiales, Universidad Complutense de
Madrid, E-28040 Madrid, Spain}

\author{Fernando Sols}
\affiliation{Departamento de F\'isica de Materiales, Universidad Complutense de
Madrid, E-28040 Madrid, Spain}

\begin{abstract}
The black-hole laser (BHL) effect is the self-amplification of Hawking radiation in the presence of a pair of horizons which act as a resonant cavity. In a flowing atomic condensate, the BHL effect arises in a finite supersonic region, where Bogoliubov-Cherenkov-Landau (BCL) radiation is coherently excited by any static perturbation. Thus, experimental attempts to produce a black-hole laser unavoidably deal with the presence of a strong BCL background, making the observation of the BHL effect still a major challenge in the analogue gravity field. Here, we perform a theoretical study of the BHL-BCL crossover using an idealized model where both phenomena can be unambiguously isolated. By drawing an analogy with an unstable pendulum, we distinguish three main regimes according to the interplay between quantum fluctuations and classical stimulation: quantum BHL, classical BHL, and BCL. Based on quite general scaling arguments, the nonlinear amplification of the initial amplitude of the quantum fluctuations up to saturation is identified as the most robust trait of a quantum BHL. A classical BHL behaves instead as a linear quantum amplifier, where the output is proportional to the input. The BCL regime also acts as a linear quantum amplifier, but its gain is exponentially smaller as compared to a classical BHL. In addition, we find that the decrease in the amplification for increasing BCL amplitude or the nonmonotonic dependence of the growth rate with respect to the background parameters are complementary signatures of black-hole lasing. We also identify interesting novel analogue phenomena such as Hawking-stimulated white-hole radiation or quantum BCL-stimulated Hawking radiation. The results of this work not only are of interest for analogue gravity, where they help to distinguish each phenomenon and to design experimental setups leading to a clear observation of the BHL effect, but they also open the prospect of finding applications of analogue concepts in quantum technologies.
\end{abstract}
\date{\today}

\maketitle

\section{Introduction}\label{sec:Intro}

As noted by Unruh \cite{Unruh1981}, the equations of motion governing the perturbations around an ideal potential flow are formally analogue to those of a massless scalar field in a curved spacetime described by the so-called acoustic metric. This discovery gave rise to the field of analogue gravity, in which inaccessible gravitational phenomena can be modeled using tabletop experiments such as atomic Bose-Einstein condensates \cite{Garay2000,Lahav2010}, water waves \cite{Weinfurtner2011,Euve2016}, nonlinear optical fibers \cite{Belgiorno2010,Drori2019}, ion rings \cite{Horstmann2010,Wittemer2019}, quantum fluids of light \cite{Carusotto2013,Nguyen2015}, or even superconducting transmon qubits \cite{Shi2023}. As a result, analogues of the dynamical Casimir effect \cite{Jaskula2012}, Sakharov oscillations \cite{Hung2013}, superradiance \cite{Torres2017}, inflation \cite{Eckel2018}, Hawking radiation \cite{deNova2019}, Unruh effect \cite{Hu2019}, quasinormal ringdown \cite{Torres2020}, backreaction \cite{Patrick2021} or cosmological particle creation \cite{Steinhauer2022a} have been observed in the laboratory. Quantum field simulators of curved spacetimes are already available in the laboratory \cite{Viermann2022}.

In this context, another remarkable phenomenon is the black-hole laser (BHL) effect \cite{Corley1999}, i.e., the self-amplification of Hawking radiation due to successive reflections between a pair of horizons, leading to the emergence of dynamical instabilities in the excitation spectrum. In an atomic condensate, the BHL effect can take place because of its superluminal dispersion relation, which allows the radiation reflected at the inner horizon to travel back to the outer one  \cite{Leonhardt2003,Barcelo2006,Jain2007,Coutant2010,Finazzi2010,Bermudez2018,Burkle2018}. Other analogue setups have been proposed to observe black-hole lasing \cite{Faccio_2012,Peloquin2016,RinconEstrada2021,Katayama2021}. 

Originally, due to its stimulated character, it was thought that the observation of the BHL effect would be a first step towards the dreamed observation of the Hawking effect. In actuality, it has been rather the opposite. This is because in a condensate the BHL effect arises in a finite supersonic region, which is energetically unstable according to the Landau criterion so any static perturbation will resonantly produce Bogoliubov-Cherenkov-Landau (BCL) radiation \cite{Carusotto2006}, the analogue of the undulation in hydraulic setups \cite{Coutant2012}. Furthermore, the BHL modes are expected to contain similar wavevectors and frequencies to those of the BCL wave, as the latter is also stimulated by the scattering of Hawking radiation at the inner horizon. Therefore, experimental attempts to isolate the BHL effect will be hindered by a background BCL signal. Indeed, the first reported observation of the BHL effect in 2014 \cite{Steinhauer2014} was later explained in terms of experimental BCL fluctuations \cite{Wang2016,Wang2017}. A mechanism of BCL-stimulated Hawking radiation was identified in 2021 \cite{Kolobov2021}, whose role in the 2014 experiment was recently confirmed \cite{Steinhauer2022}. Hence, the observation of the BHL effect still remains a major challenge in the analogue field.

As a result, it is of paramount importance to understand the  characteristic features of both the BHL and BCL mechanisms in order to eventually achieve a true BHL configuration. So far, the literature has mostly addressed the BHL-BCL crossover by directly simulating realistic setups resembling the actual experiments \cite{Tettamanti2016,Steinhauer2017,Wang2016,Wang2017,Llorente2019,Tettamanti2021,Kolobov2021,Steinhauer2022}. In this work, we adopt an alternative approach by using a simple yet highly idealized model, namely, the flat-profile model \cite{Carusotto2008,Recati2009}, where the coupling constant and external potential are perfectly matched so that the background condensate flow is homogeneous while the speed of sound is tunable at will. Albeit quite unfeasible from an experimental point of view, the theoretical simplicity of this model has allowed to clearly identify key features of several phenomena such as Hawking correlations, resonant Hawking radiation, or black-hole lasers \cite{Recati2009,deNova2014,Michel2013}, helping in this way to understand the underlying physics in more realistic scenarios of higher complexity \cite{deNova2019,Jacquet2023,deNova2021a}. 

Here, the flat-profile configuration allows us to neatly isolate the BHL and BCL contribution to the dynamics so their genuine signatures can be extracted. Using this simplified model, we report an intensive campaign of numerical simulations using the Truncated Wigner approximation \cite{Sinatra2002,Carusotto2008} to compute the dynamics of the quantum fluctuations. In particular, we focus on evaluating density-based observables, especially the density-density correlation function \cite{Carusotto2008,Recati2009}, the main tool in actual experiments  \cite{Steinhauer2014,Steinhauer2016,deNova2019,Kolobov2021}.

Dynamical instabilities are quantized like unstable harmonic oscillators \cite{Leonhardt2003,Finazzi2010,Burkle2018}. Hence, a black-hole laser is conceptually similar to an unstable pendulum. Based on this analogy, according to the interplay between quantum fluctuations and Cherenkov stimulation, where the latter plays the role of an external force, we identify three main regimes: quantum BHL, classical BHL, and BCL. In a quantum BHL, the dynamics is fully driven by the amplification of the quantum fluctuations of the lasing modes, akin to the zero-point decay of the unstable equilibrium position of a pendulum. In a classical BHL, the lasing instability is given a well-defined classical amplitude, excited by the background Cherenkov wave, akin to the case where a small kick is given to an unstable pendulum. In the BCL regime, the Cherenkov stimulation is large enough to completely dominate the dynamics, overshadowing the BHL effect, akin to applying a strong external force which overcomes gravity. 

In all cases, the dynamics eventually saturates when the system reaches certain metastable state. We find that the most characteristic signature of each regime is its efficiency in amplifying the initial amplitude of quantum fluctuations during the transient before saturation. A quantum BHL behaves as a nonlinear quantum amplifier, increasing quantum fluctuations up to the same saturation amplitude regardless of their initial strength. This is because in a quantum BHL the dynamics is also driven by quantum fluctuations themselves rather than by a background mean field. A classical BHL corresponds instead to a linear quantum amplifier, where the output is proportional to the input, with this process taking place around a well-defined mean-field trajectory which determines the saturation point. The BCL regime also provides linear quantum amplification around a background mean field, but its gain is exponentially smaller as compared to a classical BHL. This is because there is no microscopic mechanism of amplification operating, and the amplification merely originates due to the background density modulation of the BCL wave. Our analysis relies on general scaling arguments that are quite independent of the details of the model. 

The above results imply that the quantum amplification decreases with the BCL amplitude in the lasing regime, in contrast with its increasing behavior in the Cherenkov regime, providing a qualitative characterization tool. Complementarily, we propose the strongly nonmonotonic dependence of the lasing growth rate with respect to the background parameters (e.g., the cavity length or the flow speed) as another criterion to distinguish classical BHL from BCL stimulation, since the latter is expected to depend smoothly on the properties of the background flow.

Remarkably, in the process of analyzing the BHL-BCL correlation patterns, we also identify interesting analogue phenomena such as Hawking-stimulated white-hole \cite{Mayoral2011} (HSWH) radiation at the beginning of the black-hole lasing process, or BCL-stimulated Hawking radiation \cite{Kolobov2021} of a quantum origin, which can be understood as spontaneous resonant Hawking radiation \cite{Zapata2011} above a nonlinear BCL undulation that acts as a resonator.

Apart from their intrinsic interest for the analogue field, where they help to isolate the distinctive signatures of each phenomenon and to design experimental setups leading to an unambiguous demonstration of the BHL effect, the results of this work may also have a potential impact on atomtronics \cite{Amico2021} and on the more general field of quantum technologies.

The article is arranged as follows. Section \ref{sec:Model} introduces the model considered in this work while Sec. \ref{sec:Tools} discusses the tools used for the analysis. The numerical results are presented in Sec. \ref{sec:Numerical}. A global discussion on the physical significance of the findings of the paper is included in Sec. \ref{sec:Discussion}, along with a summary table containing the main results for each identified regime. Conclusions and future perspectives are outlined in Sec. \ref{sec:conclu}.

\begin{figure*}[t]
\begin{tabular}{@{}cc@{}}\stackinset{l}{0pt}{t}{0pt}{\large{(a)}}{\includegraphics[width=\columnwidth]{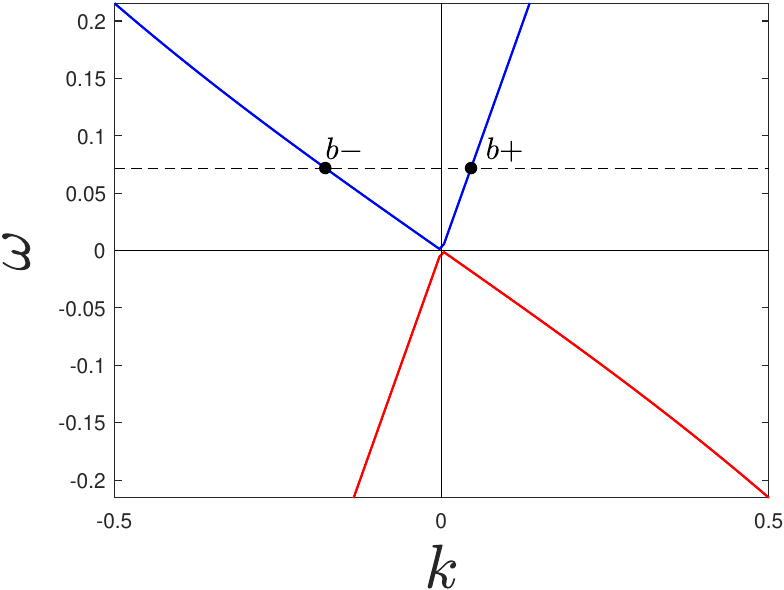}}~ &~
    \stackinset{l}{0pt}{t}{0pt}{\large{(b)}}
    {\includegraphics[width=0.98\columnwidth]{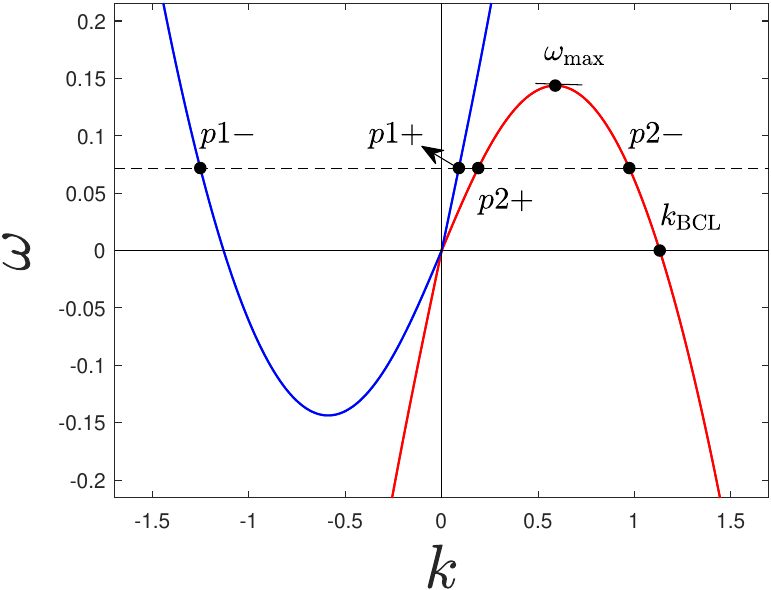}} \\ \\ \\
    \stackinset{l}{0pt}{t}{0pt}{\large{(c)}}{\includegraphics[width=\columnwidth]{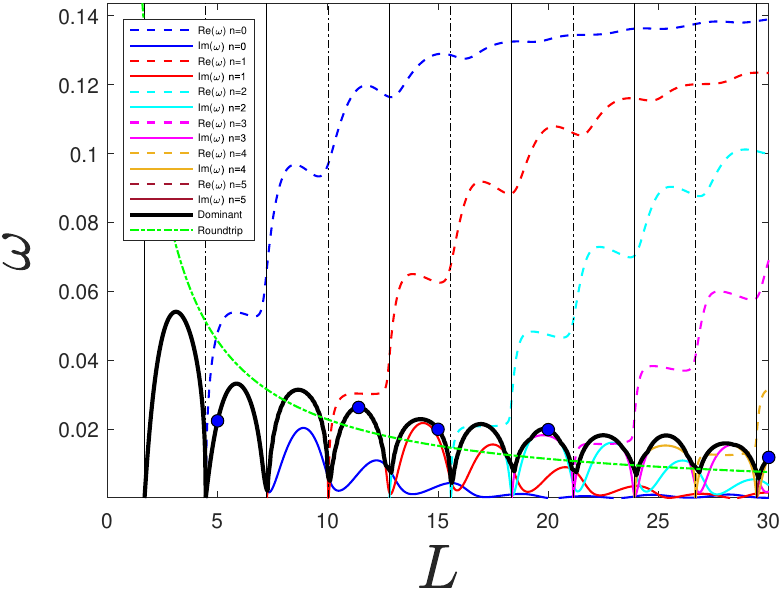}} &
    \stackinset{l}{0pt}{t}{0pt}{\large{(d)}}
    {\includegraphics[width=\columnwidth]{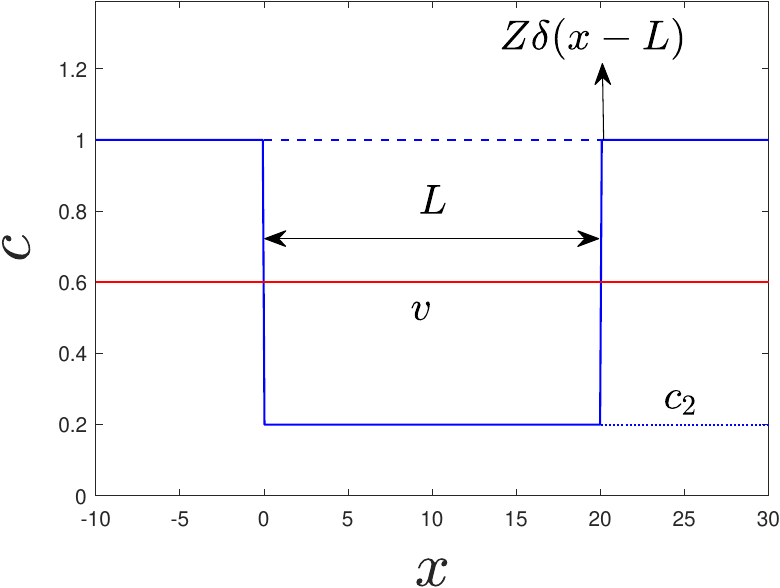}} 
\end{tabular}
\caption{Flat-profile BHL configuration with $v=0.6,~c_2=0.2$. (a) Dispersion relation in the subsonic region. The blue/red lines signal the $\pm$ branches of Eq. (\ref{eq:Dispersion}), while the $\pm$ label denotes whether the modes travel along/against the flow. (b) Dispersion relation in the supersonic region. For a certain frequency below the cutoff frequency $\omega_{\rm{max}}$, all wavevectors are purely real (horizontal dashed line). The BCL mode has zero frequency and finite wavevector $k_{\rm{BCL}}$. (c) Spectrum of dynamical instabilities as a function of the cavity length $L$. Solid (dashed) lines are the imaginary (real) part of the frequency. Vertical solid (dashed) black lines represent the critical lengths $L_n$ ($L_{n+1/2}$). Thick solid black envelope highlights the growth rate $\Gamma$ of the dominant mode. Dash-dotted green line is the inverse of the roundtrip time. Blue dots indicate the numerical values considered. (d) Schematic depiction of the model used in this work. The initial homogeneous condensate is characterized by its sound (horizontal dashed blue) and flow speed (solid red). At $t=0$, the sound speed is quenched to $c_2$ for $x>0$ (horizontal dotted blue) and a black hole is formed. At $t=t_{\rm{BCL}}>0$, a delta barrier is switched on at $x=L$ (vertical arrow), which stimulates BCL radiation. Finally, at $t=t_{\rm{BHL}}\geq t_{\rm{BCL}}$, the speed of sound is quenched back to its original value for $x>L$, giving rise to the flat-profile BHL (solid blue).}
\label{fig:BHLBCLScheme}
\end{figure*}

\section{The model}\label{sec:Model}

In order to study the BHL-BCL crossover, we use the flat-profile model as a testing ground \cite{Carusotto2008,Recati2009,deNova2016}, where both phenomena can be controlled and isolated. For times $t<0$, we consider a stationary one-dimensional homogeneous quasicondensate \cite{Menotti2002} flowing from left to right, described by a time-independent Gross-Pitaevskii (GP) wave function $\Psi_0(x)=\sqrt{n_0}e^{iqx}$. The corresponding sound and flow speeds are $c_0=\sqrt{gn_0/m}$ and $v=\hbar q/m$, with $g$ the coupling constant and $m$ the mass of the atoms. Hereafter, we set $\hbar=m=c_0=1$, and rescale the GP wave function as $\Psi(x,t) \to \sqrt{n_0}\Psi(x,t)$ so it become dimensionless. The interested reader may consult Refs. \cite{Recati2009,deNova2017b} and Refs. \cite{deNova2016,deNova2021a} for detailed discussions that follow the notation of this work on both Hawking radiation and black-hole lasers, respectively.

Quantum fluctuations of the condensate are described by the Bogoliubov-de Gennes (BdG) equations. For a homogeneous condensate, the BdG modes are given in terms of plane waves with wavevector $k$, and the corresponding dispersion relation is determined by the flow and sound speeds $v,c$ as
\begin{equation}\label{eq:Dispersion}
    \omega=vk\pm \Omega_k,~\Omega_k=\sqrt{c^2k^2+\frac{k^4}{4}},
\end{equation}
with $\Omega_k$ the usual Bogoliubov dispersion relation for a condensate at rest and $vk$ the Doppler shift. This dispersion relation yields four wavevectors for given frequency $\omega$, displaying two qualitatively different regimes, subsonic ($v<c$) or supersonic ($v>c$), as shown in upper row of Fig. \ref{fig:BHLBCLScheme}. In the subsonic case, Fig. \ref{fig:BHLBCLScheme}a, for a given frequency $\omega>0$ there are two modes with real wavevector (the other two ones are exponentially growing/decaying solutions), labeled as $b\pm$, where the $\pm$ indicates whether they propagate along/against the condensate flow, respectively. In the supersonic case, Fig. \ref{fig:BHLBCLScheme}b, for any positive frequency below the cutoff $\omega_{\rm{max}}$ all wavevectors are purely real. The $p1\pm$ modes arise from the $+$ branch in Eq. (\ref{eq:Dispersion}), while the $p2\pm$ modes arise from the $-$ branch and are the conjugates of the negative energy modes in the $+$ branch. The presence of these negative energy modes is a characteristic feature of a supersonic flow, revealing its energetic instability. Specifically, the Landau criterion establishes the appearance of a zero-frequency mode, the BCL mode, with a finite wavevector $k_{\rm{BCL}}=2\sqrt{v^2-c^2}$ computed by $vk_{\rm{BCL}}=\Omega_{k_{\rm{BCL}}}$. As a result, any static perturbation in the flow will resonantly excite BCL radiation in the condensate \cite{Carusotto2006}. This is because the BdG equations describe at the same time both linear collective motion, coherently imprinted on the condensate and accounted by linear perturbations $\delta \Psi$ of the GP wave function, and quantum quasiparticle excitations, accounted by the quantum fluctuations of the field operator $\delta \hat{\Psi}$ around the mean-field coherent expectation value. Nevertheless, we recall that energetic instability (presence of modes with negative energy) is only a necessary condition for dynamical instability (presence of modes with blowing complex frequency), but not sufficient \cite{Wu2003}.

In our configuration, we choose the initial condensate to be subsonic, $v<1$. Now, at $t=0$, we quench inhomogeneously both the external potential $W(x)$ and the coupling constant $g(x)$ so that
\begin{equation}\label{eq:FlatProfile}
    g(x)+W(x)=E_b,
\end{equation}
with $E_b$ some constant energy that can be subtracted from the Hamiltonian. In this way, $\Psi_0(x)=e^{ivx}$ remains a stationary solution of the GP equation. However, the BdG modes do experience nontrivial dynamics as the sound speed is now $c(x)=\sqrt{g(x)}$. Specifically, we choose $g(x)$ as a piecewise homogeneous function so that the condensate remains unchanged for $x<0$ and becomes supersonic for $x>0$, with a sound speed $c_2<v$. Hence, we reach a black-hole configuration (i.e., a subsonic-supersonic flow transition) with the event horizon (i.e., the subsonic-supersonic interface) placed at $x=0$. This process mimics the formation of a black hole and the subsequent production of Hawking radiation in actual experiments \cite{Steinhauer2014,Steinhauer2016,deNova2019,Kolobov2021}, which is characterized by the spontaneous emission of $b-,p2+$ modes from the event horizon into the exterior/interior (subsonic/supersonic regions) of the black hole, respectively.

At $t=t_{\rm{BCL}}>0$, an additional localized delta potential $V(x)=Z\delta(x-L)$ is switched on at $x=L>0$. Since the barrier is placed in the supersonic region, it will stimulate the emission of BCL radiation, here of wavevector $k_{\rm{BCL}}=2\sqrt{v^2-c_2^2}$, with a certain amplitude $A_{\rm{BCL}}$. This is a simple model of the BCL stimulation that occurs in the experiment, which can be previous to the birth of a second horizon. Finally, at $t=t_{\rm{BHL}}\geq t_{\rm{BCL}}$, we quench back the coupling constant for $x>L$ according to Eq. (\ref{eq:FlatProfile}) so that the condensate recovers there its subsonic character. Hence, a white-hole horizon (a supersonic-subsonic interface, the time reversal of a black-hole horizon) forms at $x=L$, and the whole structure now possesses two horizons, giving rise to a BHL configuration. Throughout this work, we consider long onset times $t_{\rm{BCL}},t_{\rm{BHL}}$ in order to mimic the presence of a stationary period of spontaneous Hawking radiation and the eventual formation of an inner horizon in the experiment \cite{Kolobov2021}. We have checked that the main conclusions of this work, concerning the BHL-BCL crossover, are rather insensitive to the transient before the BCL and BHL onsets.

Quantitatively, a black-hole laser is characterized by a discrete BdG spectrum of dynamical instabilities, computed as a function of the cavity length $L$ for the flat-profile model in Fig. \ref{fig:BHLBCLScheme}c (see Refs. \cite{Michel2013,Michel2015} for the technical details). The critical lengths $L_n$ at which a new dynamical instability emerges (vertical solid lines) can be derived analytically:

\begin{equation}\label{eq:CriticalLengths}
    L_n=L_0
    +\frac{n\pi}{\sqrt{v^2-c_2^2}},~L_0=\frac{\arctan\sqrt{\dfrac{1-v^2}{v^2-c_2^2}}}{\sqrt{v^2-c_2^2}},~n=0,1\ldots
\end{equation}
All these modes are born as degenerate unstable modes, i.e., modes with purely imaginary frequency, $\omega_n=-\omega_n^*$. For half-integer values $n+1/2$, the above equation yields the lengths $L_{n+1/2}$ at which the $n$-th unstable mode becomes oscillatory, developing a nonvanishing real part of the frequency (vertical dashed lines). The dominant mode is that with the largest growth rate, $\Gamma_n=\textrm{Im}\,\omega_n$, and determines the total growth rate $\Gamma$ of the lasing instability, $\Gamma=\max_{n}\Gamma_n$ (thick solid black envelope in Fig. \ref{fig:BHLBCLScheme}c). For short cavities, this is typically the mode with the largest value of $n$. However, as the cavity becomes larger and larger, the competition between the different unstable modes becomes stronger and stronger. 

Qualitatively, we can understand these dynamical instabilities as Hawking $p2+$ modes reflected at the white-hole horizon and scattered into $p2-$ modes that will bounce back towards the black hole, further stimulating the production of Hawking radiation and thus leading to a process of self-amplification \cite{Corley1999,Finazzi2010}. This simple argument yields a good estimation for the order of magnitude of the growth rate of the instability, $\Gamma\sim 1/\tau_{\rm{RT}}$ (green dashed line in Fig. \ref{fig:BHLBCLScheme}c), with $\tau_{\rm{RT}}$ the roundtrip time for a zero-frequency $p2+$ mode to travel back and forth between the horizons. This frequency choice is motivated by observing that the dominant mode has either zero or small real part of the frequency, which also implies that the $p2-$ wavevector involved in the lasing instability will be close to the BCL wavector. Thus, BCL stimulation and BHL self-amplification share similar short wavelengths and low frequencies, something that strongly complicates their clear distinction in real setups \cite{Steinhauer2014,Tettamanti2016,Steinhauer2017,Wang2016,Wang2017,Llorente2019,Tettamanti2021,Kolobov2021,Steinhauer2022}. Here emerges the importance of the flat-profile configuration: because of the fine-tuning condition (\ref{eq:FlatProfile}), the white hole does not further stimulate BCL radiation. Indeed, if the delta potential was not switched on, $Z=0$, there would not be any Cherenkov stimulation from the white hole and we would have the flat-profile BHL of Ref. \cite{deNova2016}. On the other hand, if the white hole was never switched on, $t_{\rm{BHL}}=\infty$, we would only have BCL stimulation. Hence, by comparing scenarios with and without white hole, and with and without Cherenkov stimulation, we can isolate the genuine BHL features from those of BCL.

Compactly, the model for the condensate dynamics developed in this section is encapsulated in the following GP equation:
\begin{eqnarray}\label{eq:GPEquation}
\nonumber i\partial_t\Psi(x,t)&=&H_{\rm{GP}}(x,t)\Psi(x,t),~\Psi(x,0)=\Psi_0(x)=e^{ivx},\\
\nonumber H_{\rm{GP}}(x,t)&=&-\frac{1}{2}\partial_x^2+g(x,t)\left[|\Psi(x,t)|^2-1\right]+V(x,t),\\
V(x,t)&=&Z\delta(x-L)\theta(t-t_{\rm{BCL}}),\\
\nonumber g(x,t)&=&1+(c_2-1)[\theta(x)\theta(t)\\
\nonumber&-&\theta(t-t_{\rm{BHL}})\theta(x-L)],
\end{eqnarray}
with $\theta$ the Heaviside function. A schematic summary of the model is presented in Fig. \ref{fig:BHLBCLScheme}d.

\section{The tools}\label{sec:Tools}

We simulate the time evolution of the condensate by numerically integrating the time-dependent GP equation (\ref{eq:GPEquation}). Quantum fluctuations are added via the Truncated Wigner approximation \cite{Sinatra2002,Carusotto2008}, which computes symmetric-ordered expectation values from ensemble averages of integrations of the GP equation using a stochastic initial condition 
\begin{equation}
\Psi_W(x,0)=[1+\delta\Psi_W(x)]e^{ivx}.
\end{equation}
In turn, $\delta\Psi_W(x)$ is given in terms of the BdG modes around the initial homogeneous condensate:
\begin{equation}
\delta\Psi_W(x)=\frac{1}{\sqrt{N}}\sum_k \alpha_{k} u_ke^{ikx}+\alpha^*_k v^*_ke^{-ikx},
\end{equation}
where $N$ is the total number of particles (we recall that in our units the density $n_0$ is factorized out from the GP wave function) and $u_k,v_k$ are the usual Bogoliubov components
\begin{equation}
u_k=\frac{\frac{k^2}{2}+\Omega_k}{\sqrt{2k^2\Omega_k}},~
v_k=\frac{\frac{k^2}{2}-\Omega_k}{\sqrt{2k^2\Omega_k}}.
\end{equation}
Thus, the quantum fluctuations $\delta\hat{\Psi}(x,t)$ of the field operator $\hat{\Psi}(x,t)$ around the mean-field condensate are accounted here by the initial condition $\delta\Psi_W(x)$, where the amplitudes $\alpha_k, \alpha^*_k$ are stochastic variables that mimic the usual phonon annihilation and creation operators $\hat{\alpha}_k,\hat{\alpha}^\dagger_{k}$, $[\hat{\alpha}_k,\hat{\alpha}^\dagger_{k'}]=\delta_{kk'}$. These classical amplitudes are sampled from the Wigner distribution of the initial equilibrium state, assumed to be the $T=0$ ground state in the comoving frame of the condensate, which is a Gaussian distribution characterized by the first and second-order momenta
\begin{eqnarray}
    \nonumber \braket{\alpha_k}&=&\braket{\hat{\alpha}_k}=0,~\braket{\alpha_{k'}\alpha_k}=\braket{\hat{\alpha}_{k'}\hat{\alpha}_k}=0,\\    \braket{\alpha^*_{k'}\alpha_k}&=&\frac{\braket{\hat{\alpha}^\dagger_{k'}\hat{\alpha}_k+\hat{\alpha}_k\hat{\alpha}^\dagger_{k'}}}{2}=\frac{\delta_{kk'}}{2}.
\end{eqnarray}
By subtracting the constant factor $\delta_{kk'}/2$ in the last equation, arising from the commutator between annihilation and creation operators, one retrieves the more usual normal-ordered expectation value $\braket{\hat{\alpha}_{k'}^\dagger\hat{\alpha}_{k}}$. Our initial condition for the quantum state eliminates the vacuum ambiguity arising in the presence of a black hole, and of dynamically unstable modes after the BHL onset \cite{Ribeiro2022}.

Regarding the observables of interest, we focus on computing the density and its correlations as they are measured in the laboratory through \textit{in situ} imaging after averaging over ensembles of repetitions of the experiment \cite{Shammass2012,Steinhauer2014,Steinhauer2016,deNova2019,Kolobov2021}. Specifically, we will compute the ensemble-averaged density from the diagonal of the first-order correlation function
\begin{equation}
     \braket{\hat{n}(x,t)}=G^{(1)}(x,t)\equiv \braket{\hat{\Psi}^\dagger(x,t)\hat{\Psi}(x,t)}.
\end{equation}
In order to characterize quantum fluctuations, we use the \textit{relative} second-order correlation function
\begin{widetext}
    \begin{equation}\label{eq:RelativeCorrelations}
   g^{(2)}(x,x',t)=\frac{\braket{\hat{\Psi}^\dagger(x,t)\hat{\Psi}^\dagger(x',t)\hat{\Psi}(x',t)\hat{\Psi}(x,t)}-\braket{\hat{\Psi}^\dagger(x,t)\hat{\Psi}(x,t)}\braket{\hat{\Psi}^\dagger(x',t)\hat{\Psi}(x',t)}}{n^2_0},
\end{equation}
\end{widetext}
where we have momentarily restored units to keep track of the different scalings. Apart from a trivial delta term, the second-order correlation function is directly given by the relative density-density correlations, measurable in experiments,
\begin{equation}
   g^{(2)}(x,x',t)=\frac{\braket{\delta\hat{n}(x,t)\delta\hat{n}(x',t)}-\braket{\hat{n}(x,t)}\delta(x-x')}{n^2_0},
\end{equation}
with $\delta\hat{n}(x,t)=\hat{n}(x,t)-\braket{\hat{n}(x,t)}$ the density fluctuations. For simplicity, we will also refer to the second-order correlation function as the density-density correlation function.

\begin{figure*}[t]
\begin{tabular}{@{}ccc@{}}\stackinset{l}{2pt}{t}{3pt}{\large{(a)}}{\includegraphics[width=0.66\columnwidth]{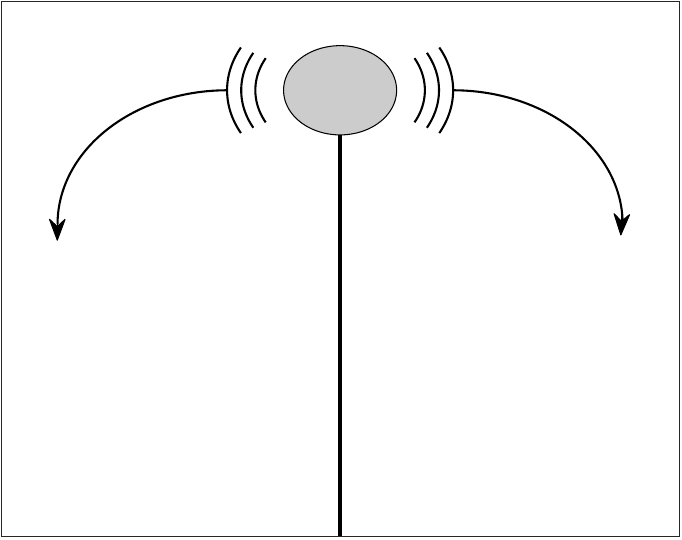}}~ &~
    \stackinset{l}{2pt}{t}{3pt}{\large{(b)}}
    {\includegraphics[width=0.66\columnwidth]{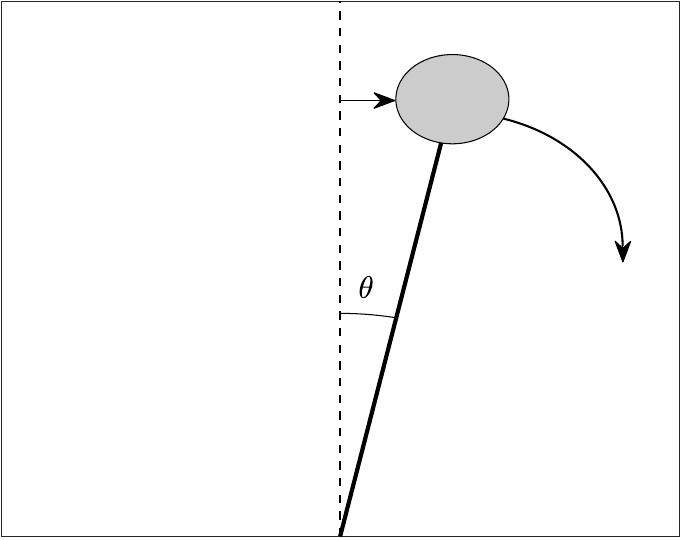}}~ &~ 
    \stackinset{l}{2pt}{t}{3pt}{\large{(c)}}{\includegraphics[width=0.66\columnwidth]{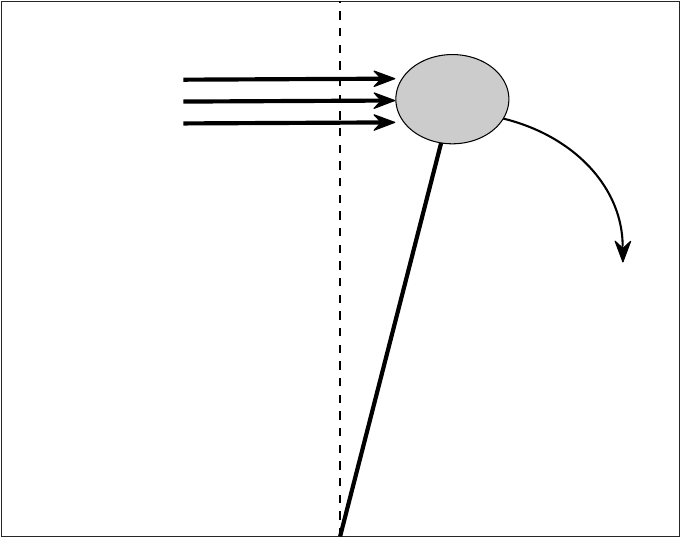}} 
\end{tabular}
\caption{Schematic depiction of an unstable pendulum. (a) Quantum BHL: The pendulum is placed at its unstable equilibrium position but quantum fluctuations make it fall. (b) Classical BHL: A small kick on the pendulum causes it to depart some small angle $\theta$ from the unstable equilibrium position, falling down with a well-defined classical trajectory as a result. (c) BCL: A strong external force (horizontal arrows) pushes the pendulum out of equilibrium, dominating the dynamics instead of gravity.}
\label{fig:Pendula}
\end{figure*}

By expanding the density operator in terms of $\delta\hat{\Psi}(x,t)$ around the initial homogeneous flowing condensate,
\begin{eqnarray}\label{eq:Density expansion}
    \hat{n}(x,t)&=&n_0+\delta\hat{n}^{(1)}(x,t)+\delta\hat{n}^{(2)}(x,t),\\
    \nonumber \delta\hat{n}^{(1)}(x,t)&=&\sqrt{n_0}\hat{\chi}(x,t),~\hat{\chi}(x,t)=\delta\hat{\Psi}(x,t)+\delta\hat{\Psi}^\dagger(x,t),\\
    \nonumber \delta\hat{n}^{(2)}(x,t)&=&\delta\hat{\Psi}^\dagger(x,t)\delta\hat{\Psi}(x,t),
\end{eqnarray}
we find that to lowest order in the quantum fluctuations
\begin{equation}\label{eq:RelativeCorrelationsBdG}
   g^{(2)}(x,x',t)\simeq \frac{\braket{\hat{\chi}(x,t)\hat{\chi}(x',t)}-\delta(x-x')}{n_0}.
\end{equation}
Since $\braket{\hat{\chi}(x,t)\hat{\chi}(x',t)}$ has units of inverse length, the \textit{normalized} density-density correlation function
\begin{equation}\label{eq:NormalizedCorrelations}
    G^{(2)}(x,x',t)\equiv n_0\xi_0 g^{(2)}(x,x',t),~\xi_0=\frac{\hbar}{mc_0},
\end{equation}
is a dimensionless function that does not depend explicitly on the density $n_0$ in the BdG approximation, just implicitly via the healing length $\xi_0$ as the density solely intervenes in the dynamics at this level through the term $gn_0=\hbar^2/m\xi^2_0$. Specifically, for the initial equilibrium state at $t\leq 0$, the density-density correlation function can be evaluated analytically by \cite{Deuar2003,Larre2012}
\begin{eqnarray}\label{eq:antibunching}
    G^{(2)}(x,x',t)&=&G\left(\frac{x-x'}{\xi_0}\right),\\
    \nonumber G(z)&=&-\frac{1}{\pi z}\int^\infty_0\mathrm{d}q~\frac{\sin 2 q z}{(1+q^2)^{\frac{3}{2}}},
\end{eqnarray}
with $G(0)=-2/\pi$. This is the celebrated antibunching phenomenon, resulting from the repulsive nature of the interactions. 

When switching back to our system of units, where $\delta\hat{\Psi}$ is relative to the condensate amplitude $\sqrt{n_0}$, the above dimensional arguments dictate that the typical amplitude of the quantum fluctuations $A_{\rm{QF}}$ scales as 
\begin{equation}\label{eq:QFAmplitude}
     \delta\hat{\Psi}\sim A_{\rm{QF}}\sim \frac{1}{\sqrt{n_0\xi_0}}\ll 1.
\end{equation}
The condition $n_0\xi_0\gg 1$ is required so that the relative density fluctuations are small, $g^{(2)}\sim (n_0\xi_0)^{-1}\ll 1$, and thus our one-dimensional quasicondensate description is valid \cite{Mora2003}.
Although, strictly speaking, $\xi_0=1$ in our units, we will write the complete dimensionless factor $n_0\xi_0$ to stress its physical meaning as the regulator of the strength of the quantum fluctuations. 

From the previous considerations, one can anticipate three different regimes for the dynamics after the BHL formation depending on the amplitudes of the background Cherenkov wave, $A_{\rm{BCL}}$, and of the quantum fluctuations, $A_{\rm{QF}}$:

\begin{enumerate}
    \item Quantum BHL: $A_{\rm{BCL}}\ll A_{\rm{QF}}\ll 1$. The BHL instability is triggered by quantum fluctuations (e.g., no barrier is placed, $A_{\rm{BCL}}=0$), and the dynamics is driven by the zero-point motion of the quasiparticle vacuum.
    
    \item Classical BHL: $A_{\rm{QF}}\ll A_{\rm{BCL}}\ll 1$. The BHL amplification still dominates the dynamics but the seed of the instability is now the classical amplitude of the BCL wave in the condensate, leading to a well-defined mean-field trajectory. 
    
    \item BCL: $A_{\rm{QF}}\ll A_{\rm{BCL}}\sim 1$. The amplitude of the BCL stimulation is highly nonlinear and dominates the mean-field dynamics towards the saturation regime.
\end{enumerate}

Since a BHL behaves as an unstable harmonic oscillator \cite{Leonhardt2003,Finazzi2010,Burkle2018}, we can qualitatively understand these regimes using an analogy with an unstable pendulum, Fig. \ref{fig:Pendula}. The unstable equilibrium position is equivalent to the GP wave function of the flat-profile BHL. However, due to zero-point motion, this configuration is unstable at the quantum level, and then the pendulum falls (Fig. \ref{fig:Pendula}a). This is akin to a quantum BHL. We can give a deterministic amplitude to the pendulum with a small kick (Fig. \ref{fig:Pendula}b). The pendulum departs some small angle $\theta$ from its equilibrium position and consequently falls following a well-defined classical trajectory. This is akin to a classical BHL, where the angle $\theta$ plays here the role of the Cherenkov amplitude $A_{\rm{BCL}}$ that seeds the BHL instability. If a strong enough force is applied (Fig. \ref{fig:Pendula}c), it will drive the pendulum motion outside its unstable equilibrium position instead of gravity, as in the case where BCL stimulation dominates the dynamics over the BHL mechanism.

\section{Numerical results}\label{sec:Numerical}

In this section we present the main results of the work, where we compute the time evolution of the condensate and its quantum fluctuations (see Refs. \cite{deNova2016,deNova2022} for the technical details about the numerical techniques employed in this work). As reference background BHL parameters, we choose for the moment those of Fig. \ref{fig:BHLBCLScheme}d: $v=0.6$, $c_2=0.2$, and $L=20$. This cavity contains $4$ unstable lasing modes, with the $n=2$ mode expected to be the dominant one, leading to a growth rate $\Gamma=\Gamma_2\approx 0.02$. Regarding the specific parameters involved in the Truncated Wigner method, we take $L_{t}\approx 1885$ as the total length of the numerical grid (periodic boundary conditions are imposed) and $N=10^{7}$ as the number of particles, with $n_0=N/L_t$ the condensate density. This yields $n_0\xi_0\approx 5\cdot 10^3\gg 1$. The number of modes is $N_m=3000\ll N$, corresponding to a momentum cut-off $|k|<5$, much larger than the typical wavevectors involved in the dynamics. Ensemble averages are evaluated after $1000$ Monte Carlo simulations, although good convergence is already found for the main BHL and BCL features after few hundreds of simulations. Finally, appropriate factors are subtracted to retrieve the required normal-ordered expectation values; their explicit form is discussed in Ref. \cite{Jacquet2022}.

In our model, the Cherenkov amplitude $A_{\rm{BCL}}$ is directly controlled by the barrier strength $Z$. The initial amplitude of the quantum fluctuations $A_{\rm{QF}}$ can be adjusted by a dimensionless control parameter $\lambda$ in the following way: we increase the initial density $n_0$ as $n_0\to\lambda n_0$ while changing the coupling constant so that $gn_0$ is kept constant (in particular, $gn_0=1$ in our units). By doing so, the mean-field dynamics remains the same but now $n_0\xi_0\to \lambda n_0\xi_0$. Thinking in more physical terms, this is like 
modifying the condensate density by changing the number of atoms while tuning the coupling constant using Feshbach resonances \cite{Chin2010}. As a result, the amplitude of the quantum fluctuations in our units will scale as
\begin{equation}
     A_{\rm{QF}}\sim \frac{1}{\sqrt{\lambda}},
\end{equation}
so $\lambda$ can be regarded as a control parameter of the strength of the quantum fluctuations. For brevity, we will simply refer to $\lambda^{-1}$ as the quantum strength. Thus, the interplay between the quantum and Cherenkov seeding will be explored via the parameters $Z,\lambda$, manipulating also the onset times $t_{\rm{BCL}},t_{\rm{BHL}}$ to further isolate the contribution of each effect. 

Since we are mainly concerned about the conceptual differences between the BHL and BCL mechanisms, we will restrict the simulations to intermediate times when the system reaches the saturation regime. This saturation can be associated to a certain stationary GP solution of the nonlinear spectrum \cite{Michel2013}. This solution is in turn also metastable, with a lifetime much longer than that of the initial BHL configuration, and it will eventually collapse \cite{Michel2015,deNova2016}. For sufficiently long times, the system will either reach the true ground state or the CES (Continuous Emission of Solitons) state \cite{deNova2016,deNova2021a,deNova2022}. However, the study of such long-time dynamics is beyond the scope of this work.

\begin{figure*}
\begin{tabular}{@{}cccc@{}}
    \stackinset{l}{0pt}{t}{0pt}{(a)}{\includegraphics[width=0.25\textwidth]{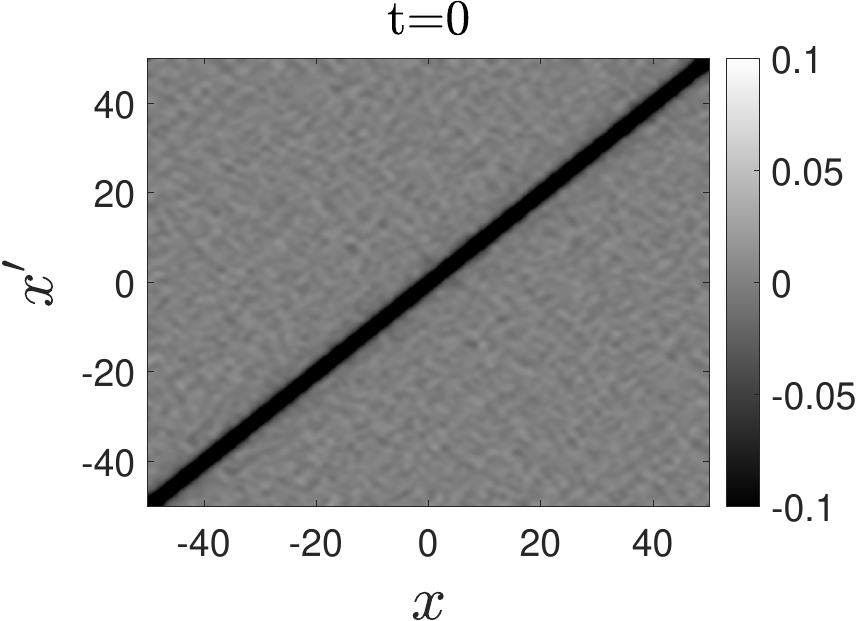}} 
    & \stackinset{l}{0pt}{t}{0pt}{(b)}{\includegraphics[width=0.25\textwidth]{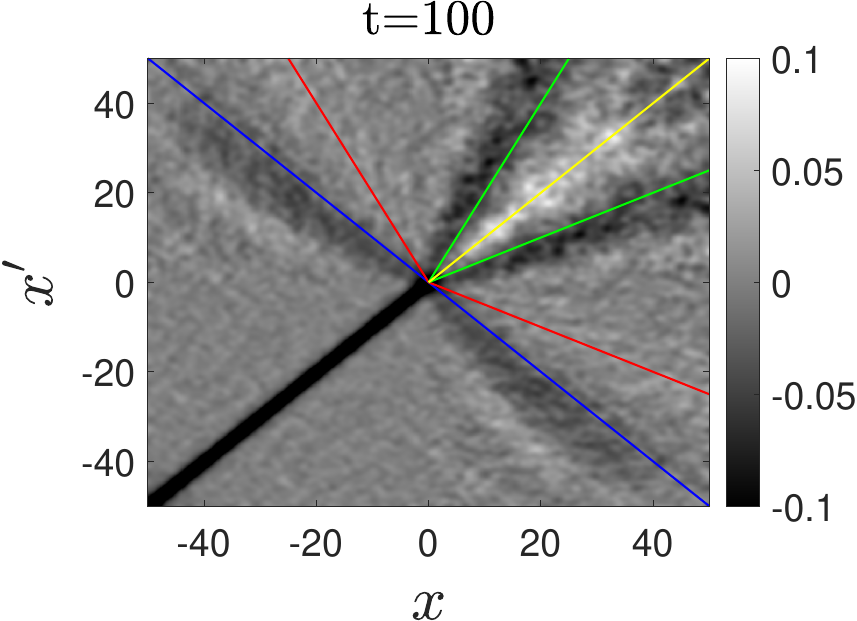}}& 
    \stackinset{l}{0pt}{t}{0pt}{(c)}{\includegraphics[width=0.25\textwidth]{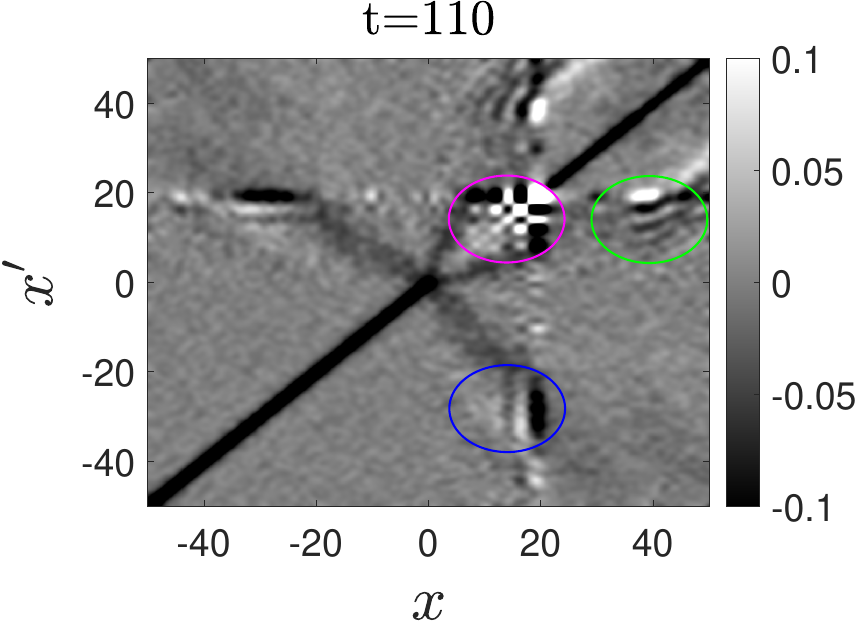}} &
    \stackinset{l}{0pt}{t}{0pt}{(d)}{\includegraphics[width=0.24\textwidth]{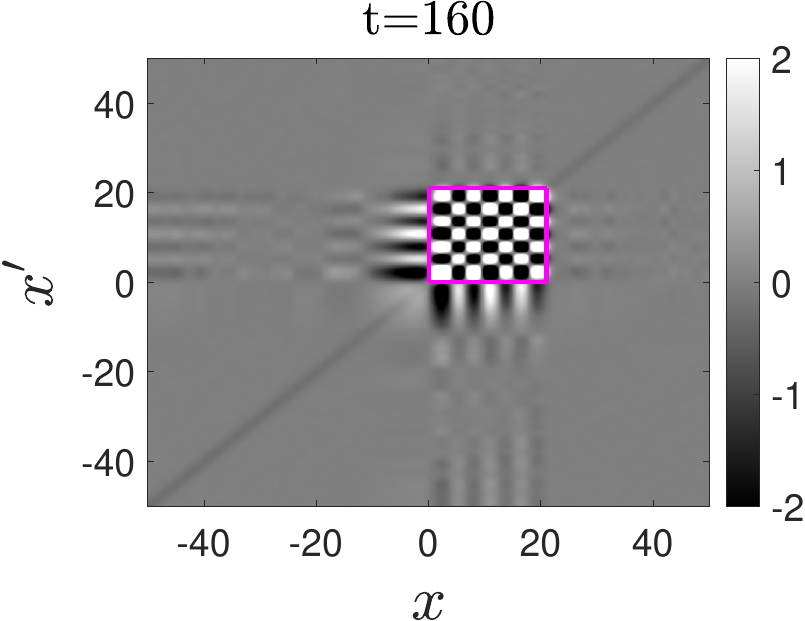}}
    \\ \\
    \stackinset{l}{0pt}{t}{0pt}{(e)}{\includegraphics[width=0.25\textwidth]{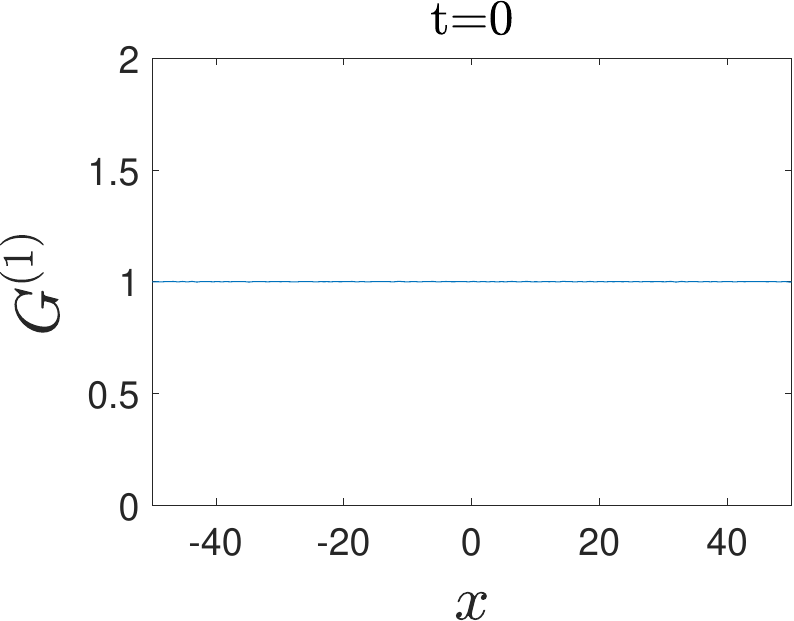}} 
    & \stackinset{l}{0pt}{t}{0pt}{(f)}{\includegraphics[width=0.25\textwidth]{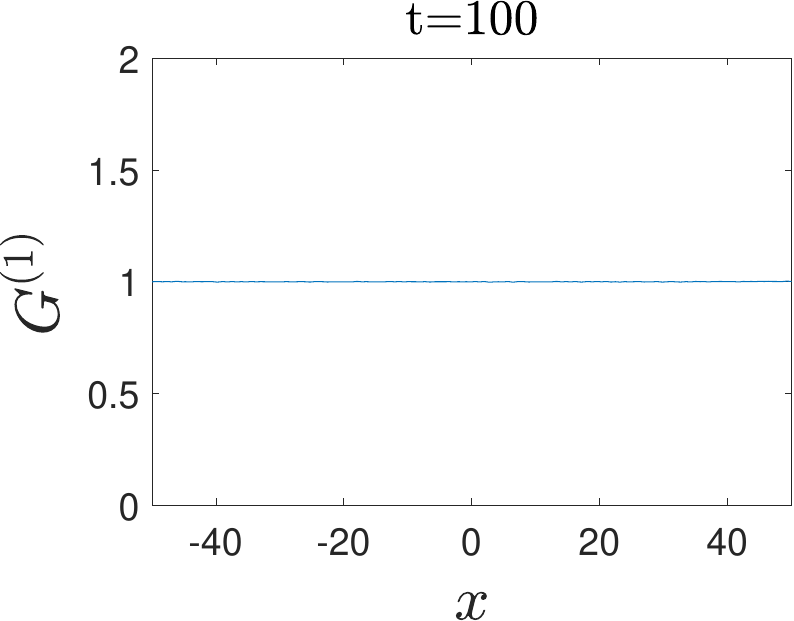}}& 
    \stackinset{l}{0pt}{t}{0pt}{(g)}{\includegraphics[width=0.25\textwidth]{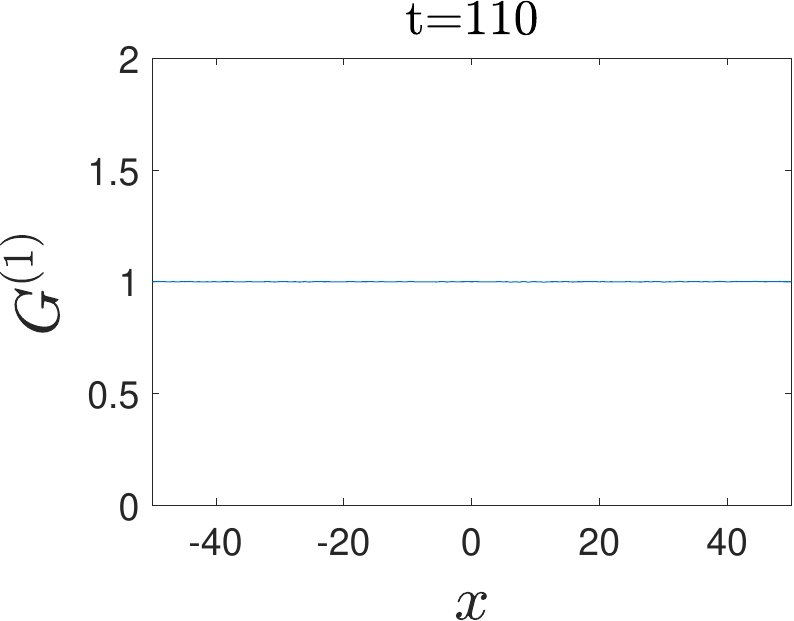}} &
    \stackinset{l}{0pt}{t}{0pt}{(h)}{\includegraphics[width=0.25\textwidth]{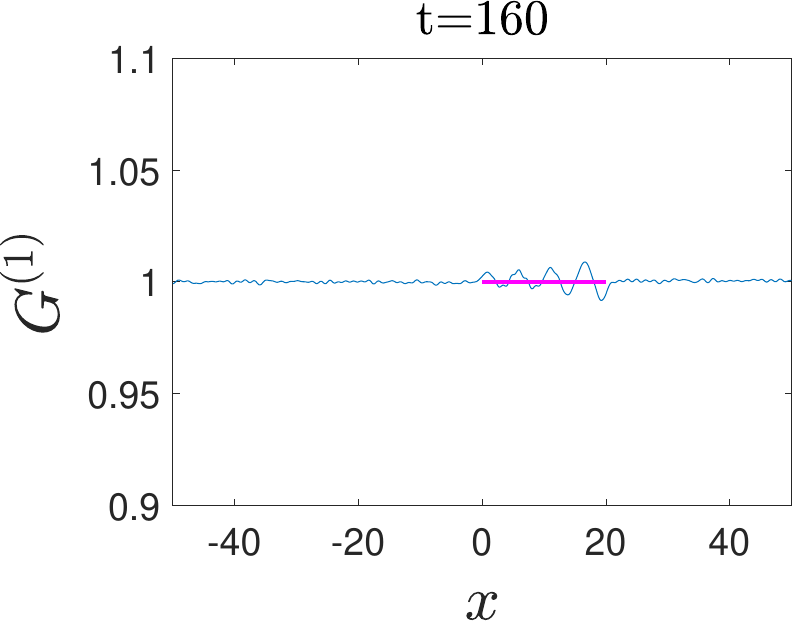}}
    \\ \\
     \stackinset{l}{0pt}{t}{0pt}{(i)}{\includegraphics[width=0.25\textwidth]{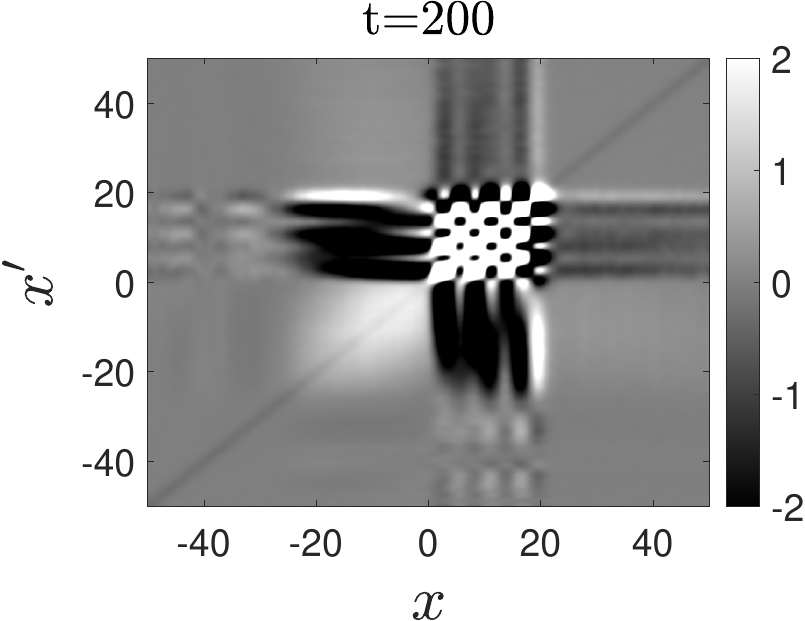}} 
    & \stackinset{l}{0pt}{t}{0pt}{(j)}{\includegraphics[width=0.25\textwidth]{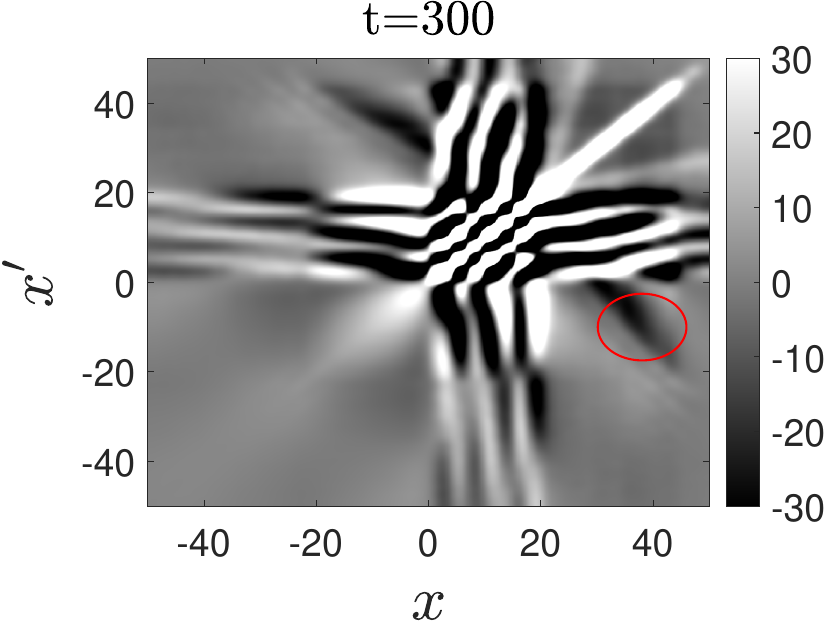}}& 
    \stackinset{l}{0pt}{t}{0pt}{(k)}{\includegraphics[width=0.25\textwidth]{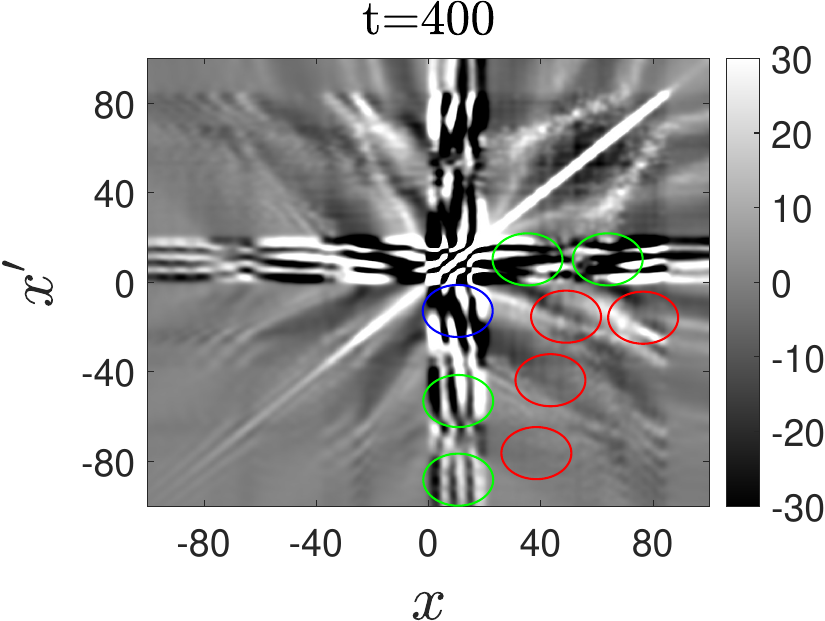}} &
    \stackinset{l}{0pt}{t}{0pt}{(l)}{\includegraphics[width=0.25\textwidth]{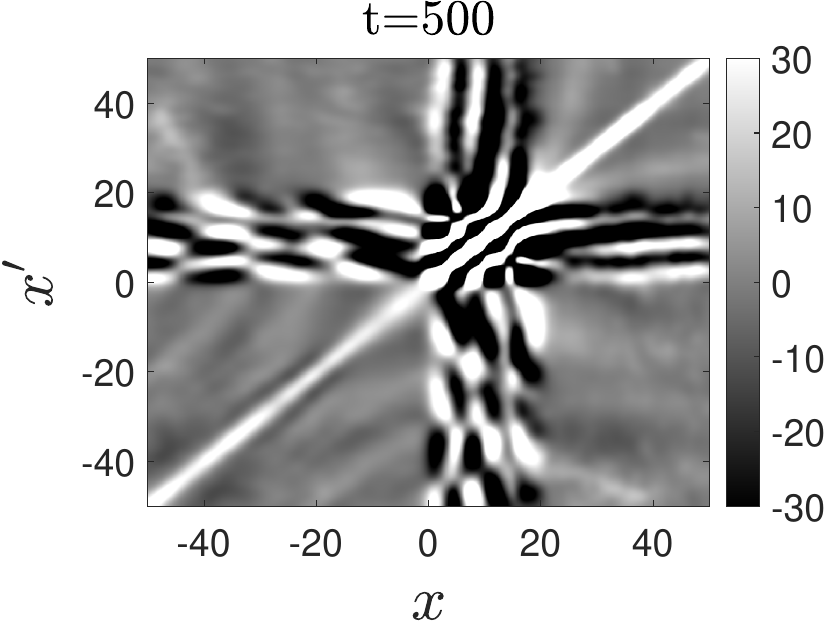}}
    \\ \\
    \stackinset{l}{0pt}{t}{0pt}{(m)}{\includegraphics[width=0.25\textwidth]{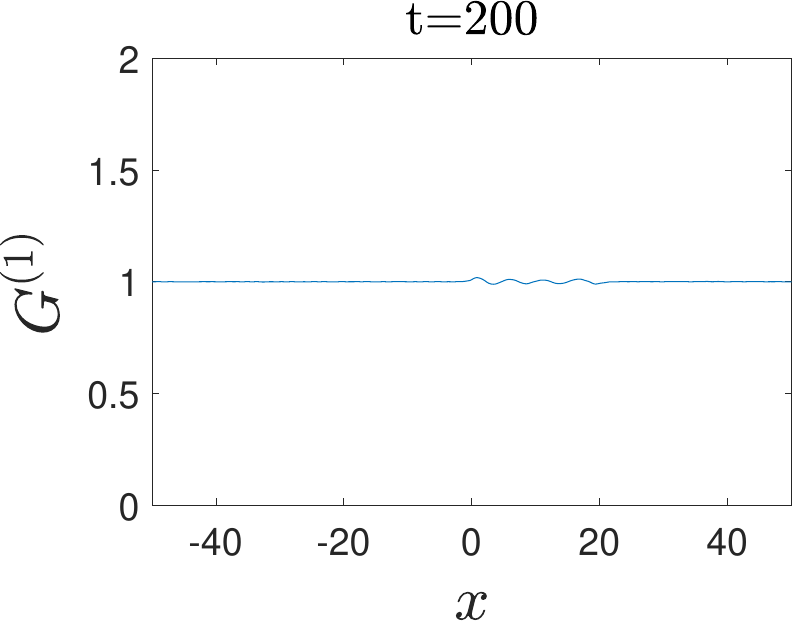}} 
    & \stackinset{l}{0pt}{t}{0pt}{(n)}{\includegraphics[width=0.25\textwidth]{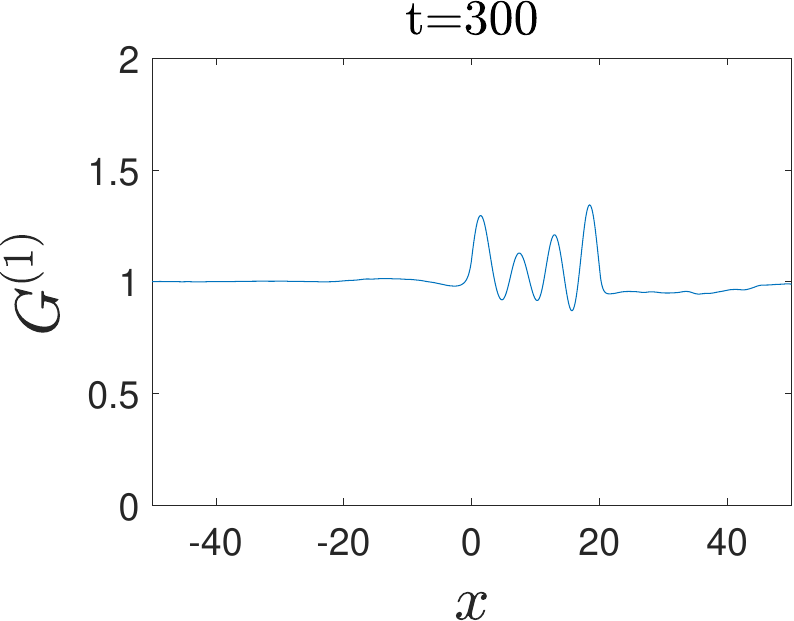}}& 
    \stackinset{l}{0pt}{t}{0pt}{(o)}{\includegraphics[width=0.25\textwidth]{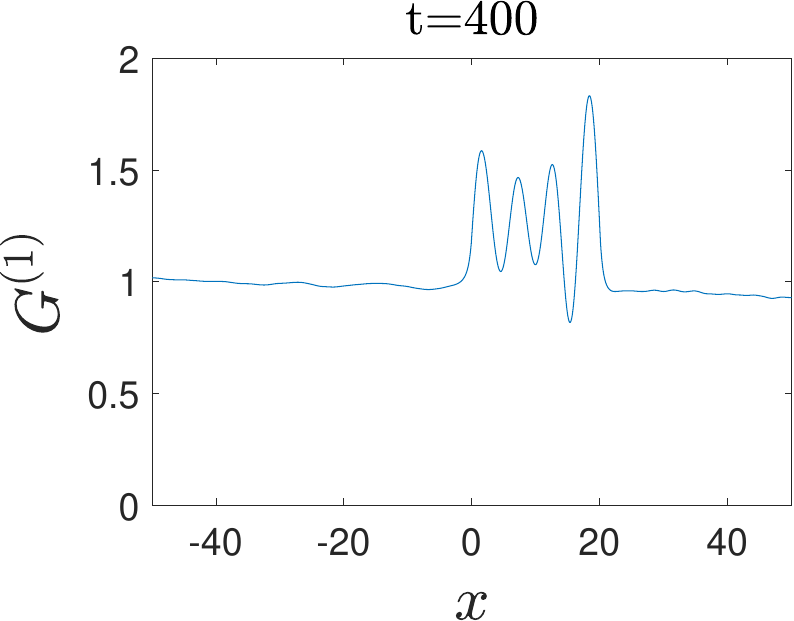}} &
    \stackinset{l}{0pt}{t}{0pt}{(p)}{\includegraphics[width=0.25\textwidth]{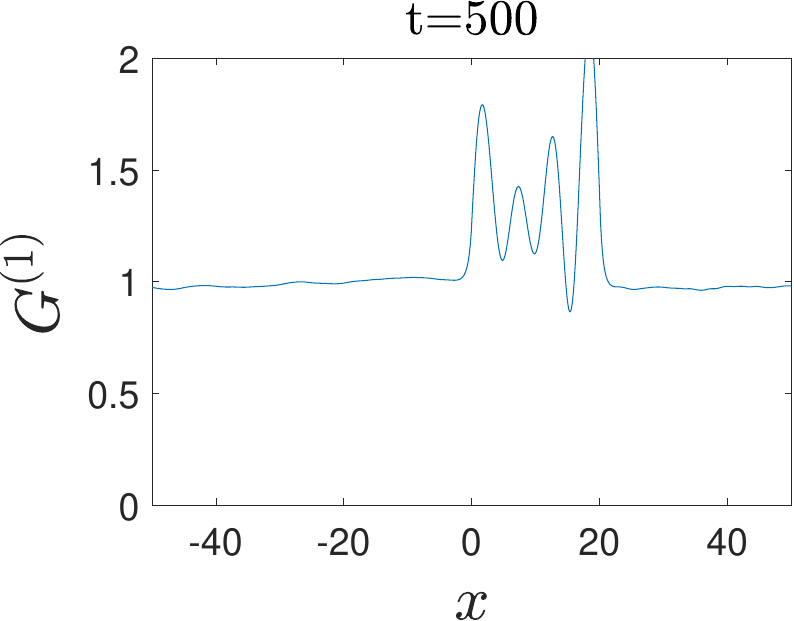}}
\end{tabular}
\caption{Time evolution of the density-density correlation function $G^{(2)}(x,x',t)$ [(a)-(d), (i)-(l)] and the ensemble-averaged density $G^{(1)}(x,t)$ [(e)-(h), (m)-(p)] for a purely quantum BHL ($Z=0$) with background parameters $v=0.6,c_2=0.2,L=20$. The quantum strength is set to $\lambda=1$. The black hole is switched on at $t=0$ and the white hole is switched on at $t=t_{\rm{BHL}}=100$. The time for each snapshot is indicated above the panel. Different color curves highlight several features discussed in the main text.}
\label{fig:QBHLTime}
\end{figure*}

\begin{figure*}
\begin{tabular}{@{}cccc@{}}
      \stackinset{l}{0pt}{t}{0pt}{(a)}{\includegraphics[width=0.25\textwidth]{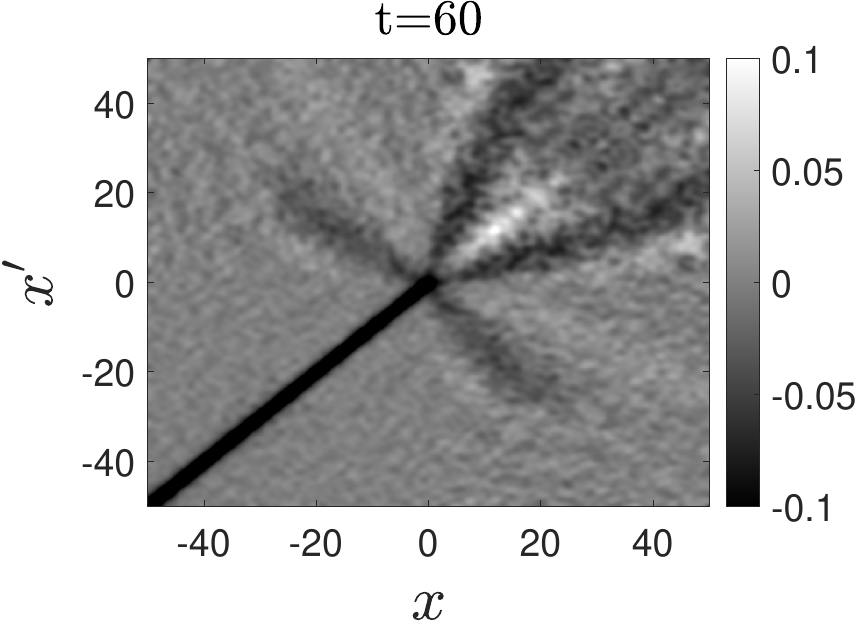}} 
    & \stackinset{l}{0pt}{t}{0pt}{(b)}{\includegraphics[width=0.25\textwidth]{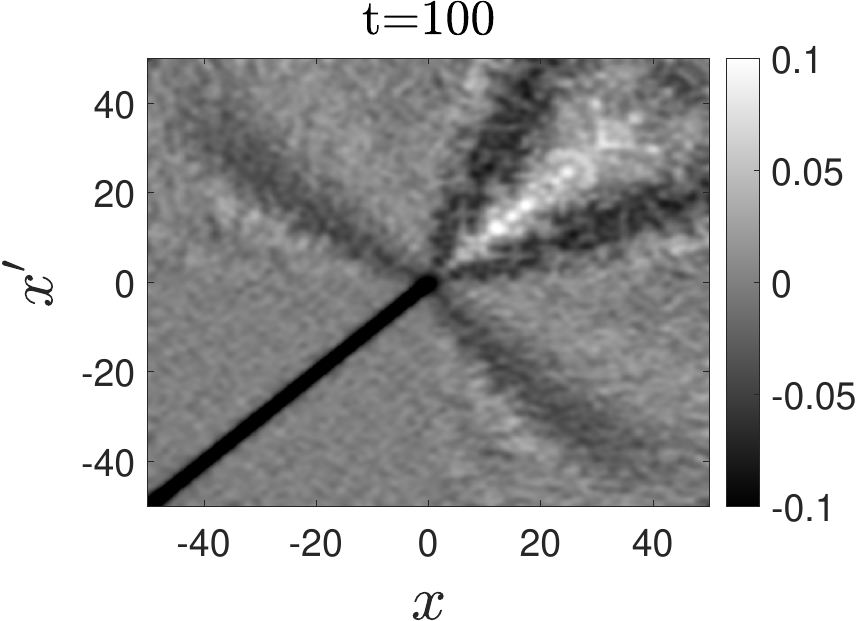}}& 
    \stackinset{l}{0pt}{t}{0pt}{(c)}{\includegraphics[width=0.25\textwidth]{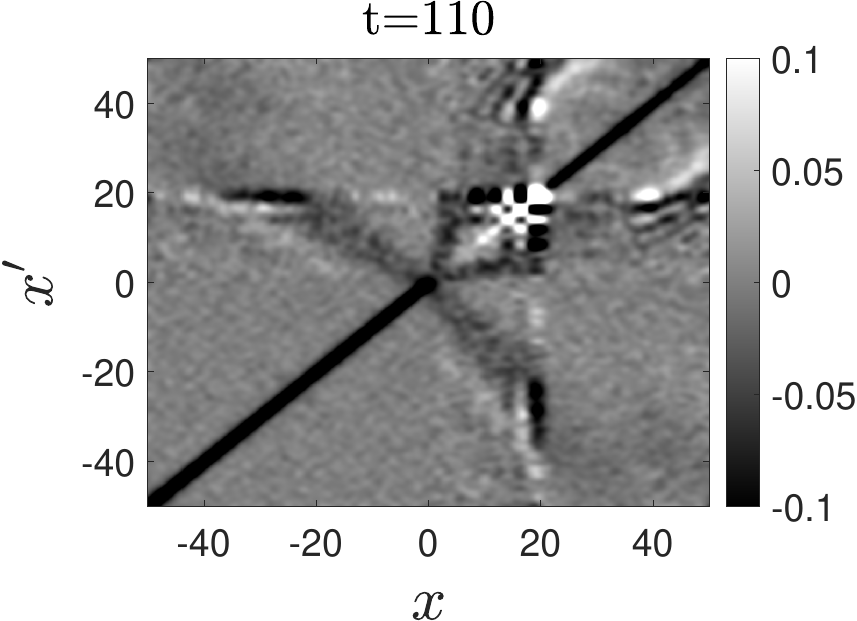}} &
    \stackinset{l}{0pt}{t}{0pt}{(d)}{\includegraphics[width=0.24\textwidth]{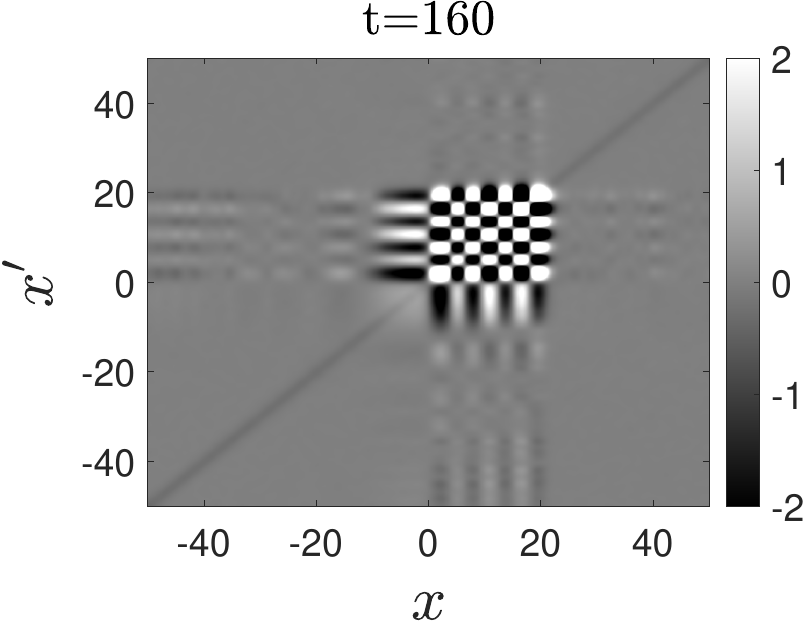}}
    \\ \\
    \stackinset{l}{0pt}{t}{0pt}{(e)}{\includegraphics[width=0.25\textwidth]{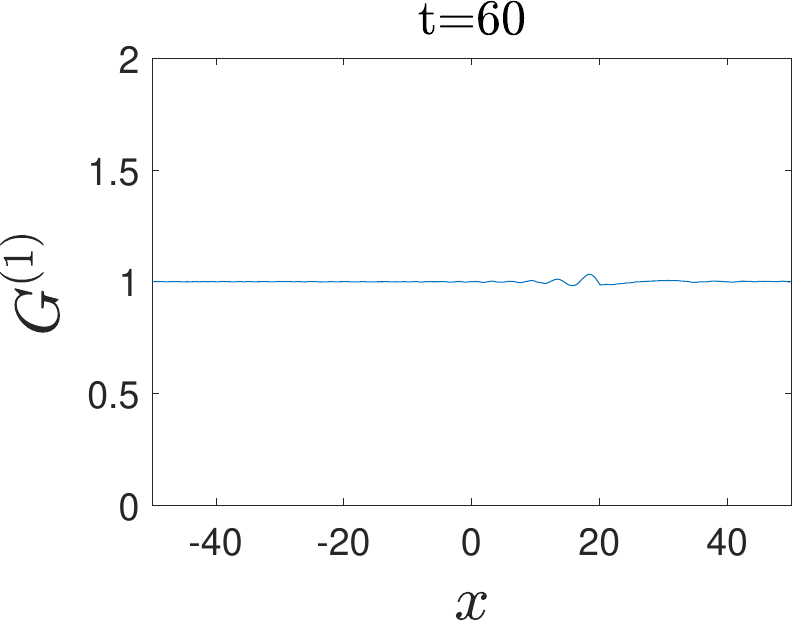}} 
    & \stackinset{l}{0pt}{t}{0pt}{(f)}{\includegraphics[width=0.25\textwidth]{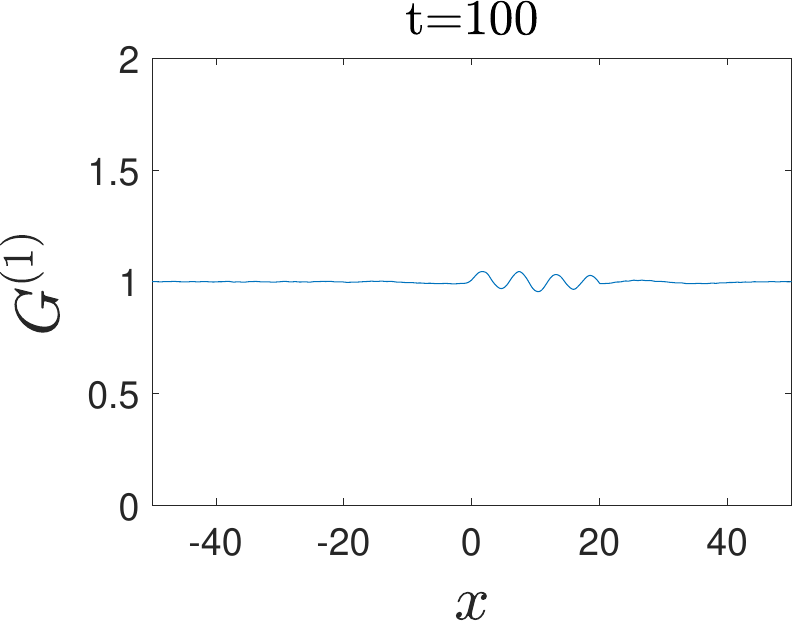}}& 
    \stackinset{l}{0pt}{t}{0pt}{(g)}{\includegraphics[width=0.25\textwidth]{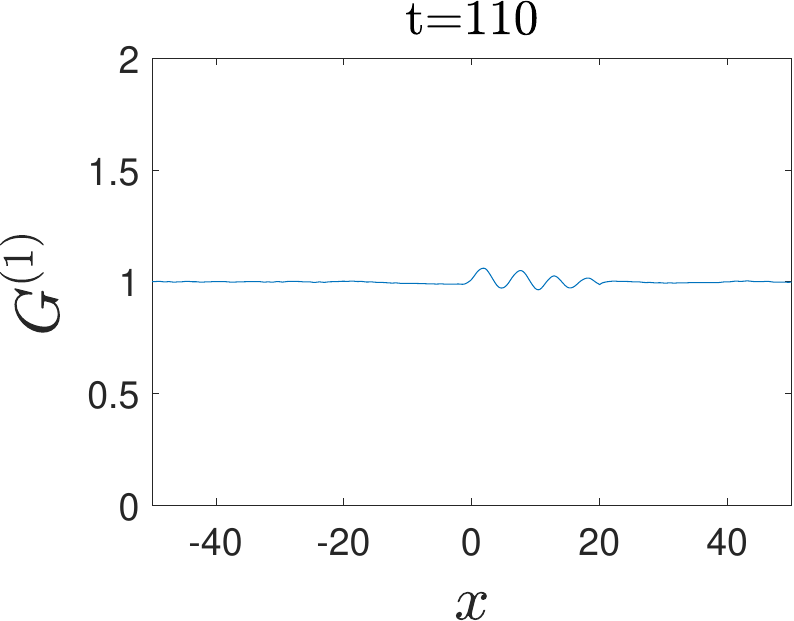}} &
    \stackinset{l}{0pt}{t}{0pt}{(h)}{\includegraphics[width=0.25\textwidth]{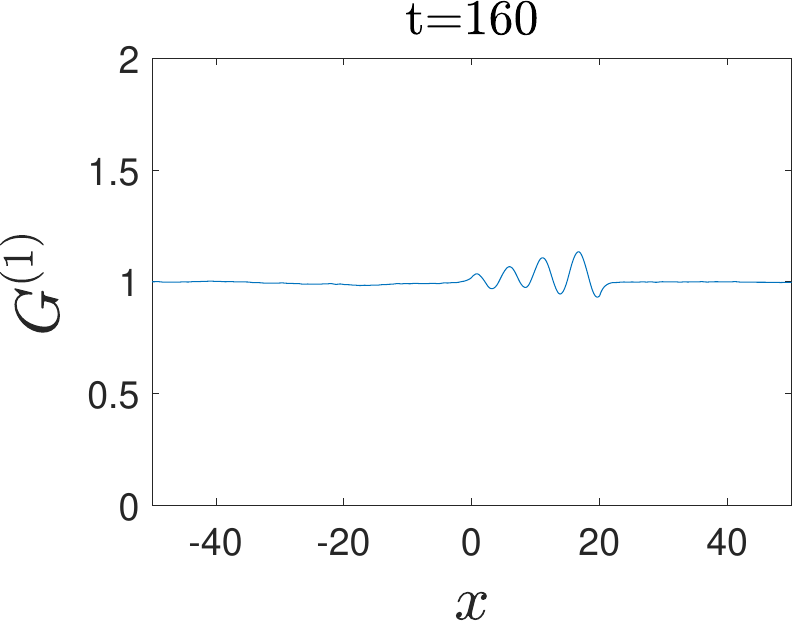}}
    \\ \\
     \stackinset{l}{0pt}{t}{0pt}{(i)}{\includegraphics[width=0.25\textwidth]{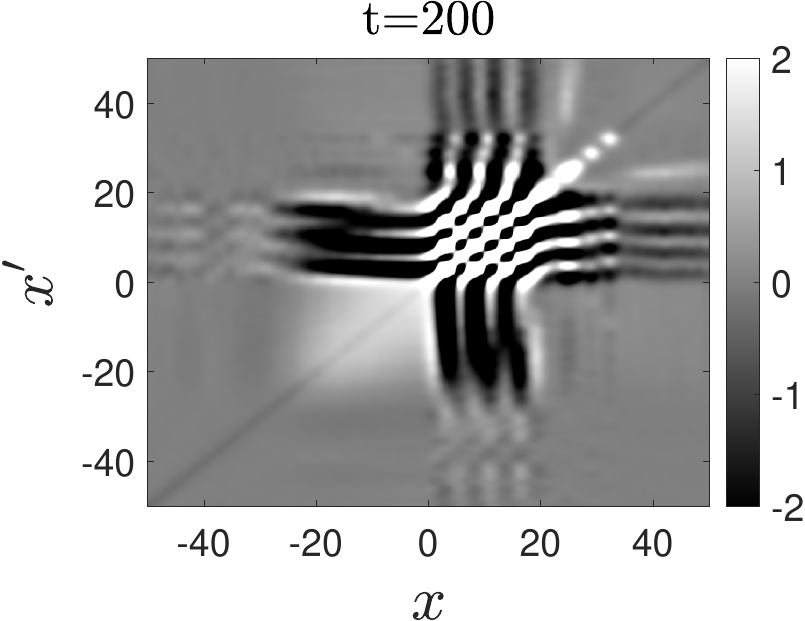}} 
    & \stackinset{l}{0pt}{t}{0pt}{(j)}{\includegraphics[width=0.25\textwidth]{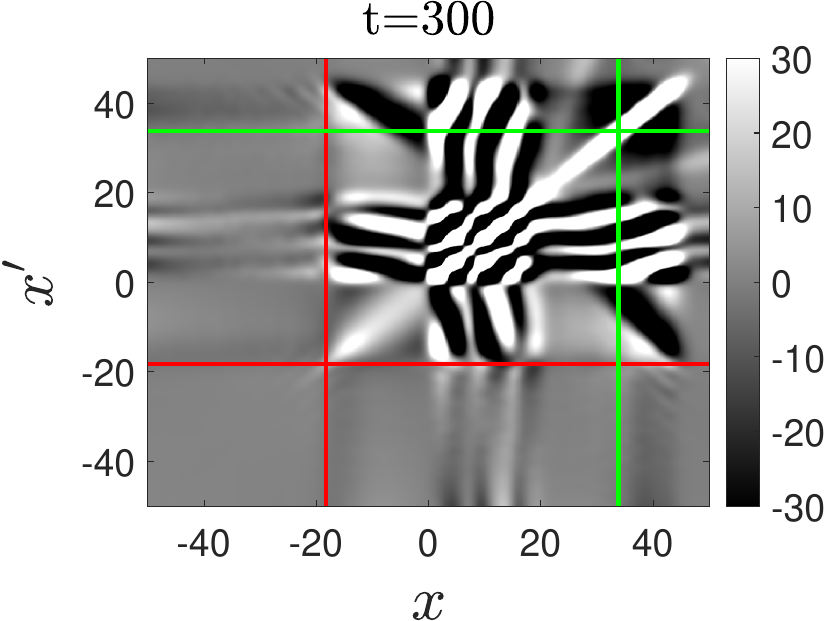}}& 
    \stackinset{l}{0pt}{t}{0pt}{(k)}{\includegraphics[width=0.25\textwidth]{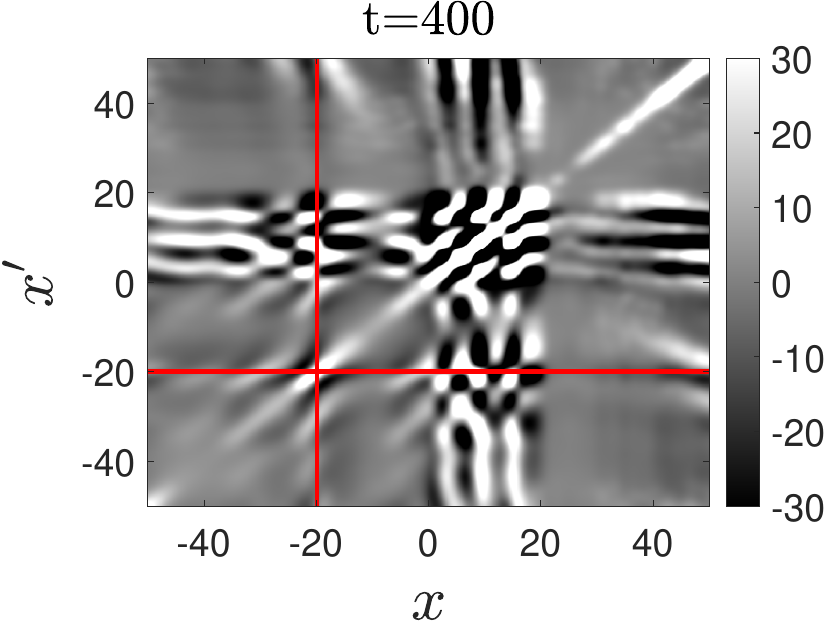}} &
    \stackinset{l}{0pt}{t}{0pt}{(l)}{\includegraphics[width=0.25\textwidth]{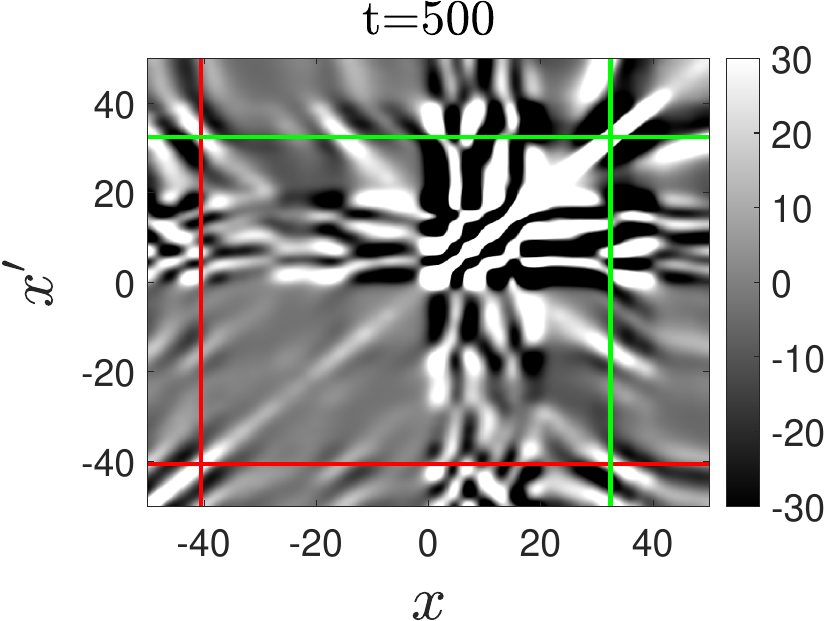}}
    \\ \\
    \stackinset{l}{0pt}{t}{0pt}{(m)}{\includegraphics[width=0.25\textwidth]{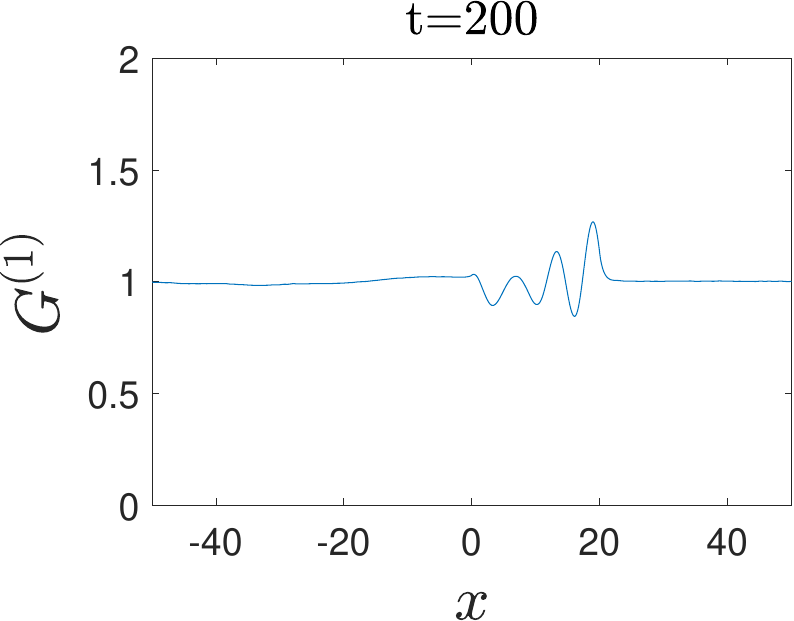}} 
    & \stackinset{l}{0pt}{t}{0pt}{(n)}{\includegraphics[width=0.25\textwidth]{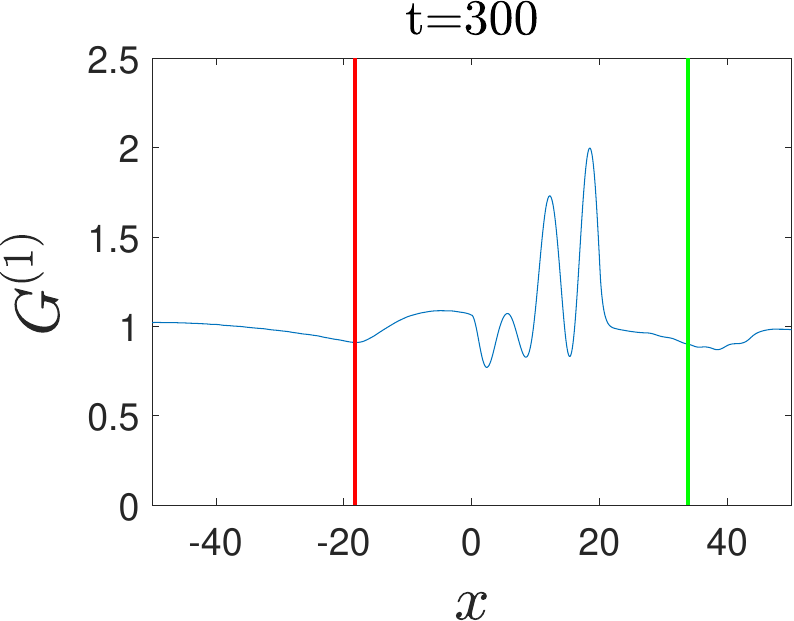}}& 
    \stackinset{l}{0pt}{t}{0pt}{(o)}{\includegraphics[width=0.25\textwidth]{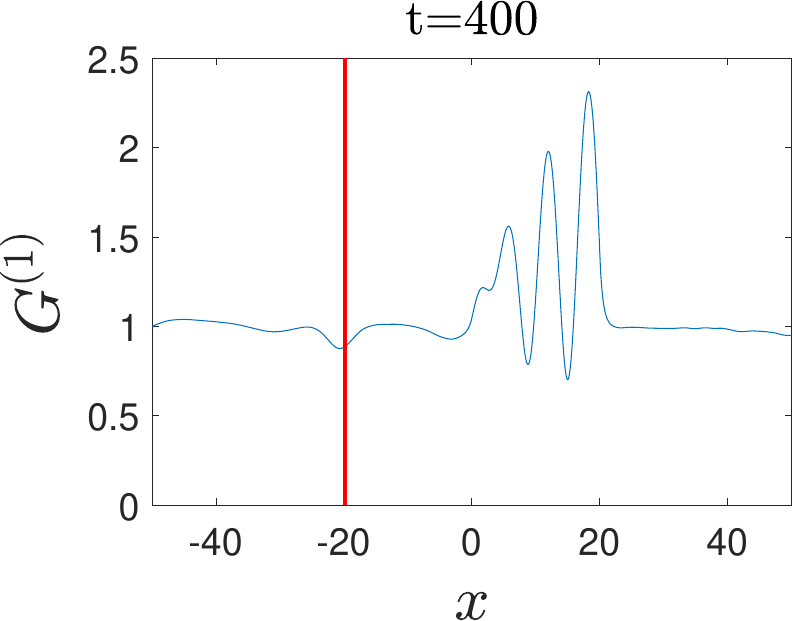}} &
    \stackinset{l}{0pt}{t}{0pt}{(p)}{\includegraphics[width=0.25\textwidth]{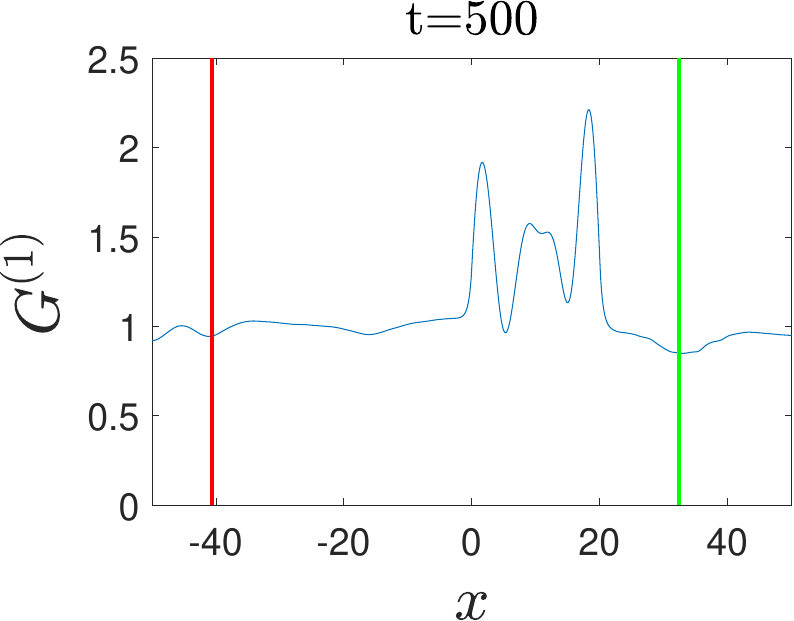}}
\end{tabular}
\caption{Same as Fig. \ref{fig:QBHLTime} but now a delta barrier of amplitude $Z=0.01$ is placed at $x=L=20$ and switched on at $t=t_{\rm{BCL}}=50$. Solid red and green lines mark the expected positions of solitons from the mean-field trajectory; see also Fig. \ref{fig:CBHLTime1000}.}
\label{fig:CBHLTime}
\end{figure*}

\begin{figure*}
\begin{tabular}{@{}cccc@{}}
      \stackinset{l}{0pt}{t}{0pt}{(a)}{\includegraphics[width=0.25\textwidth]{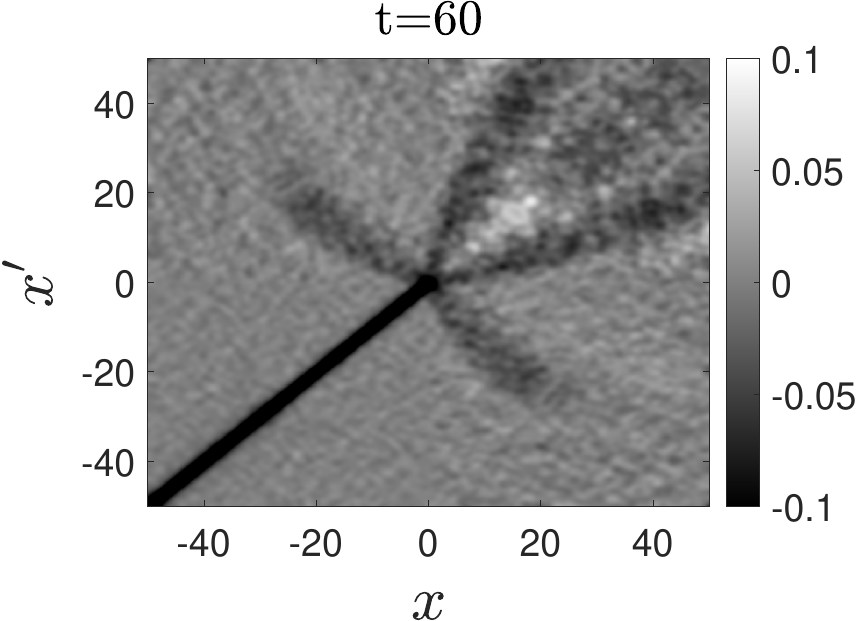}} 
    & \stackinset{l}{0pt}{t}{0pt}{(b)}{\includegraphics[width=0.25\textwidth]{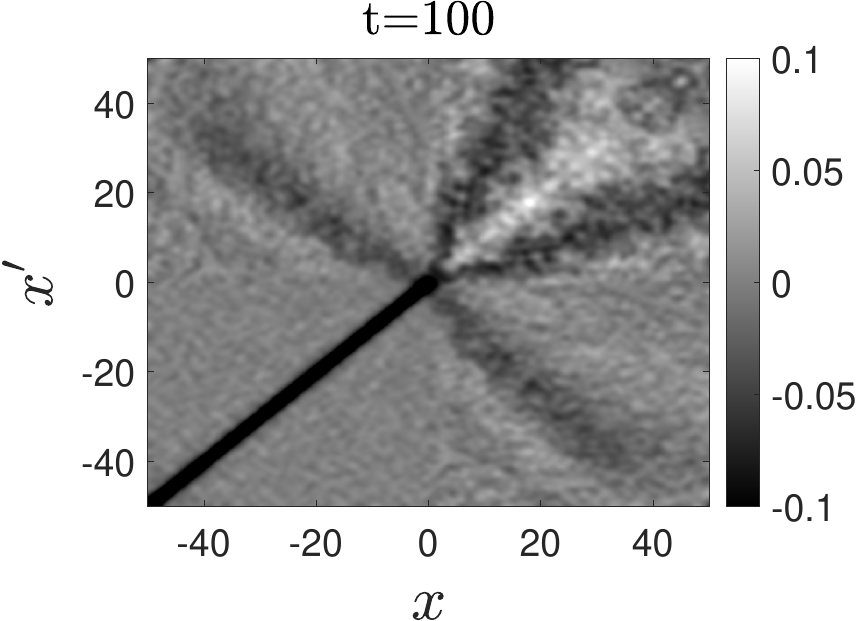}}& 
    \stackinset{l}{0pt}{t}{0pt}{(c)}{\includegraphics[width=0.25\textwidth]{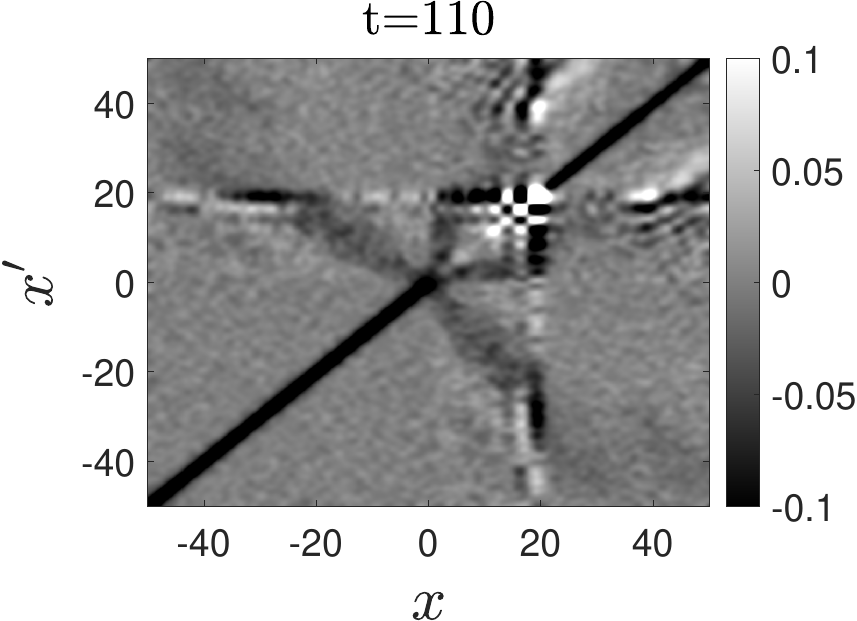}} &
    \stackinset{l}{0pt}{t}{0pt}{(d)}{\includegraphics[width=0.24\textwidth]{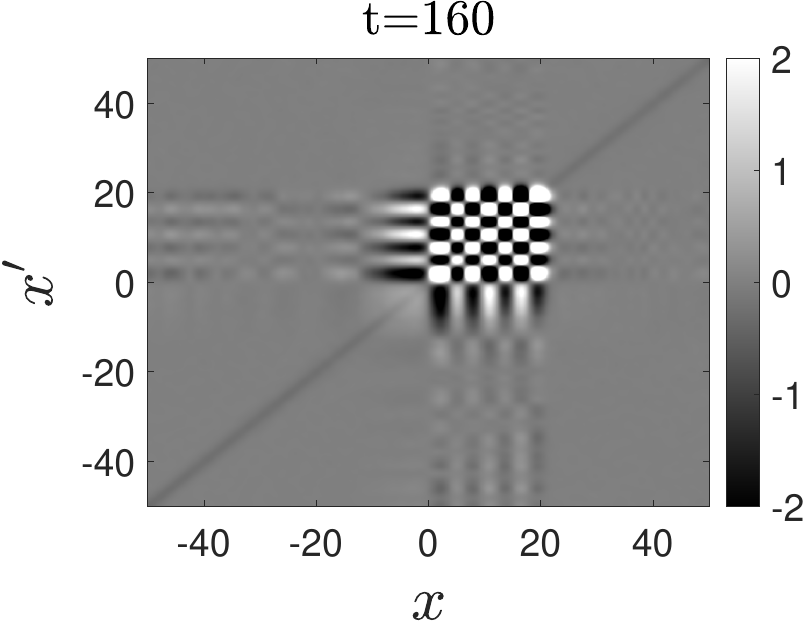}}
    \\ \\
    \stackinset{l}{0pt}{t}{0pt}{(e)}{\includegraphics[width=0.25\textwidth]{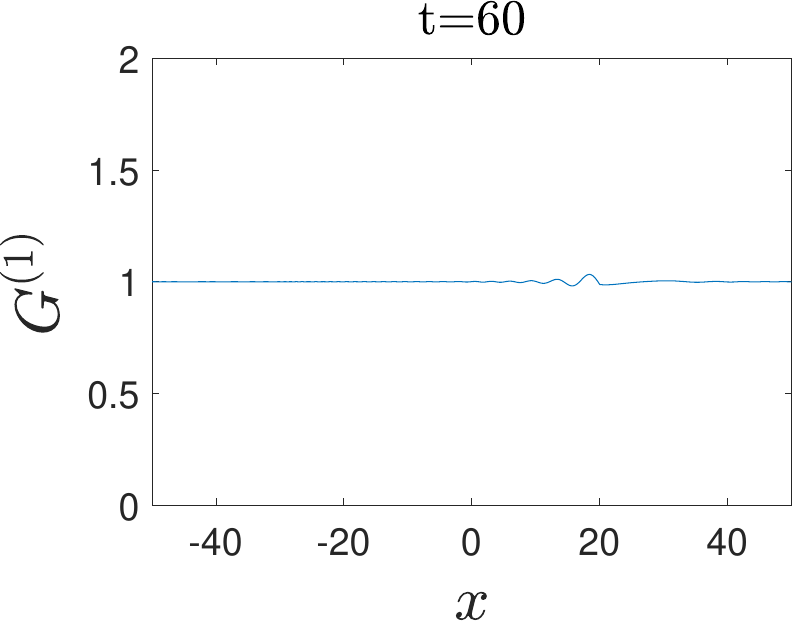}} 
    & \stackinset{l}{0pt}{t}{0pt}{(f)}{\includegraphics[width=0.25\textwidth]{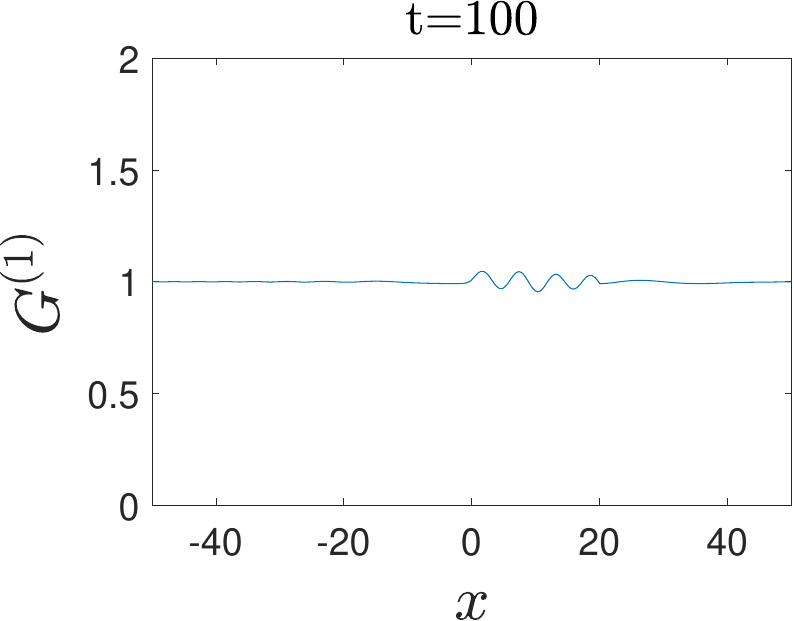}} & 
    \stackinset{l}{0pt}{t}{0pt}{(g)}{\includegraphics[width=0.25\textwidth]{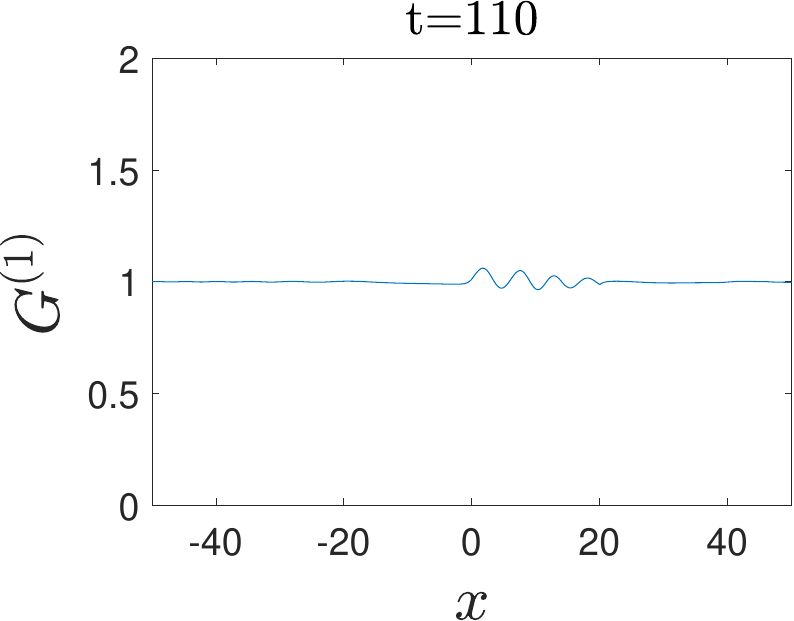}} &
    \stackinset{l}{0pt}{t}{0pt}{(h)}{\includegraphics[width=0.25\textwidth]{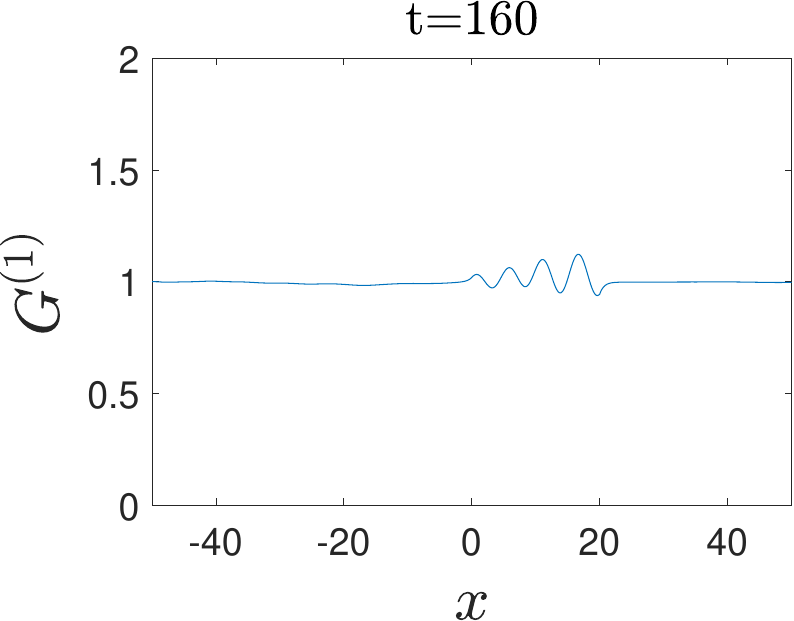}}
    \\ \\
     \stackinset{l}{0pt}{t}{0pt}{(i)}{\includegraphics[width=0.25\textwidth]{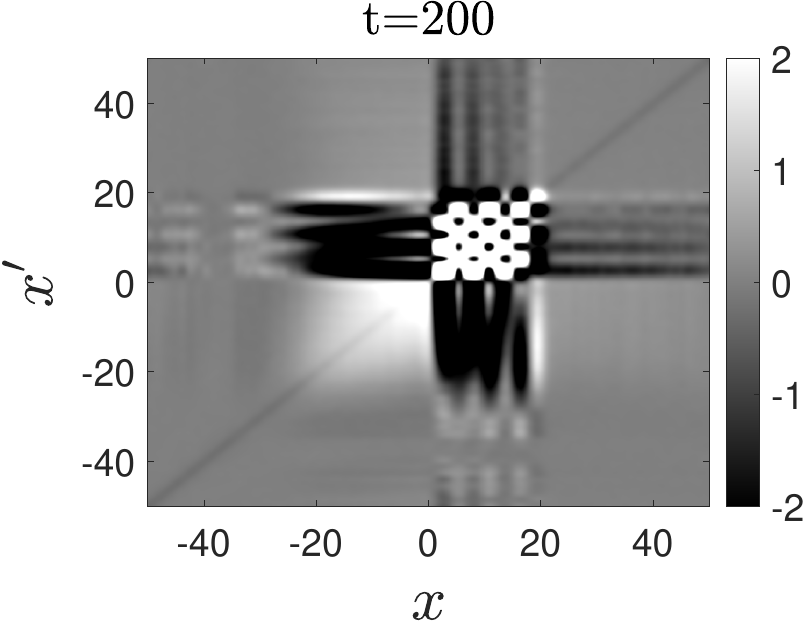}} 
    & \stackinset{l}{0pt}{t}{0pt}{(j)}{\includegraphics[width=0.25\textwidth]{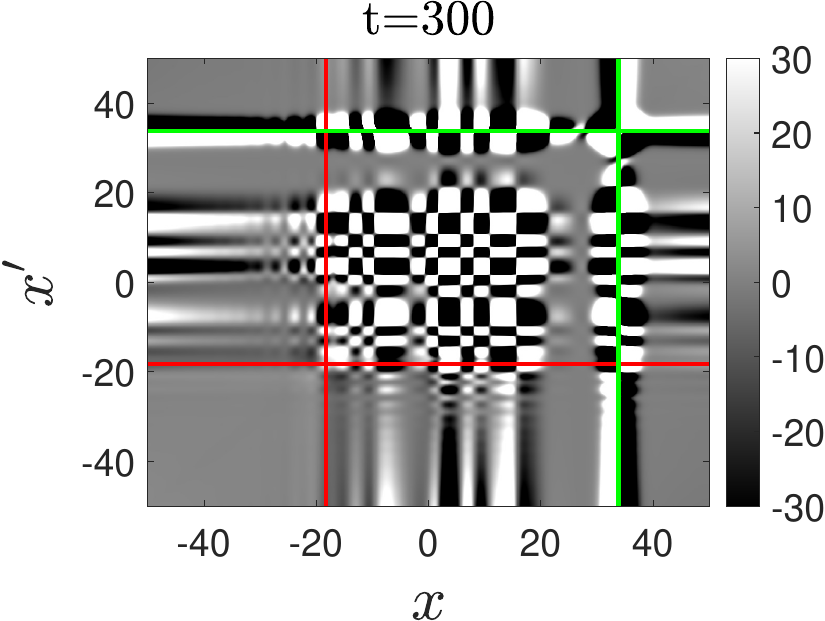}} & 
    \stackinset{l}{0pt}{t}{0pt}{(k)}{\includegraphics[width=0.25\textwidth]{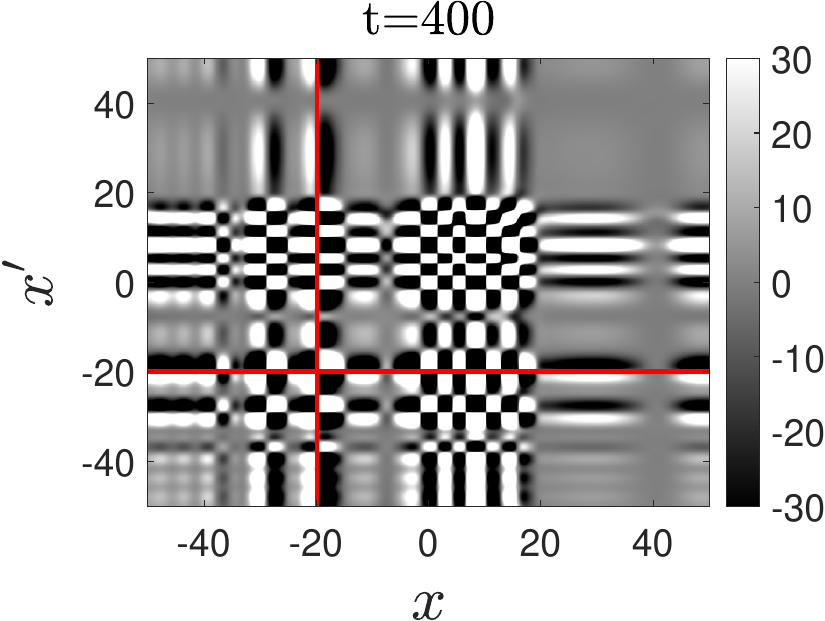}} &
    \stackinset{l}{0pt}{t}{0pt}{(l)}{\includegraphics[width=0.25\textwidth]{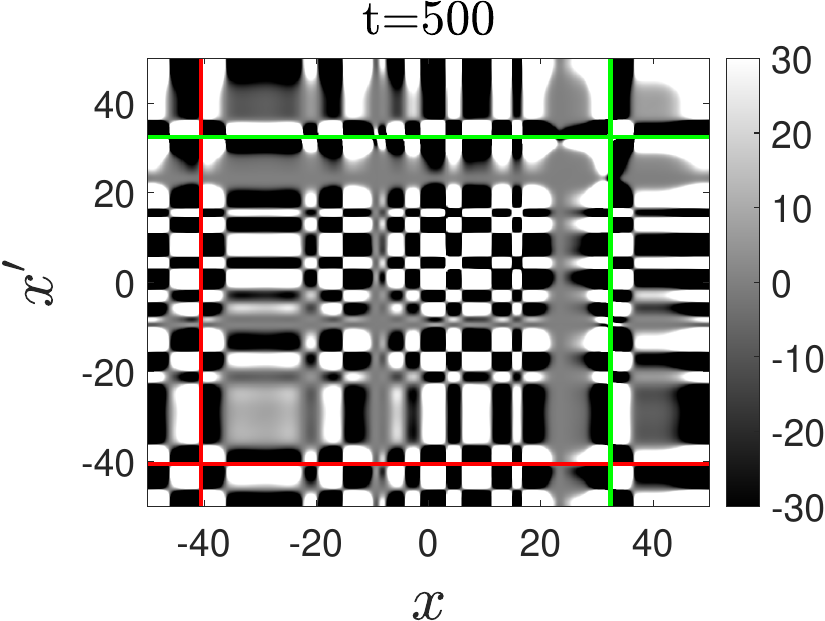}}
    \\ \\
    \stackinset{l}{0pt}{t}{0pt}{(m)}{\includegraphics[width=0.25\textwidth]{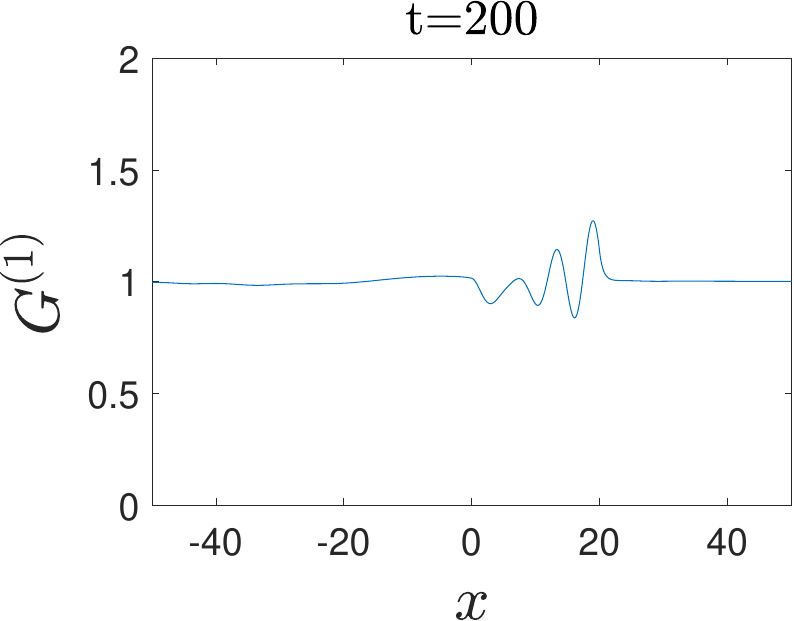}} 
    & \stackinset{l}{0pt}{t}{0pt}{(n)}{\includegraphics[width=0.25\textwidth]{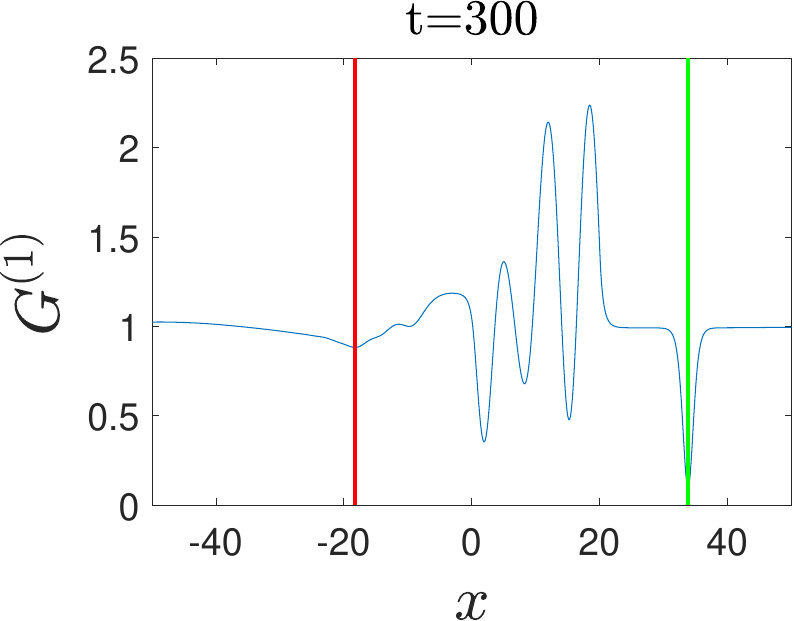}} & 
    \stackinset{l}{0pt}{t}{0pt}{(o)}{\includegraphics[width=0.25\textwidth]{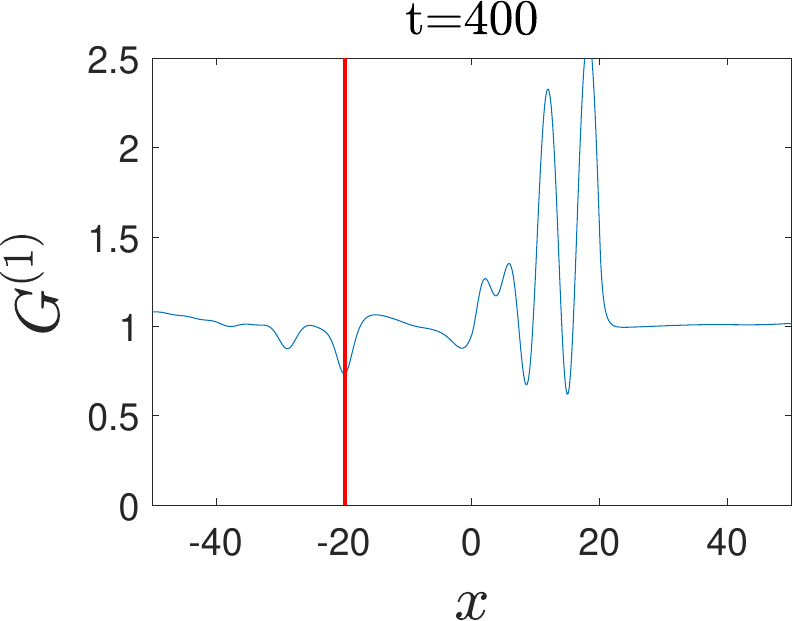}} &
    \stackinset{l}{0pt}{t}{0pt}{(p)}{\includegraphics[width=0.25\textwidth]{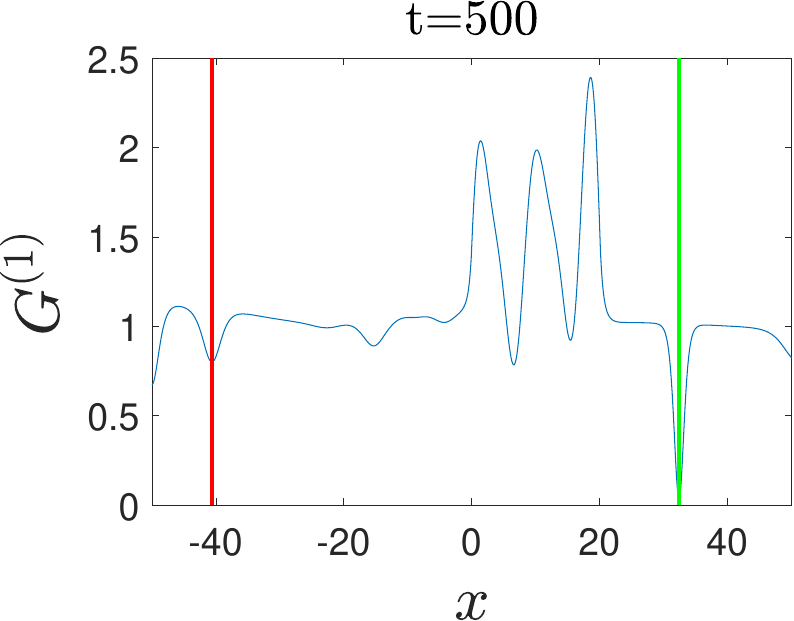}}
\end{tabular}
\caption{Same as Fig. \ref{fig:CBHLTime} but now $\lambda=1000$.}
\label{fig:CBHLTime1000}
\end{figure*}

\begin{figure*}
\begin{tabular}{@{}cccc@{}}
      \stackinset{l}{0pt}{t}{0pt}{(a)}{\includegraphics[width=0.25\textwidth]{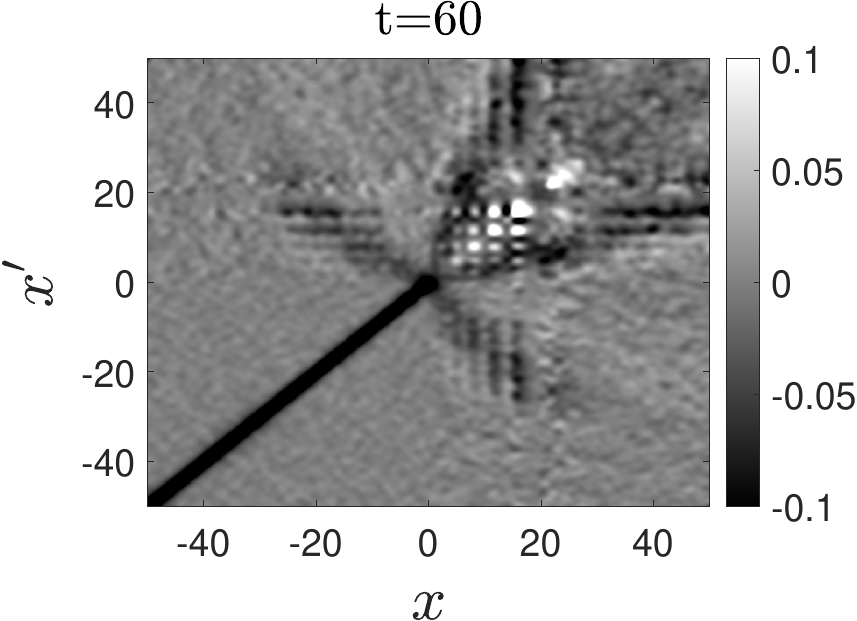}} 
    & \stackinset{l}{0pt}{t}{0pt}{(b)}{\includegraphics[width=0.25\textwidth]{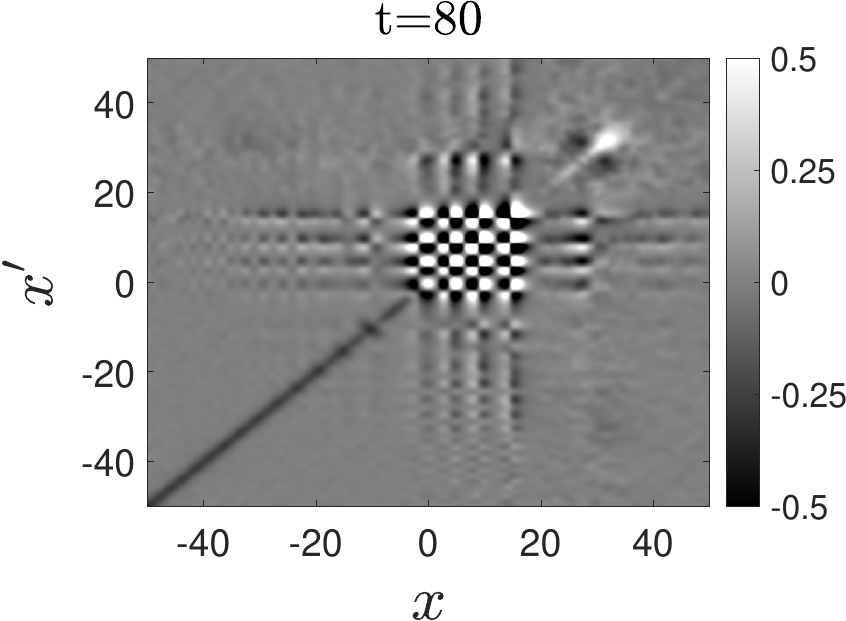}} & 
    \stackinset{l}{0pt}{t}{0pt}{(c)}{\includegraphics[width=0.25\textwidth]{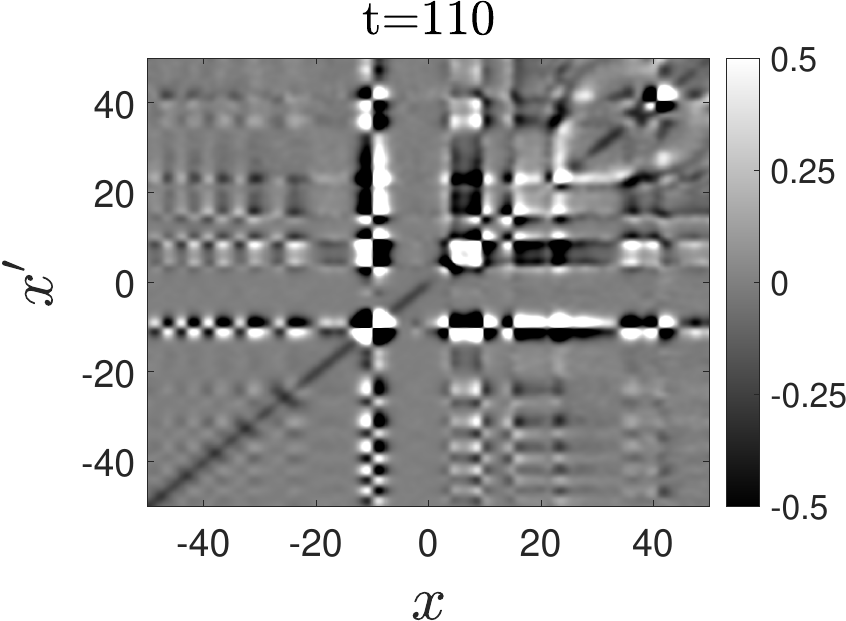}} &
    \stackinset{l}{0pt}{t}{0pt}{(d)}{\includegraphics[width=0.25\textwidth]{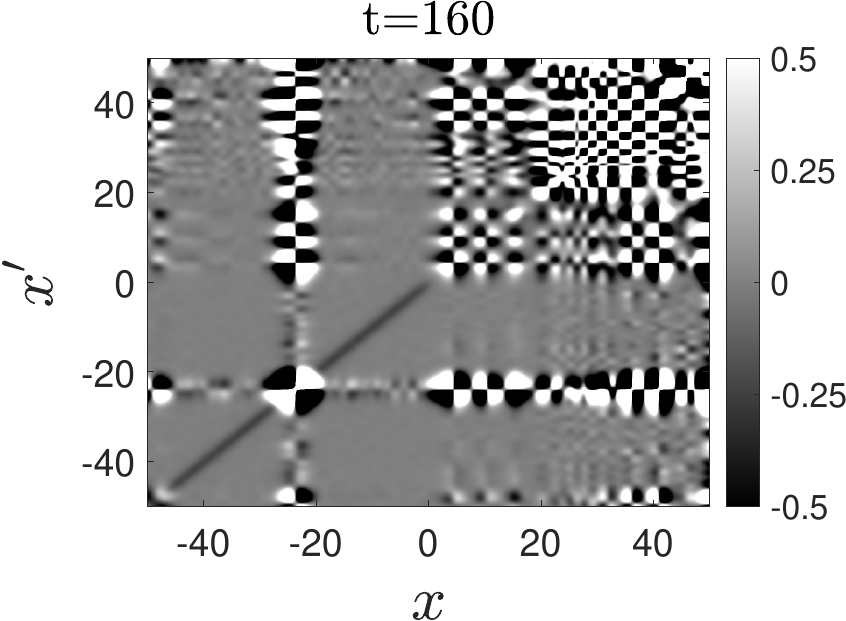}}
    \\ \\
    \stackinset{l}{0pt}{t}{0pt}{(e)}{\includegraphics[width=0.25\textwidth]{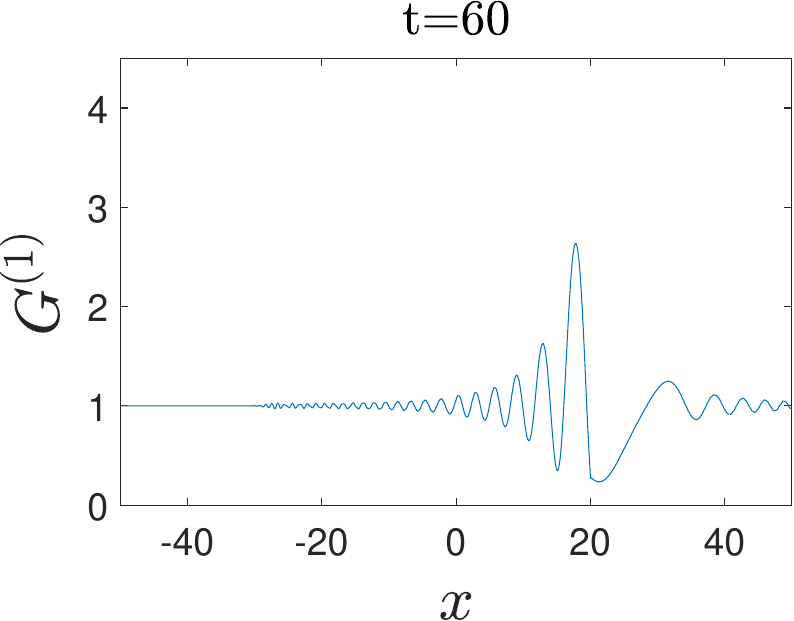}} 
    & \stackinset{l}{0pt}{t}{0pt}{(f)}{\includegraphics[width=0.25\textwidth]{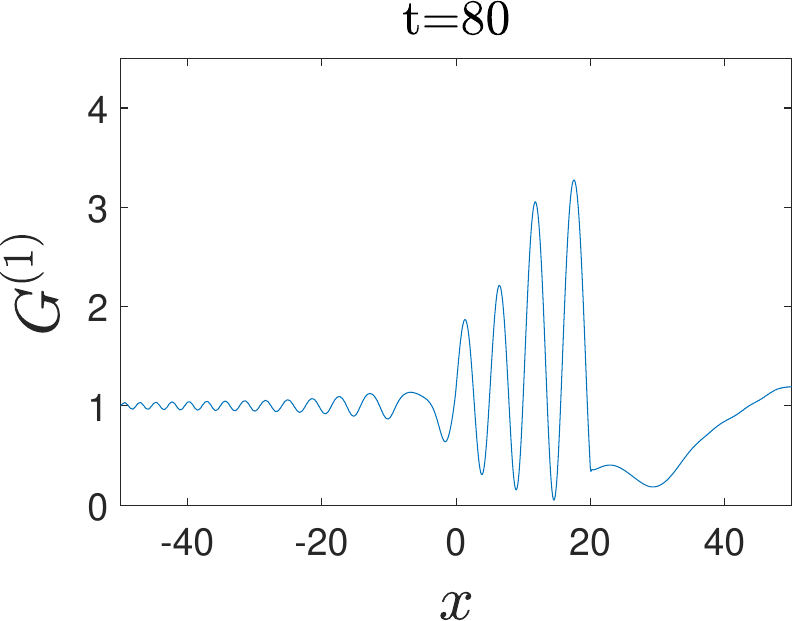}} & 
    \stackinset{l}{0pt}{t}{0pt}{(g)}{\includegraphics[width=0.25\textwidth]{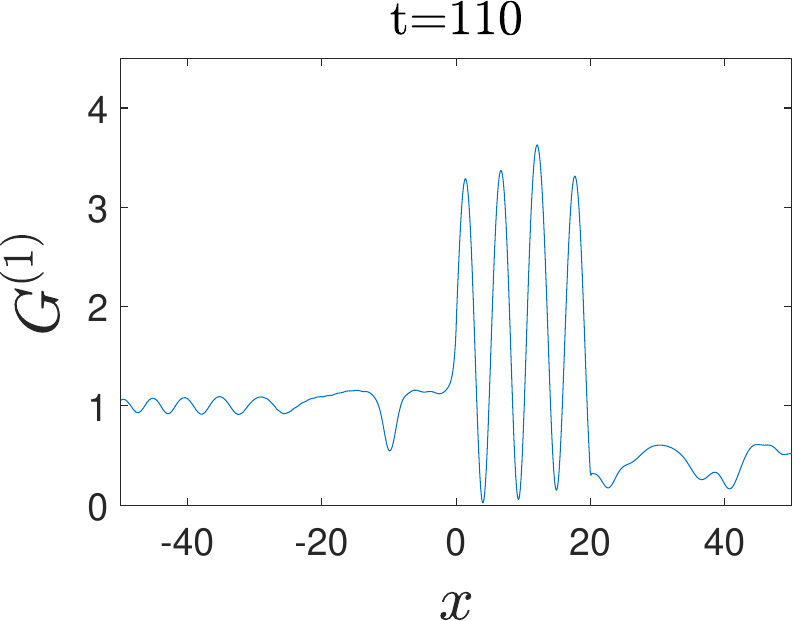}} &
    \stackinset{l}{0pt}{t}{0pt}{(h)}{\includegraphics[width=0.25\textwidth]{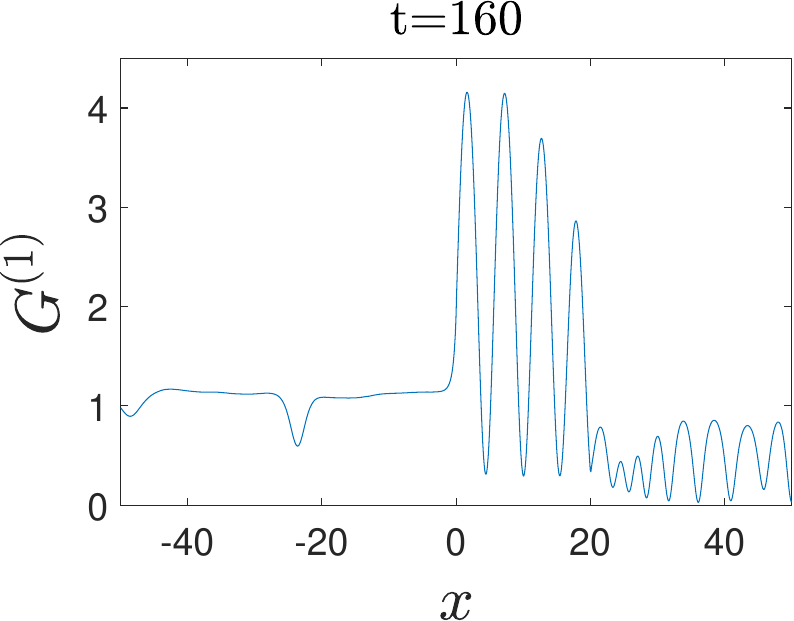}}
    \\ \\
     \stackinset{l}{0pt}{t}{0pt}{(i)}{\includegraphics[width=0.25\textwidth]{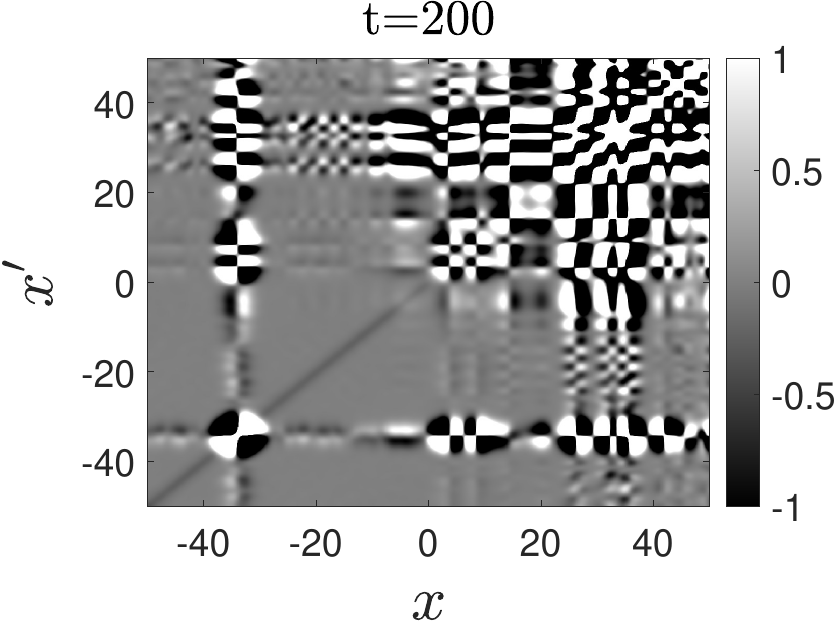}} 
    & \stackinset{l}{0pt}{t}{0pt}{(j)}{\includegraphics[width=0.25\textwidth]{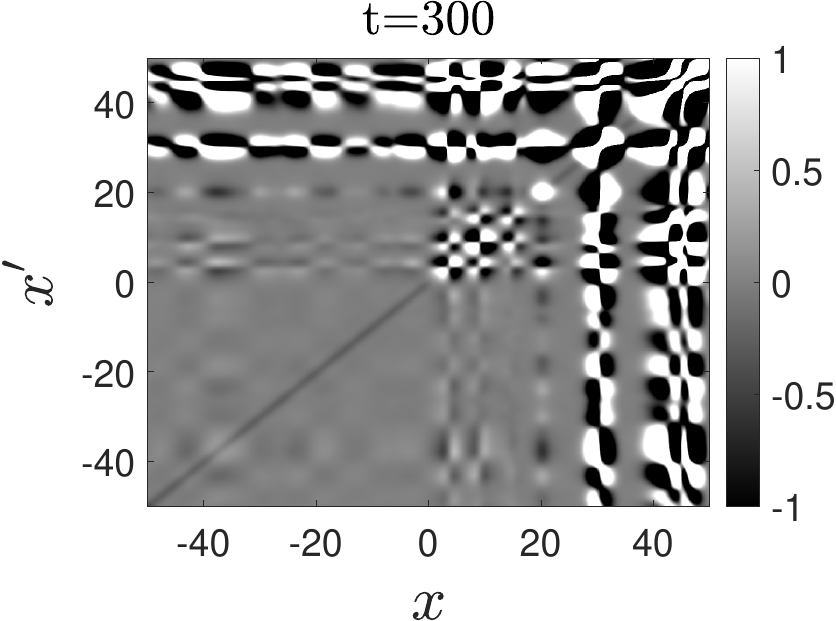}} & 
    \stackinset{l}{0pt}{t}{0pt}{(k)}{\includegraphics[width=0.25\textwidth]{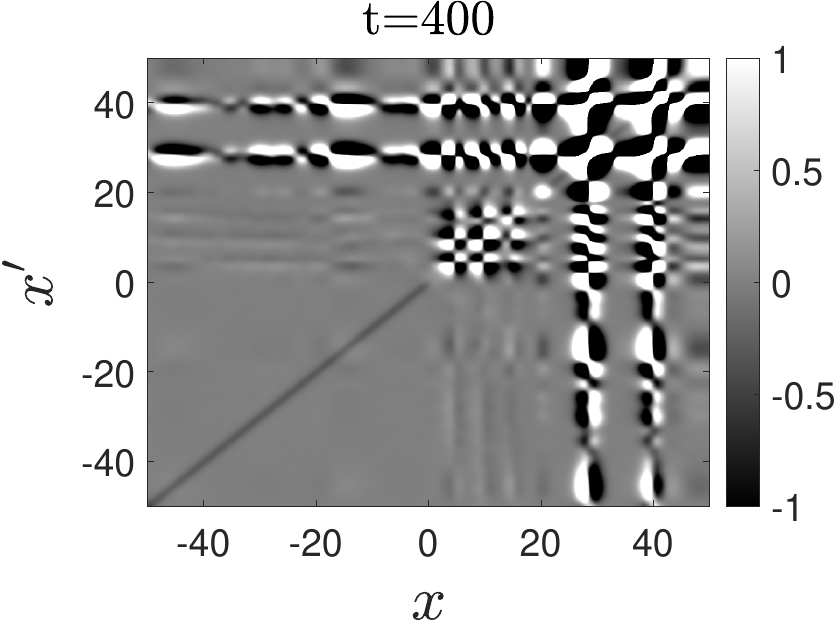}} &
    \stackinset{l}{0pt}{t}{0pt}{(l)}{\includegraphics[width=0.25\textwidth]{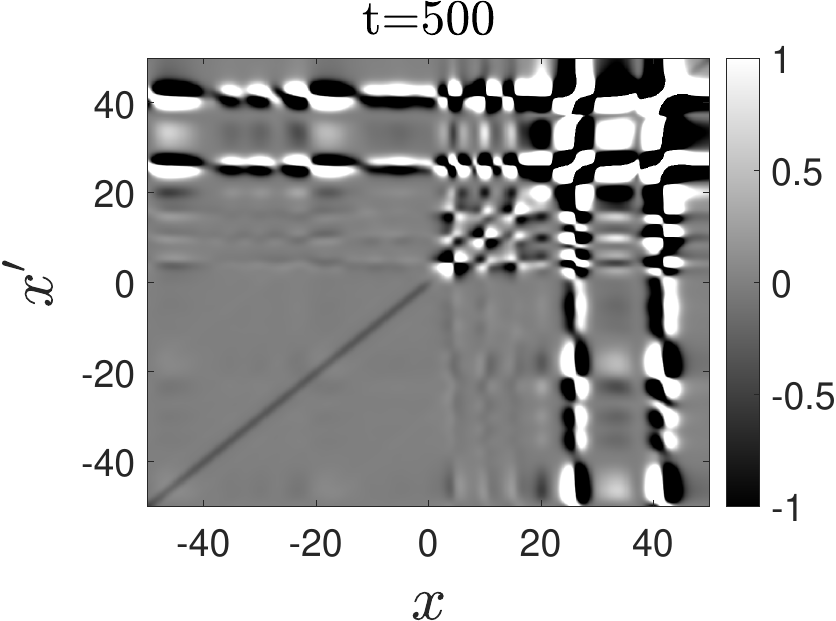}}
    \\ \\
    \stackinset{l}{0pt}{t}{0pt}{(m)}{\includegraphics[width=0.25\textwidth]{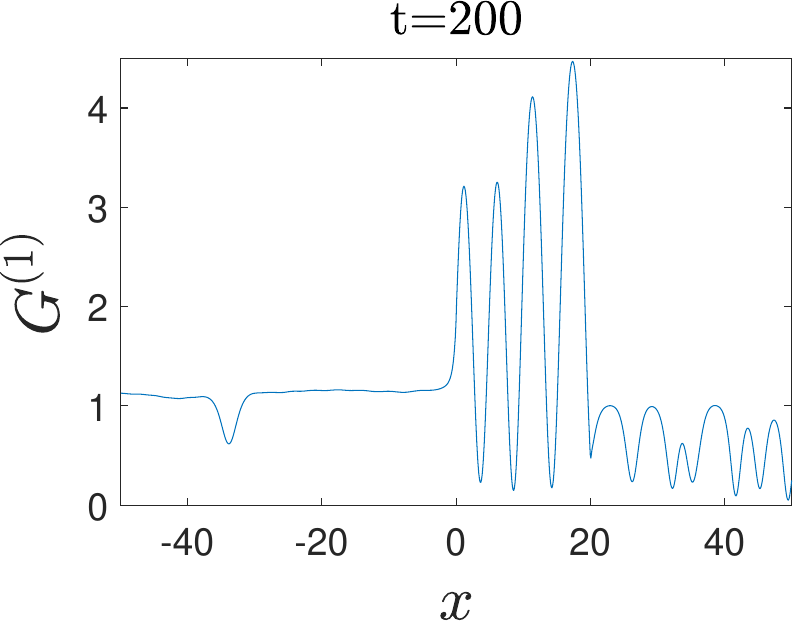}} 
    & \stackinset{l}{0pt}{t}{0pt}{(n)}{\includegraphics[width=0.25\textwidth]{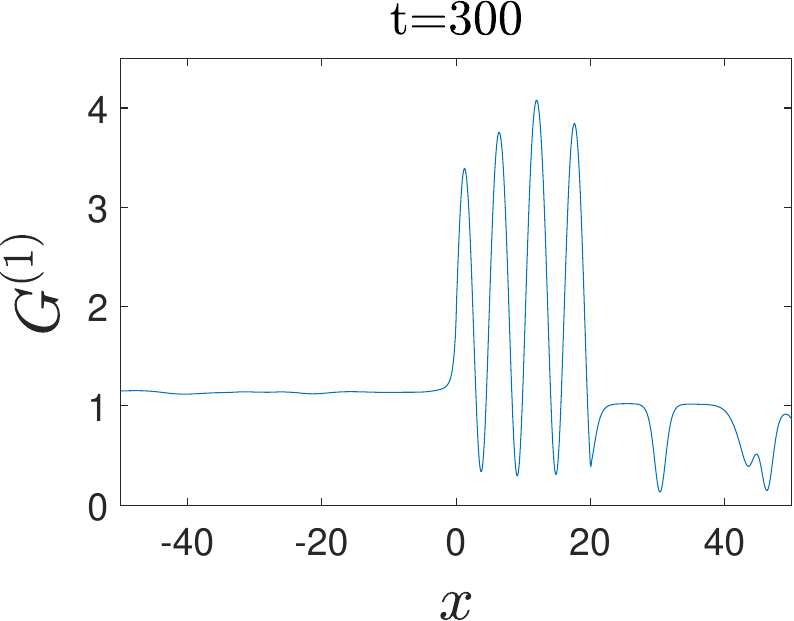}} & 
    \stackinset{l}{0pt}{t}{0pt}{(o)}{\includegraphics[width=0.25\textwidth]{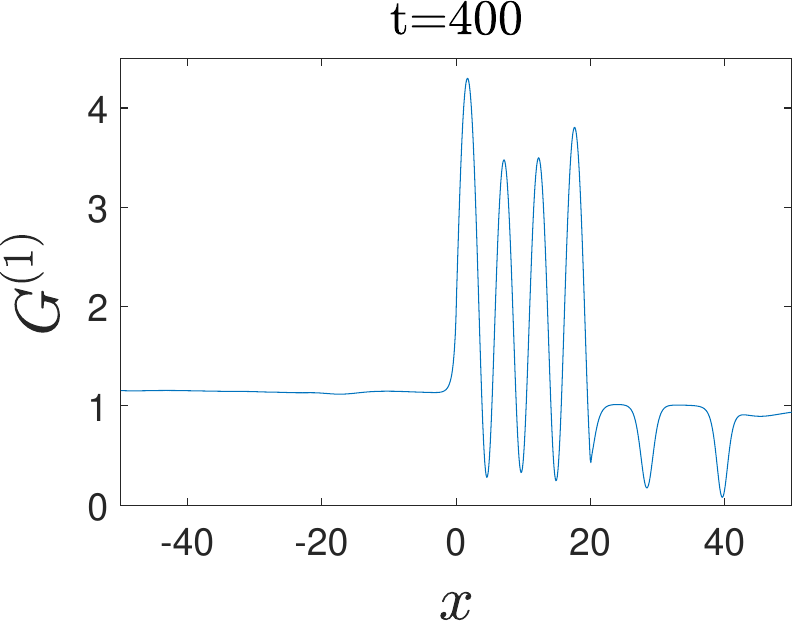}} &
    \stackinset{l}{0pt}{t}{0pt}{(p)}{\includegraphics[width=0.25\textwidth]{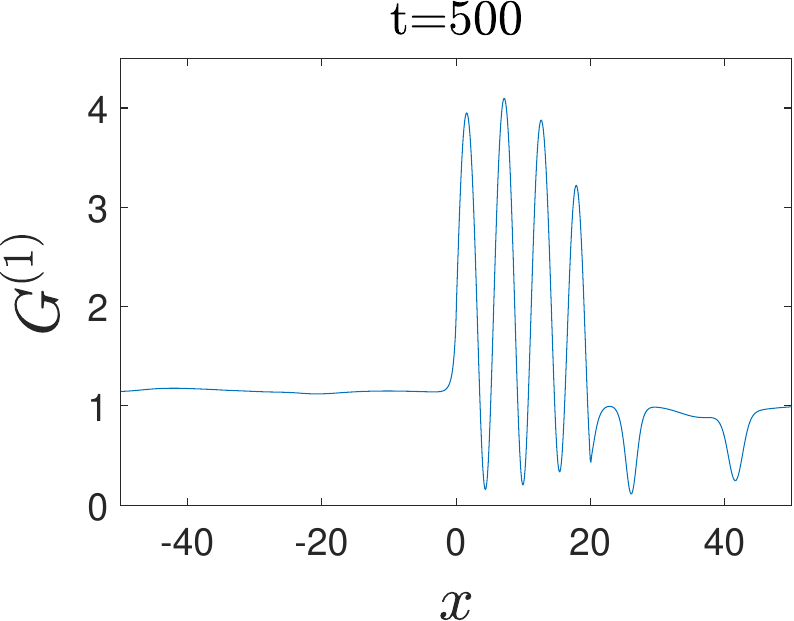}}
\end{tabular}
\caption{Same as Fig. \ref{fig:CBHLTime} but now $Z=0.75$.}
\label{fig:BCLTime}
\end{figure*}

\begin{figure*}
\begin{tabular}{@{}cccc@{}}
      \stackinset{l}{0pt}{t}{0pt}{(a)}{\includegraphics[width=0.25\textwidth]{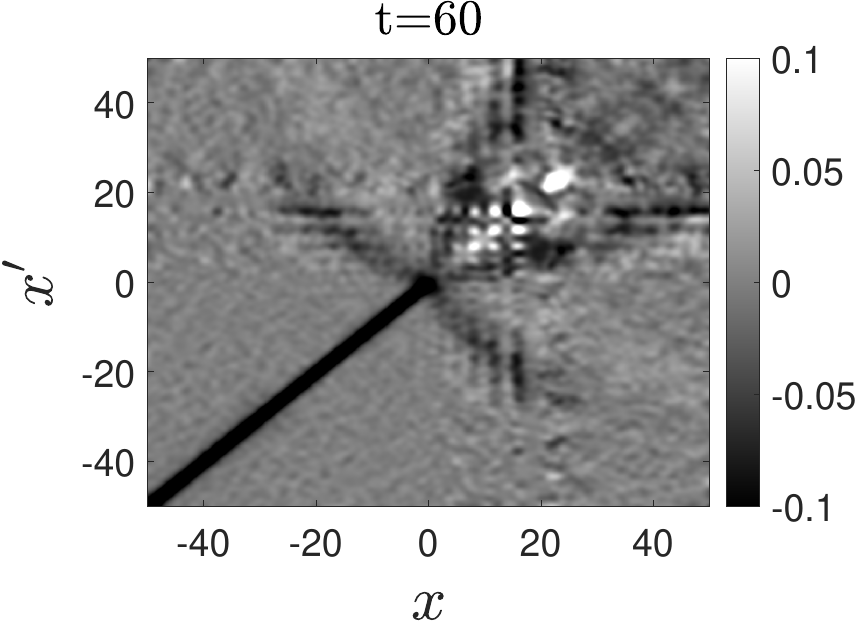}} 
    & \stackinset{l}{0pt}{t}{0pt}{(b)}{\includegraphics[width=0.25\textwidth]{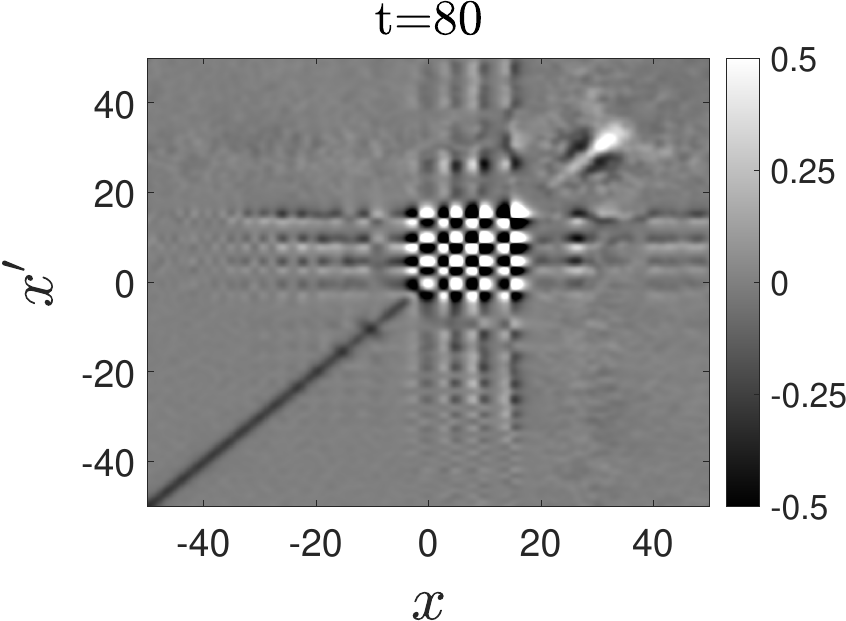}}& 
    \stackinset{l}{0pt}{t}{0pt}{(c)}{\includegraphics[width=0.25\textwidth]{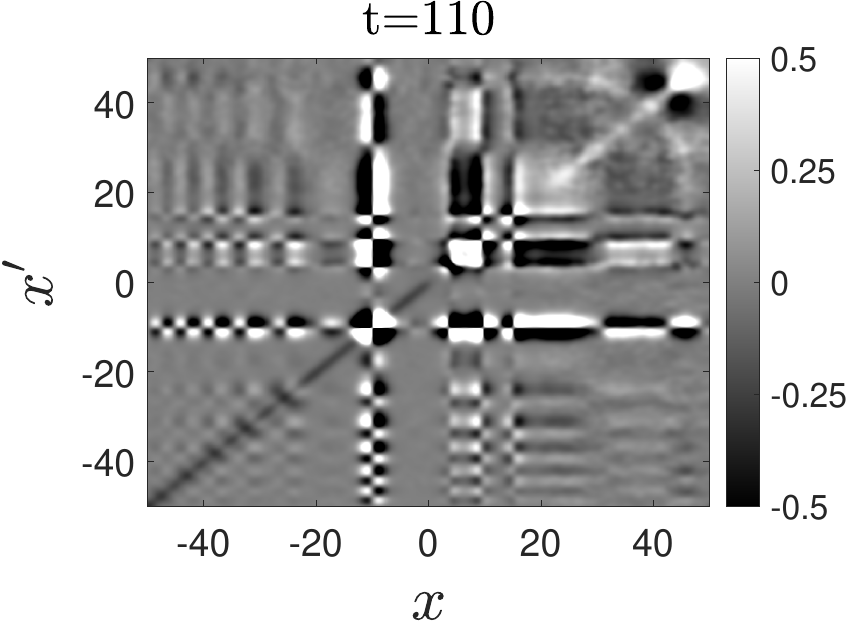}} &
    \stackinset{l}{0pt}{t}{0pt}{(d)}{\includegraphics[width=0.25\textwidth]{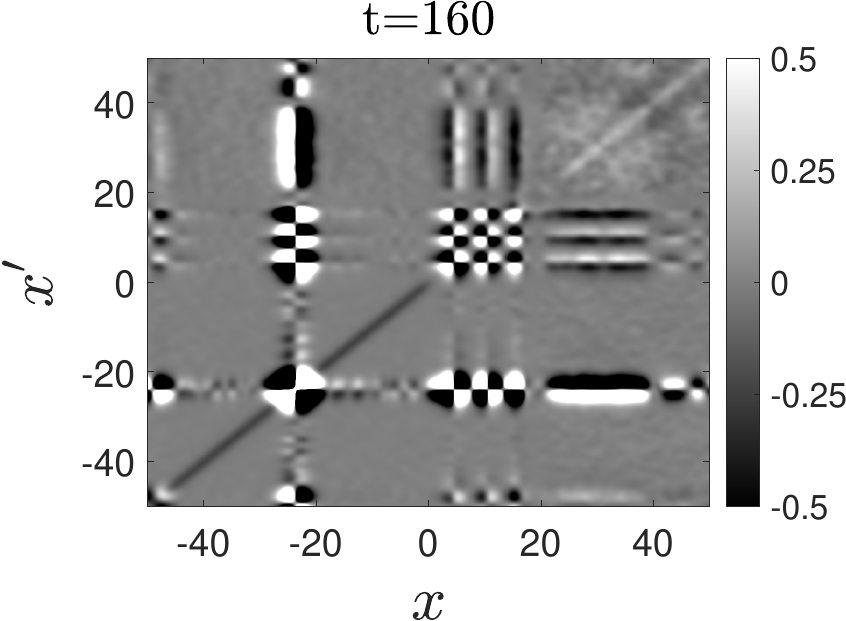}}
    \\ \\
    \stackinset{l}{0pt}{t}{0pt}{(e)}{\includegraphics[width=0.25\textwidth]{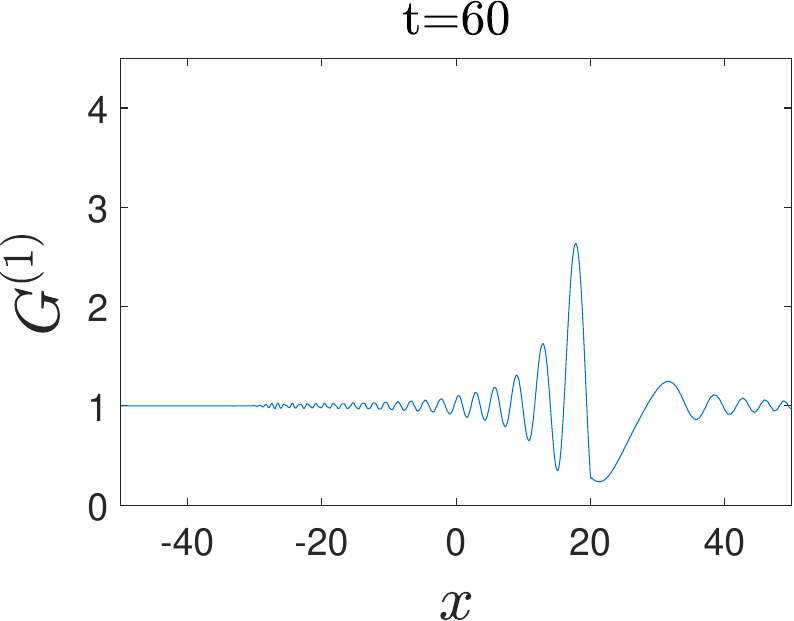}} 
    & \stackinset{l}{0pt}{t}{0pt}{(f)}{\includegraphics[width=0.25\textwidth]{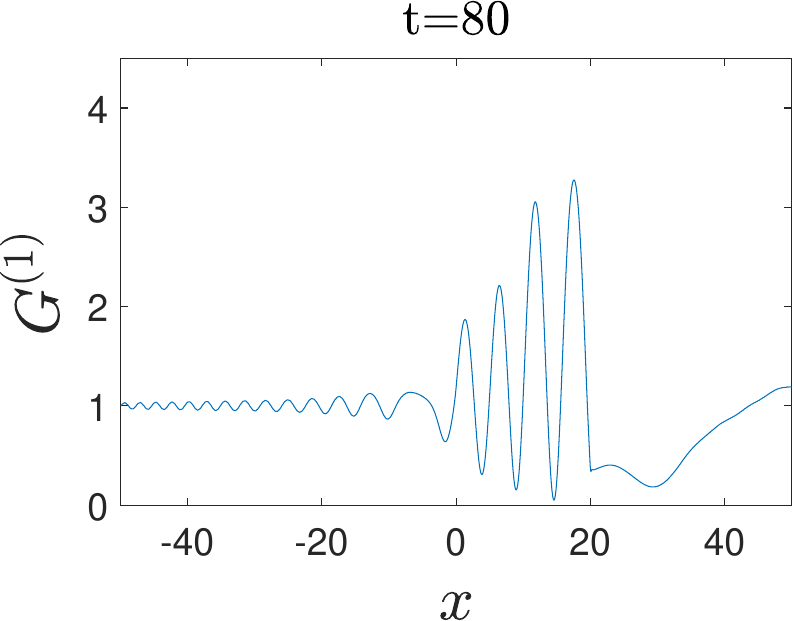}}& 
    \stackinset{l}{0pt}{t}{0pt}{(g)}{\includegraphics[width=0.25\textwidth]{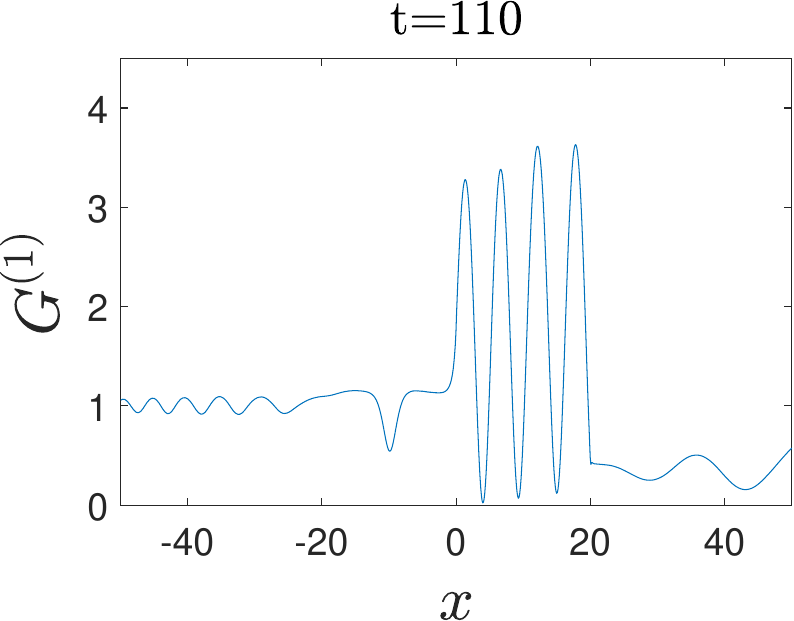}} &
    \stackinset{l}{0pt}{t}{0pt}{(h)}{\includegraphics[width=0.25\textwidth]{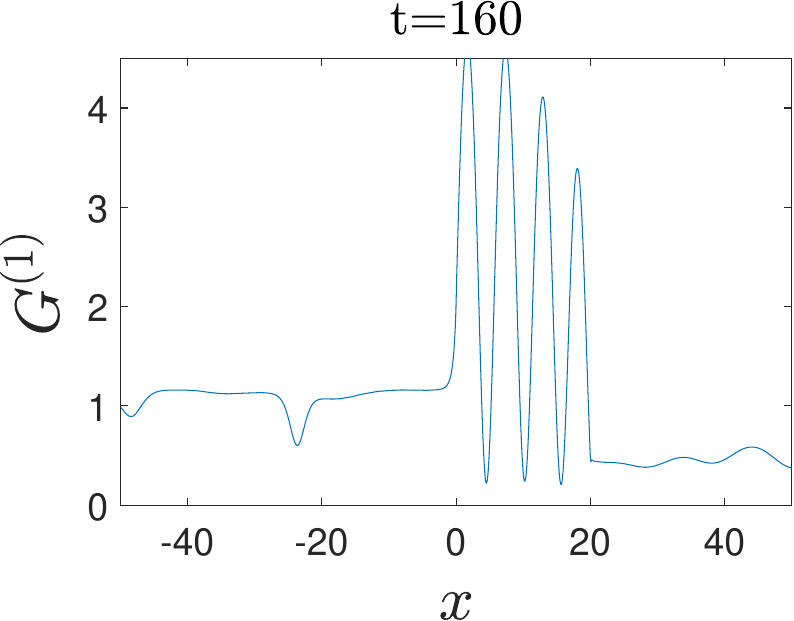}}
    \\ \\
     \stackinset{l}{0pt}{t}{0pt}{(i)}{\includegraphics[width=0.25\textwidth]{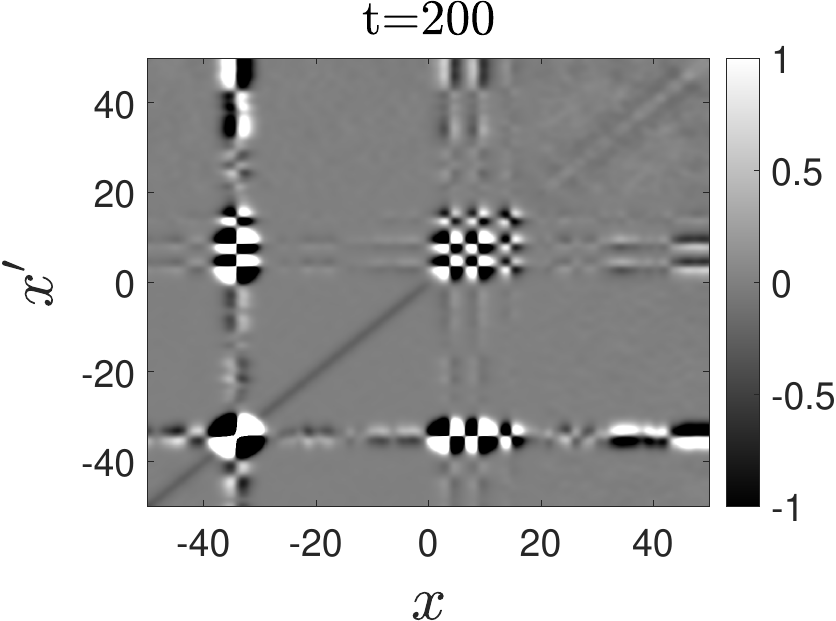}} 
    & \stackinset{l}{0pt}{t}{0pt}{(j)}{\includegraphics[width=0.25\textwidth]{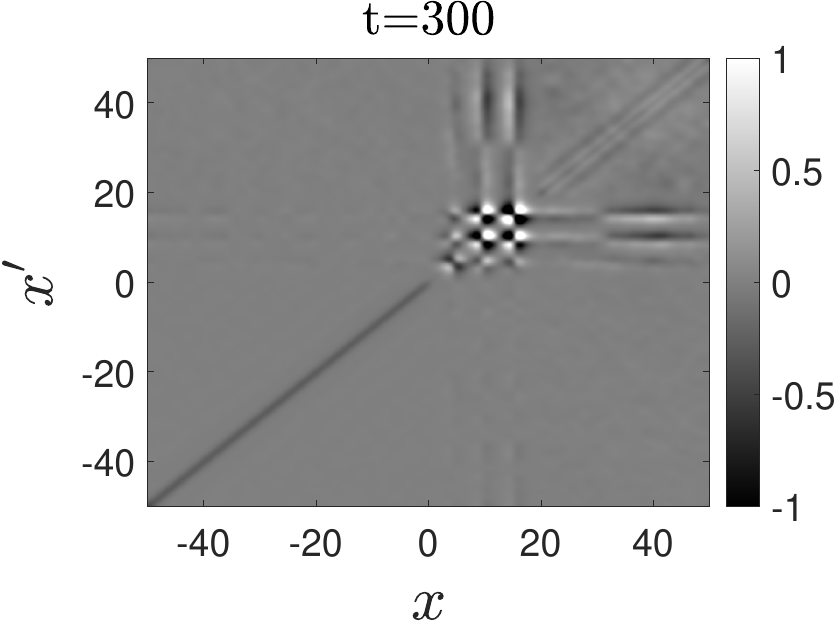}}& 
    \stackinset{l}{0pt}{t}{0pt}{(k)}{\includegraphics[width=0.25\textwidth]{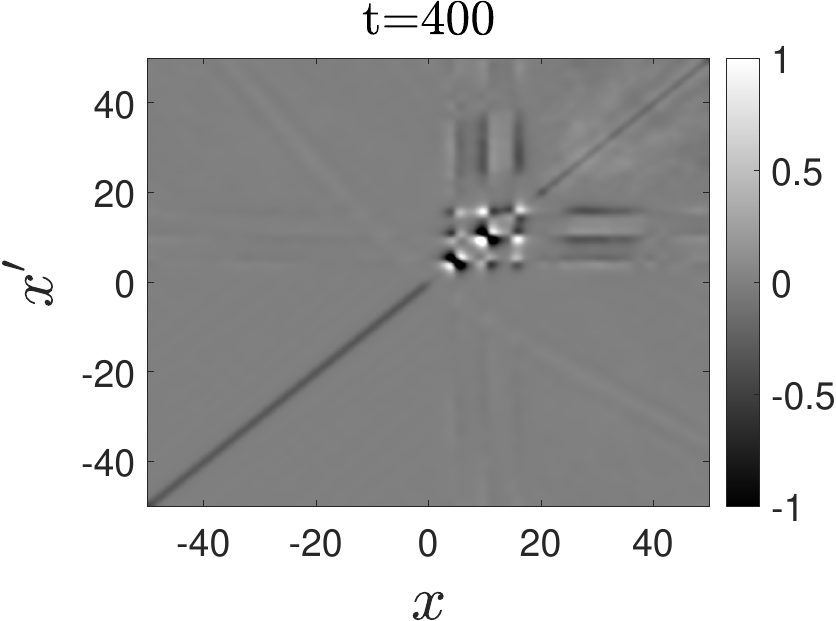}} &
    \stackinset{l}{0pt}{t}{0pt}{(l)}{\includegraphics[width=0.25\textwidth]{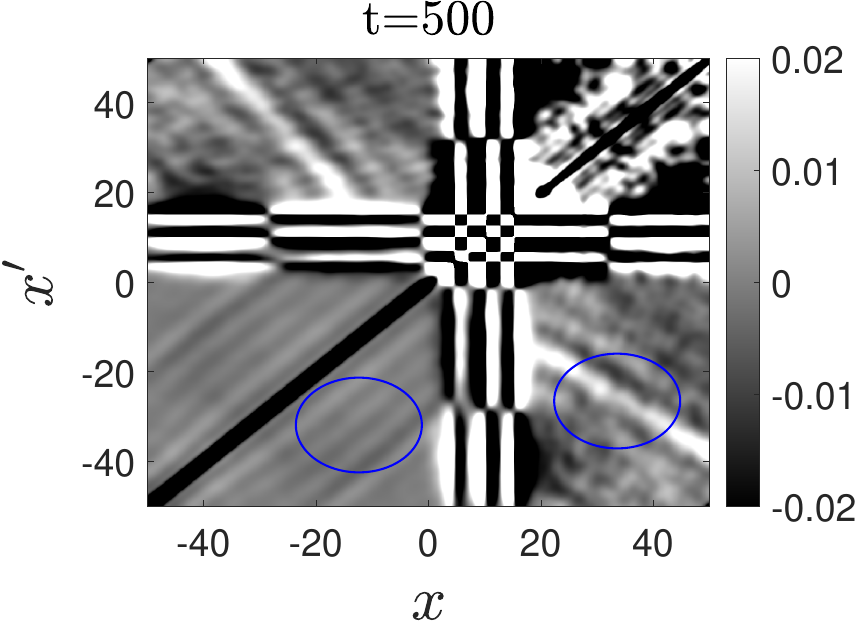}}
    \\ \\
    \stackinset{l}{0pt}{t}{0pt}{(m)}{\includegraphics[width=0.25\textwidth]{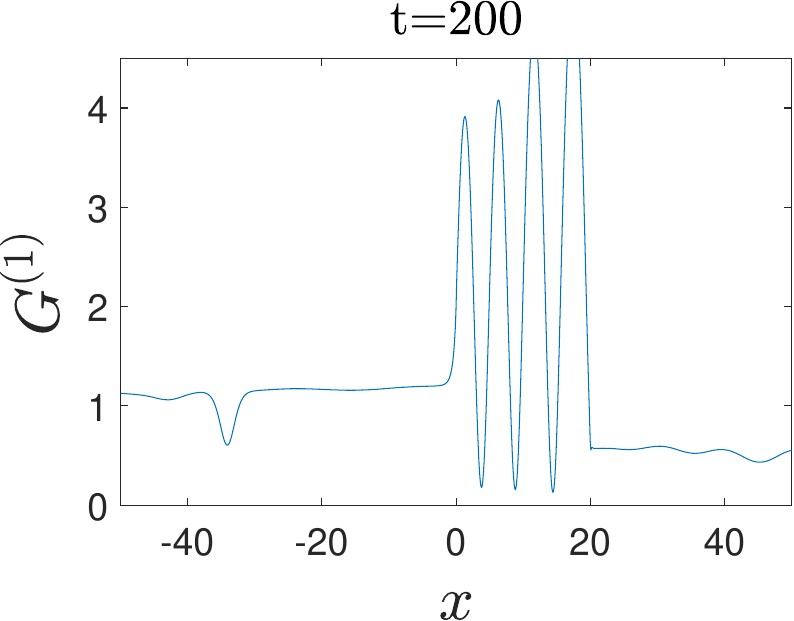}} 
    & \stackinset{l}{0pt}{t}{0pt}{(n)}{\includegraphics[width=0.25\textwidth]{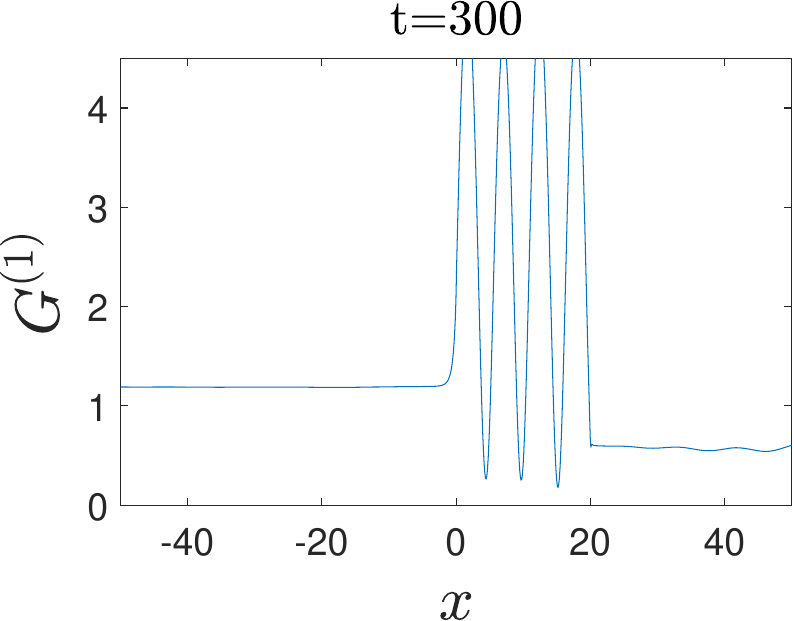}}& 
    \stackinset{l}{0pt}{t}{0pt}{(o)}{\includegraphics[width=0.25\textwidth]{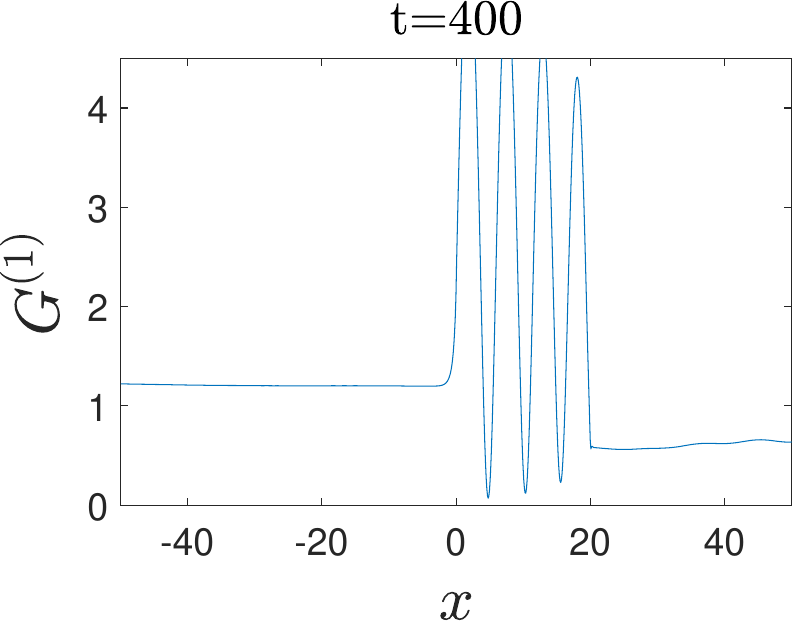}} &
    \stackinset{l}{0pt}{t}{0pt}{(p)}{\includegraphics[width=0.25\textwidth]{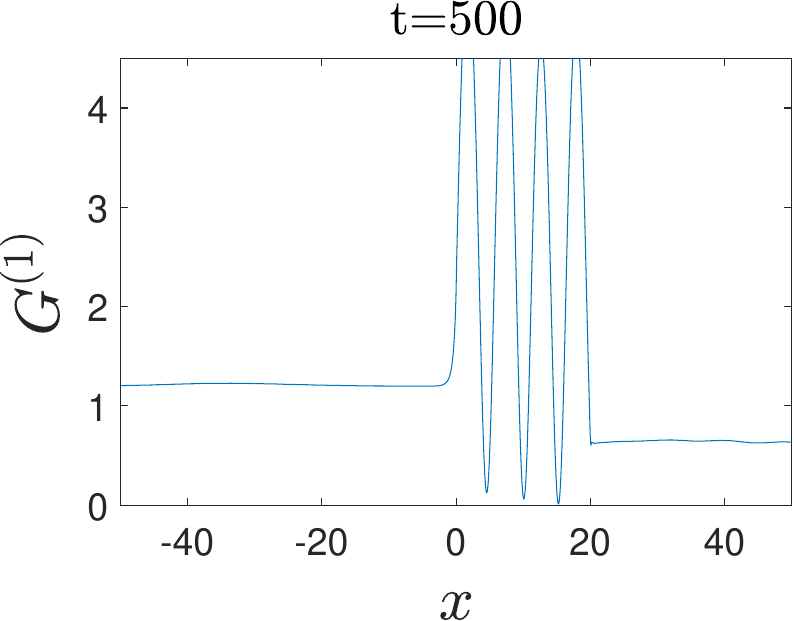}}
\end{tabular}
\caption{Same as Fig. \ref{fig:BCLTime} but now only BCL stimulation is present, $t_{\rm{BHL}}=\infty$. Blue circles in panel (l) highlight monochromatic features arising from quantum BCL-stimulated Hawking radiation, as discussed in the main text.}
\label{fig:BCLPuroTime}
\end{figure*}

\begin{figure*}[t]
\begin{tabular}{@{}ccc@{}}\stackinset{l}{0pt}{t}{0pt}{\large{(a)}}{\includegraphics[width=0.66\columnwidth]{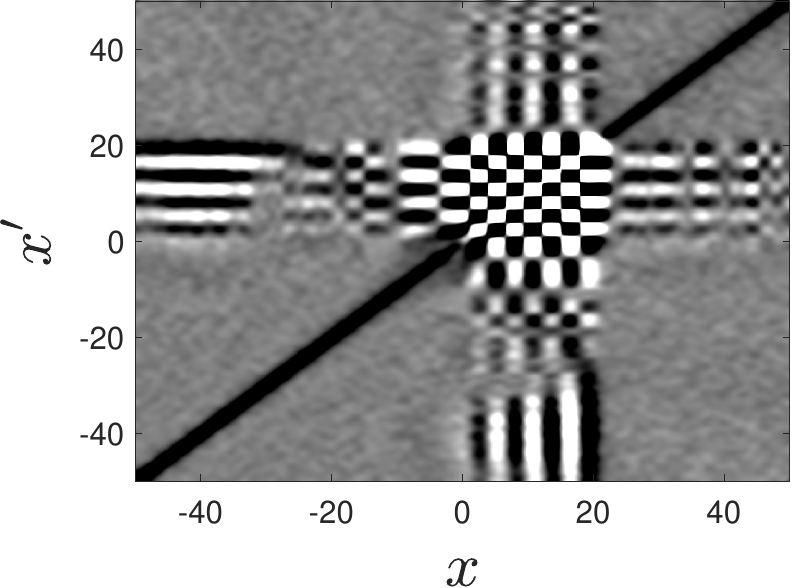}} &~
    \stackinset{l}{0pt}{t}{0pt}{\large{(b)}}
    {\includegraphics[width=0.66\columnwidth]{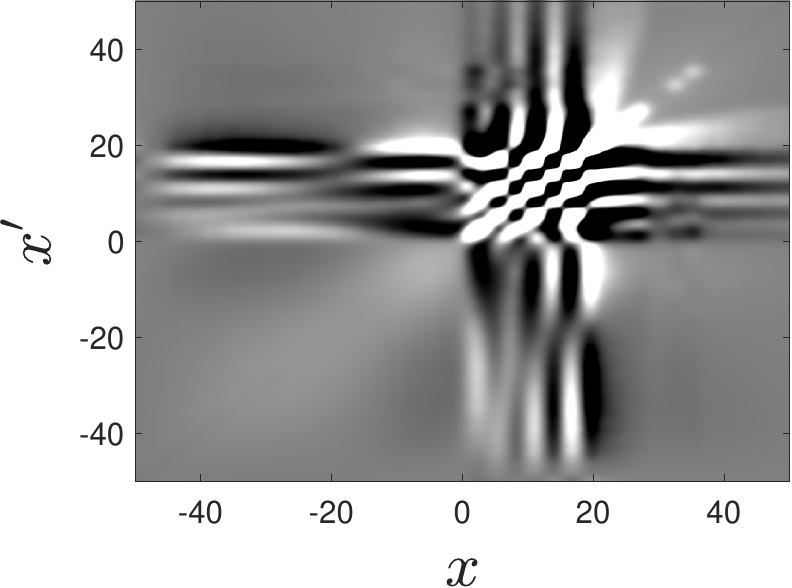}} &~
    \stackinset{l}{0pt}{t}{0pt}{\large{(c)}}
    {\includegraphics[width=0.66\columnwidth]{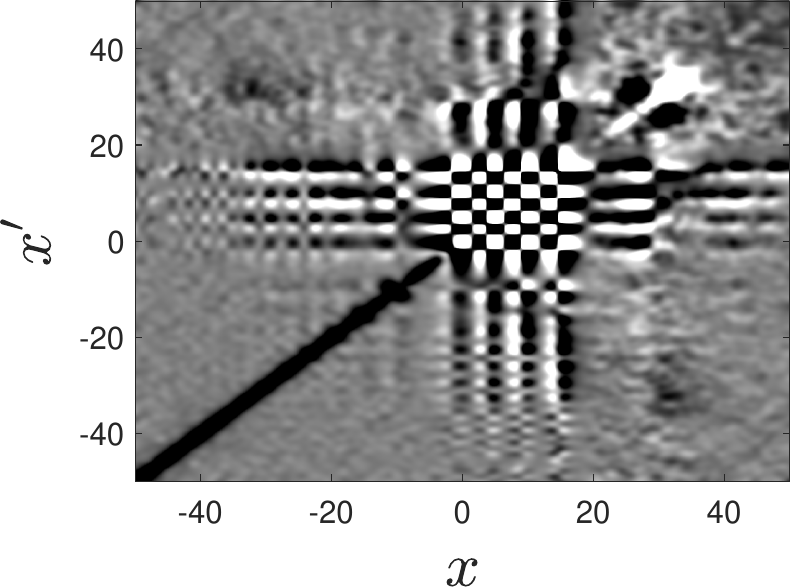}} \\ \\ \\
    \stackinset{l}{0pt}{t}{0pt}{\large{(d)}}{\includegraphics[width=0.66\columnwidth]{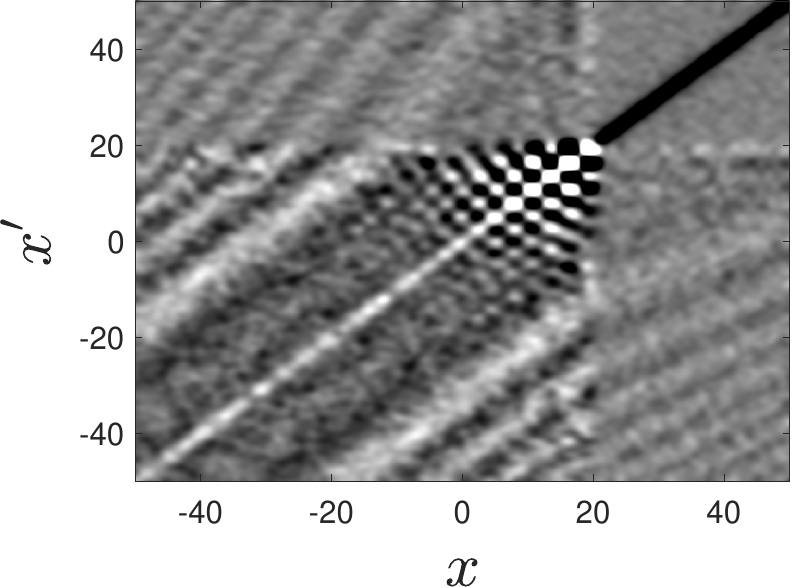}} &
    \stackinset{l}{0pt}{t}{0pt}{\large{(e)}}
    {\includegraphics[width=0.66\columnwidth]{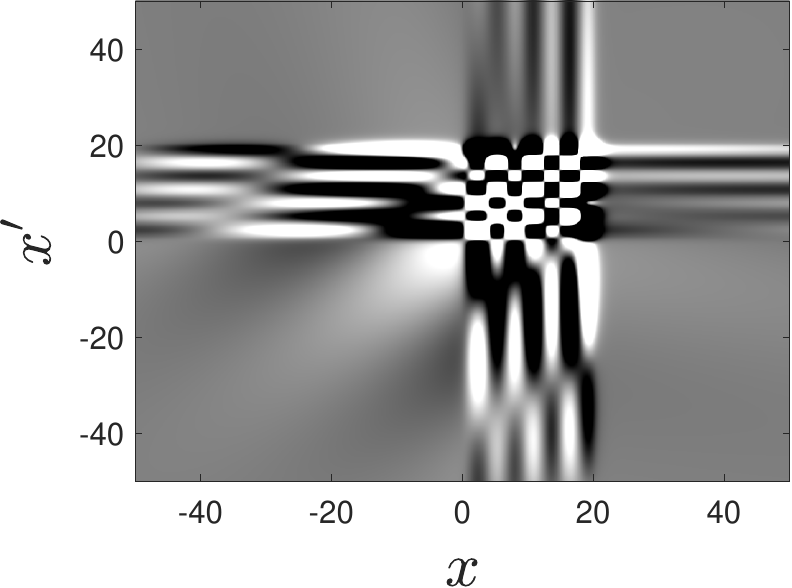}} &
    \stackinset{l}{0pt}{t}{0pt}{\large{(f)}}
    {\includegraphics[width=0.66\columnwidth]{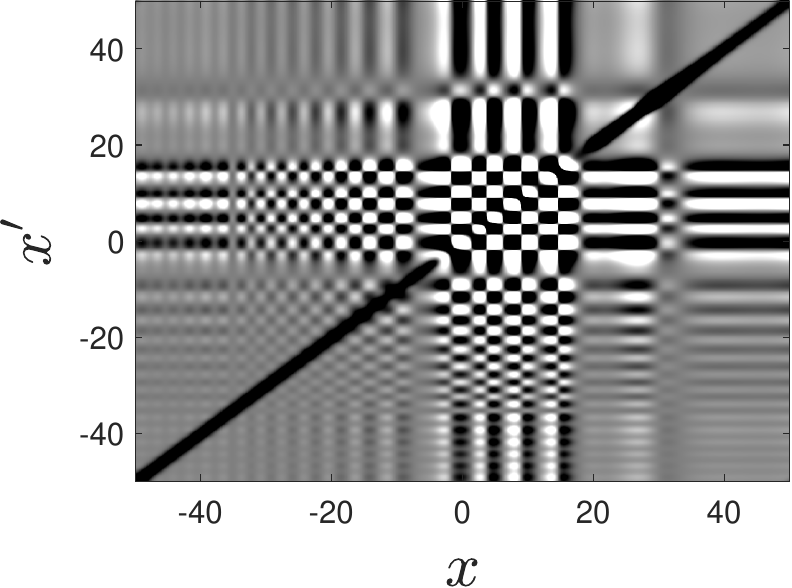}} 
\end{tabular}
\caption{Density-density correlation function $G^{(2)}(x,x')$ for different configurations. The scale is not shown since we are only interested in the qualitative spatial structure of the correlation patterns. (a) Snapshot at $t=140$ of the simulation in Fig. \ref{fig:QBHLTime}.
(b) Same as (a) but for $t=250$. (c) Snapshot at $t=80$ of the simulation in Fig. \ref{fig:BCLTime}. (d) Same as (a) but without black hole. (e) Spatial correlation pattern arising solely from the dominant mode in (b), computed in the BdG approximation by taking Eq. (\ref{eq:QuantumDominant}) and setting $\braket{\hat{\alpha}^\dagger_I\hat{\alpha}_I}=1/2,~\braket{\hat{\alpha}_I\hat{\alpha}_I}=0$. (f) Same as (c) but now replacing quantum fluctuations by particle-number fluctuations. Specifically, we take as initial condition in the Truncated Wigner method $\Psi_W(x,0)=\sqrt{1+\delta n}\,e^{ivx}$, with $\delta n$ a Gaussian variable with zero mean and $\sqrt{\braket{\delta n^2}}=0.001$.}
\label{fig:BHLBCLExplan}
\end{figure*}

\subsection{Correlation patterns}\label{subsec:Pattern}

We begin by studying the time evolution of the density-density correlation function $G^{(2)}(x,x',t)$ and the corresponding ensemble-averaged density $G^{(1)}(x,t)$, displayed in Figs. \ref{fig:QBHLTime}-\ref{fig:BCLPuroTime} for different values of $Z,\lambda$. We focus on the qualitative correlation patterns exhibited, leaving the quantitative analysis for the following subsections.

In Fig. \ref{fig:QBHLTime}, we present the case of a purely quantum BHL as there is no BCL stimulation ($Z=0$), so the dynamics is fully driven by quantum fluctuations. At $t=0$, Fig. \ref{fig:QBHLTime}a, the homogeneous condensate is at equilibrium and $G^{(2)}(x,x')$ displays the expected antibunching along the main diagonal $x=x'$, Eq. (\ref{eq:antibunching}). Then, the black hole is switched on and the emission of Hawking radiation begins, as revealed by the emergence of the celebrated Hawking moustache (marked by a blue line in Fig. \ref{fig:QBHLTime}b), stemming from the correlations between the Hawking $b-$ and the partner $p2+$ modes \cite{Carusotto2008}. Along them, we observe the correlations between the Hawking and the $p1+$ modes \cite{Recati2009} (red line), and the correlations between the $p1+$ and the $p2+$ modes (green line), which represent the bosonic analogue of the Andreev reflection \cite{Zapata2009a}. Self-correlations of both the $p1+$ and $p2+$ modes can be also seen along the main diagonal (yellow line). All these lines are predicted following the usual hydrodynamic approximation \cite{Recati2009}. Since there is no BCL stimulation, the background condensate remains homogeneous, as shown by Fig. \ref{fig:QBHLTime}f.

At $t=t_{\rm{BHL}}=100$, the white hole is switched on and a BHL configuration is reached. At early times after the BHL onset, Fig. \ref{fig:QBHLTime}c, the presence of an inner horizon is revealed by fringe patterns in the supersonic-upstream (blue circle) and supersonic-downstream (green circle) regions. They result from the correlation between the Hawking (blue) or Andreev (green) modes and their partner $p2+$ modes, now reflected at the inner horizon as $p1-,p2-$ modes with large wavevector close to $k_{\rm{BCL}}$. On the other hand, the white band of self-correlations along the main diagonal vanishes away as the downstream region becomes subsonic again. In turn, the correlation between the reflected $p1-,p2-$ modes gives rise to a checkerboard pattern inside the lasing cavity (magenta circle), which is the white-hole analogue of the Andreev correlations since here the outgoing supersonic modes have large wavevector for low frequencies. 

Physically, we can understand all these features as white-hole radiation \cite{Mayoral2011} stimulated by the scattering of the Hawking radiation emitted from the black hole. We denote this phenomenon as Hawking-stimulated white-hole radiation or, more compactly, HSWH radiation. We compare HSWH radiation with the spontaneous white-hole radiation that would arise in the absence of a black hole in the leftmost column of Fig. \ref{fig:BHLBCLExplan}. Specifically, Fig. \ref{fig:BHLBCLExplan}a is a snapshot of the density-density correlation function at $t=140$, while Fig. \ref{fig:BHLBCLExplan}d shows the same result without switching on the black hole at $t=0$, so only the white hole placed at $x=L$ is present. We observe that a checkerboard pattern also emerges from the white-hole horizon because of the scattering of vacuum fluctuations. The checkerboard amplitude is expected to grow logarithmically in time \cite{Mayoral2011}. However, spontaneous white-hole radiation lacks the scions of the Hawking and Andreev correlations (blue and green circles in Fig. \ref{fig:QBHLTime}c), which arise between the white-hole horizon and distant regions along the Hawking and Andreev bands almost immediately after the white-hole onset, and thus can be regarded as the most distinctive signature of HSWH radiation. We note that the fringes parallel to the main diagonal in Fig. \ref{fig:BHLBCLExplan}d are a transient feature, arising due to the dynamical Casimir effect induced by the sudden change of the coupling constant in the upstream region \cite{Mayoral2011}.

At this stage, the continuous spectrum of spontaneous Hawking radiation contributes to the white-hole stimulation. As time goes by, the discrete nature of the unstable lasing spectrum enters in place. In particular, the dominant mode begins to overshadow the remaining fluctuations and to drive the dynamics, Figs. \ref{fig:QBHLTime}d,h,i,m. This can be seen by the emergence of a strong checkerboard pattern in the supersonic-supersonic correlations (magenta square of Fig. \ref{fig:QBHLTime}d), and of an incipient ripple in the supersonic density (magenta line of Fig. \ref{fig:QBHLTime}h; notice the change of scale). 

Eventually, the dominant mode reaches a large amplitude where nonlinear effects are crucial, Figs. \ref{fig:QBHLTime}j-l, n-p. This is singularly illustrated by Fig. \ref{fig:QBHLTime}k, where we have enlarged the spatial span of the plot. We observe the emergence of a highly enhanced moustache (blue circle) along with series of parallel stripes both upstream and downstream (green circles), resulting from the monochromatic character of the instability. These features are strong enough to extend beyond the supersonic region and give rise to upstream-downstream correlations (red circles). In this way, the localized dark band (red circle) in Fig. \ref{fig:QBHLTime}j which is parallel to the moustache is just a transient feature before reaching the monochromatic regime, resembling the relation between the multimode and the monochromatic stimulated periods in the experiment \cite{Kolobov2021}. The checkerboard pattern also reaches a large amplitude while the density profile shows a peaked structure whose amplitude is saturated, corresponding to the $n=3$ nonlinear GP solution \cite{Michel2013,deNova2016}. 

Therefore, the BHL dynamics can be mainly understood in terms of the dynamics of a single mode. In order to further validate this interpretation, we compute the contribution to the density-density correlation function from the dominant mode by using the corresponding BdG wave function. The result is depicted in Fig. \ref{fig:BHLBCLExplan}e, where we compare it with the correlation function at an intermediate time $t=250$, Fig. \ref{fig:BHLBCLExplan}b, when the dominant mode stands well above the remaining lasing modes but the system is not fully yet in the saturation regime. A good agreement is found between both correlation patterns; most of the blurring in Fig. \ref{fig:BHLBCLExplan}b can be attributed to nonlinear effects (see discussion in the following paragraphs). 

In Fig. \ref{fig:CBHLTime}, we now switch on a weak delta barrier ($Z=0.01$) at $t=t_{\rm{BCL}}=50$, placed at the eventual position of the inner horizon so it stimulates a small BCL amplitude. The dynamics for times $t<t_{\rm{BCL}}$ remains the same. For early times after the BCL and BHL onsets, Figs. \ref{fig:CBHLTime}a-h, we see that the only noticeable difference is the emergence of a small ripple in the density profile arising from the background BCL wave. On the opposite side, precisely because of its deterministic classical nature, the BCL wave does not show up in the correlation function.

The presence of an initial classical amplitude modifies the dynamics for later times. The first effect is that the nonlinear regime is reached sooner, Figs. \ref{fig:CBHLTime}i,m. The second effect is that several features of the correlation function become distorted as compared to the purely quantum case, Figs. \ref{fig:CBHLTime}j-l, n-p.

More information can be inferred if we reduce the quantum strength by setting $\lambda=1000$, Fig. \ref{fig:CBHLTime1000}. The early dynamics, Figs. \ref{fig:CBHLTime1000}a-h, is almost identical. However, the nonlinear regime is dramatically altered, Figs. \ref{fig:CBHLTime1000}i-p. Both the density profile and the correlation function are less smoothed and more peaked, losing most of the quantum BHL features of Fig. \ref{fig:QBHLTime}. In addition, the density profile now clearly displays soliton emission (solid red and green lines), which was unobserved in the previous cases. These solitons also show up in the correlation function as sharp features since they carry a large localized density depletion. We can understand the disappearance of solitons for strong quantum fluctuations as follows. In each individual trajectory of the Truncated Wigner ensemble, solitons are always emitted \cite{deNova2016}. In a quantum BHL, Fig. \ref{fig:QBHLTime}, the dynamics is solely driven by the amplification of quantum fluctuations, mainly by that of the amplitude of the dominant mode. The quantum nature of this amplitude leads to strong variations between different trajectories, and in particular between the positions of the emitted solitons. Since both $G^{(1)},G^{(2)}$ are obtained from ensemble averages, solitons are washed out by the averaging process. In contrast, in the classical BHL of Fig. \ref{fig:CBHLTime1000}, the dominant mode has an initial well-defined amplitude, from which the instability develops. Thus, all the trajectories of the ensemble amount to small quantum fluctuations around the deterministic classical trajectory given by Eq. (\ref{eq:GPEquation}). When averaging over the ensemble, the mean-field trajectory is then recovered, which neatly displays solitons. Figure \ref{fig:CBHLTime} represents a limiting case where both quantum and classical BHL are competing. The mean-field dynamics is driven deterministically by the BHL amplification of the classical BCL seed. Nonetheless, the lasing cavity also amplifies quantum fluctuations, which become sufficiently strong in the saturation regime to blur the sharp soliton features when computing ensemble averages. These arguments also explain the peaked checkerboard pattern within the lasing cavity in the saturation regime of Fig. \ref{fig:CBHLTime1000}, which now can be understood as fluctuations around the highly nonlinear ripple in the background mean-field density, in contrast to the blurred checkerboard for stronger quantum fluctuations in Figs. \ref{fig:QBHLTime},\ref{fig:CBHLTime}.

In Fig. \ref{fig:BCLTime}, instead of diluting quantum fluctuations, we introduce a strong potential barrier which generates a large BCL amplitude. This is already observed soon after the onset of the barrier, Figs. \ref{fig:BCLTime}e,f. Moreover, the Cherenkov wave now does show up in the density-density correlation function as a checkerboard pattern, Figs. \ref{fig:BCLTime}a,b, contradicting the naive intuition that it should not appear there. The answer to this apparent paradox is that here the large BCL amplitude cannot be regarded as a linear BdG mode on top of a uniform condensate, but instead it strongly backreacts onto the mean-field background around which quantum fluctuations evolve. The large modulation of the background density is translated into a checkerboard pattern in the density-density correlation function, similarly to the saturation regime of Fig. \ref{fig:CBHLTime1000}. The white-hole onset barely alters the dynamics, except for the emission of trains of solitons into the downstream region.

We further confirm that the checkerboard structure arises due to the strong BCL modulation of the mean-field background in the rightmost column of Fig. \ref{fig:BHLBCLExplan}, where we compare our results with a simulation that only includes particle-number fluctuations \cite{Wang2017}. A good agreement is observed between the correlation patterns in the cavity region, revealing the same underlying mechanism since the spatial structure in the case of particle-number fluctuations can only emerge due to the background envelope.

Finally, in Fig. \ref{fig:BCLPuroTime}, we analyze the same case of Fig. \ref{fig:BCLTime} but without white hole, $t_{\rm{BHL}}=\infty$, so the dynamics is purely driven by BCL stimulation. We observe that the evolution is essentially the same, especially in the cavity and the upstream region. The only significant difference arises in the downstream region, which now is supersonic and does not support trains of solitons. Thus, we conclude that the main correlation patterns of Fig. \ref{fig:BCLTime} can be unambiguously attributed to BCL stimulation. 

Interestingly, due to the supersonic character of the downstream region, at late times we can observe monochromatic features arising from quantum BCL-stimulated Hawking radiation \cite{Kolobov2021} (blue circles in Fig. \ref{fig:BCLPuroTime}l), where quantum fluctuations around the nonlinear mean-field BCL background stimulate Hawking radiation. Actually, since the BCL modulation acts as a resonant cavity, we can also understand this phenomenon as spontaneous resonant Hawking radiation \cite{Zapata2011}. 

\begin{figure*}
\begin{tabular}{@{}ccccc@{}}
      \stackinset{l}{0pt}{t}{0pt}{\footnotesize{(a)}}{\includegraphics[width=0.2\textwidth]{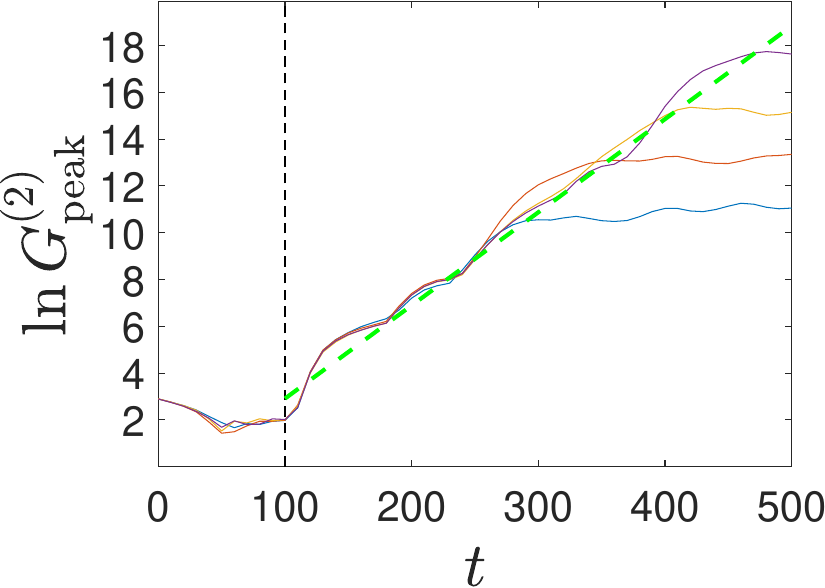}} 
    & \stackinset{l}{0pt}{t}{0pt}{\footnotesize{(b)}}{\includegraphics[width=0.2\textwidth]{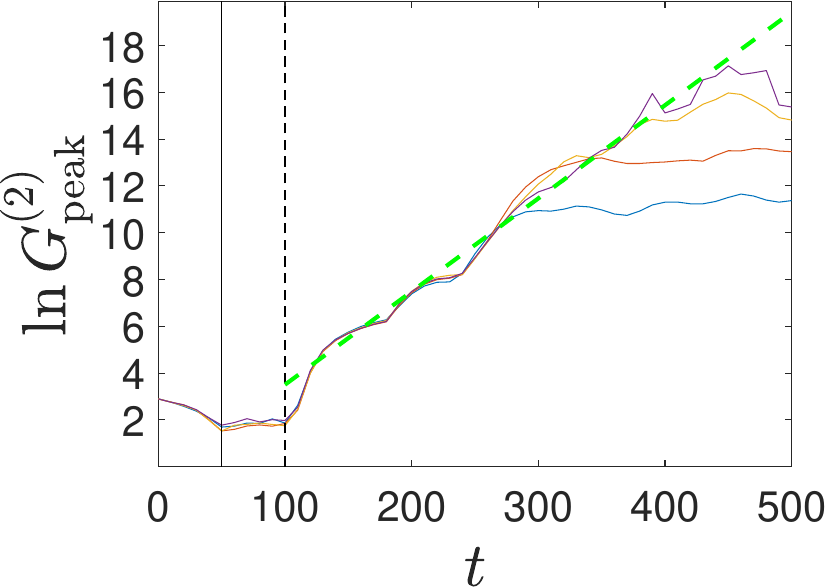}} &
    \stackinset{l}{0pt}{t}{0pt}{\footnotesize{(c)}}{\includegraphics[width=0.2\textwidth]{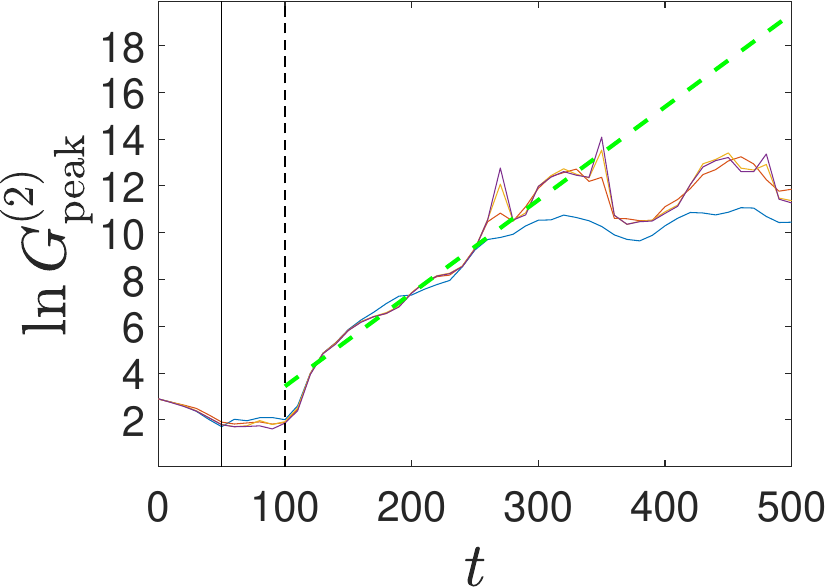}} &
    \stackinset{l}{0pt}{t}{0pt}{\footnotesize{(d)}}{\includegraphics[width=0.2\textwidth]{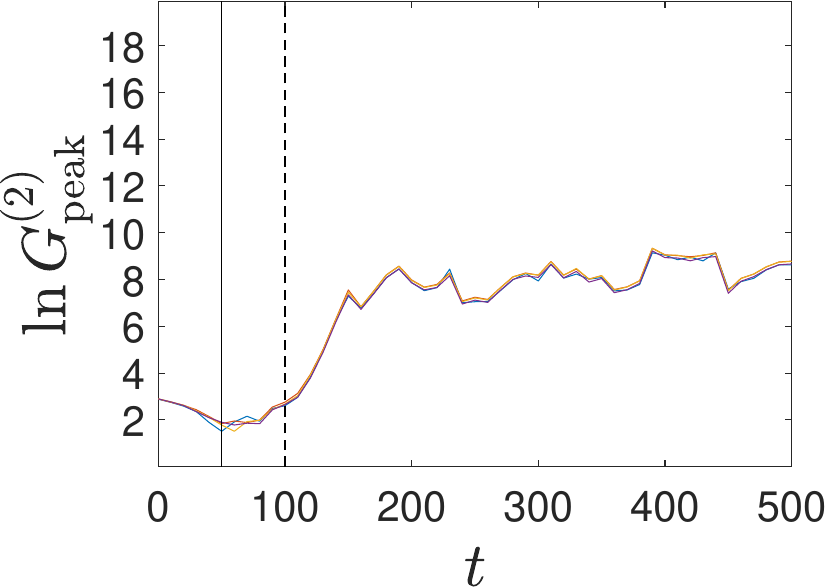}}  &
    \stackinset{l}{0pt}{t}{0pt}{\footnotesize{(e)}}{\includegraphics[width=0.2\textwidth]{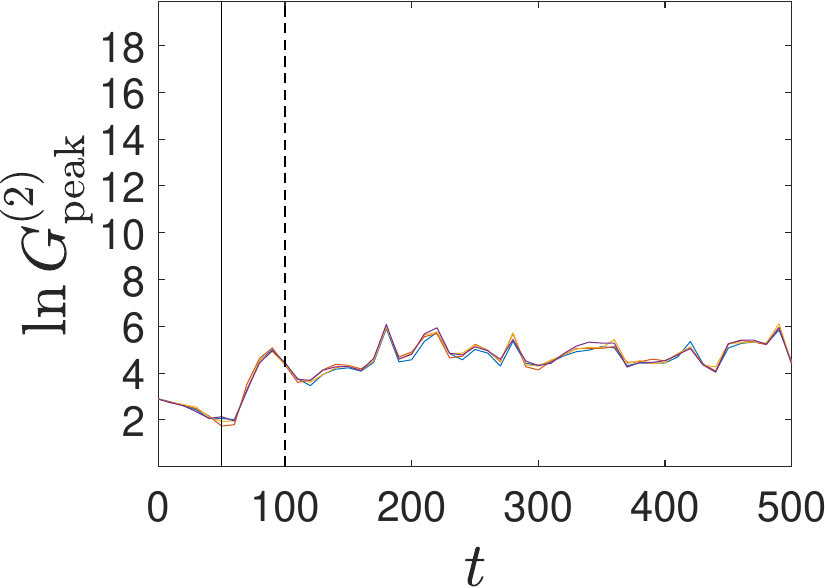}}\\
    \stackinset{l}{0pt}{t}{0pt}{\footnotesize{(f)}}{\includegraphics[width=0.2\textwidth]{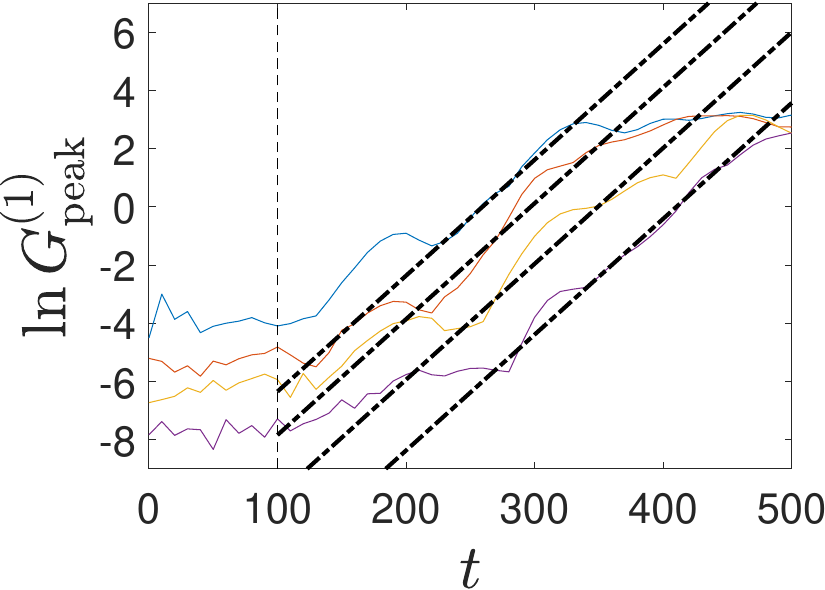}} 
    & \stackinset{l}{0pt}{t}{0pt}{\footnotesize{(g)}}{\includegraphics[width=0.2\textwidth]{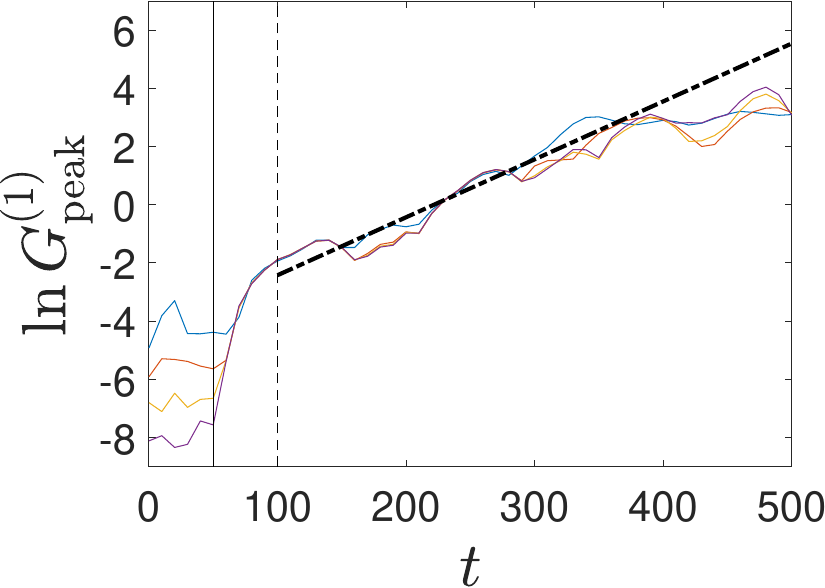}} &
    \stackinset{l}{0pt}{t}{0pt}{\footnotesize{(h)}}{\includegraphics[width=0.2\textwidth]{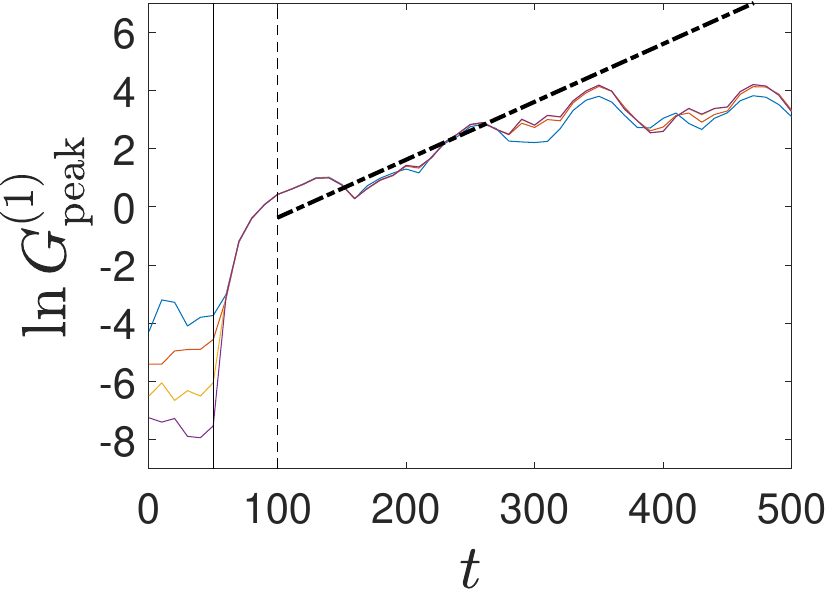}} &
    \stackinset{l}{0pt}{t}{0pt}{\footnotesize{(i)}}{\includegraphics[width=0.2\textwidth]{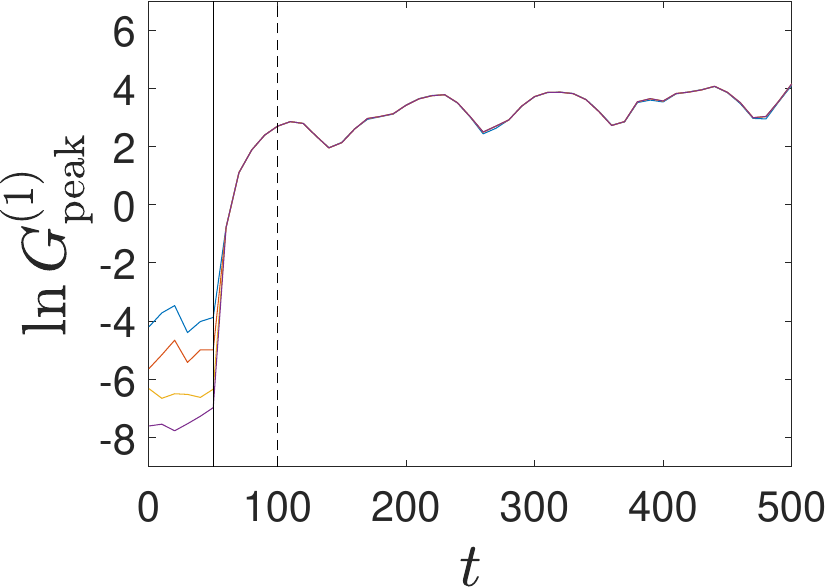}}  &
    \stackinset{l}{0pt}{t}{0pt}{\footnotesize{(j)}}{\includegraphics[width=0.2\textwidth]{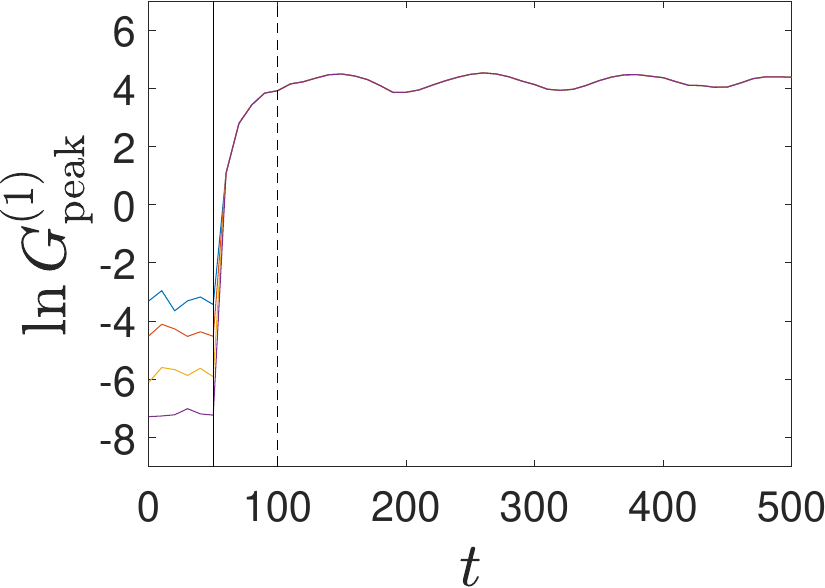}} 
\end{tabular}
\caption{Time evolution of $G^{(n)}_{\rm{peak}}$ for a BHL configuration with $v=0.6$, $c_2=0.2$, $L=20$. The vertical solid (dashed) line signals $t_{\rm{BCL}}$ ($t_{\rm{BHL}}$). Columns, from left to right, correspond to $Z=0,0.001,0.01,0.1,0.75$, respectively. Blue, red, orange, purple solid lines represent $\lambda=1,10,100,1000$. (a)-(e) $G^{(2)}_{\rm{peak}}(t)$. Dashed green lines in (a)-(c) represent the theoretical prediction $\ln G^{(2)}_{\rm{peak}}\sim 2\Gamma t$, Eqs. (\ref{eq:Z2prediction}), (\ref{eq:Z2Breaking}).  (f)-(j) $G^{(1)}_{\rm{peak}}(t)$. Dash-dotted black lines represent the theoretical prediction $\ln G^{(1)}_{\rm{peak}}\sim 2\Gamma t$ in (f), Eq. (\ref{eq:Z2prediction}), and $\ln G^{(1)}_{\rm{peak}}\sim \Gamma t$ in (g)-(h), Eq. (\ref{eq:Z2Breaking}).}
\label{fig:SaturationQF}
\end{figure*}

\subsection{Quantum versus classical BHL}\label{subsec:QuantumClassical}

Once we have identified the main correlation patterns, we proceed to perform a quantitative analysis. Specifically, we evaluate the Fourier transform of the correlation functions inside the lasing cavity (magenta lines in Figs. \ref{fig:QBHLTime}d,h). As figures of merit, we choose
\begin{eqnarray}
     \nonumber G^{(2)}_{\rm{peak}}(t)&\equiv& \max_{k,k'} |G^{(2)}(k,k',t)|,\\
     G^{(1)}_{\rm{peak}}(t)&\equiv& \max_{k} |G^{(1)}(k,t)-G^{(1)}(0,t)|,
\end{eqnarray}
where in the last line we are subtracting the background homogeneous contribution from the condensate. These observables are expected to capture the main dynamics of the BHL and BCL regimes because both involve modes with well-defined wavevector within the lasing cavity. In particular, $G^{(1)}_{\rm{peak}}$ represents the amplitude of the density ripple and $G^{(2)}_{\rm{peak}}$ that of the checkerboard pattern, which are characteristic features of both regimes from their very beginning.

We display the time evolution of $G^{(2)}_{\rm{peak}}$ (upper row) and $G^{(1)}_{\rm{peak}}$ (lower row) in Fig. \ref{fig:SaturationQF}, where the columns correspond to $Z=0,0.001,0.01,0.1,0.75$, respectively. Inside each panel, blue, red, orange, purple curves represent $\lambda=1,10,100,1000$. Vertical solid (dashed) lines signal the onset of the BCL (BHL) mechanisms.

We first focus on the case of a purely quantum BHL, Figs. \ref{fig:SaturationQF}a,f, where we note that the blue curve corresponds to the simulation of Fig. \ref{fig:QBHLTime}. We observe that, before the BHL onset, the amplitude of the density-density correlations is essentially constant (note the logarithmic scale) and independent of $\lambda$. This last property arises precisely because of the use of the normalized correlation function of Eq. (\ref{eq:NormalizedCorrelations}). On the other hand, the perturbations around the homogeneous background in the average density do depend on $\lambda$ since $\braket{\delta\hat{n}^{(1)}(x,t)}=0$ and hence
\begin{equation}\label{eq:Density}
    \braket{\hat{n}(x,t)}-n_0=\braket{\delta\hat{n}^{(2)}(x,t)}=\braket{\delta\hat{\Psi}^\dagger(x,t)\delta\hat{\Psi}(x,t)}\sim \lambda^{-1},
\end{equation}
so $G^{(1)}_{\textrm{peak}}\sim \lambda^{-1}$.

After the BHL onset, both the average density and the density-density correlations exponentially increase. Specifically, since the dominant mode drives the dynamics, in this regime we expect the quantum fluctuations to be \textit{grosso modo} described by

\begin{equation}\label{eq:QuantumDominant}
    \delta\hat{\Psi}(x,t)\simeq \frac{e^{\Gamma t}}{\sqrt{\lambda n_0\xi_0}}\left[u_I(x)e^{-i\gamma t}\hat{\alpha}_I+v^*_I(x)e^{i\gamma t}\hat{\alpha}^\dagger_I\right],
\end{equation}
with the time $t$ measured here starting from the BHL onset, $u_I,v_I$ the BdG components of the dominant mode, $\gamma$ the real part of its frequency, and $\hat{\alpha}_I$ its amplitude. We recall that, since dynamically unstable modes have zero norm according to the BdG inner product, their amplitudes do not behave as annihilation operators but they instead commute with their conjugate, $[\hat{\alpha}_I,\hat{\alpha}^\dagger_I]=0$ \cite{Finazzi2010}. In a quantum BHL, the phase of the amplitude of the dominant mode is expected to be random and hence we can neglect $\braket{\hat{\alpha}_I\hat{\alpha}_I}\simeq 0$. This assumption yields that $G^{(1)}_{\textrm{peak}},G^{(2)}_{\textrm{peak}}$ behave as
\begin{eqnarray}\label{eq:QBHLGrowth}
 \nonumber G^{(1)}_{\textrm{peak}}(t)&\sim& \braket{\delta\hat{n}^{(2)}}\sim \frac{e^{2\Gamma t}}{\lambda},\\
    G^{(2)}_{\textrm{peak}}(t)&\sim& \lambda \braket{\delta\hat{n}^{(1)}\delta\hat{n}^{(1)}}\sim e^{2\Gamma t}.
\end{eqnarray}
Hence, for a quantum BHL, 
\begin{equation}\label{eq:Z2prediction}
    \ln G^{(n)}_{\textrm{peak}}(t)\sim 2\Gamma t, ~n=1,2,
\end{equation}
since $G^{(1)}_{\textrm{peak}}(t)$ also scales quadratically in the field fluctuations because the $\mathbb{Z}_2$ symmetry of a purely quantum BHL sets $\braket{\delta\hat{\Psi}(x,t)}=0$ \cite{Michel2015}. Using the pendulum picture, we can understand the $\mathbb{Z}_2$ symmetry as that of the quantum fluctuations around the initial unstable position. 

The prediction of Eq. (\ref{eq:Z2prediction}) is depicted in dashed green, dash-dotted black lines in Figs. \ref{fig:SaturationQF}a,f, respectively, where $\Gamma$ is obtained from Fig. \ref{fig:BHLBCLScheme}c. A good agreement with the numerical results is observed, extending the qualitative agreement between the spatial correlation patterns present in the central column of Fig. \ref{fig:BHLBCLExplan}. Nevertheless, this approximation necessarily fails at early times after the BHL onset, when the whole spectrum of Hawking radiation contributes to the checkerboard amplitude through HSWH radiation, and at late times, when nonlinear effects become crucial. Both limits give rise to deviations from the dashed green slope. In particular, the exponential growth ceases when the system reaches the saturation regime. There, both $G^{(1)}_{\textrm{peak}},G^{(2)}_{\textrm{peak}}$ saturate to the values $G^{(1)}_{\textrm{sat}},G^{(2)}_{\textrm{sat}}$ (latest times in Fig. \ref{fig:SaturationQF}). Using these values, we can define the saturation time $t_{\textrm{sat}}$ as the time when $G^{(2)}_{\textrm{peak}}$ reaches $G^{(2)}_{\textrm{sat}}$. Technically, we extract $G^{(1)}_{\textrm{sat}},G^{(2)}_{\textrm{sat}}$ from the simulations by averaging in time once in the saturation regime, and take $t_{\textrm{sat}}$ as the time needed for $G^{(2)}_{\textrm{peak}}(t)$ to reach the value $G^{(2)}_{\textrm{peak}}(t=0)+0.9[G^{(2)}_{\textrm{sat}}-G^{(2)}_{\textrm{peak}}(t=0)]$. We have checked that the main conclusions are quite insensitive to the specific details of these definitions.

To capture the underlying physics governing these processes, we use a crude model in which Eq. (\ref{eq:QuantumDominant}) governs the whole lasing period until the saturation regime, estimating in this way $G^{(2)}_{\textrm{sat}}$ and $t_{\textrm{sat}}$.

\begin{figure}[t]
\begin{tabular}{@{}cc@{}}
      \stackinset{l}{0pt}{t}{0pt}{\scriptsize{(a)}}{\includegraphics[width=0.5\columnwidth]{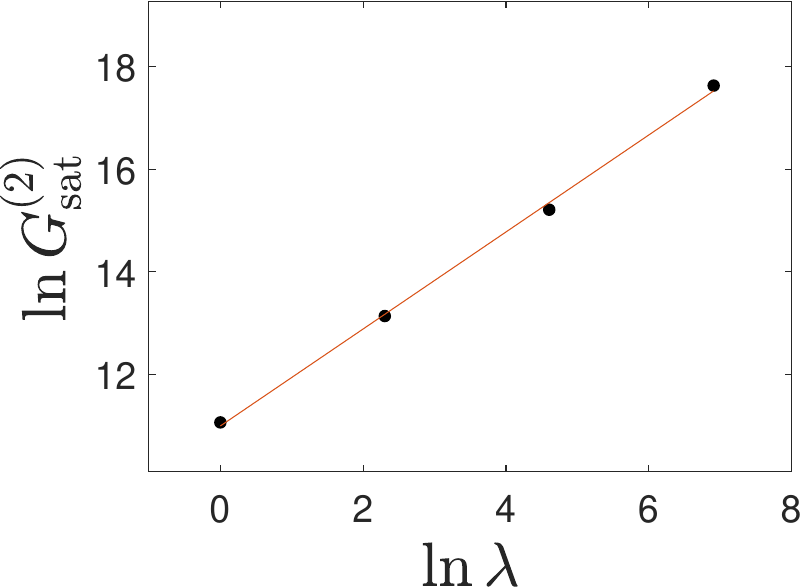}} 
    & \stackinset{l}{0pt}{t}{0pt}{\scriptsize{(b)}}{\includegraphics[width=0.5\columnwidth]{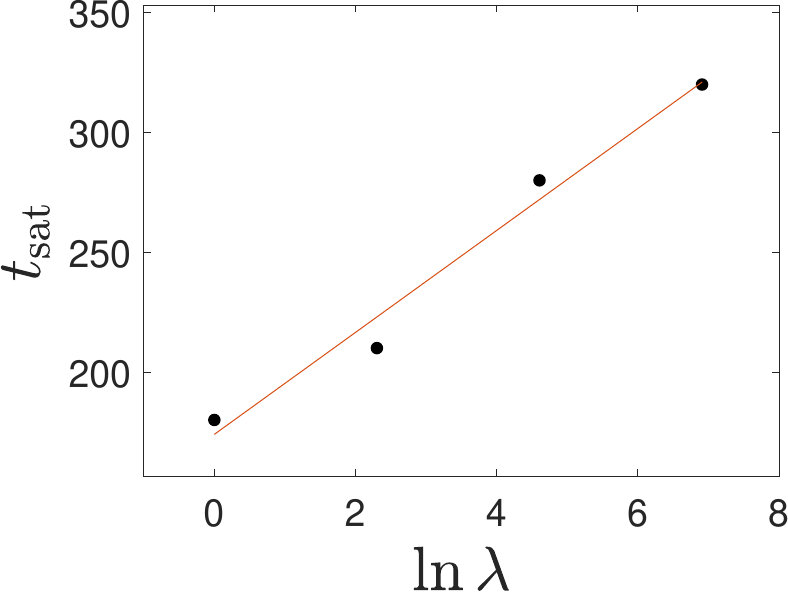}}\\ \stackinset{l}{0pt}{t}{0pt}{\scriptsize{(c)}}{\includegraphics[width=0.5\columnwidth]{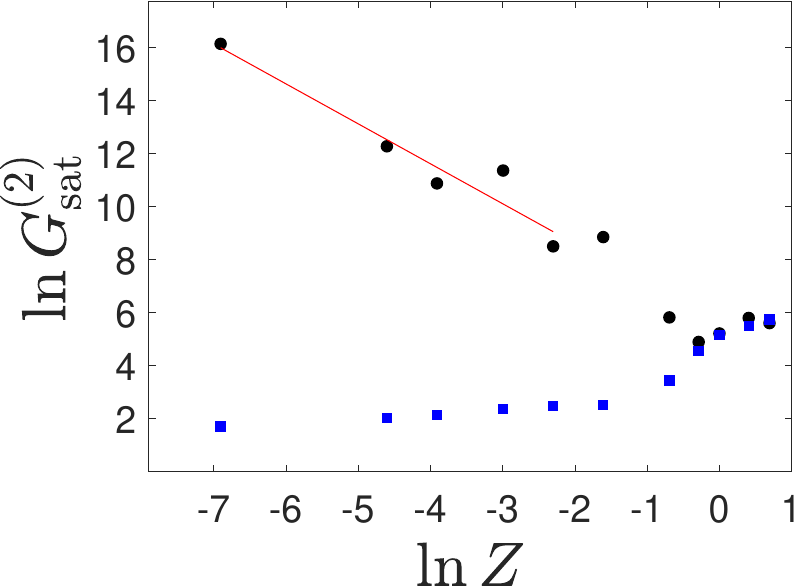}} 
    & \stackinset{l}{0pt}{t}{0pt}{\scriptsize{(d)}}{\includegraphics[width=0.5\columnwidth]{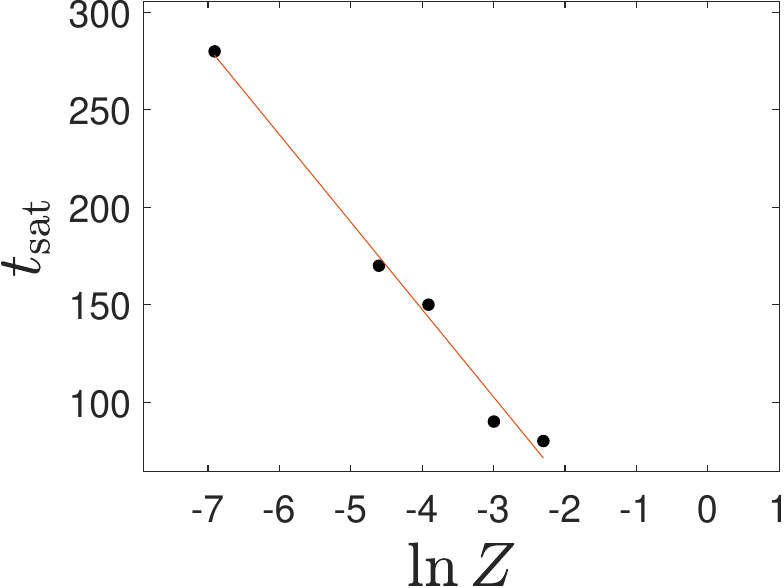}}
\end{tabular}
\caption{Saturation values for a BHL with $v=0.6$, $c_2=0.2$, $L=20$. Red lines represent linear fits to the black dots. (a) Saturation amplitude $G^{(2)}_{\textrm{sat}}$ as a function of the quantum strength $\lambda^{-1}$ for a quantum BHL with $Z=0$, Fig. \ref{fig:SaturationQF}a. (b) Saturation time $t_{\rm{sat}}$ for the simulations in (a). (c) $G^{(2)}_{\textrm{sat}}$ as a function of the barrier strength $Z$ for fixed $\lambda=1000$. Black dots correspond to $t_{\rm{BHL}}=100$ and blue squares to pure BCL stimulation, $t_{\rm{BHL}}=\infty$, with $t_{\rm{BCL}}=50$. The fit is restricted to small values of $Z$. (d) $t_{\rm{sat}}$ for the points fitted in (c).}
\label{fig:SaturationAmplitudesBHL}
\end{figure}

For a quantum BHL, Fig. \ref{fig:SaturationQF} shows that $G^{(1)}_{\textrm{sat}}$ does not depend on $\lambda$ but $G^{(2)}_{\textrm{sat}}$ does, which is precisely the opposite behavior to that of $G^{(1)}_{\textrm{peak}},G^{(2)}_{\textrm{peak}}$ before saturation. This is because  the saturation regime is reached when quantum fluctuations have grown enough to enter the nonlinear regime,
\begin{equation}\label{eq:SaturatedAmplitude}
    \delta\hat{n}\sim 1.
\end{equation}
Our crude model then predicts
\begin{eqnarray}
 \nonumber G^{(1)}_{\textrm{sat}}&\sim& 1\sim \frac{e^{2\Gamma t_{\textrm{sat}}}}{\lambda},\\
    G^{(2)}_{\textrm{sat}}&\sim& \lambda \sim e^{2\Gamma t_{\textrm{sat}}},
\end{eqnarray}
so
\begin{eqnarray}
    \ln G^{(2)}_{\textrm{sat}}\sim \ln \lambda,~~t_{\textrm{sat}}\sim \frac{\ln \lambda}{2\Gamma}.
\end{eqnarray}
We represent both $G^{(2)}_{\textrm{sat}},t_{\textrm{sat}}$ as a function of $\lambda$ for a purely quantum BHL in Figs. \ref{fig:SaturationAmplitudesBHL}a,b, finding good agreement with the expected scaling. 

In the regime of classical BHL, the $\mathbb{Z}_2$ symmetry is broken by the BCL stimulation. The Cherenkov wave is a coherent undulation above the homogeneous background, accounted by linear perturbations of the GP wave function $\braket{\delta\hat{\Psi}(x,t)}=\delta\Psi(x,t)\sim A_{\rm{BCL}}\neq 0$. This is translated into a classical version of Eq. (\ref{eq:QuantumDominant}) that yields
\begin{equation}\label{eq:Z2BreakingLinear}
   G^{(1)}_{\textrm{peak}}(t)\sim \braket{\delta\hat{n}^{(1)}}\sim A_{\rm{BCL}}e^{\Gamma t}\cos(\gamma t+\delta),
\end{equation}
with $\delta$ some phase. In the pendulum picture, the amplitude $A_{\rm{BCL}}$ is the equivalent of the initial angle $\theta$ which breaks the $\mathbb{Z}_2$ symmetry of the unstable equilibrium position. As a result, a classical BHL can be understood as a well-defined classical trajectory plus small quantum fluctuations around it. Consequently, the average density does not depend here on $\lambda$.

Precisely because of its classical deterministic character, at the linear level the BCL amplitude does not show up in the density-density correlation function and $G^{(2)}_{\textrm{peak}}(t)$ still follows Eq. (\ref{eq:QBHLGrowth}). Thus, the $\mathbb{Z}_2$ symmetry-breaking implies now
\begin{equation}\label{eq:Z2Breaking}
    \ln G^{(1)}_{\textrm{peak}}(t)\sim \Gamma t,~~\ln G^{(2)}_{\textrm{peak}}(t)\sim 2\Gamma t.
\end{equation}
These predictions are depicted in dashed green, dash-dotted black lines in Figs. \ref{fig:SaturationQF}b-c, g-h, respectively, where good agreement with the numerical results is again observed. The small oscillations observed in $G^{(1)}_{\textrm{peak}}(t)$ result from the sinusoidal term in Eq. (\ref{eq:Z2BreakingLinear}), arising from the oscillatory component of the frequency of the dominant mode.

The saturation regime is reached now when the mean-field amplitude, given here by $G^{(1)}_{\textrm{peak}}$, grows up to that of a nonlinear stationary GP solution. In our crude model, the saturation of a classical BHL is then determined by the condition
\begin{equation}\label{eq:SaturationTimeCBHL}
    G^{(1)}_{\textrm{sat}}\sim 1 \sim A_{\rm{BCL}}e^{\Gamma t_{\textrm{sat}}}.
\end{equation}
Therefore, the saturation time is predicted to behave as
\begin{equation}
    t_{\textrm{sat}}\sim - \frac{\ln A_{\rm{BCL}}}{\Gamma}\sim - \frac{\ln Z}{\Gamma},
\end{equation} 
since we are in the regime of small $A_{\rm{BCL}}$ where it is linear in the barrier amplitude $Z$, $A_{\rm{BCL}}\sim Z$. Accordingly, the saturation amplitude for the quantum fluctuations will be given by $G^{(2)}_{\textrm{sat}}\sim e^{2\Gamma t_{\textrm{sat}}}$ so
\begin{equation}\label{eq:SaturationAmplitudeCBHL}
    \ln G^{(2)}_{\textrm{sat}}\sim 2\Gamma t_{\textrm{sat}}\sim -2\ln Z.
\end{equation}
We represent both magnitudes as a function of the delta strength $Z$ in Figs. \ref{fig:SaturationAmplitudesBHL}c,d, finding good agreement with the predicted scalings in the regime of validity of small $Z$. The decrease of both magnitudes with $A_{\rm{BCL}}$ (controlled by the barrier strength $Z$) is quite intuitive: the closer the system is to the saturation regime, the less time is needed to reach it. Since the amplification of quantum fluctuations mainly occurs during the lasing time, a shorter saturation time is translated into a smaller amplification factor $e^{2\Gamma t_{\textrm{sat}}}$.

\begin{figure}[t]
\begin{tabular}{@{}cc@{}}
      \stackinset{l}{0pt}{t}{0pt}{\scriptsize{(a)}}{\includegraphics[width=0.5\columnwidth]{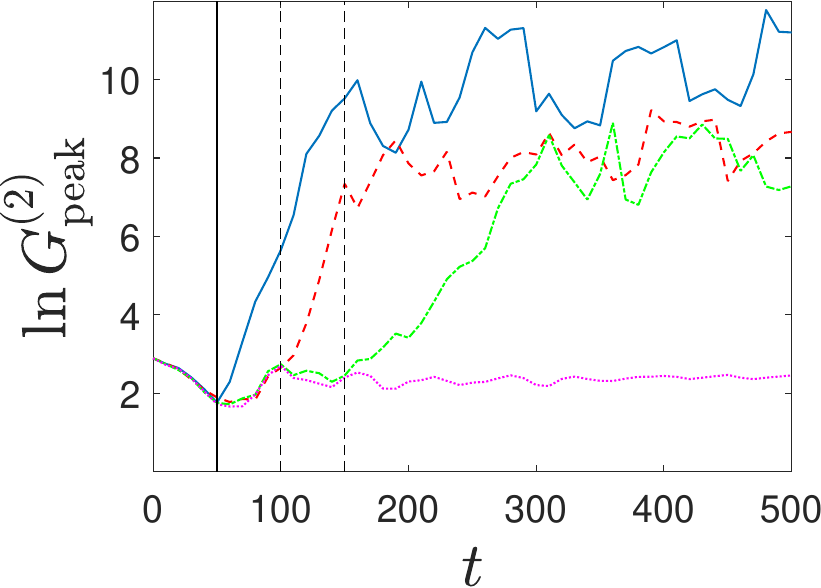}} 
    & \stackinset{l}{0pt}{t}{0pt}{\scriptsize{(b)}}{\includegraphics[width=0.5\columnwidth]{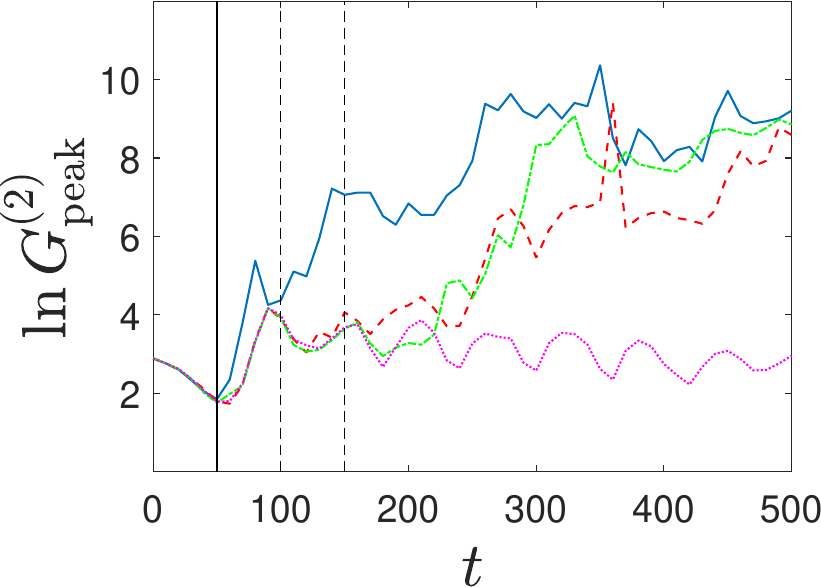}}\\
    \stackinset{l}{0pt}{t}{0pt}{\scriptsize{(c)}}{\includegraphics[width=0.5\columnwidth]{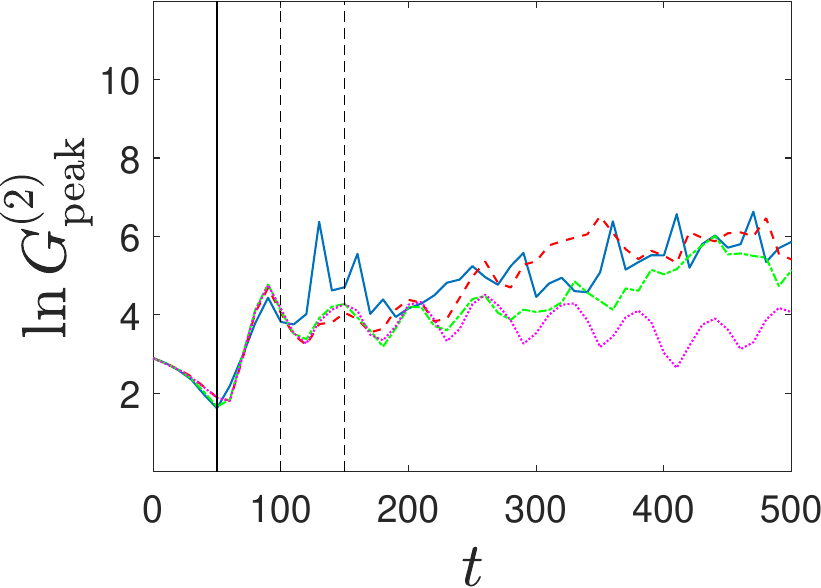}} 
    & \stackinset{l}{0pt}{t}{0pt}{\scriptsize{(d)}}{\includegraphics[width=0.5\columnwidth]{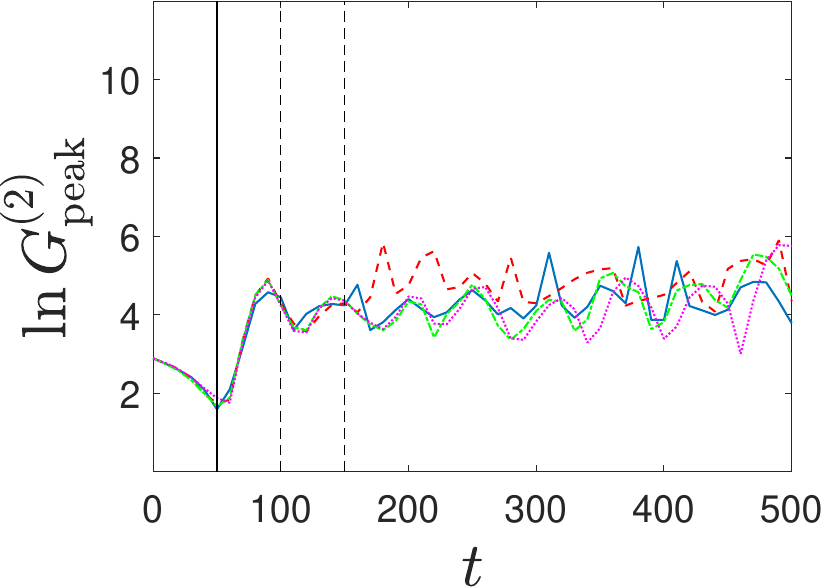}}
\end{tabular}
\caption{Time evolution of $G^{(2)}_{\rm{peak}}$ for different BHL onsets $t_{\rm{BHL}}=50$ (solid blue), $t_{\rm{BHL}}=100$ (dashed red), $t_{\rm{BHL}}=150$ (dash-dotted green), and $t_{\rm{BHL}}=\infty$ (dotted magenta). The BHL parameters are $v=0.6$, $c_2=0.2$, $L=20$, and the quantum strength is set to $\lambda=1000$. Vertical solid (dashed) lines signal the BCL (BHL) onsets, where the former is fixed to $t_{\rm{BCL}}=50$. (a)-(d) Delta strength values $Z=0.1,0.3,0.5,0.75$, respectively.}
\label{fig:SaturationBHLBCLPlots}
\end{figure}

\begin{figure*}
\begin{tabular}{@{}ccccc@{}}
      \stackinset{l}{0pt}{t}{0pt}{\footnotesize{(a)}}{\includegraphics[width=0.2\textwidth]{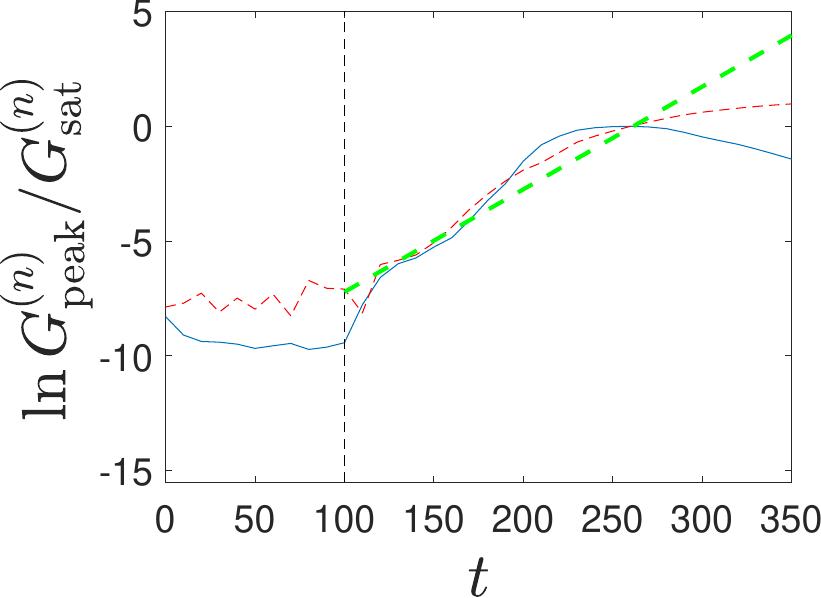}} 
    & \stackinset{l}{0pt}{t}{0pt}{\footnotesize{(b)}}{\includegraphics[width=0.2\textwidth]{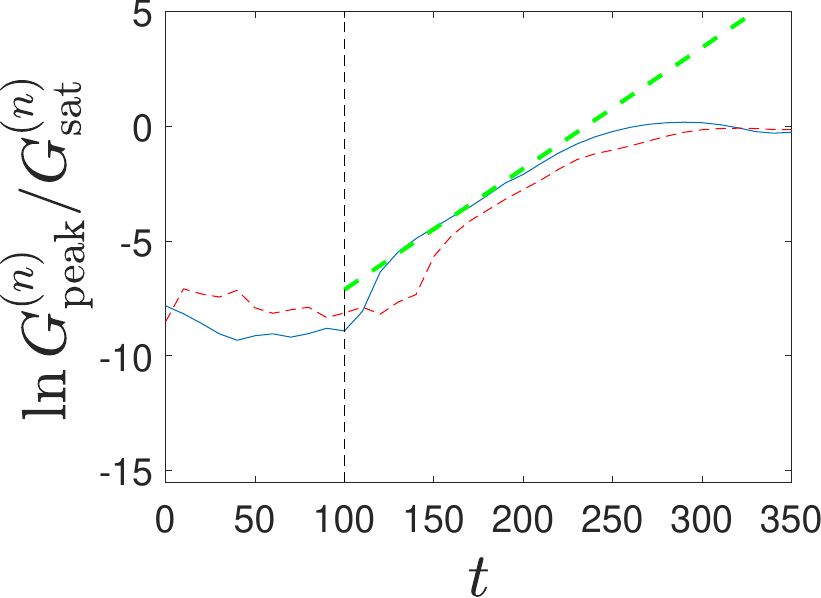}} &
    \stackinset{l}{0pt}{t}{0pt}{\footnotesize{(c)}}{\includegraphics[width=0.2\textwidth]{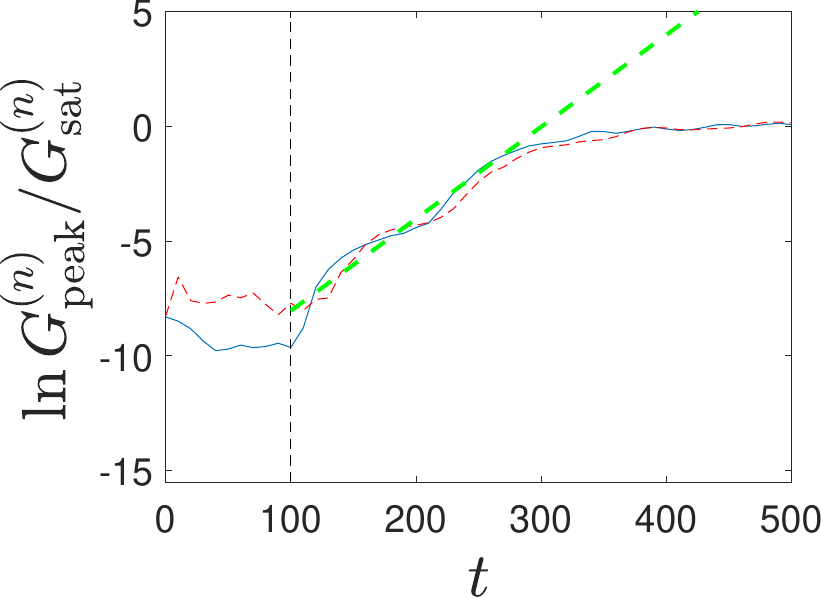}} &
    \stackinset{l}{0pt}{t}{0pt}{\footnotesize{(d)}}{\includegraphics[width=0.2\textwidth]{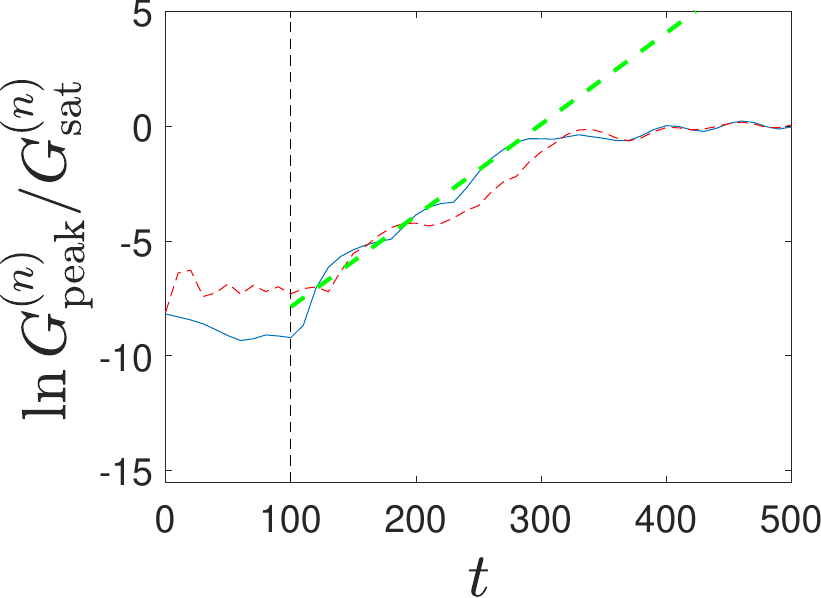}}  &
    \stackinset{l}{0pt}{t}{0pt}{\footnotesize{(e)}}{\includegraphics[width=0.2\textwidth]{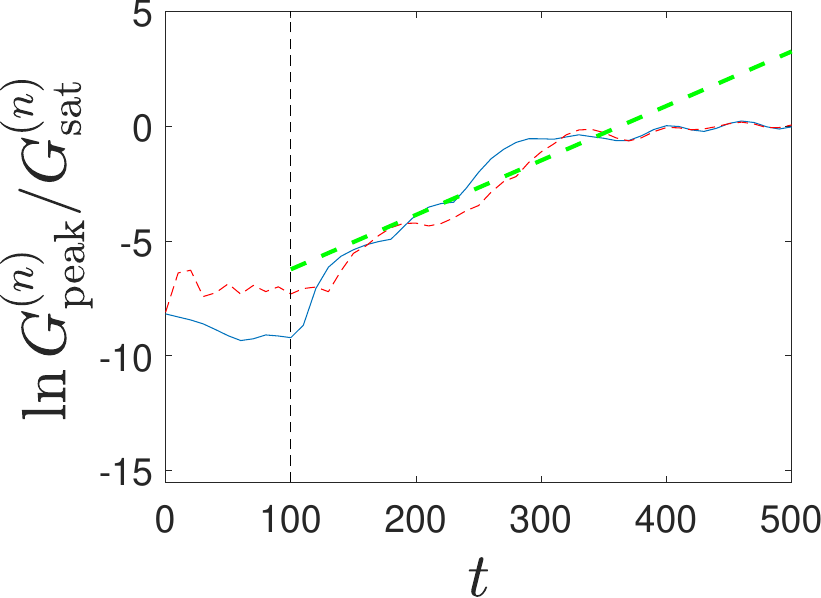}}\\
    \stackinset{l}{0pt}{t}{0pt}{\footnotesize{(f)}}{\includegraphics[width=0.2\textwidth]{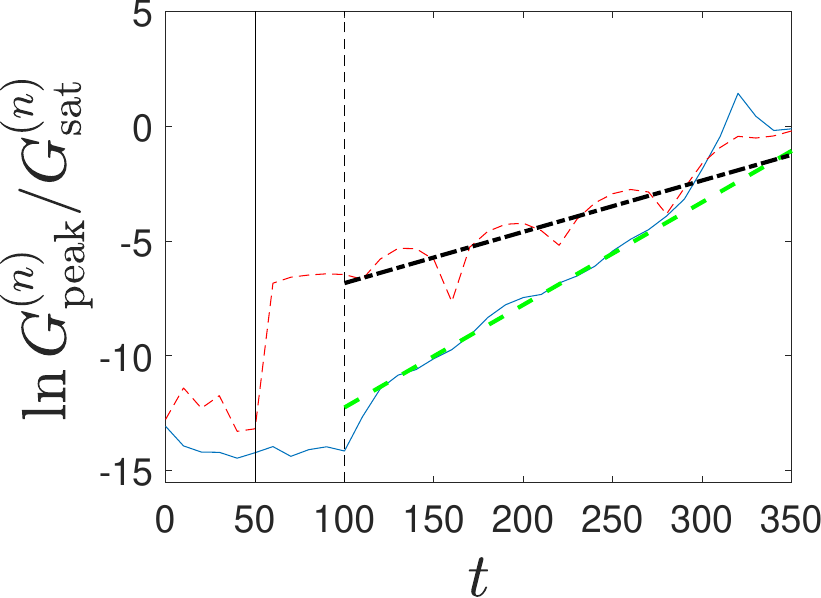}} 
    & \stackinset{l}{0pt}{t}{0pt}{\footnotesize{(g)}}{\includegraphics[width=0.2\textwidth]{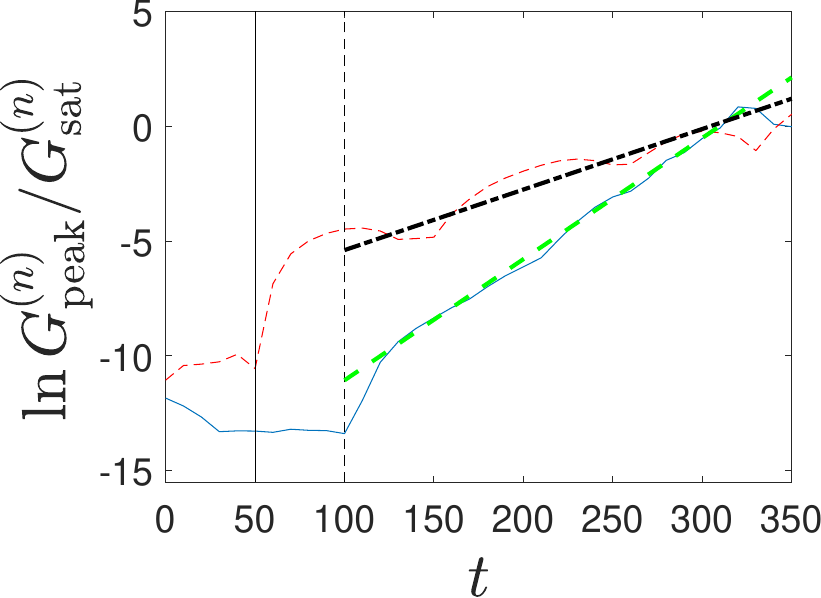}} &
    \stackinset{l}{0pt}{t}{0pt}{\footnotesize{(h)}}{\includegraphics[width=0.2\textwidth]{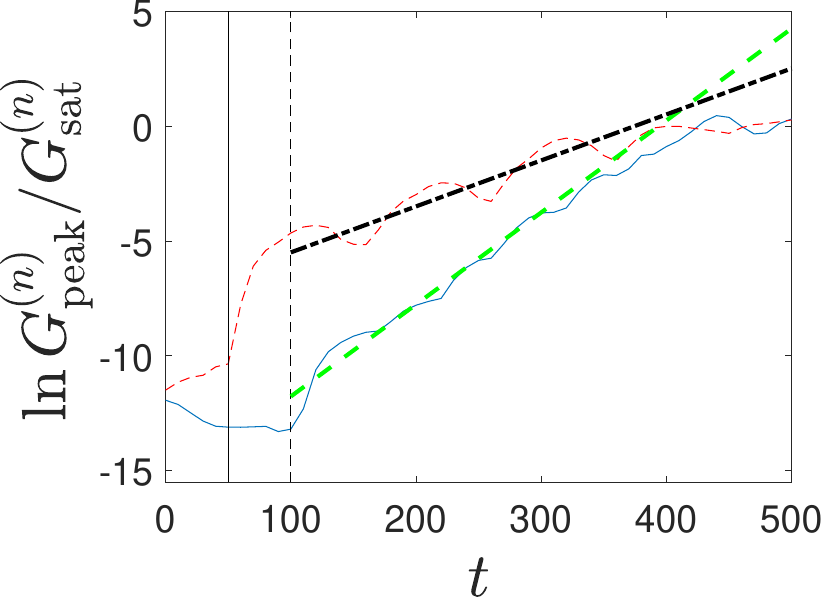}} &
    \stackinset{l}{0pt}{t}{0pt}{\footnotesize{(i)}}{\includegraphics[width=0.2\textwidth]{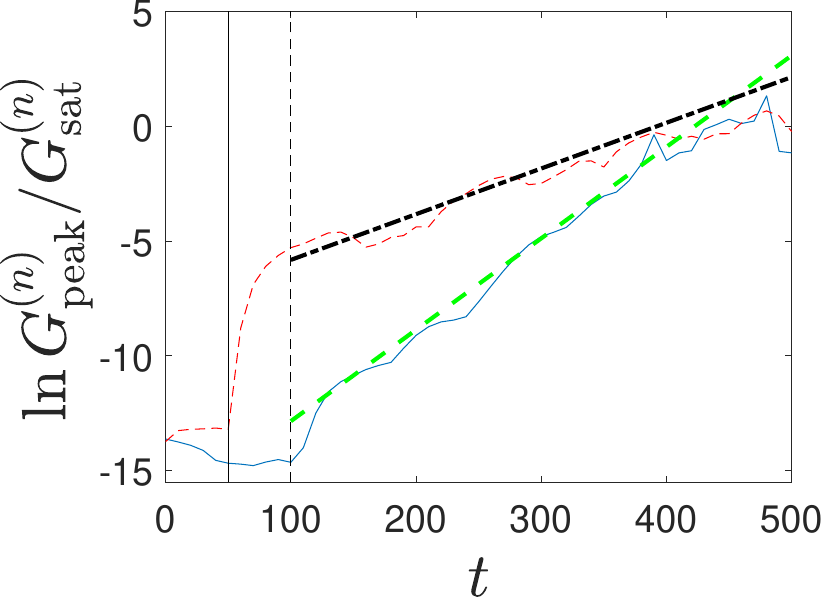}}  &
    \stackinset{l}{0pt}{t}{0pt}{\footnotesize{(j)}}{\includegraphics[width=0.2\textwidth]{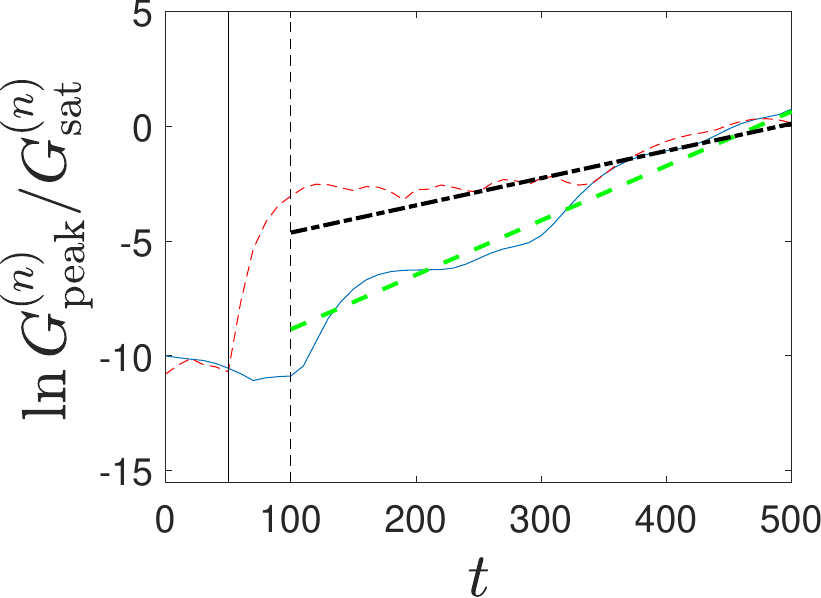}} \\
    \stackinset{l}{0pt}{t}{0pt}{\footnotesize{(k)}}{\includegraphics[width=0.2\textwidth]{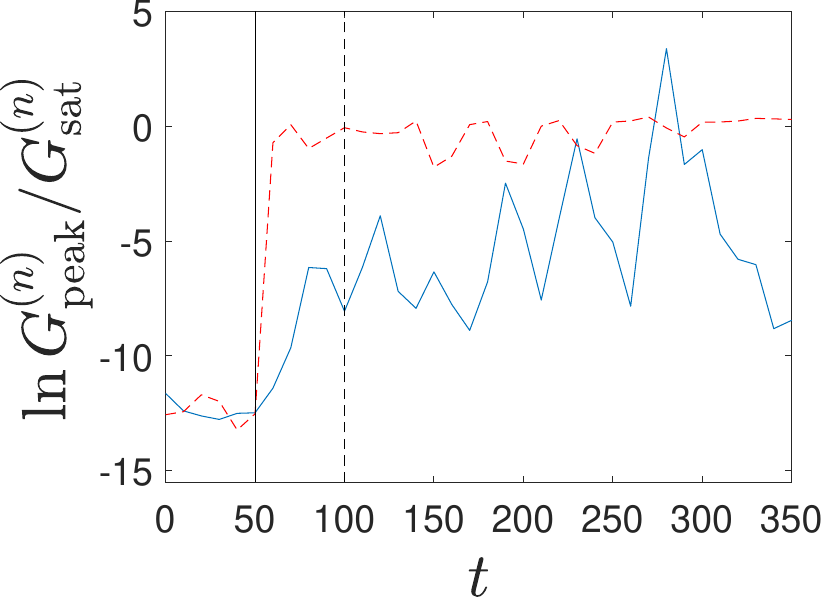}} 
    & \stackinset{l}{0pt}{t}{0pt}{\footnotesize{(l)}}{\includegraphics[width=0.2\textwidth]{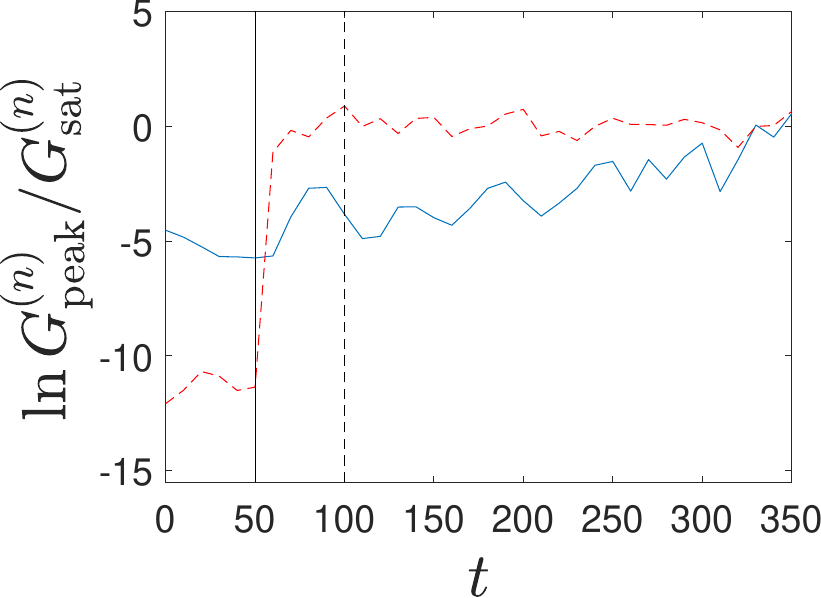}} &
    \stackinset{l}{0pt}{t}{0pt}{\footnotesize{(m)}}{\includegraphics[width=0.2\textwidth]{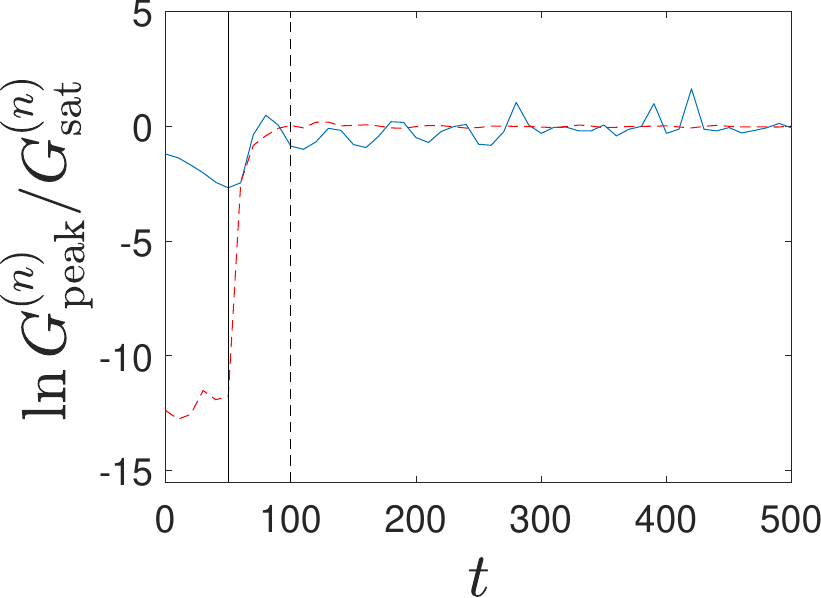}} &
    \stackinset{l}{0pt}{t}{0pt}{\footnotesize{(n)}}{\includegraphics[width=0.2\textwidth]{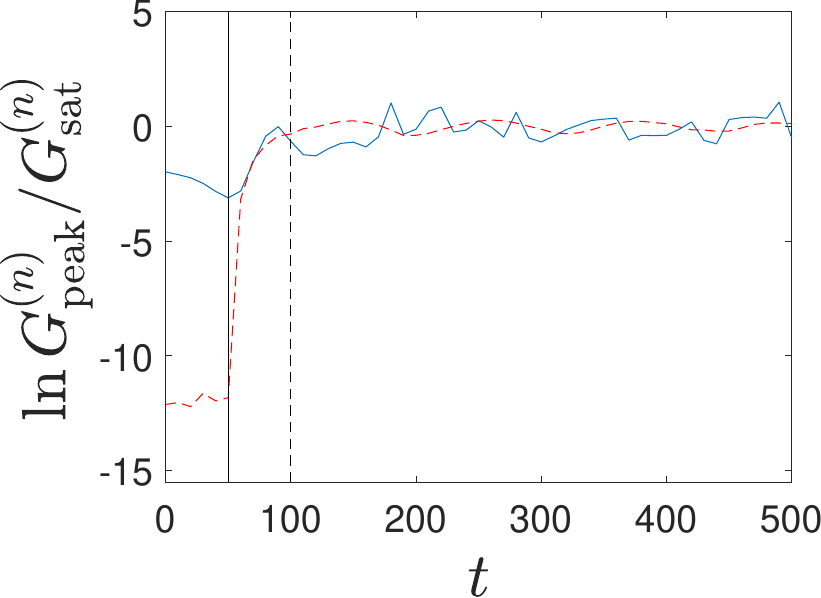}}  &
    \stackinset{l}{0pt}{t}{0pt}{\footnotesize{(o)}}{\includegraphics[width=0.2\textwidth]{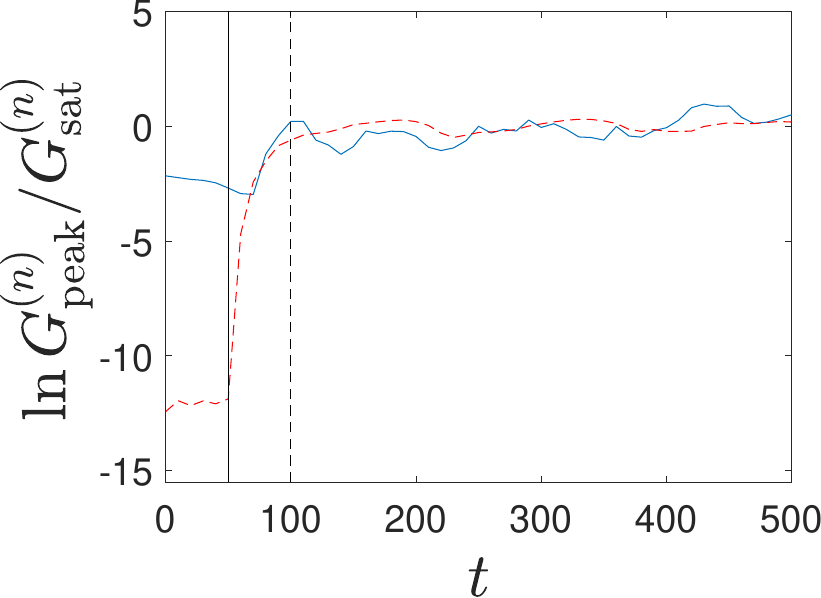}}
\end{tabular}
\caption{Time evolution of $G^{(n)}_{\rm{peak}}/G^{(n)}_{\rm{sat}}$ for $n=1$ (dashed red) and $n=2$ (solid blue), for fixed $v=0.6$, $c_2=0.2$. The vertical solid (dashed) line signals $t_{\rm{BCL}}$ ($t_{\rm{BHL}}$). Columns, from left to right, correspond to $L=5,11.4,15,20,30$, respectively. (a)-(e) Quantum BHL with $Z=0$ and $\lambda=1$. Dashed green line shows the prediction $\ln G^{(n)}_{\rm{peak}}\sim 2\Gamma t$, Eq. (\ref{eq:Z2prediction}).  (f)-(j) Classical BHL with $Z=0.001$ and $\lambda=1000$. Dashed green, dash-dotted black lines represent, respectively, the theoretical predictions $\ln G^{(2)}_{\rm{peak}}\sim 2\Gamma t$, $\ln G^{(1)}_{\rm{peak}}\sim \Gamma t$, Eq. (\ref{eq:Z2Breaking}). (k)-(o) BCL regime with $Z=0.75$ and $\lambda=1000$.}
\label{fig:SaturationCorrelationCavity}
\end{figure*}

\subsection{Classical BHL versus BCL}\label{subsec:CBHLBCL}

We now address the distinction between classical BHL and BCL, Figs. \ref{fig:SaturationQF}d-e, i-j. Due to the strong BCL stimulation, leading to a well-defined mean-field trajectory, the results here are essentially independent of $\lambda$. We first notice remarkable qualitative differences between the fourth and fifth column of Fig. \ref{fig:SaturationQF}. In Fig. \ref{fig:SaturationQF}d, the density-density correlation function is quite insensitive to the rather large BCL amplitude, Fig. \ref{fig:SaturationQF}i. However, once the white hole is switched on, a large amplification of quantum fluctuations is observed again, although with a different growth rate as compared to the expected BHL one, dashed green lines in Figs. \ref{fig:SaturationQF}a-c. Two main factors contribute to this behavior: (i) the starting point for the BHL amplification in Fig. \ref{fig:SaturationQF}d is not anymore some linear BCL wave on top of a homogeneous stationary condensate, where the BdG spectrum of Fig. \ref{fig:BHLBCLScheme}c is still expected to be valid; and (ii) even if the linear BdG approximation was valid, the saturation time is so short that there is no room for the dominant mode to stand above. On the other hand, Fig. \ref{fig:SaturationQF}e shows some amplification of the quantum fluctuations even when the BCL mechanism is operating alone, and the BHL onset barely affects the dynamics.

In order to further understand these features, we play with the BCL and BHL onsets in Figs. \ref{fig:SaturationBHLBCLPlots}a-d where, for the corresponding values of $Z=0.1,0.3,0.5,0.75$, we display the time evolution of $G^{(2)}_{\rm{peak}}$. Solid blue, dashed red, and dash-dotted green lines in each panel correspond to BHL onset times $t_{\rm{BHL}}=50,100,150$, respectively, while the dotted magenta line represents the case where there is no white hole, $t_{\rm{BHL}}=\infty$. The BCL onset and the quantum strength are fixed to $t_{\rm{BCL}}=50$ and $\lambda=1000$.

In Fig. \ref{fig:SaturationBHLBCLPlots}a, the amplification of $G^{(2)}_{\rm{peak}}$ can be unambiguously attributed to the presence of a white hole. However, this effect is attenuated as we further increase $Z$. Quantitatively, for increasing $Z$, the BCL contribution to the checkerboard pattern increases while the saturation amplitude $G^{(2)}_{\rm{sat}}$ decreases. These trends are confirmed by Fig. \ref{fig:SaturationAmplitudesBHL}c, where we depict the saturation amplitude $G^{(2)}_{\rm{sat}}$ for $t_{\rm{BHL}}=100$ ($t_{\rm{BHL}}=\infty$) as a function of $Z$ using black dots (blue squares). 

Both trends are in fact in close relationship. We have seen in the previous section that, in the classical regime, saturation is determined by the moment in which the mean-field density reaches $G^{(1)}_{\rm{sat}}$. Moreover, the second row of Fig. \ref{fig:SaturationQF} shows that the dependence of $G^{(1)}_{\rm{sat}}$ on $Z$ is very mild, even in the highly nonlinear regime of $Z\sim 1$. Therefore, the saturation time $t_{\rm{sat}}$ is essentially fixed by the starting point of the mean-field dynamics, in turn given by the initial BCL amplitude $A_{\rm{BCL}}$.  Hence, we explain the decrease of $G^{(2)}_{\rm{sat}}$ for increasing $Z$ by extending the arguments leading to Eqs. (\ref{eq:SaturationTimeCBHL})-(\ref{eq:SaturationAmplitudeCBHL}) to the nonlinear regime. 

At the same time (see Figs. \ref{fig:BCLTime}-\ref{fig:BCLPuroTime} and ensuing discussion), if the BCL undulation is highly nonlinear, we can no longer understand it as a perturbation but rather as a new mean-field background over which fluctuations evolve. The sharp peaked structure of the BCL wave, alternating density maxima and minima, results in the emergence of a checkerboard pattern in the correlation function, whose origin is completely different from that from a black-hole laser, Figs. \ref{fig:QBHLTime}-\ref{fig:CBHLTime1000}, arising from the exponential amplification of the quantum fluctuations of the lasing modes. Accordingly, the checkerboard amplitude from pure BCL stimulation behaves very differently; it weakly increases with $Z$ (blue squares in Fig. \ref{fig:SaturationAmplitudesBHL}c) and is exponentially smaller than its lasing counterpart (black dots). 

For sufficiently strong BCL stimulation (last points in Fig. \ref{fig:SaturationAmplitudesBHL}c), both saturation amplitudes converge, indicating the dominance of the BCL mechanism over the BHL one. In the BCL regime, the effect of the BHL onset is small, and the main physics can be understood in terms of the large background Cherenkov wave. The density ripple saturates here to the BCL amplitude squared, $G^{(1)}_{\rm{sat}}\sim A^2_{\rm{BCL}}$. Regarding the correlation function, the trend is reversed and the saturated checkerboard amplitude $G^{(2)}_{\rm{sat}}$ now grows with $Z$, or equivalently, with $A_{\rm{BCL}}$. Specifically, we expect $G^{(2)}_{\rm{sat}}=F(A_{\rm{BCL}})$, with $F$ some increasing function of $A_{\rm{BCL}}$ that also depends on the parameters determining the background flow. Finally, due to the rapid growth of the BCL wave, $t_{\rm{sat}}$ will be essentially limited by the time $\tau_{\rm{BCL}}<\tau_{\rm{RT}}$ needed for the BCL wave to travel towards the black hole and extend over the full cavity, $t_{\rm{sat}}\gtrsim \tau_{\rm{BCL}}$.

\subsection{Mean-field parameters}\label{subsec:MFParameteres}

We analyze in this section the role of the background parameters determining the BHL configuration. For the sake of clarity, we focus on variations of the cavity length $L$ and the flow speed $v$, which can be both controlled in the experiment: the cavity length is related to the depth of the waterfall potential and the flow speed to its velocity  \cite{Steinhauer2014}.

The spectrum of dynamical instabilities as a function of $L$ is shown in Fig. \ref{fig:BHLBCLScheme}c, where blue dots mark the numerical values considered in this section. In Fig. \ref{fig:SaturationCorrelationCavity} we jointly depict the time evolution of $G_{\rm{peak}}^{(1)}$ (dashed red) and $G_{\rm{peak}}^{(2)}$ (solid blue), both normalized to their corresponding saturation values, where columns correspond to $L=5,11.4,15,20,30$ and rows to $Z=0,0.001,0.75$, respectively. In this way, each row matches one of the three regimes discussed here: quantum BHL, classical BHL, and BCL. Dashed green and dash-dotted black lines in the first two rows represent theoretical fits from the predictions of Sec. \ref{subsec:QuantumClassical}, in good agreement with the numerical data. 

The first column of Fig. \ref{fig:SaturationCorrelationCavity} addresses a short cavity where only one nondegenerate unstable mode is present. Short cavities are translated into short roundtrip times as $\tau^{-1}_{\rm{RT}}\sim 1/L$, and hence large growth rates can be expected. This is what is seen Fig. \ref{fig:SaturationCorrelationCavity}a, where the dominant mode governs the dynamics almost from the very beginning of the lasing period (note also the shorter span in the time axis). In the latest times of the simulation, the system eventually reaches the ground state as no other nonlinear solutions are present. Since the ground state is dynamically stable, the enhanced quantum fluctuations flow away from the cavity, reducing the magnitude of $G_{\rm{peak}}^{(2)}$. The nondegenerate character of the unstable mode is clearly shown in Fig. \ref{fig:SaturationCorrelationCavity}f, where neat oscillations in the density growth are observed, as predicted by Eq. (\ref{eq:Z2BreakingLinear}). Finally, the BCL regime in Fig. \ref{fig:SaturationCorrelationCavity}k presents large periodic oscillations in its saturation regime because the system reaches a CES state, induced by the strong potential barrier.

\begin{figure}
    \includegraphics[width=\columnwidth]{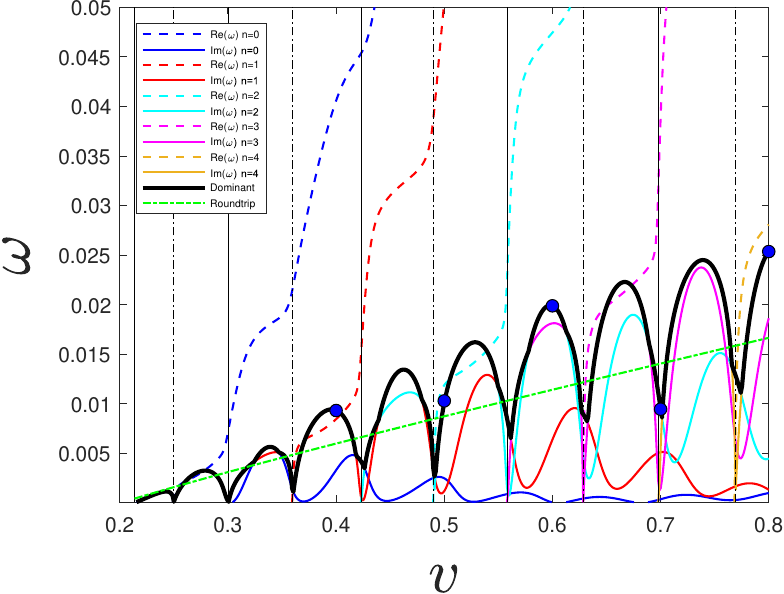}
    \caption{Spectrum of dynamical instabilities as a function of the flow speed $v$ for a flat-profile BHL configuration with $c_2=0.2,~L=20$. Solid (dashed) lines are the imaginary (real) part of the frequency. Vertical solid (dashed) black lines represent the critical velocities $v_n$ ($v_{n+1/2}$). Thick solid black envelope highlights the growth rate $\Gamma$ of the dominant mode. Dash-dotted green line is the inverse of the roundtrip time. Blue dots indicate the numerical values considered.}
    \label{fig:SpectrumV}
\end{figure}

\begin{figure*}
\begin{tabular}{@{}ccccc@{}}
      \stackinset{l}{0pt}{t}{0pt}{\footnotesize{(a)}}{\includegraphics[width=0.2\textwidth]{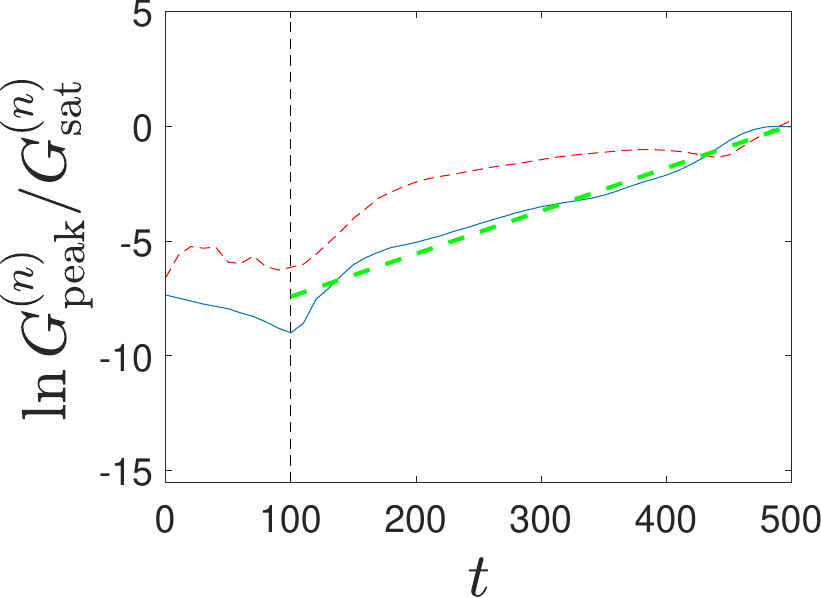}} 
    & \stackinset{l}{0pt}{t}{0pt}{\footnotesize{(b)}}{\includegraphics[width=0.2\textwidth]{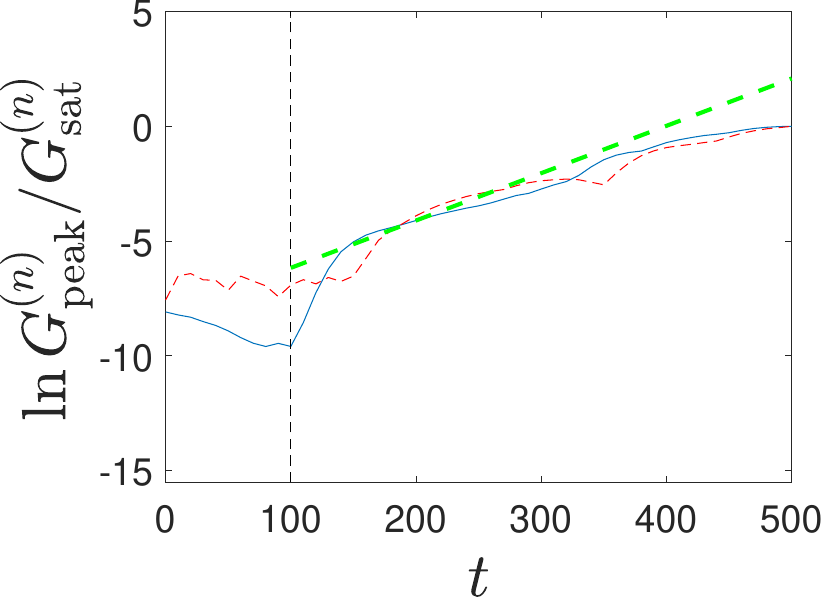}} &
    \stackinset{l}{0pt}{t}{0pt}{\footnotesize{(c)}}{\includegraphics[width=0.2\textwidth]{G2G1X20Puro}} &
    \stackinset{l}{0pt}{t}{0pt}{\footnotesize{(d)}}{\includegraphics[width=0.2\textwidth]{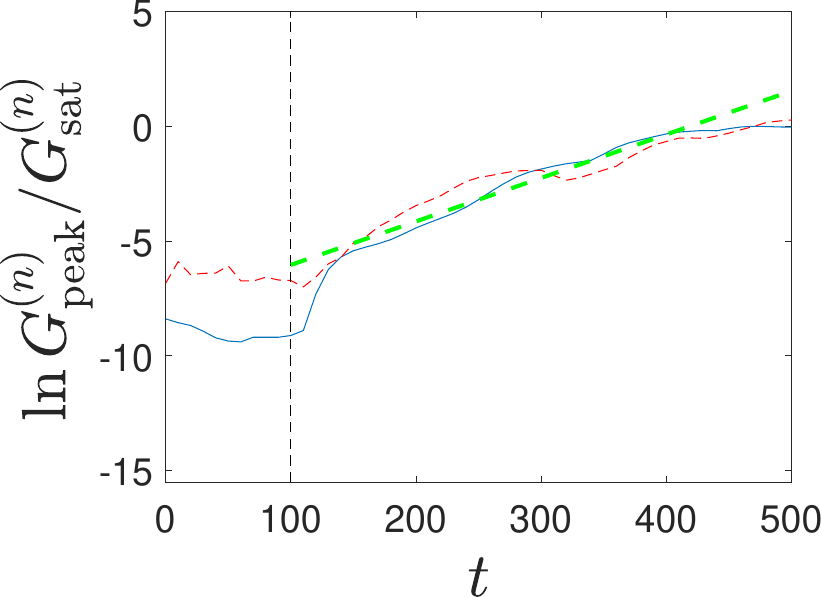}}  &
    \stackinset{l}{0pt}{t}{0pt}{\footnotesize{(e)}}{\includegraphics[width=0.2\textwidth]{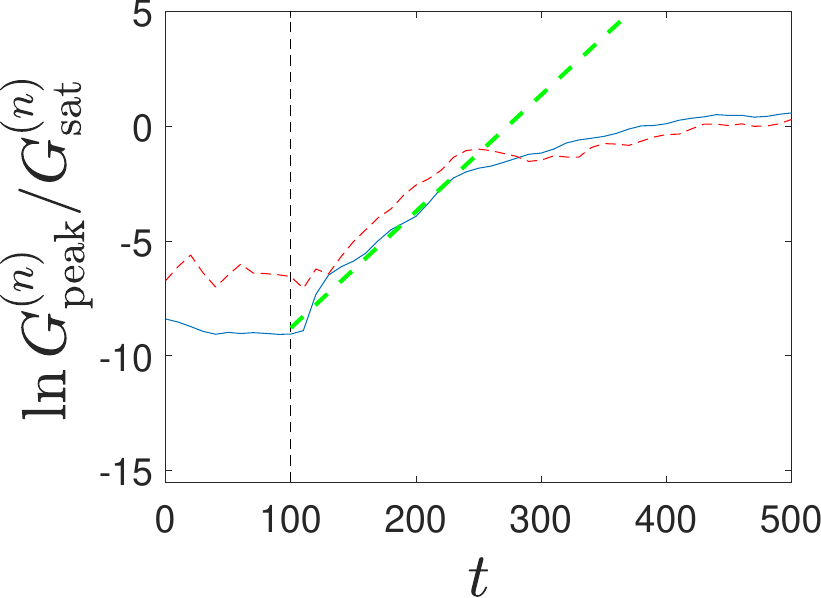}}\\
    \stackinset{l}{0pt}{t}{0pt}{\footnotesize{(f)}}{\includegraphics[width=0.2\textwidth]{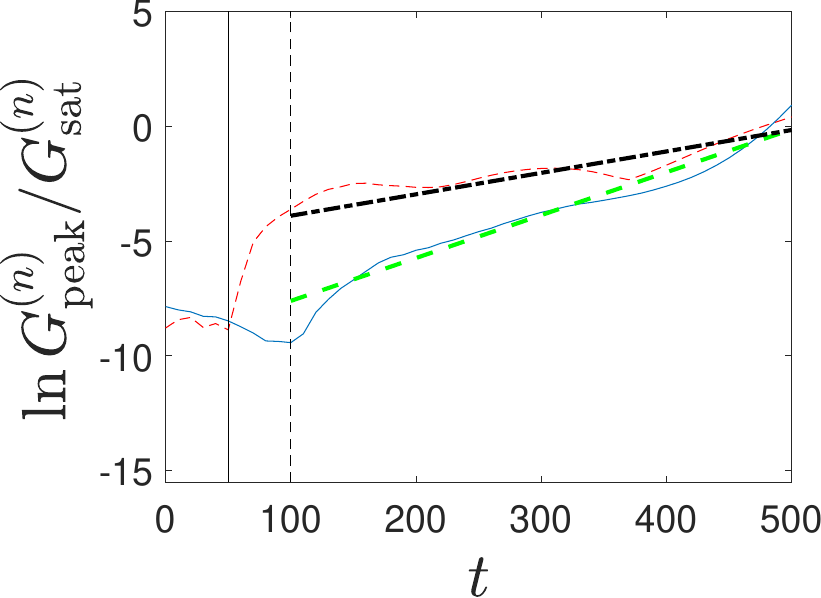}} 
    & \stackinset{l}{0pt}{t}{0pt}{\footnotesize{(g)}}{\includegraphics[width=0.2\textwidth]{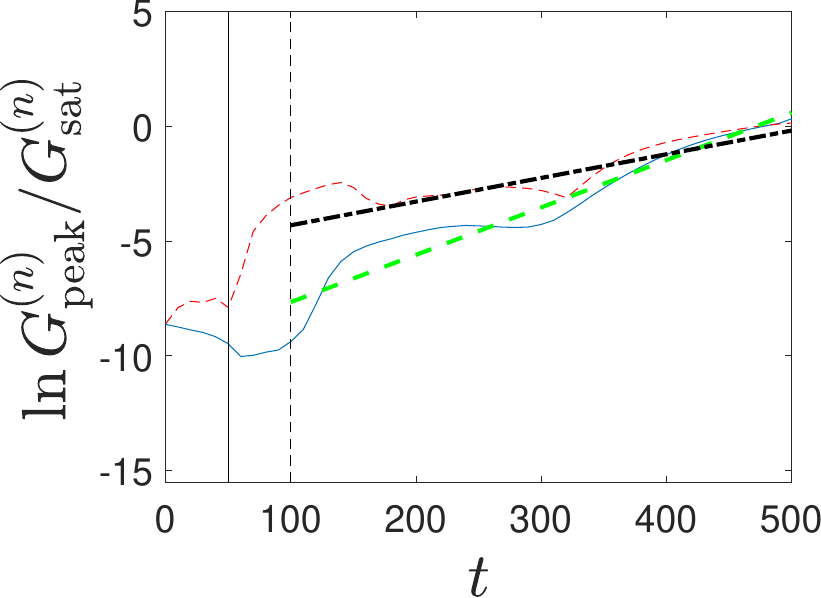}} &
    \stackinset{l}{0pt}{t}{0pt}{\footnotesize{(h)}}{\includegraphics[width=0.2\textwidth]{G2G1X20BHL}} &
    \stackinset{l}{0pt}{t}{0pt}{\footnotesize{(i)}}{\includegraphics[width=0.2\textwidth]{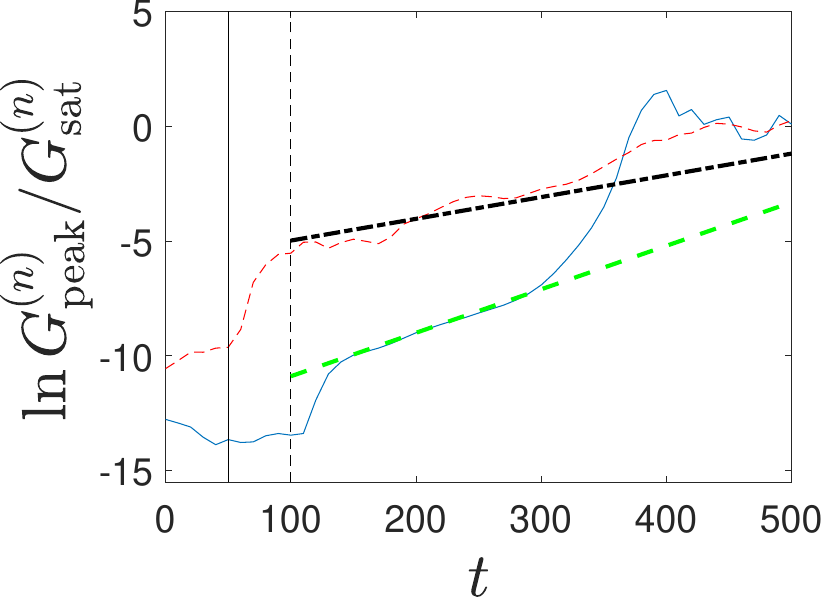}}  &
    \stackinset{l}{0pt}{t}{0pt}{\footnotesize{(j)}}{\includegraphics[width=0.2\textwidth]{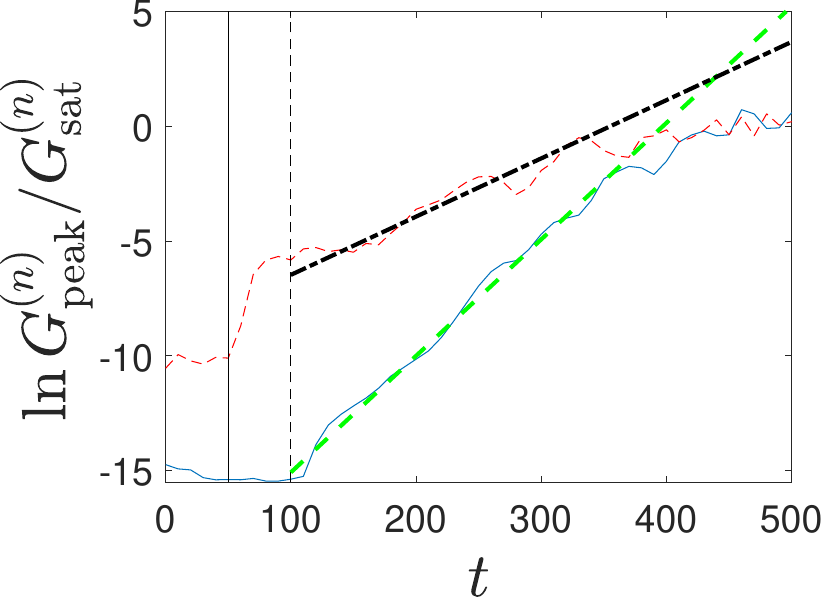}} \\
    \stackinset{l}{0pt}{t}{0pt}{\footnotesize{(k)}}{\includegraphics[width=0.2\textwidth]{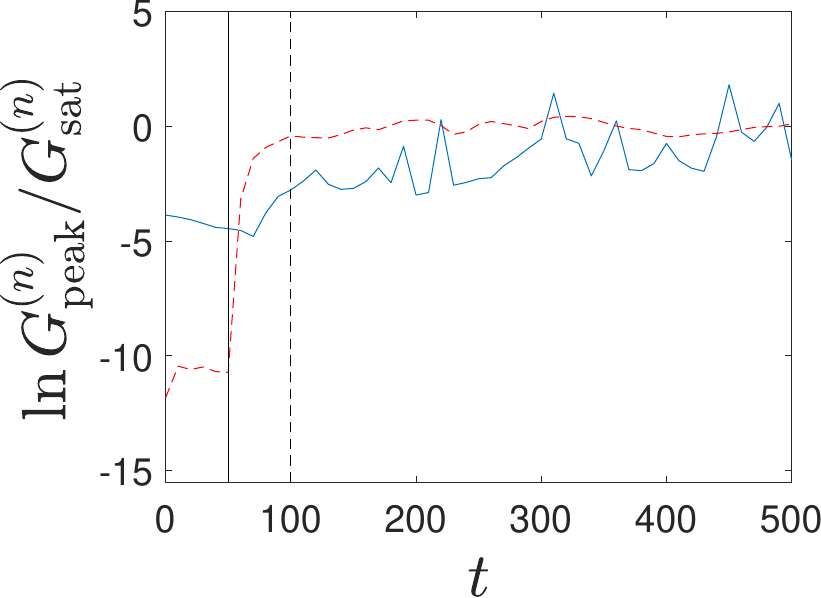}} 
    & \stackinset{l}{0pt}{t}{0pt}{\footnotesize{(l)}}{\includegraphics[width=0.2\textwidth]{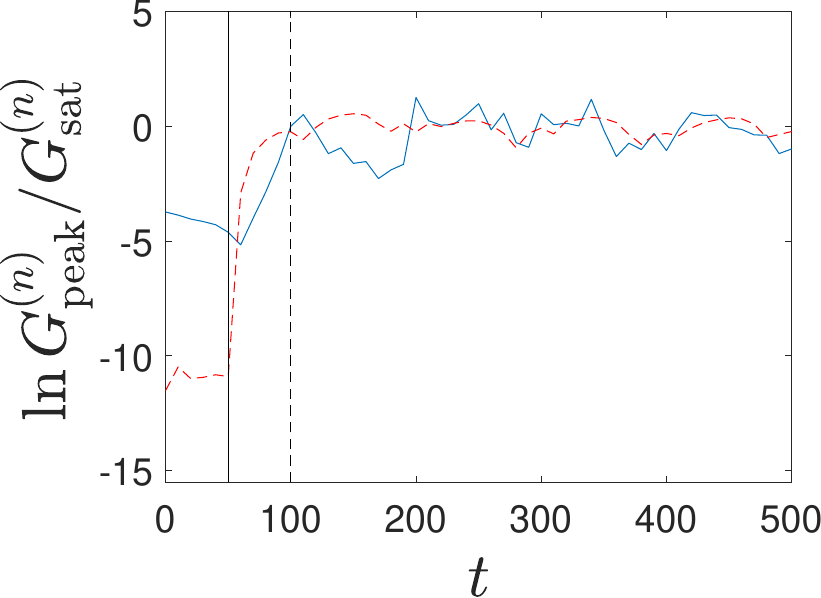}} &
    \stackinset{l}{0pt}{t}{0pt}{\footnotesize{(m)}}{\includegraphics[width=0.2\textwidth]{G2G1X20BCL}} &
    \stackinset{l}{0pt}{t}{0pt}{\footnotesize{(n)}}{\includegraphics[width=0.2\textwidth]{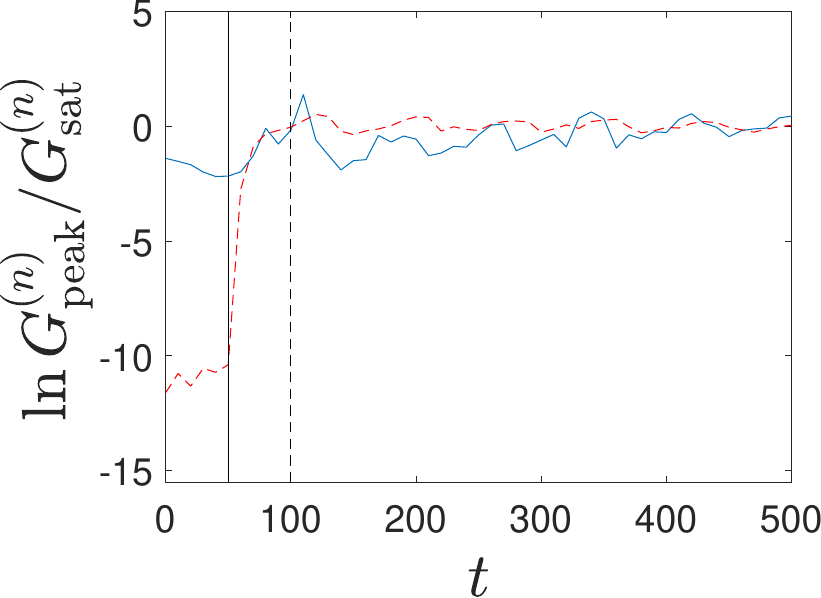}}  &
    \stackinset{l}{0pt}{t}{0pt}{\footnotesize{(o)}}{\includegraphics[width=0.2\textwidth]{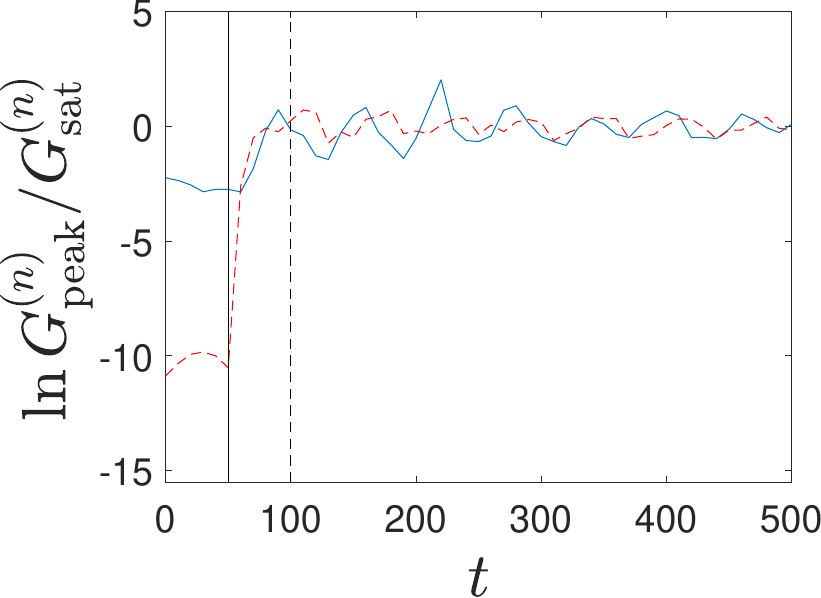}}
\end{tabular}
\caption{Same as Fig. \ref{fig:SaturationCorrelationCavity} but now the cavity length is fixed to $L=20$ and the columns correspond to $v=0.4,0.5,0.6,0.7,0.8$, respectively.}
\label{fig:SaturationCorrelationVelocity}
\end{figure*}

It must be noted that the roundtrip time only provides an estimation of the actual growth rate. From Fig. \ref{fig:BHLBCLScheme}c, we expect a highly nonmonotonic dependence of $\Gamma$ with respect to the cavity length. This is indeed observed in the second column of Fig. \ref{fig:SaturationCorrelationCavity}, where although the cavity is more than twice longer, the growth rate is even larger than in the first column. The oscillation period of the density for classical BHL, Fig. \ref{fig:SaturationCorrelationCavity}g, is also larger, in agreement with the prediction from Fig. \ref{fig:BHLBCLScheme}c.

For sufficiently long cavities, the qualitative trends converge, as shown by the last three columns of Fig. \ref{fig:SaturationCorrelationCavity}. In the linear regime, this is because there is an increasing number of lasing modes and we are closer to the ideal long-cavity limit where the usual WKB prescriptions become exact \cite{Finazzi2010}. In particular, the nonmonotonic behavior of the growth rate is attenuated, and the real part of the frequency of the dominant mode tends to zero, which translates into an increasing period of the oscillations in the density growth, Figs. \ref{fig:SaturationCorrelationCavity}h-j. In the saturation regime, the trends converge because, once there, the system typically oscillates around the nonlinear GP solution labeled with the largest $n$ available, which is highly metastable with a long lifetime \cite{Michel2015,deNova2016}. The more regular behavior of the nonlinear regime suggests that the predicted scalings for the saturation regime, namely Fig. \ref{fig:SaturationAmplitudesBHL}, are more accurate in the long-cavity limit. 

With respect to the flow speed, the spectrum of dynamical instabilities as a function of $v$ is shown in Fig. \ref{fig:SpectrumV}, where blue dots mark again the numerical values considered. We see that now the roundtrip time is reduced for increasing $v$; in the limit of large supersonic Mach number it scales as $\tau^{-1}_{\rm{RT}}\simeq 2v/L$, resulting in an increasing estimated growth rate. However, once more, nonmonotonic oscillations of the actual growth rate are observed, with increasing amplitude. In particular, the critical velocity $v_n$ at which a new unstable mode emerges is given by the adapted version of Eq. (\ref{eq:CriticalLengths}):
\begin{equation}\label{eq:CriticalVs}
    L=\frac{\arctan\sqrt{\dfrac{1-v_n^2}{v_n^2-c_2^2}}}{\sqrt{v_n^2-c_2^2}}+\frac{n\pi}{\sqrt{v_n^2-c_2^2}},~n=0,1\ldots
\end{equation}
For half-integer values $n+1/2$, the above equation yields the velocities $v_{n+1/2}$ at which the $n$-th unstable mode becomes nondegenerate.

In Fig. \ref{fig:SaturationCorrelationVelocity}, we perform the same analysis of Fig. \ref{fig:SaturationCorrelationCavity} but now the cavity length is fixed to $L=20$ and each column corresponds to $v=0.4,0.5,0.6,0.7,0.8$, respectively. Once more, a good agreement with the theoretical predictions is found. The oscillation period of the density in the classical BHL regime also follows the expected behavior. Perhaps the most remarkable result is the high degree of nonmonotonicity revealed by the fourth column, $v=0.7$, where the growth rate is halved during the lasing regime as compared to the adjacent columns $v=0.6,0.8$, as correctly predicted by Fig. \ref{fig:SpectrumV}.

\section{Discussion}\label{sec:Discussion}

We proceed here to critically discuss the results of the previous section from a global perspective. We first analyzed in Sec. \ref{subsec:Pattern} the characteristic spatial patterns of each regime (quantum BHL, classical BHL, and BCL) in the ensemble average density $G^{(1)}$ and in the normalized density-density correlation function $G^{(2)}$. For a quantum BHL, at early times we observe HSWH radiation that results from the scattering of the partner modes of the Hawking effect at the inner horizon. Due to the existence of low frequency and large wavevector modes in the lasing cavity, a checkerboard pattern emerges in the density-density correlations, similarly to the usual white-hole radiation stimulated from vacuum fluctuations. However, here these modes are also correlated with the originally emitted Hawking and Andreev modes, now far away from the outer horizon, yielding the distinctive signature of HSWH radiation. All these correlations arise from a continuous spectrum. In contrast, at late times, a monochromatic pattern is observed in both $G^{(1)},~G^{(2)}$, associated to the exponential growth and subsequent saturation of the single dominant unstable mode. This is translated into an enhanced checkerboard pattern and parallel stripes in the subsonic-supersonic and upstream-downstream correlations, as well as a large density ripple within the lasing cavity. 

The classical BHL regime differs from the quantum one (i) at early times, by the presence of a density ripple due to the background BCL stimulation; (ii) at late times, by the presence of sharp features, including soliton emission. The BCL regime is characterized by a strong ripple that is quite insensitive to the BHL onset, displaying a checkerboard pattern from the very beginning of the BCL stimulation.

A note of caution is placed here: the use of any of the above signatures as a conclusive smoking gun in a real experiment can be very problematic. For instance, the neat presence of sharp features (e.g., soliton emission) is indeed an evidence of classical behavior, either BHL or BCL. However, their absence does not exclude these phenomena because run-to-run experimental variations can also wash out these features in the ensemble average. The presence of HSWH radiation is neither an unambiguous signature since it can perfectly arise from an inner horizon that will eventually stimulate enough BCL radiation to overshadow the BHL amplification, as in actual experiments \cite{Steinhauer2014,Kolobov2021}.

Even more confusing can be the presence of a monochromatic pattern in the correlation function, which naively could be attributed to the underlying operation of a quantum BHL, because that can also be generated by experimental variations of BCL-stimulated Hawking radiation, as observed in the 2021 experiment \cite{Kolobov2021}. 

Actually, even the presence of a checkerboard pattern, common to all phenomena, is of a very different nature depending on the dominant mechanism. In a BHL, it arises at early times from HSWH radiation, and at late times from the dominant lasing mode. Once in the saturation regime, the blurring of the checkerboard is characteristic of quantum BHL. In contrast, in the BCL regime, the checkerboard is originated by the large Cherenkov amplitude, whose nonlinear peaked structure acts as a background density modulation over which quantum fluctuations evolve. Nevertheless, experimental variations can also blur a neat checkerboard arising from either classical BHL or BCL stimulation. Remarkably, the intricacy of this problem is captured by Fig. \ref{fig:BHLBCLExplan}, where three different phenomena (HSWH radiation, exponential amplification of the single dominant mode, BCL background modulation) give rise to a checkerboard pattern.

Sections \ref{subsec:QuantumClassical}, \ref{subsec:CBHLBCL} introduced a more robust characterization of each phenomenon, based on a quantitative analysis of the time evolution of the amplitude of the density ripple in $G^{(1)}$ and of the checkerboard pattern in $G^{(2)}$. In the context of the quantum-classical transition of a black-hole laser, we analyzed and confirmed the role of the $\mathbb{Z}_2$ symmetry, first predicted in Ref. \cite{Michel2015}. However, this symmetry is very fragile and a minor BCL stimulation can break it, as seen in Fig. \ref{fig:SaturationQF}b (see also central row of Figs. \ref{fig:SaturationCorrelationCavity}, \ref{fig:SaturationCorrelationVelocity}). In turn, this symmetry breaking gives an oscillatory character to the growth of the density ripple, in contrast with the BCL undulation, by definition a zero-frequency mode. Again, this difference is neither a conclusive distinguishability criterion because in experiments the BCL wave can acquire a finite frequency in the black-hole rest frame due to the Doppler shift induced by the recoil of the inner horizon \cite{Wang2016,Kolobov2021}.

Nevertheless, even when the $\mathbb{Z}_2$ symmetry is broken, signatures of quantum BHL can still be present, revealed as a dependence of the saturation amplitude $G^{(2)}_{\textrm{sat}}$ on the initial strength of the quantum fluctuations, determined here by the dimensionless parameter $\lambda$; see blue and red curves in Fig. \ref{fig:SaturationQF}b and blue curve in Fig. \ref{fig:SaturationQF}c. In fact, the blue and purple curves in Fig. \ref{fig:SaturationQF}c correspond to the simulations of Figs. \ref{fig:CBHLTime}, \ref{fig:CBHLTime1000}, respectively, which displayed strong qualitative differences in their correlation patterns. 

Based on the above discussion, we identify the exponential amplification of the initial quantum fluctuations as the most robust trait of the BHL effect. Specifically, we characterize the BHL-BCL crossover using very general scaling arguments that allow us to quantify the efficiency of the system as a quantum amplifier. We measure the amplitude of the quantum fluctuations through the \textit{relative} density-density correlation function $g^{(2)}$ of Eq. (\ref{eq:RelativeCorrelations}). In the initial state, we can compute analytically the Fourier transform of the correlation function [see Eq. (\ref{eq:antibunching}) and ensuing discussion] from:
\begin{eqnarray}
    \nonumber g^{(2)}(x,x',t=0)&=&\frac{1}{2\pi n_0\xi_0}\int^{\infty}_{-\infty}\mathrm{d}k~[S_0(k)-1]e^{ik(x-x')}\\
    \nonumber  S_0(k)&=&|u_k+v_k|^2=\frac{k^2}{2\Omega_k}=\frac{|k|}{\sqrt{4+k^2}},\\ 
\end{eqnarray}
where $S_0(k)$ is the zero-temperature static structure factor. Thus, at $t=0$, the peak of the Fourier transform is at $k=0$ and satisfies, apart from some proportionality factor,
\begin{equation}
    g_{\rm{peak}}^{(2)}(t=0)\propto \frac{1}{n_0\xi_0}\ll 1. 
\end{equation}
This is the input of the quantum amplifier. The output is the saturation value $g_{\rm{sat}}^{(2)}$. Then, we can quantify the gain of the quantum amplifier as the dimensionless ratio
\begin{equation}
    \mathcal{G}\equiv \frac{g_{\rm{sat}}^{(2)}}{g_{\rm{peak}}^{(2)}(t=0)}\propto n_0\xi_0 g_{\rm{sat}}^{(2)}= G_{\rm{sat}}^{(2)}. 
\end{equation}
Therefore, the saturation value of the \textit{normalized} correlation function $G_{\rm{sat}}^{(2)}$ is a good measure of the gain of the quantum amplifier. 

In the case of a quantum BHL, the saturation regime is characterized by the condition 
\begin{equation}
    g_{\rm{sat}}^{(2)}\sim 1,
\end{equation}
when density fluctuations become nonlinear, of the order of the condensate density itself. This means that a quantum BHL behaves as a nonlinear quantum amplifier since it amplifies the initial quantum fluctuations up to the same saturation amplitude, with an output not proportional to the input:
\begin{equation}
    \mathcal{G}_{\rm{QBHL}}\sim n_0\xi_0.
\end{equation}
In other words, the gain depends on the input amplitude $1/n_0\xi_0$, as already revealed by the $\lambda$-dependence observed in Fig. \ref{fig:SaturationAmplitudesBHL}a.

\begin{table*}[t]
\begin{tabular}[c]{|c|c|c|c|c|c|c|}
\hline
~ & $G^{(1)}_{\rm{peak}}(t)$ & $G^{(2)}_{\rm{peak}}(t)$ & $G^{(1)}_{\rm{sat}}$ & $G^{(2)}_{\rm{sat}}$ & $t_{\rm{sat}}$ & Monotonic \\
\hline
Quantum BHL & $\sim e^{2\Gamma t}/n_0\xi_0 $ & $\sim e^{2\Gamma t}$ & $\sim 1$ & $\sim n_0\xi_0 $ &  $\sim \ln n_0\xi_0 /2\Gamma$  & No\\
\hline
Classical BHL  & $\sim A_{\rm{BCL}}e^{\Gamma t}\cos(\gamma t+\delta)$ & $\sim e^{2\Gamma t}$  & $\sim 1$ & $\sim e^{2\Gamma t_{\rm{sat}}} \sim A_{\rm{BCL}}^{-2}$ &  $\sim -\ln A_{\rm{BCL}}/\Gamma$ & No\\
\hline
BCL  & ----- & ----- & $\sim A^2_{\rm{BCL}}$ & $F(A_{\rm{BCL}}) $ &  $\gtrsim \tau_{\rm{BCL}}$ & Yes\\
\hline
\end{tabular}
\caption{Summary of the main results for the three regimes discussed in this work: quantum BHL, classical BHL, and BCL. There is no analytical prediction for $G^{(n)}_{\rm{peak}}(t)$ in the BCL regime due to its highly nonlinear behavior. $F(A_{\rm{BCL}})$ is an increasing function of  $A_{\rm{BCL}}$ and $\tau_{\rm{BCL}}$ is the time that it takes the BCL wave to reach the black hole. The column ``Monotonic'' indicates a monotonic dependence on the background parameters of the flow. }
\label{table}
\end{table*}

In the case of a classical BHL, the saturation regime is independent of the initial quantum fluctuations, with the saturation time $t_{\rm{sat}}$ being determined by the mean-field dynamics, in turn governed by the exponential amplification of the initial BCL seed. This is translated into a saturation value [see Eq. (\ref{eq:SaturationAmplitudeCBHL})]
\begin{equation}
    g_{\rm{sat}}^{(2)}\sim \frac{e^{2\Gamma t_{\rm{sat}}}}{n_0\xi_0}.
\end{equation}
Hence, a classical BHL is a linear quantum amplifier, with a gain factor 
\begin{equation}
    \mathcal{G}_{\rm{CBHL}}\sim e^{2\Gamma t_{\rm{sat}}}
\end{equation}
independent of the input and exponentially large in the saturation time $t_{\rm{sat}}$. Thus, we can also regard $t_{\rm{sat}}$ as the lasing time during which the exponential amplification of quantum fluctuations takes place. As a result, this time typically decreases with the BCL amplitude $A_{\rm{BCL}}$ as the system then starts closer to saturation. 

In the BCL regime, linear quantum amplification occurs since the saturation properties are also independent of the initial quantum strength, Figs. \ref{fig:SaturationQF}d-e. However, the quantum gain 
\begin{equation}
    \mathcal{G}_{\rm{BCL}}\propto G_{\rm{sat}}^{(2)}=F(A_{\rm{BCL}})
\end{equation}
is exponentially smaller as compared to that of a classical BHL, Fig. \ref{fig:SaturationAmplitudesBHL}c. This results from the lack of a microscopic mechanism of exponential amplification, and the enhancement of quantum fluctuations stems just from the large BCL modulation of the mean-field density. This also implies that the function $F(A_{\rm{BCL}})$ determining the gain increases with $A_{\rm{BCL}}$, in stark contrast with the decrease expected for lasing amplification. Thus, the dependence of the gain with respect to the BCL amplitude can be used to distinguish between classical BHL and BCL.

The relation between the existence of classical trajectories and linear quantum amplification results from the fact that the quantum dynamics around a well-defined mean-field background is linear, governed by the BdG equations. Nonlinear amplification precisely emerges when quantum fluctuations backreact onto the condensate, as in a quantum BHL, and the linear BdG approximation no longer holds. Therefore, any significant deviation from linear amplification, i.e., any dependence of the gain on $n_0\xi_0$ represents a signature of quantum BHL, as shown in Figs. \ref{fig:SaturationQF}a-c.

Further understanding on how each amplifier works can be extracted from Figs. \ref{fig:SaturationCorrelationCavity}, \ref{fig:SaturationCorrelationVelocity}. Before saturation, the dynamics of $G_{\rm{peak}}^{(1)},G_{\rm{peak}}^{(2)}$ are strongly coupled for quantum BHL and BCL, while they become uncorrelated for classical BHL. The explanation behind this observation relies on the different mechanisms governing the dynamics of each magnitude. In a quantum BHL, the evolution of both $G_{\rm{peak}}^{(1)},G_{\rm{peak}}^{(2)}$ is mainly determined by the same mechanism: the amplification of the quantum amplitude of the unstable mode. In a classical BHL, $G_{\rm{peak}}^{(1)}$ is determined by the growth of the initial classical BCL amplitude while $G_{\rm{peak}}^{(2)}$ still represents the amplification of the quantum fluctuations of the unstable mode. The coupling of both magnitudes is again retrieved in the BCL regime, where the nonlinear amplitude of the Cherenkov wave that saturates $G_{\rm{peak}}^{(1)}$ is also responsible for the amplification in $G_{\rm{peak}}^{(2)}$. 

This analysis yields an important conclusion: the BCL mechanism requires a large nonlinear mean-field amplitude to show up in the correlation function, in contrast to a classical BHL, where mean-field and quantum dynamics (characterized by $G_{\rm{peak}}^{(1)},G_{\rm{peak}}^{(2)}$, respectively) are decoupled. This implies that, for a classical BHL, one can have a large checkerboard amplification which is not translated into a large ripple in the density profile, as shown in Figs. \ref{fig:CBHLTime}, \ref{fig:CBHLTime1000}. Actually, this statement also applies to a purely quantum BHL due to the $\mathbb{Z}_2$-symmetry suppression of the ripple growth (see Fig. \ref{fig:QBHLTime}d).

Hence, a joint analysis of the behavior of $G_{\rm{peak}}^{(1)},G_{\rm{peak}}^{(2)}$ and a quantitative characterization of the quantum gain can help to distinguish BHL from BCL in experimental setups in a robust way, regardless of the specific details of the configuration. Moreover, the study can be supplemented by varying the background parameters: any nonmonotonic behavior of the growth rate of both $G_{\rm{peak}}^{(1)},G_{\rm{peak}}^{(2)}$ further hints at the operation of the BHL effect because the BCL dependence should be smooth due to its zero-frequency nature. Indeed, nonmonotonicity is a quite general feature of resonant structures, not only in analogue setups \cite{Zapata2011}. This general qualitative analysis can be always complemented with a more specific quantitative one based on the estimation of the roundtrip time and the expected BCL signal \cite{Wang2016,Wang2017,Kolobov2021,Steinhauer2022}.

To conclude, we summarize the main results for each regime in Table \ref{table}, namely the behavior of $G^{(n)}_{\rm{peak}}(t)$, the dependence of the saturation parameters on the quantum and BCL amplitudes, $1/n_0\xi_0$ and $A_{\rm{BCL}}$, and the monotonicity with respect to the background parameters.

\section{Conclusions and outlook}\label{sec:conclu}

In this work, we have studied the BHL-BCL crossover using the idealized flat-profile model, which allows to unambiguously distinguish the contribution of each mechanism to the dynamics. By drawing an analogy with an unstable pendulum, we have identified three main regimes depending on the interplay between quantum fluctuations and classical stimulation: quantum BHL, where the dynamics is driven by the exponential amplification of vacuum fluctuations; classical BHL, where the lasing instability has a well-defined coherent amplitude induced by the background Cherenkov wave; and BCL, where Cherenkov stimulation governs the evolution until saturation.

General scaling arguments allow us to characterize each regime according to its behavior as amplifier of the initial quantum fluctuations. In this way, a quantum BHL is revealed as a nonlinear quantum amplifier, which takes quantum fluctuations up to the same saturation amplitude regardless of their initial strength. A classical BHL behaves instead as a linear quantum amplifier, where the output is proportional to the input and the gain is exponentially large in the lasing time. The BCL regime also acts as linear quantum amplifier, but its nature is very different since the amplification arises from the strong background modulation induced by the BCL wave, and not from a microscopic amplification mechanism. This is translated into an exponentially smaller gain as compared to a classical BHL. 

In order to clearly distinguish between the BHL and the BCL mechanisms in experiments, the measurement of the quantum gain can be complemented with a detailed study of the density ripple and the checkerboard pattern, including their joint behavior, the dependence of their saturation amplitudes with respect to the BCL strength, and the monotonicity of their growth rate with respect to the background parameters.

From an analogue gravity perspective, our work neatly isolates the BHL and BCL characteristic traits, providing practical tools for the unambiguous identification of the BHL effect in future experiments. Furthermore, our analysis suggests that the most reachable target is a classical BHL, where the background Cherenkov wave is sufficiently attenuated to become the seed of the BHL amplification instead of overshadowing it. At the same time, one can aim at optimizing the background mean-field configuration in order to maximize the lasing growth rate. Using the pendulum picture, the idea is to minimize the external force, so that it just gives a small amplitude to the pendulum, and maximize the effect of gravity. Once a classical BHL is achieved, the next challenge is the observation of the quantum BHL effect by further reducing the BCL background while increasing the amplitude of the quantum fluctuations by approaching the interacting regime $n_0\xi_0\gtrsim 1$. Nevertheless, a dedicated analysis of realistic experimental setups, including effects such as temperature, inhomogeneity and nonstationarity of the background, or experimental variations, is left for future work.


The results of this study can be also useful for other analogue setups in which low-frequency undulations similar to the Cherenkov wave compete with the BHL effect \cite{Coutant2014,Bossard2023}. Moreover, our model provides an ideal testing ground for the study of quantum \cite{Balbinot2005a,Baak2022,Butera2023} and classical \cite{Patrick2021} backreaction within the quantum and classical BHL regimes here described, respectively. In addition, the identification of novel phenomena such as HSWH radiation or quantum BCL-stimulated Hawking radiation is also relevant for the analogue community.

From a more global perspective, the characterization of a BHL as a quantum amplifier can be of interest for atomtronics and quantum transport. In general, the achievement and control of a stationary regime of spontaneous emission of Hawking radiation \cite{Kolobov2021}, the emergence of a spontaneous many-body Floquet state with the subsequent formation of a continuous time crystal in the long-time regime of a black-hole laser \cite{deNova2022}, the behavior of an optical lattice as a low-pass filter of Hawking radiation \cite{deNova2014a,deNova2017b}, and the quantum amplification here identified open the prospect of using gravitational analogues to also investigate condensed-matter phenomena and potential applications in quantum technologies. 

\acknowledgments
We thank D. Berm\'udez, S. Butera, I. Carusotto, Uwe R. Fischer, M. Jacquet, C. Lobo, S. Weinfurtner, and I. Zapata for valuable discussions. This project has received funding from European Union's Horizon 2020 research and innovation programme under the Marie Sk\l{}odowska-Curie Grant Agreement No. 847635, from Spain's Agencia Estatal de Investigaci\'on through Grants No. FIS2017-84368-P and No. PID2022-139288NB-I00, and from Universidad Complutense de Madrid through Grant No. FEI-EU-19-12.

\bibliography{Hawking}

\begin{thebibliography}{76}%
\makeatletter
\providecommand \@ifxundefined [1]{%
 \@ifx{#1\undefined}
}%
\providecommand \@ifnum [1]{%
 \ifnum #1\expandafter \@firstoftwo
 \else \expandafter \@secondoftwo
 \fi
}%
\providecommand \@ifx [1]{%
 \ifx #1\expandafter \@firstoftwo
 \else \expandafter \@secondoftwo
 \fi
}%
\providecommand \natexlab [1]{#1}%
\providecommand \enquote  [1]{``#1''}%
\providecommand \bibnamefont  [1]{#1}%
\providecommand \bibfnamefont [1]{#1}%
\providecommand \citenamefont [1]{#1}%
\providecommand \href@noop [0]{\@secondoftwo}%
\providecommand \href [0]{\begingroup \@sanitize@url \@href}%
\providecommand \@href[1]{\@@startlink{#1}\@@href}%
\providecommand \@@href[1]{\endgroup#1\@@endlink}%
\providecommand \@sanitize@url [0]{\catcode `\\12\catcode `\$12\catcode
  `\&12\catcode `\#12\catcode `\^12\catcode `\_12\catcode `\%12\relax}%
\providecommand \@@startlink[1]{}%
\providecommand \@@endlink[0]{}%
\providecommand \url  [0]{\begingroup\@sanitize@url \@url }%
\providecommand \@url [1]{\endgroup\@href {#1}{\urlprefix }}%
\providecommand \urlprefix  [0]{URL }%
\providecommand \Eprint [0]{\href }%
\providecommand \doibase [0]{http://dx.doi.org/}%
\providecommand \selectlanguage [0]{\@gobble}%
\providecommand \bibinfo  [0]{\@secondoftwo}%
\providecommand \bibfield  [0]{\@secondoftwo}%
\providecommand \translation [1]{[#1]}%
\providecommand \BibitemOpen [0]{}%
\providecommand \bibitemStop [0]{}%
\providecommand \bibitemNoStop [0]{.\EOS\space}%
\providecommand \EOS [0]{\spacefactor3000\relax}%
\providecommand \BibitemShut  [1]{\csname bibitem#1\endcsname}%
\let\auto@bib@innerbib\@empty
\bibitem [{\citenamefont {Unruh}(1981)}]{Unruh1981}%
  \BibitemOpen
  \bibfield  {author} {\bibinfo {author} {\bibfnamefont {W~G}\ \bibnamefont
  {Unruh}},\ }\bibfield  {title} {\enquote {\bibinfo {title} {{Experimental
  Black-Hole Evaporation?}}}\ }\href {\doibase 10.1103/PhysRevLett.46.1351}
  {\bibfield  {journal} {\bibinfo  {journal} {Phys. Rev. Lett.}\ }\textbf
  {\bibinfo {volume} {46}},\ \bibinfo {pages} {1351--1353} (\bibinfo {year}
  {1981})}\BibitemShut {NoStop}%
\bibitem [{\citenamefont {Garay}\ \emph {et~al.}(2000)\citenamefont {Garay},
  \citenamefont {Anglin}, \citenamefont {Cirac},\ and\ \citenamefont
  {Zoller}}]{Garay2000}%
  \BibitemOpen
  \bibfield  {author} {\bibinfo {author} {\bibfnamefont {L~J}\ \bibnamefont
  {Garay}}, \bibinfo {author} {\bibfnamefont {J~R}\ \bibnamefont {Anglin}},
  \bibinfo {author} {\bibfnamefont {J~I}\ \bibnamefont {Cirac}}, \ and\
  \bibinfo {author} {\bibfnamefont {P}~\bibnamefont {Zoller}},\ }\bibfield
  {title} {\enquote {\bibinfo {title} {{Sonic Analog of Gravitational Black
  Holes in Bose-Einstein Condensates}},}\ }\href {\doibase
  10.1103/PhysRevLett.85.4643} {\bibfield  {journal} {\bibinfo  {journal}
  {Phys. Rev. Lett.}\ }\textbf {\bibinfo {volume} {85}},\ \bibinfo {pages}
  {4643--4647} (\bibinfo {year} {2000})}\BibitemShut {NoStop}%
\bibitem [{\citenamefont {Lahav}\ \emph {et~al.}(2010)\citenamefont {Lahav},
  \citenamefont {Itah}, \citenamefont {Blumkin}, \citenamefont {Gordon},
  \citenamefont {Rinott}, \citenamefont {Zayats},\ and\ \citenamefont
  {Steinhauer}}]{Lahav2010}%
  \BibitemOpen
  \bibfield  {author} {\bibinfo {author} {\bibfnamefont {Oren}\ \bibnamefont
  {Lahav}}, \bibinfo {author} {\bibfnamefont {Amir}\ \bibnamefont {Itah}},
  \bibinfo {author} {\bibfnamefont {Alex}\ \bibnamefont {Blumkin}}, \bibinfo
  {author} {\bibfnamefont {Carmit}\ \bibnamefont {Gordon}}, \bibinfo {author}
  {\bibfnamefont {Shahar}\ \bibnamefont {Rinott}}, \bibinfo {author}
  {\bibfnamefont {Alona}\ \bibnamefont {Zayats}}, \ and\ \bibinfo {author}
  {\bibfnamefont {Jeff}\ \bibnamefont {Steinhauer}},\ }\bibfield  {title}
  {\enquote {\bibinfo {title} {{Realization of a Sonic Black Hole Analog in a
  Bose-Einstein Condensate}},}\ }\href {\doibase
  10.1103/PhysRevLett.105.240401} {\bibfield  {journal} {\bibinfo  {journal}
  {Phys. Rev. Lett.}\ }\textbf {\bibinfo {volume} {105}},\ \bibinfo {pages}
  {240401} (\bibinfo {year} {2010})}\BibitemShut {NoStop}%
\bibitem [{\citenamefont {Weinfurtner}\ \emph {et~al.}(2011)\citenamefont
  {Weinfurtner}, \citenamefont {Tedford}, \citenamefont {Penrice},
  \citenamefont {Unruh},\ and\ \citenamefont {Lawrence}}]{Weinfurtner2011}%
  \BibitemOpen
  \bibfield  {author} {\bibinfo {author} {\bibfnamefont {Silke}\ \bibnamefont
  {Weinfurtner}}, \bibinfo {author} {\bibfnamefont {Edmund~W.}\ \bibnamefont
  {Tedford}}, \bibinfo {author} {\bibfnamefont {Matthew C.~J.}\ \bibnamefont
  {Penrice}}, \bibinfo {author} {\bibfnamefont {William~G.}\ \bibnamefont
  {Unruh}}, \ and\ \bibinfo {author} {\bibfnamefont {Gregory~A.}\ \bibnamefont
  {Lawrence}},\ }\bibfield  {title} {\enquote {\bibinfo {title} {{Measurement
  of Stimulated Hawking Emission in an Analogue System}},}\ }\href {\doibase
  10.1103/PhysRevLett.106.021302} {\bibfield  {journal} {\bibinfo  {journal}
  {Phys. Rev. Lett.}\ }\textbf {\bibinfo {volume} {106}},\ \bibinfo {pages}
  {021302} (\bibinfo {year} {2011})}\BibitemShut {NoStop}%
\bibitem [{\citenamefont {Euv\'e}\ \emph {et~al.}(2016)\citenamefont {Euv\'e},
  \citenamefont {Michel}, \citenamefont {Parentani}, \citenamefont {Philbin},\
  and\ \citenamefont {Rousseaux}}]{Euve2016}%
  \BibitemOpen
  \bibfield  {author} {\bibinfo {author} {\bibfnamefont {L.-P.}\ \bibnamefont
  {Euv\'e}}, \bibinfo {author} {\bibfnamefont {F.}~\bibnamefont {Michel}},
  \bibinfo {author} {\bibfnamefont {R.}~\bibnamefont {Parentani}}, \bibinfo
  {author} {\bibfnamefont {T.~G.}\ \bibnamefont {Philbin}}, \ and\ \bibinfo
  {author} {\bibfnamefont {G.}~\bibnamefont {Rousseaux}},\ }\bibfield  {title}
  {\enquote {\bibinfo {title} {{Observation of Noise Correlated by the Hawking
  Effect in a Water Tank}},}\ }\href {\doibase 10.1103/PhysRevLett.117.121301}
  {\bibfield  {journal} {\bibinfo  {journal} {Phys. Rev. Lett.}\ }\textbf
  {\bibinfo {volume} {117}},\ \bibinfo {pages} {121301} (\bibinfo {year}
  {2016})}\BibitemShut {NoStop}%
\bibitem [{\citenamefont {Belgiorno}\ \emph {et~al.}(2010)\citenamefont
  {Belgiorno}, \citenamefont {Cacciatori}, \citenamefont {Clerici},
  \citenamefont {Gorini}, \citenamefont {Ortenzi}, \citenamefont {Rizzi},
  \citenamefont {Rubino}, \citenamefont {Sala},\ and\ \citenamefont
  {Faccio}}]{Belgiorno2010}%
  \BibitemOpen
  \bibfield  {author} {\bibinfo {author} {\bibfnamefont {F}~\bibnamefont
  {Belgiorno}}, \bibinfo {author} {\bibfnamefont {S~L}\ \bibnamefont
  {Cacciatori}}, \bibinfo {author} {\bibfnamefont {M}~\bibnamefont {Clerici}},
  \bibinfo {author} {\bibfnamefont {V}~\bibnamefont {Gorini}}, \bibinfo
  {author} {\bibfnamefont {G}~\bibnamefont {Ortenzi}}, \bibinfo {author}
  {\bibfnamefont {L}~\bibnamefont {Rizzi}}, \bibinfo {author} {\bibfnamefont
  {E}~\bibnamefont {Rubino}}, \bibinfo {author} {\bibfnamefont {V~G}\
  \bibnamefont {Sala}}, \ and\ \bibinfo {author} {\bibfnamefont
  {D}~\bibnamefont {Faccio}},\ }\bibfield  {title} {\enquote {\bibinfo {title}
  {{Hawking Radiation from Ultrashort Laser Pulse Filaments}},}\ }\href
  {\doibase 10.1103/PhysRevLett.105.203901} {\bibfield  {journal} {\bibinfo
  {journal} {Phys. Rev. Lett.}\ }\textbf {\bibinfo {volume} {105}},\ \bibinfo
  {pages} {203901} (\bibinfo {year} {2010})}\BibitemShut {NoStop}%
\bibitem [{\citenamefont {Drori}\ \emph {et~al.}(2019)\citenamefont {Drori},
  \citenamefont {Rosenberg}, \citenamefont {Berm\'udez}, \citenamefont
  {Silberberg},\ and\ \citenamefont {Leonhardt}}]{Drori2019}%
  \BibitemOpen
  \bibfield  {author} {\bibinfo {author} {\bibfnamefont {Jonathan}\
  \bibnamefont {Drori}}, \bibinfo {author} {\bibfnamefont {Yuval}\ \bibnamefont
  {Rosenberg}}, \bibinfo {author} {\bibfnamefont {David}\ \bibnamefont
  {Berm\'udez}}, \bibinfo {author} {\bibfnamefont {Yaron}\ \bibnamefont
  {Silberberg}}, \ and\ \bibinfo {author} {\bibfnamefont {Ulf}\ \bibnamefont
  {Leonhardt}},\ }\bibfield  {title} {\enquote {\bibinfo {title} {{Observation
  of Stimulated Hawking Radiation in an Optical Analogue}},}\ }\href {\doibase
  10.1103/PhysRevLett.122.010404} {\bibfield  {journal} {\bibinfo  {journal}
  {Phys. Rev. Lett.}\ }\textbf {\bibinfo {volume} {122}},\ \bibinfo {pages}
  {010404} (\bibinfo {year} {2019})}\BibitemShut {NoStop}%
\bibitem [{\citenamefont {Horstmann}\ \emph {et~al.}(2010)\citenamefont
  {Horstmann}, \citenamefont {Reznik}, \citenamefont {Fagnocchi},\ and\
  \citenamefont {Cirac}}]{Horstmann2010}%
  \BibitemOpen
  \bibfield  {author} {\bibinfo {author} {\bibfnamefont {B}~\bibnamefont
  {Horstmann}}, \bibinfo {author} {\bibfnamefont {B}~\bibnamefont {Reznik}},
  \bibinfo {author} {\bibfnamefont {S}~\bibnamefont {Fagnocchi}}, \ and\
  \bibinfo {author} {\bibfnamefont {J~I}\ \bibnamefont {Cirac}},\ }\bibfield
  {title} {\enquote {\bibinfo {title} {{Hawking Radiation from an Acoustic
  Black Hole on an Ion Ring}},}\ }\href {\doibase
  10.1103/PhysRevLett.104.250403} {\bibfield  {journal} {\bibinfo  {journal}
  {Phys. Rev. Lett.}\ }\textbf {\bibinfo {volume} {104}},\ \bibinfo {pages}
  {250403} (\bibinfo {year} {2010})}\BibitemShut {NoStop}%
\bibitem [{\citenamefont {Wittemer}\ \emph {et~al.}(2019)\citenamefont
  {Wittemer}, \citenamefont {Hakelberg}, \citenamefont {Kiefer}, \citenamefont
  {Schr\"oder}, \citenamefont {Fey}, \citenamefont {Sch\"utzhold},
  \citenamefont {Warring},\ and\ \citenamefont {Schaetz}}]{Wittemer2019}%
  \BibitemOpen
  \bibfield  {author} {\bibinfo {author} {\bibfnamefont {Matthias}\
  \bibnamefont {Wittemer}}, \bibinfo {author} {\bibfnamefont {Frederick}\
  \bibnamefont {Hakelberg}}, \bibinfo {author} {\bibfnamefont {Philip}\
  \bibnamefont {Kiefer}}, \bibinfo {author} {\bibfnamefont {Jan-Philipp}\
  \bibnamefont {Schr\"oder}}, \bibinfo {author} {\bibfnamefont {Christian}\
  \bibnamefont {Fey}}, \bibinfo {author} {\bibfnamefont {Ralf}\ \bibnamefont
  {Sch\"utzhold}}, \bibinfo {author} {\bibfnamefont {Ulrich}\ \bibnamefont
  {Warring}}, \ and\ \bibinfo {author} {\bibfnamefont {Tobias}\ \bibnamefont
  {Schaetz}},\ }\bibfield  {title} {\enquote {\bibinfo {title} {Phonon pair
  creation by inflating quantum fluctuations in an ion trap},}\ }\href
  {\doibase 10.1103/PhysRevLett.123.180502} {\bibfield  {journal} {\bibinfo
  {journal} {Phys. Rev. Lett.}\ }\textbf {\bibinfo {volume} {123}},\ \bibinfo
  {pages} {180502} (\bibinfo {year} {2019})}\BibitemShut {NoStop}%
\bibitem [{\citenamefont {Carusotto}\ and\ \citenamefont
  {Ciuti}(2013)}]{Carusotto2013}%
  \BibitemOpen
  \bibfield  {author} {\bibinfo {author} {\bibfnamefont {Iacopo}\ \bibnamefont
  {Carusotto}}\ and\ \bibinfo {author} {\bibfnamefont {Cristiano}\ \bibnamefont
  {Ciuti}},\ }\bibfield  {title} {\enquote {\bibinfo {title} {Quantum fluids of
  light},}\ }\href {\doibase 10.1103/RevModPhys.85.299} {\bibfield  {journal}
  {\bibinfo  {journal} {Rev. Mod. Phys.}\ }\textbf {\bibinfo {volume} {85}},\
  \bibinfo {pages} {299--366} (\bibinfo {year} {2013})}\BibitemShut {NoStop}%
\bibitem [{\citenamefont {Nguyen}\ \emph {et~al.}(2015)\citenamefont {Nguyen},
  \citenamefont {Gerace}, \citenamefont {Carusotto}, \citenamefont {Sanvitto},
  \citenamefont {Galopin}, \citenamefont {Lema\^{i}tre}, \citenamefont
  {Sagnes}, \citenamefont {Bloch},\ and\ \citenamefont {Amo}}]{Nguyen2015}%
  \BibitemOpen
  \bibfield  {author} {\bibinfo {author} {\bibfnamefont {H.~S.}\ \bibnamefont
  {Nguyen}}, \bibinfo {author} {\bibfnamefont {D.}~\bibnamefont {Gerace}},
  \bibinfo {author} {\bibfnamefont {I.}~\bibnamefont {Carusotto}}, \bibinfo
  {author} {\bibfnamefont {D.}~\bibnamefont {Sanvitto}}, \bibinfo {author}
  {\bibfnamefont {E.}~\bibnamefont {Galopin}}, \bibinfo {author} {\bibfnamefont
  {A.}~\bibnamefont {Lema\^{i}tre}}, \bibinfo {author} {\bibfnamefont
  {I.}~\bibnamefont {Sagnes}}, \bibinfo {author} {\bibfnamefont
  {J.}~\bibnamefont {Bloch}}, \ and\ \bibinfo {author} {\bibfnamefont
  {A.}~\bibnamefont {Amo}},\ }\bibfield  {title} {\enquote {\bibinfo {title}
  {Acoustic black hole in a stationary hydrodynamic flow of microcavity
  polaritons},}\ }\href {\doibase 10.1103/PhysRevLett.114.036402} {\bibfield
  {journal} {\bibinfo  {journal} {Phys. Rev. Lett.}\ }\textbf {\bibinfo
  {volume} {114}},\ \bibinfo {pages} {036402} (\bibinfo {year}
  {2015})}\BibitemShut {NoStop}%
\bibitem [{\citenamefont {Shi}\ \emph {et~al.}(2023)\citenamefont {Shi},
  \citenamefont {Yang}, \citenamefont {Xiang}, \citenamefont {Ge},
  \citenamefont {Li}, \citenamefont {Wang}, \citenamefont {Huang},
  \citenamefont {Tian}, \citenamefont {Song}, \citenamefont {Zheng} \emph
  {et~al.}}]{Shi2023}%
  \BibitemOpen
  \bibfield  {author} {\bibinfo {author} {\bibfnamefont {Yun-Hao}\ \bibnamefont
  {Shi}}, \bibinfo {author} {\bibfnamefont {Run-Qiu}\ \bibnamefont {Yang}},
  \bibinfo {author} {\bibfnamefont {Zhongcheng}\ \bibnamefont {Xiang}},
  \bibinfo {author} {\bibfnamefont {Zi-Yong}\ \bibnamefont {Ge}}, \bibinfo
  {author} {\bibfnamefont {Hao}\ \bibnamefont {Li}}, \bibinfo {author}
  {\bibfnamefont {Yong-Yi}\ \bibnamefont {Wang}}, \bibinfo {author}
  {\bibfnamefont {Kaixuan}\ \bibnamefont {Huang}}, \bibinfo {author}
  {\bibfnamefont {Ye}~\bibnamefont {Tian}}, \bibinfo {author} {\bibfnamefont
  {Xiaohui}\ \bibnamefont {Song}}, \bibinfo {author} {\bibfnamefont {Dongning}\
  \bibnamefont {Zheng}},  \emph {et~al.},\ }\bibfield  {title} {\enquote
  {\bibinfo {title} {{Quantum simulation of Hawking radiation and curved
  spacetime with a superconducting on-chip black hole}},}\ }\href
  {https://www.nature.com/articles/s41467-023-39064-6} {\bibfield  {journal}
  {\bibinfo  {journal} {Nature Communications}\ }\textbf {\bibinfo {volume}
  {14}},\ \bibinfo {pages} {3263} (\bibinfo {year} {2023})}\BibitemShut
  {NoStop}%
\bibitem [{\citenamefont {Jaskula}\ \emph {et~al.}(2012)\citenamefont
  {Jaskula}, \citenamefont {Partridge}, \citenamefont {Bonneau}, \citenamefont
  {Lopes}, \citenamefont {Ruaudel}, \citenamefont {Boiron},\ and\ \citenamefont
  {Westbrook}}]{Jaskula2012}%
  \BibitemOpen
  \bibfield  {author} {\bibinfo {author} {\bibfnamefont {J.-C.}\ \bibnamefont
  {Jaskula}}, \bibinfo {author} {\bibfnamefont {G.~B.}\ \bibnamefont
  {Partridge}}, \bibinfo {author} {\bibfnamefont {M.}~\bibnamefont {Bonneau}},
  \bibinfo {author} {\bibfnamefont {R.}~\bibnamefont {Lopes}}, \bibinfo
  {author} {\bibfnamefont {J.}~\bibnamefont {Ruaudel}}, \bibinfo {author}
  {\bibfnamefont {D.}~\bibnamefont {Boiron}}, \ and\ \bibinfo {author}
  {\bibfnamefont {C.~I.}\ \bibnamefont {Westbrook}},\ }\bibfield  {title}
  {\enquote {\bibinfo {title} {{Acoustic Analog to the Dynamical Casimir Effect
  in a Bose-Einstein Condensate}},}\ }\href {\doibase
  10.1103/PhysRevLett.109.220401} {\bibfield  {journal} {\bibinfo  {journal}
  {Phys. Rev. Lett.}\ }\textbf {\bibinfo {volume} {109}},\ \bibinfo {pages}
  {220401} (\bibinfo {year} {2012})}\BibitemShut {NoStop}%
\bibitem [{\citenamefont {Hung}\ \emph {et~al.}(2013)\citenamefont {Hung},
  \citenamefont {Gurarie},\ and\ \citenamefont {Chin}}]{Hung2013}%
  \BibitemOpen
  \bibfield  {author} {\bibinfo {author} {\bibfnamefont {Chen-Lung}\
  \bibnamefont {Hung}}, \bibinfo {author} {\bibfnamefont {Victor}\ \bibnamefont
  {Gurarie}}, \ and\ \bibinfo {author} {\bibfnamefont {Cheng}\ \bibnamefont
  {Chin}},\ }\bibfield  {title} {\enquote {\bibinfo {title} {{From Cosmology to
  Cold Atoms: Observation of Sakharov Oscillations in a Quenched Atomic
  Superfluid}},}\ }\href {\doibase 10.1126/science.1237557} {\bibfield
  {journal} {\bibinfo  {journal} {Science}\ }\textbf {\bibinfo {volume}
  {341}},\ \bibinfo {pages} {1213--1215} (\bibinfo {year} {2013})}\BibitemShut
  {NoStop}%
\bibitem [{\citenamefont {Torres}\ \emph {et~al.}(2017)\citenamefont {Torres},
  \citenamefont {Patrick}, \citenamefont {Coutant}, \citenamefont {Richartz},
  \citenamefont {Tedford},\ and\ \citenamefont {Weinfurtner}}]{Torres2017}%
  \BibitemOpen
  \bibfield  {author} {\bibinfo {author} {\bibfnamefont {Theo}\ \bibnamefont
  {Torres}}, \bibinfo {author} {\bibfnamefont {Sam}\ \bibnamefont {Patrick}},
  \bibinfo {author} {\bibfnamefont {Antonin}\ \bibnamefont {Coutant}}, \bibinfo
  {author} {\bibfnamefont {Mauricio}\ \bibnamefont {Richartz}}, \bibinfo
  {author} {\bibfnamefont {Edmund~W}\ \bibnamefont {Tedford}}, \ and\ \bibinfo
  {author} {\bibfnamefont {Silke}\ \bibnamefont {Weinfurtner}},\ }\bibfield
  {title} {\enquote {\bibinfo {title} {Rotational superradiant scattering in a
  vortex flow},}\ }\href {https://www.nature.com/articles/nphys4151} {\bibfield
   {journal} {\bibinfo  {journal} {Nature Physics}\ }\textbf {\bibinfo {volume}
  {13}},\ \bibinfo {pages} {833--836} (\bibinfo {year} {2017})}\BibitemShut
  {NoStop}%
\bibitem [{\citenamefont {Eckel}\ \emph {et~al.}(2018)\citenamefont {Eckel},
  \citenamefont {Kumar}, \citenamefont {Jacobson}, \citenamefont {Spielman},\
  and\ \citenamefont {Campbell}}]{Eckel2018}%
  \BibitemOpen
  \bibfield  {author} {\bibinfo {author} {\bibfnamefont {S.}~\bibnamefont
  {Eckel}}, \bibinfo {author} {\bibfnamefont {A.}~\bibnamefont {Kumar}},
  \bibinfo {author} {\bibfnamefont {T.}~\bibnamefont {Jacobson}}, \bibinfo
  {author} {\bibfnamefont {I.~B.}\ \bibnamefont {Spielman}}, \ and\ \bibinfo
  {author} {\bibfnamefont {G.~K.}\ \bibnamefont {Campbell}},\ }\bibfield
  {title} {\enquote {\bibinfo {title} {{A Rapidly Expanding Bose-Einstein
  Condensate: An Expanding Universe in the Lab}},}\ }\href {\doibase
  10.1103/PhysRevX.8.021021} {\bibfield  {journal} {\bibinfo  {journal} {Phys.
  Rev. X}\ }\textbf {\bibinfo {volume} {8}},\ \bibinfo {pages} {021021}
  (\bibinfo {year} {2018})}\BibitemShut {NoStop}%
\bibitem [{\citenamefont {de~Nova}\ \emph {et~al.}(2019)\citenamefont
  {de~Nova}, \citenamefont {Golubkov}, \citenamefont {Kolobov},\ and\
  \citenamefont {Steinhauer}}]{deNova2019}%
  \BibitemOpen
  \bibfield  {author} {\bibinfo {author} {\bibfnamefont {J.~R.~M.}\
  \bibnamefont {de~Nova}}, \bibinfo {author} {\bibfnamefont {K.}~\bibnamefont
  {Golubkov}}, \bibinfo {author} {\bibfnamefont {V.~I.}\ \bibnamefont
  {Kolobov}}, \ and\ \bibinfo {author} {\bibfnamefont {J.}~\bibnamefont
  {Steinhauer}},\ }\bibfield  {title} {\enquote {\bibinfo {title} {{Observation
  of thermal Hawking radiation and its temperature in an analogue black
  hole}},}\ }\href {\doibase 10.1038/s41586-019-1241-0} {\bibfield  {journal}
  {\bibinfo  {journal} {Nature}\ }\textbf {\bibinfo {volume} {569}},\ \bibinfo
  {pages} {688} (\bibinfo {year} {2019})}\BibitemShut {NoStop}%
\bibitem [{\citenamefont {Hu}\ \emph {et~al.}(2019)\citenamefont {Hu},
  \citenamefont {Feng}, \citenamefont {Zhang},\ and\ \citenamefont
  {Chin}}]{Hu2019}%
  \BibitemOpen
  \bibfield  {author} {\bibinfo {author} {\bibfnamefont {Jiazhong}\
  \bibnamefont {Hu}}, \bibinfo {author} {\bibfnamefont {Lei}\ \bibnamefont
  {Feng}}, \bibinfo {author} {\bibfnamefont {Zhendong}\ \bibnamefont {Zhang}},
  \ and\ \bibinfo {author} {\bibfnamefont {Cheng}\ \bibnamefont {Chin}},\
  }\bibfield  {title} {\enquote {\bibinfo {title} {{Quantum simulation of Unruh
  radiation}},}\ }\href {https://www.nature.com/articles/s41567-019-0537-1}
  {\bibfield  {journal} {\bibinfo  {journal} {Nature Physics}\ }\textbf
  {\bibinfo {volume} {15}},\ \bibinfo {pages} {785--789} (\bibinfo {year}
  {2019})}\BibitemShut {NoStop}%
\bibitem [{\citenamefont {Torres}\ \emph {et~al.}(2020)\citenamefont {Torres},
  \citenamefont {Patrick}, \citenamefont {Richartz},\ and\ \citenamefont
  {Weinfurtner}}]{Torres2020}%
  \BibitemOpen
  \bibfield  {author} {\bibinfo {author} {\bibfnamefont {Theo}\ \bibnamefont
  {Torres}}, \bibinfo {author} {\bibfnamefont {Sam}\ \bibnamefont {Patrick}},
  \bibinfo {author} {\bibfnamefont {Maur\'{\i}cio}\ \bibnamefont {Richartz}}, \
  and\ \bibinfo {author} {\bibfnamefont {Silke}\ \bibnamefont {Weinfurtner}},\
  }\bibfield  {title} {\enquote {\bibinfo {title} {Quasinormal mode
  oscillations in an analogue black hole experiment},}\ }\href {\doibase
  10.1103/PhysRevLett.125.011301} {\bibfield  {journal} {\bibinfo  {journal}
  {Phys. Rev. Lett.}\ }\textbf {\bibinfo {volume} {125}},\ \bibinfo {pages}
  {011301} (\bibinfo {year} {2020})}\BibitemShut {NoStop}%
\bibitem [{\citenamefont {Patrick}\ \emph {et~al.}(2021)\citenamefont
  {Patrick}, \citenamefont {Goodhew}, \citenamefont {Gooding},\ and\
  \citenamefont {Weinfurtner}}]{Patrick2021}%
  \BibitemOpen
  \bibfield  {author} {\bibinfo {author} {\bibfnamefont {Sam}\ \bibnamefont
  {Patrick}}, \bibinfo {author} {\bibfnamefont {Harry}\ \bibnamefont
  {Goodhew}}, \bibinfo {author} {\bibfnamefont {Cisco}\ \bibnamefont
  {Gooding}}, \ and\ \bibinfo {author} {\bibfnamefont {Silke}\ \bibnamefont
  {Weinfurtner}},\ }\bibfield  {title} {\enquote {\bibinfo {title}
  {Backreaction in an analogue black hole experiment},}\ }\href {\doibase
  10.1103/PhysRevLett.126.041105} {\bibfield  {journal} {\bibinfo  {journal}
  {Phys. Rev. Lett.}\ }\textbf {\bibinfo {volume} {126}},\ \bibinfo {pages}
  {041105} (\bibinfo {year} {2021})}\BibitemShut {NoStop}%
\bibitem [{\citenamefont {Steinhauer}\ \emph {et~al.}(2022)\citenamefont
  {Steinhauer}, \citenamefont {Abuzarli}, \citenamefont {Aladjidi},
  \citenamefont {Bienaim{\'e}}, \citenamefont {Piekarski}, \citenamefont {Liu},
  \citenamefont {Giacobino}, \citenamefont {Bramati},\ and\ \citenamefont
  {Glorieux}}]{Steinhauer2022a}%
  \BibitemOpen
  \bibfield  {author} {\bibinfo {author} {\bibfnamefont {Jeff}\ \bibnamefont
  {Steinhauer}}, \bibinfo {author} {\bibfnamefont {Murad}\ \bibnamefont
  {Abuzarli}}, \bibinfo {author} {\bibfnamefont {Tangui}\ \bibnamefont
  {Aladjidi}}, \bibinfo {author} {\bibfnamefont {Tom}\ \bibnamefont
  {Bienaim{\'e}}}, \bibinfo {author} {\bibfnamefont {Clara}\ \bibnamefont
  {Piekarski}}, \bibinfo {author} {\bibfnamefont {Wei}\ \bibnamefont {Liu}},
  \bibinfo {author} {\bibfnamefont {Elisabeth}\ \bibnamefont {Giacobino}},
  \bibinfo {author} {\bibfnamefont {Alberto}\ \bibnamefont {Bramati}}, \ and\
  \bibinfo {author} {\bibfnamefont {Quentin}\ \bibnamefont {Glorieux}},\
  }\bibfield  {title} {\enquote {\bibinfo {title} {Analogue cosmological
  particle creation in an ultracold quantum fluid of light},}\ }\href
  {https://www.nature.com/articles/s41467-022-30603-1} {\bibfield  {journal}
  {\bibinfo  {journal} {Nat. Commun.}\ }\textbf {\bibinfo {volume} {13}},\
  \bibinfo {pages} {2890} (\bibinfo {year} {2022})}\BibitemShut {NoStop}%
\bibitem [{\citenamefont {Viermann}\ \emph {et~al.}(2022)\citenamefont
  {Viermann}, \citenamefont {Sparn}, \citenamefont {Liebster}, \citenamefont
  {Hans}, \citenamefont {Kath}, \citenamefont {Parra-L{\'o}pez}, \citenamefont
  {Tolosa-Sime{\'o}n}, \citenamefont {S{\'a}nchez-Kuntz}, \citenamefont {Haas},
  \citenamefont {Strobel} \emph {et~al.}}]{Viermann2022}%
  \BibitemOpen
  \bibfield  {author} {\bibinfo {author} {\bibfnamefont {Celia}\ \bibnamefont
  {Viermann}}, \bibinfo {author} {\bibfnamefont {Marius}\ \bibnamefont
  {Sparn}}, \bibinfo {author} {\bibfnamefont {Nikolas}\ \bibnamefont
  {Liebster}}, \bibinfo {author} {\bibfnamefont {Maurus}\ \bibnamefont {Hans}},
  \bibinfo {author} {\bibfnamefont {Elinor}\ \bibnamefont {Kath}}, \bibinfo
  {author} {\bibfnamefont {{\'A}lvaro}\ \bibnamefont {Parra-L{\'o}pez}},
  \bibinfo {author} {\bibfnamefont {Mireia}\ \bibnamefont {Tolosa-Sime{\'o}n}},
  \bibinfo {author} {\bibfnamefont {Natalia}\ \bibnamefont
  {S{\'a}nchez-Kuntz}}, \bibinfo {author} {\bibfnamefont {Tobias}\ \bibnamefont
  {Haas}}, \bibinfo {author} {\bibfnamefont {Helmut}\ \bibnamefont {Strobel}},
  \emph {et~al.},\ }\bibfield  {title} {\enquote {\bibinfo {title} {Quantum
  field simulator for dynamics in curved spacetime},}\ }\href
  {https://www.nature.com/articles/s41586-022-05313-9} {\bibfield  {journal}
  {\bibinfo  {journal} {Nature}\ }\textbf {\bibinfo {volume} {611}},\ \bibinfo
  {pages} {260--264} (\bibinfo {year} {2022})}\BibitemShut {NoStop}%
\bibitem [{\citenamefont {Corley}\ and\ \citenamefont
  {Jacobson}(1999)}]{Corley1999}%
  \BibitemOpen
  \bibfield  {author} {\bibinfo {author} {\bibfnamefont {Steven}\ \bibnamefont
  {Corley}}\ and\ \bibinfo {author} {\bibfnamefont {Ted}\ \bibnamefont
  {Jacobson}},\ }\bibfield  {title} {\enquote {\bibinfo {title} {Black hole
  lasers},}\ }\href {\doibase 10.1103/PhysRevD.59.124011} {\bibfield  {journal}
  {\bibinfo  {journal} {Phys. Rev. D}\ }\textbf {\bibinfo {volume} {59}},\
  \bibinfo {pages} {124011} (\bibinfo {year} {1999})}\BibitemShut {NoStop}%
\bibitem [{\citenamefont {Leonhardt}\ \emph {et~al.}(2003)\citenamefont
  {Leonhardt}, \citenamefont {Kiss},\ and\ \citenamefont
  {\"{O}hberg}}]{Leonhardt2003}%
  \BibitemOpen
  \bibfield  {author} {\bibinfo {author} {\bibfnamefont {U}~\bibnamefont
  {Leonhardt}}, \bibinfo {author} {\bibfnamefont {T}~\bibnamefont {Kiss}}, \
  and\ \bibinfo {author} {\bibfnamefont {P}~\bibnamefont {\"{O}hberg}},\
  }\bibfield  {title} {\enquote {\bibinfo {title} {{Theory of elementary
  excitations in unstable Bose-Einstein condensates and the instability of
  sonic horizons}},}\ }\href {\doibase 10.1103/PhysRevA.67.033602} {\bibfield
  {journal} {\bibinfo  {journal} {Phys. Rev. A}\ }\textbf {\bibinfo {volume}
  {67}},\ \bibinfo {pages} {33602} (\bibinfo {year} {2003})}\BibitemShut
  {NoStop}%
\bibitem [{\citenamefont {Barcel\'o}\ \emph {et~al.}(2006)\citenamefont
  {Barcel\'o}, \citenamefont {Cano}, \citenamefont {Garay},\ and\ \citenamefont
  {Jannes}}]{Barcelo2006}%
  \BibitemOpen
  \bibfield  {author} {\bibinfo {author} {\bibfnamefont {C.}~\bibnamefont
  {Barcel\'o}}, \bibinfo {author} {\bibfnamefont {A.}~\bibnamefont {Cano}},
  \bibinfo {author} {\bibfnamefont {L.~J.}\ \bibnamefont {Garay}}, \ and\
  \bibinfo {author} {\bibfnamefont {G.}~\bibnamefont {Jannes}},\ }\bibfield
  {title} {\enquote {\bibinfo {title} {{Stability analysis of sonic horizons in
  Bose-Einstein condensates}},}\ }\href {\doibase 10.1103/PhysRevD.74.024008}
  {\bibfield  {journal} {\bibinfo  {journal} {Phys. Rev. D}\ }\textbf {\bibinfo
  {volume} {74}},\ \bibinfo {pages} {024008} (\bibinfo {year}
  {2006})}\BibitemShut {NoStop}%
\bibitem [{\citenamefont {Jain}\ \emph {et~al.}(2007)\citenamefont {Jain},
  \citenamefont {Bradley},\ and\ \citenamefont {Gardiner}}]{Jain2007}%
  \BibitemOpen
  \bibfield  {author} {\bibinfo {author} {\bibfnamefont {P}~\bibnamefont
  {Jain}}, \bibinfo {author} {\bibfnamefont {A~S}\ \bibnamefont {Bradley}}, \
  and\ \bibinfo {author} {\bibfnamefont {C}~\bibnamefont {Gardiner}},\
  }\bibfield  {title} {\enquote {\bibinfo {title} {{Quantum de Laval nozzle:
  Stability and quantum dynamics of sonic horizons in a toroidally trapped Bose
  gas containing a superflow}},}\ }\href {\doibase 10.1103/PhysRevA.76.023617}
  {\bibfield  {journal} {\bibinfo  {journal} {Phys. Rev. A}\ }\textbf {\bibinfo
  {volume} {76}},\ \bibinfo {pages} {23617} (\bibinfo {year}
  {2007})}\BibitemShut {NoStop}%
\bibitem [{\citenamefont {Coutant}\ and\ \citenamefont
  {Parentani}(2010)}]{Coutant2010}%
  \BibitemOpen
  \bibfield  {author} {\bibinfo {author} {\bibfnamefont {Antonin}\ \bibnamefont
  {Coutant}}\ and\ \bibinfo {author} {\bibfnamefont {Renaud}\ \bibnamefont
  {Parentani}},\ }\bibfield  {title} {\enquote {\bibinfo {title} {{Black hole
  lasers, a mode analysis}},}\ }\href {\doibase 10.1103/PhysRevD.81.084042}
  {\bibfield  {journal} {\bibinfo  {journal} {Phys. Rev. D}\ }\textbf {\bibinfo
  {volume} {81}},\ \bibinfo {pages} {84042} (\bibinfo {year}
  {2010})}\BibitemShut {NoStop}%
\bibitem [{\citenamefont {Finazzi}\ and\ \citenamefont
  {Parentani}(2010)}]{Finazzi2010}%
  \BibitemOpen
  \bibfield  {author} {\bibinfo {author} {\bibfnamefont {S}~\bibnamefont
  {Finazzi}}\ and\ \bibinfo {author} {\bibfnamefont {R}~\bibnamefont
  {Parentani}},\ }\bibfield  {title} {\enquote {\bibinfo {title} {{Black hole
  lasers in Bose-Einstein condensates}},}\ }\href
  {http://stacks.iop.org/1367-2630/12/i=9/a=095015} {\bibfield  {journal}
  {\bibinfo  {journal} {New J. Phys.}\ }\textbf {\bibinfo {volume} {12}},\
  \bibinfo {pages} {095015} (\bibinfo {year} {2010})}\BibitemShut {NoStop}%
\bibitem [{\citenamefont {Berm\'{u}dez}\ and\ \citenamefont
  {Leonhardt}(2019)}]{Bermudez2018}%
  \BibitemOpen
  \bibfield  {author} {\bibinfo {author} {\bibfnamefont {David}\ \bibnamefont
  {Berm\'{u}dez}}\ and\ \bibinfo {author} {\bibfnamefont {Ulf}\ \bibnamefont
  {Leonhardt}},\ }\bibfield  {title} {\enquote {\bibinfo {title} {{Resonant
  Hawking radiation as an instability}},}\ }\href {\doibase
  10.1088/1361-6382/aaf435} {\bibfield  {journal} {\bibinfo  {journal}
  {Classical and Quantum Gravity}\ }\textbf {\bibinfo {volume} {36}},\ \bibinfo
  {pages} {024001} (\bibinfo {year} {2019})}\BibitemShut {NoStop}%
\bibitem [{\citenamefont {B\"urkle}\ \emph {et~al.}(2018)\citenamefont
  {B\"urkle}, \citenamefont {Gaidoukov},\ and\ \citenamefont
  {Anglin}}]{Burkle2018}%
  \BibitemOpen
  \bibfield  {author} {\bibinfo {author} {\bibfnamefont {R}~\bibnamefont
  {B\"urkle}}, \bibinfo {author} {\bibfnamefont {A}~\bibnamefont {Gaidoukov}},
  \ and\ \bibinfo {author} {\bibfnamefont {J~R}\ \bibnamefont {Anglin}},\
  }\bibfield  {title} {\enquote {\bibinfo {title} {Quasi-steady radiation of
  sound from turbulent sonic ergoregions},}\ }\href {\doibase
  10.1088/1367-2630/aad7ed} {\bibfield  {journal} {\bibinfo  {journal} {New
  Journal of Physics}\ }\textbf {\bibinfo {volume} {20}},\ \bibinfo {pages}
  {083020} (\bibinfo {year} {2018})}\BibitemShut {NoStop}%
\bibitem [{\citenamefont {Faccio}\ \emph {et~al.}(2012)\citenamefont {Faccio},
  \citenamefont {Arane}, \citenamefont {Lamperti},\ and\ \citenamefont
  {Leonhardt}}]{Faccio_2012}%
  \BibitemOpen
  \bibfield  {author} {\bibinfo {author} {\bibfnamefont {Daniele}\ \bibnamefont
  {Faccio}}, \bibinfo {author} {\bibfnamefont {Tal}\ \bibnamefont {Arane}},
  \bibinfo {author} {\bibfnamefont {Marco}\ \bibnamefont {Lamperti}}, \ and\
  \bibinfo {author} {\bibfnamefont {Ulf}\ \bibnamefont {Leonhardt}},\
  }\bibfield  {title} {\enquote {\bibinfo {title} {Optical black hole
  lasers},}\ }\href {\doibase 10.1088/0264-9381/29/22/224009} {\bibfield
  {journal} {\bibinfo  {journal} {Classical and Quantum Gravity}\ }\textbf
  {\bibinfo {volume} {29}},\ \bibinfo {pages} {224009} (\bibinfo {year}
  {2012})}\BibitemShut {NoStop}%
\bibitem [{\citenamefont {Peloquin}\ \emph {et~al.}(2016)\citenamefont
  {Peloquin}, \citenamefont {Euv\'e}, \citenamefont {Philbin},\ and\
  \citenamefont {Rousseaux}}]{Peloquin2016}%
  \BibitemOpen
  \bibfield  {author} {\bibinfo {author} {\bibfnamefont {C\'edric}\
  \bibnamefont {Peloquin}}, \bibinfo {author} {\bibfnamefont {L\'eo-Paul}\
  \bibnamefont {Euv\'e}}, \bibinfo {author} {\bibfnamefont {Thomas}\
  \bibnamefont {Philbin}}, \ and\ \bibinfo {author} {\bibfnamefont {Germain}\
  \bibnamefont {Rousseaux}},\ }\bibfield  {title} {\enquote {\bibinfo {title}
  {Analog wormholes and black hole laser effects in hydrodynamics},}\ }\href
  {\doibase 10.1103/PhysRevD.93.084032} {\bibfield  {journal} {\bibinfo
  {journal} {Phys. Rev. D}\ }\textbf {\bibinfo {volume} {93}},\ \bibinfo
  {pages} {084032} (\bibinfo {year} {2016})}\BibitemShut {NoStop}%
\bibitem [{\citenamefont {Rinc{\'o}n-Estrada}\ and\ \citenamefont
  {Berm{\'u}dez}(2021)}]{RinconEstrada2021}%
  \BibitemOpen
  \bibfield  {author} {\bibinfo {author} {\bibfnamefont {Juan~David}\
  \bibnamefont {Rinc{\'o}n-Estrada}}\ and\ \bibinfo {author} {\bibfnamefont
  {David}\ \bibnamefont {Berm{\'u}dez}},\ }\bibfield  {title} {\enquote
  {\bibinfo {title} {Instabilities in an optical black-hole laser},}\ }\href
  {\doibase https://doi.org/10.1002/andp.202000239} {\bibfield  {journal}
  {\bibinfo  {journal} {Annalen der Physik}\ }\textbf {\bibinfo {volume}
  {533}},\ \bibinfo {pages} {2000239} (\bibinfo {year} {2021})}\BibitemShut
  {NoStop}%
\bibitem [{\citenamefont {Katayama}(2021)}]{Katayama2021}%
  \BibitemOpen
  \bibfield  {author} {\bibinfo {author} {\bibfnamefont {Haruna}\ \bibnamefont
  {Katayama}},\ }\bibfield  {title} {\enquote {\bibinfo {title}
  {Quantum-circuit black hole lasers},}\ }\href
  {https://www.nature.com/articles/s41598-021-98456-0} {\bibfield  {journal}
  {\bibinfo  {journal} {Scientific Reports}\ }\textbf {\bibinfo {volume}
  {11}},\ \bibinfo {pages} {19137} (\bibinfo {year} {2021})}\BibitemShut
  {NoStop}%
\bibitem [{\citenamefont {Carusotto}\ \emph {et~al.}(2006)\citenamefont
  {Carusotto}, \citenamefont {Hu}, \citenamefont {Collins},\ and\ \citenamefont
  {Smerzi}}]{Carusotto2006}%
  \BibitemOpen
  \bibfield  {author} {\bibinfo {author} {\bibfnamefont {I.}~\bibnamefont
  {Carusotto}}, \bibinfo {author} {\bibfnamefont {S.~X.}\ \bibnamefont {Hu}},
  \bibinfo {author} {\bibfnamefont {L.~A.}\ \bibnamefont {Collins}}, \ and\
  \bibinfo {author} {\bibfnamefont {A.}~\bibnamefont {Smerzi}},\ }\bibfield
  {title} {\enquote {\bibinfo {title} {{Bogoliubov-\ifmmode \check{C}\else
  \v{C}\fi{}erenkov Radiation in a Bose-Einstein Condensate Flowing against an
  Obstacle}},}\ }\href {\doibase 10.1103/PhysRevLett.97.260403} {\bibfield
  {journal} {\bibinfo  {journal} {Phys. Rev. Lett.}\ }\textbf {\bibinfo
  {volume} {97}},\ \bibinfo {pages} {260403} (\bibinfo {year}
  {2006})}\BibitemShut {NoStop}%
\bibitem [{\citenamefont {Coutant}\ \emph {et~al.}(2012)\citenamefont
  {Coutant}, \citenamefont {Fabbri}, \citenamefont {Parentani}, \citenamefont
  {Balbinot},\ and\ \citenamefont {Anderson}}]{Coutant2012}%
  \BibitemOpen
  \bibfield  {author} {\bibinfo {author} {\bibfnamefont {Antonin}\ \bibnamefont
  {Coutant}}, \bibinfo {author} {\bibfnamefont {Alessandro}\ \bibnamefont
  {Fabbri}}, \bibinfo {author} {\bibfnamefont {Renaud}\ \bibnamefont
  {Parentani}}, \bibinfo {author} {\bibfnamefont {Roberto}\ \bibnamefont
  {Balbinot}}, \ and\ \bibinfo {author} {\bibfnamefont {Paul~R.}\ \bibnamefont
  {Anderson}},\ }\bibfield  {title} {\enquote {\bibinfo {title} {Hawking
  radiation of massive modes and undulations},}\ }\href {\doibase
  10.1103/PhysRevD.86.064022} {\bibfield  {journal} {\bibinfo  {journal} {Phys.
  Rev. D}\ }\textbf {\bibinfo {volume} {86}},\ \bibinfo {pages} {064022}
  (\bibinfo {year} {2012})}\BibitemShut {NoStop}%
\bibitem [{\citenamefont {Steinhauer}(2014)}]{Steinhauer2014}%
  \BibitemOpen
  \bibfield  {author} {\bibinfo {author} {\bibfnamefont {J.}~\bibnamefont
  {Steinhauer}},\ }\bibfield  {title} {\enquote {\bibinfo {title} {{Observation
  of self-amplifying Hawking radiation in an analog black hole laser}},}\
  }\href {http://dx.doi.org/10.1038/nphys3104} {\bibfield  {journal} {\bibinfo
  {journal} {Nature Physics}\ }\textbf {\bibinfo {volume} {10}},\ \bibinfo
  {pages} {864} (\bibinfo {year} {2014})}\BibitemShut {NoStop}%
\bibitem [{\citenamefont {Wang}\ \emph
  {et~al.}(2017{\natexlab{a}})\citenamefont {Wang}, \citenamefont {Jacobson},
  \citenamefont {Edwards},\ and\ \citenamefont {Clark}}]{Wang2016}%
  \BibitemOpen
  \bibfield  {author} {\bibinfo {author} {\bibfnamefont {Yi-Hsieh}\
  \bibnamefont {Wang}}, \bibinfo {author} {\bibfnamefont {Ted}\ \bibnamefont
  {Jacobson}}, \bibinfo {author} {\bibfnamefont {Mark}\ \bibnamefont
  {Edwards}}, \ and\ \bibinfo {author} {\bibfnamefont {Charles~W.}\
  \bibnamefont {Clark}},\ }\bibfield  {title} {\enquote {\bibinfo {title}
  {{Mechanism of stimulated Hawking radiation in a laboratory Bose-Einstein
  condensate}},}\ }\href {\doibase 10.1103/PhysRevA.96.023616} {\bibfield
  {journal} {\bibinfo  {journal} {Phys. Rev. A}\ }\textbf {\bibinfo {volume}
  {96}},\ \bibinfo {pages} {023616} (\bibinfo {year}
  {2017}{\natexlab{a}})}\BibitemShut {NoStop}%
\bibitem [{\citenamefont {Wang}\ \emph
  {et~al.}(2017{\natexlab{b}})\citenamefont {Wang}, \citenamefont {Jacobson},
  \citenamefont {Edwards},\ and\ \citenamefont {Clark}}]{Wang2017}%
  \BibitemOpen
  \bibfield  {author} {\bibinfo {author} {\bibfnamefont {Yi-Hsieh}\
  \bibnamefont {Wang}}, \bibinfo {author} {\bibfnamefont {Ted}\ \bibnamefont
  {Jacobson}}, \bibinfo {author} {\bibfnamefont {Mark}\ \bibnamefont
  {Edwards}}, \ and\ \bibinfo {author} {\bibfnamefont {Charles~W.}\
  \bibnamefont {Clark}},\ }\bibfield  {title} {\enquote {\bibinfo {title}
  {{Induced density correlations in a sonic black hole condensate}},}\ }\href
  {\doibase 10.21468/SciPostPhys.3.3.022} {\bibfield  {journal} {\bibinfo
  {journal} {SciPost Phys.}\ }\textbf {\bibinfo {volume} {3}},\ \bibinfo
  {pages} {022} (\bibinfo {year} {2017}{\natexlab{b}})}\BibitemShut {NoStop}%
\bibitem [{\citenamefont {Kolobov}\ \emph {et~al.}(2021)\citenamefont
  {Kolobov}, \citenamefont {Golubkov}, \citenamefont {de~Nova},\ and\
  \citenamefont {Steinhauer}}]{Kolobov2021}%
  \BibitemOpen
  \bibfield  {author} {\bibinfo {author} {\bibfnamefont {Victor~I}\
  \bibnamefont {Kolobov}}, \bibinfo {author} {\bibfnamefont {Katrine}\
  \bibnamefont {Golubkov}}, \bibinfo {author} {\bibfnamefont {Juan
  Ram{\'o}n~Mu{\~n}oz}\ \bibnamefont {de~Nova}}, \ and\ \bibinfo {author}
  {\bibfnamefont {Jeff}\ \bibnamefont {Steinhauer}},\ }\bibfield  {title}
  {\enquote {\bibinfo {title} {{Observation of stationary spontaneous Hawking
  radiation and the time evolution of an analogue black hole}},}\ }\href
  {https://www.nature.com/articles/s41567-020-01076-0} {\bibfield  {journal}
  {\bibinfo  {journal} {Nature Physics}\ }\textbf {\bibinfo {volume} {17}},\
  \bibinfo {pages} {362--367} (\bibinfo {year} {2021})}\BibitemShut {NoStop}%
\bibitem [{\citenamefont {Steinhauer}(2022)}]{Steinhauer2022}%
  \BibitemOpen
  \bibfield  {author} {\bibinfo {author} {\bibfnamefont {Jeff}\ \bibnamefont
  {Steinhauer}},\ }\bibfield  {title} {\enquote {\bibinfo {title}
  {{Confirmation of stimulated Hawking radiation, but not of black hole
  lasing}},}\ }\href {\doibase 10.1103/PhysRevD.106.102007} {\bibfield
  {journal} {\bibinfo  {journal} {Phys. Rev. D}\ }\textbf {\bibinfo {volume}
  {106}},\ \bibinfo {pages} {102007} (\bibinfo {year} {2022})}\BibitemShut
  {NoStop}%
\bibitem [{\citenamefont {Tettamanti}\ \emph {et~al.}(2016)\citenamefont
  {Tettamanti}, \citenamefont {Cacciatori}, \citenamefont {Parola},\ and\
  \citenamefont {Carusotto}}]{Tettamanti2016}%
  \BibitemOpen
  \bibfield  {author} {\bibinfo {author} {\bibfnamefont {M.}~\bibnamefont
  {Tettamanti}}, \bibinfo {author} {\bibfnamefont {S.~L.}\ \bibnamefont
  {Cacciatori}}, \bibinfo {author} {\bibfnamefont {A.}~\bibnamefont {Parola}},
  \ and\ \bibinfo {author} {\bibfnamefont {I.}~\bibnamefont {Carusotto}},\
  }\bibfield  {title} {\enquote {\bibinfo {title} {Numerical study of a recent
  black-hole lasing experiment},}\ }\href
  {http://stacks.iop.org/0295-5075/114/i=6/a=60011} {\bibfield  {journal}
  {\bibinfo  {journal} {EPL (Europhysics Letters)}\ }\textbf {\bibinfo {volume}
  {114}},\ \bibinfo {pages} {60011} (\bibinfo {year} {2016})}\BibitemShut
  {NoStop}%
\bibitem [{\citenamefont {Steinhauer}\ and\ \citenamefont
  {de~Nova}(2017)}]{Steinhauer2017}%
  \BibitemOpen
  \bibfield  {author} {\bibinfo {author} {\bibfnamefont {Jeff}\ \bibnamefont
  {Steinhauer}}\ and\ \bibinfo {author} {\bibfnamefont {J.~R.~M.}\ \bibnamefont
  {de~Nova}},\ }\bibfield  {title} {\enquote {\bibinfo {title} {Self-amplifying
  hawking radiation and its background: A numerical study},}\ }\href {\doibase
  10.1103/PhysRevA.95.033604} {\bibfield  {journal} {\bibinfo  {journal} {Phys.
  Rev. A}\ }\textbf {\bibinfo {volume} {95}},\ \bibinfo {pages} {033604}
  (\bibinfo {year} {2017})}\BibitemShut {NoStop}%
\bibitem [{\citenamefont {Llorente}\ and\ \citenamefont
  {Plata}(2019)}]{Llorente2019}%
  \BibitemOpen
  \bibfield  {author} {\bibinfo {author} {\bibfnamefont {J~M~Gomez}\
  \bibnamefont {Llorente}}\ and\ \bibinfo {author} {\bibfnamefont
  {J}~\bibnamefont {Plata}},\ }\bibfield  {title} {\enquote {\bibinfo {title}
  {{Black-hole lasing in Bose{\textendash}Einstein condensates: analysis of the
  role of the dynamical instabilities in a nonstationary setup}},}\ }\href
  {\doibase 10.1088/1361-6455/ab0bcb} {\bibfield  {journal} {\bibinfo
  {journal} {J. Phys. B: At. Mol. Opt. Phys}\ }\textbf {\bibinfo {volume}
  {52}},\ \bibinfo {pages} {075004} (\bibinfo {year} {2019})}\BibitemShut
  {NoStop}%
\bibitem [{\citenamefont {Tettamanti}\ \emph {et~al.}(2021)\citenamefont
  {Tettamanti}, \citenamefont {Carusotto},\ and\ \citenamefont
  {Parola}}]{Tettamanti2021}%
  \BibitemOpen
  \bibfield  {author} {\bibinfo {author} {\bibfnamefont {M.}~\bibnamefont
  {Tettamanti}}, \bibinfo {author} {\bibfnamefont {I.}~\bibnamefont
  {Carusotto}}, \ and\ \bibinfo {author} {\bibfnamefont {A.}~\bibnamefont
  {Parola}},\ }\bibfield  {title} {\enquote {\bibinfo {title} {On the role of
  interactions in trans-sonically flowing atomic condensates},}\ }\href
  {\doibase 10.1209/0295-5075/133/20002} {\bibfield  {journal} {\bibinfo
  {journal} {Europhysics Letters}\ }\textbf {\bibinfo {volume} {133}},\
  \bibinfo {pages} {20002} (\bibinfo {year} {2021})}\BibitemShut {NoStop}%
\bibitem [{\citenamefont {Carusotto}\ \emph {et~al.}(2008)\citenamefont
  {Carusotto}, \citenamefont {Fagnocchi}, \citenamefont {Recati}, \citenamefont
  {Balbinot},\ and\ \citenamefont {Fabbri}}]{Carusotto2008}%
  \BibitemOpen
  \bibfield  {author} {\bibinfo {author} {\bibfnamefont {Iacopo}\ \bibnamefont
  {Carusotto}}, \bibinfo {author} {\bibfnamefont {Serena}\ \bibnamefont
  {Fagnocchi}}, \bibinfo {author} {\bibfnamefont {Alessio}\ \bibnamefont
  {Recati}}, \bibinfo {author} {\bibfnamefont {Roberto}\ \bibnamefont
  {Balbinot}}, \ and\ \bibinfo {author} {\bibfnamefont {Alessandro}\
  \bibnamefont {Fabbri}},\ }\bibfield  {title} {\enquote {\bibinfo {title}
  {{Numerical observation of Hawking radiation from acoustic black holes in
  atomic Bose-Einstein condensates}},}\ }\href {\doibase
  10.1088/1367-2630/10/10/103001} {\bibfield  {journal} {\bibinfo  {journal}
  {New J. Phys.}\ }\textbf {\bibinfo {volume} {10}},\ \bibinfo {pages} {103001}
  (\bibinfo {year} {2008})}\BibitemShut {NoStop}%
\bibitem [{\citenamefont {Recati}\ \emph {et~al.}(2009)\citenamefont {Recati},
  \citenamefont {Pavloff},\ and\ \citenamefont {Carusotto}}]{Recati2009}%
  \BibitemOpen
  \bibfield  {author} {\bibinfo {author} {\bibfnamefont {A.}~\bibnamefont
  {Recati}}, \bibinfo {author} {\bibfnamefont {N}~\bibnamefont {Pavloff}}, \
  and\ \bibinfo {author} {\bibfnamefont {I}~\bibnamefont {Carusotto}},\
  }\bibfield  {title} {\enquote {\bibinfo {title} {{Bogoliubov theory of
  acoustic Hawking radiation in Bose-Einstein condensates}},}\ }\href {\doibase
  10.1103/PhysRevA.80.043603} {\bibfield  {journal} {\bibinfo  {journal} {Phys.
  Rev. A}\ }\textbf {\bibinfo {volume} {80}},\ \bibinfo {pages} {43603}
  (\bibinfo {year} {2009})}\BibitemShut {NoStop}%
\bibitem [{\citenamefont {de~Nova}\ \emph
  {et~al.}(2014{\natexlab{a}})\citenamefont {de~Nova}, \citenamefont {Sols},\
  and\ \citenamefont {Zapata}}]{deNova2014}%
  \BibitemOpen
  \bibfield  {author} {\bibinfo {author} {\bibfnamefont {J.~R.~M.}\
  \bibnamefont {de~Nova}}, \bibinfo {author} {\bibfnamefont {F.}~\bibnamefont
  {Sols}}, \ and\ \bibinfo {author} {\bibfnamefont {I.}~\bibnamefont
  {Zapata}},\ }\bibfield  {title} {\enquote {\bibinfo {title} {{Violation of
  Cauchy-Schwarz inequalities by spontaneous Hawking radiation in resonant
  boson structures}},}\ }\href {\doibase 10.1103/PhysRevA.89.043808} {\bibfield
   {journal} {\bibinfo  {journal} {Phys. Rev. A}\ }\textbf {\bibinfo {volume}
  {89}},\ \bibinfo {pages} {043808} (\bibinfo {year}
  {2014}{\natexlab{a}})}\BibitemShut {NoStop}%
\bibitem [{\citenamefont {Michel}\ and\ \citenamefont
  {Parentani}(2013)}]{Michel2013}%
  \BibitemOpen
  \bibfield  {author} {\bibinfo {author} {\bibfnamefont {Florent}\ \bibnamefont
  {Michel}}\ and\ \bibinfo {author} {\bibfnamefont {Renaud}\ \bibnamefont
  {Parentani}},\ }\bibfield  {title} {\enquote {\bibinfo {title} {{Saturation
  of black hole lasers in Bose-Einstein condensates}},}\ }\href {\doibase
  10.1103/PhysRevD.88.125012} {\bibfield  {journal} {\bibinfo  {journal} {Phys.
  Rev. D}\ }\textbf {\bibinfo {volume} {88}},\ \bibinfo {pages} {125012}
  (\bibinfo {year} {2013})}\BibitemShut {NoStop}%
\bibitem [{\citenamefont {Jacquet}\ \emph {et~al.}(2023)\citenamefont
  {Jacquet}, \citenamefont {Giacomelli}, \citenamefont {Valnais}, \citenamefont
  {Joly}, \citenamefont {Claude}, \citenamefont {Giacobino}, \citenamefont
  {Glorieux}, \citenamefont {Carusotto},\ and\ \citenamefont
  {Bramati}}]{Jacquet2023}%
  \BibitemOpen
  \bibfield  {author} {\bibinfo {author} {\bibfnamefont {M.~J.}\ \bibnamefont
  {Jacquet}}, \bibinfo {author} {\bibfnamefont {L.}~\bibnamefont {Giacomelli}},
  \bibinfo {author} {\bibfnamefont {Q.}~\bibnamefont {Valnais}}, \bibinfo
  {author} {\bibfnamefont {M.}~\bibnamefont {Joly}}, \bibinfo {author}
  {\bibfnamefont {F.}~\bibnamefont {Claude}}, \bibinfo {author} {\bibfnamefont
  {E.}~\bibnamefont {Giacobino}}, \bibinfo {author} {\bibfnamefont
  {Q.}~\bibnamefont {Glorieux}}, \bibinfo {author} {\bibfnamefont
  {I.}~\bibnamefont {Carusotto}}, \ and\ \bibinfo {author} {\bibfnamefont
  {A.}~\bibnamefont {Bramati}},\ }\bibfield  {title} {\enquote {\bibinfo
  {title} {Quantum vacuum excitation of a quasinormal mode in an analog model
  of black hole spacetime},}\ }\href {\doibase 10.1103/PhysRevLett.130.111501}
  {\bibfield  {journal} {\bibinfo  {journal} {Phys. Rev. Lett.}\ }\textbf
  {\bibinfo {volume} {130}},\ \bibinfo {pages} {111501} (\bibinfo {year}
  {2023})}\BibitemShut {NoStop}%
\bibitem [{\citenamefont {de~Nova}\ \emph {et~al.}(2021)\citenamefont
  {de~Nova}, \citenamefont {Palacios}, \citenamefont {Carusotto},\ and\
  \citenamefont {Sols}}]{deNova2021a}%
  \BibitemOpen
  \bibfield  {author} {\bibinfo {author} {\bibfnamefont {J.~R.~M.}\
  \bibnamefont {de~Nova}}, \bibinfo {author} {\bibfnamefont {P.~F.}\
  \bibnamefont {Palacios}}, \bibinfo {author} {\bibfnamefont {I.}~\bibnamefont
  {Carusotto}}, \ and\ \bibinfo {author} {\bibfnamefont {F.}~\bibnamefont
  {Sols}},\ }\bibfield  {title} {\enquote {\bibinfo {title} {Long time
  universality of black-hole lasers},}\ }\href {\doibase
  10.1088/1367-2630/abdce2} {\bibfield  {journal} {\bibinfo  {journal} {New
  Journal of Physics}\ }\textbf {\bibinfo {volume} {23}},\ \bibinfo {pages}
  {023040} (\bibinfo {year} {2021})}\BibitemShut {NoStop}%
\bibitem [{\citenamefont {Sinatra}\ \emph {et~al.}(2002)\citenamefont
  {Sinatra}, \citenamefont {Lobo},\ and\ \citenamefont {Castin}}]{Sinatra2002}%
  \BibitemOpen
  \bibfield  {author} {\bibinfo {author} {\bibfnamefont {Alice}\ \bibnamefont
  {Sinatra}}, \bibinfo {author} {\bibfnamefont {Carlos}\ \bibnamefont {Lobo}},
  \ and\ \bibinfo {author} {\bibfnamefont {Yvan}\ \bibnamefont {Castin}},\
  }\bibfield  {title} {\enquote {\bibinfo {title} {{The truncated Wigner method
  for Bose-condensed gases: limits of validity and applications}},}\ }\href
  {http://stacks.iop.org/0953-4075/35/i=17/a=301} {\bibfield  {journal}
  {\bibinfo  {journal} {J. Phys. B: At. Mol. Opt. Phys}\ }\textbf {\bibinfo
  {volume} {35}},\ \bibinfo {pages} {3599} (\bibinfo {year}
  {2002})}\BibitemShut {NoStop}%
\bibitem [{\citenamefont {Steinhauer}(2016)}]{Steinhauer2016}%
  \BibitemOpen
  \bibfield  {author} {\bibinfo {author} {\bibfnamefont {J.}~\bibnamefont
  {Steinhauer}},\ }\bibfield  {title} {\enquote {\bibinfo {title} {{Observation
  of quantum Hawking radiation and its entanglement in an analogue black
  hole}},}\ }\href {http://dx.doi.org/10.1038/nphys3863} {\bibfield  {journal}
  {\bibinfo  {journal} {Nature Physics}\ }\textbf {\bibinfo {volume} {12}},\
  \bibinfo {pages} {959} (\bibinfo {year} {2016})}\BibitemShut {NoStop}%
\bibitem [{\citenamefont {Mayoral}\ \emph {et~al.}(2011)\citenamefont
  {Mayoral}, \citenamefont {Recati}, \citenamefont {Fabbri}, \citenamefont
  {Parentani}, \citenamefont {Balbinot},\ and\ \citenamefont
  {Carusotto}}]{Mayoral2011}%
  \BibitemOpen
  \bibfield  {author} {\bibinfo {author} {\bibfnamefont {Carlos}\ \bibnamefont
  {Mayoral}}, \bibinfo {author} {\bibfnamefont {Alessio}\ \bibnamefont
  {Recati}}, \bibinfo {author} {\bibfnamefont {Alessandro}\ \bibnamefont
  {Fabbri}}, \bibinfo {author} {\bibfnamefont {Renaud}\ \bibnamefont
  {Parentani}}, \bibinfo {author} {\bibfnamefont {Roberto}\ \bibnamefont
  {Balbinot}}, \ and\ \bibinfo {author} {\bibfnamefont {Iacopo}\ \bibnamefont
  {Carusotto}},\ }\bibfield  {title} {\enquote {\bibinfo {title} {{Acoustic
  white holes in flowing atomic Bose-Einstein condensates}},}\ }\href {\doibase
  10.1088/1367-2630/13/2/025007} {\bibfield  {journal} {\bibinfo  {journal}
  {New Journal of Physics}\ }\textbf {\bibinfo {volume} {13}},\ \bibinfo
  {pages} {025007} (\bibinfo {year} {2011})}\BibitemShut {NoStop}%
\bibitem [{\citenamefont {Zapata}\ \emph {et~al.}(2011)\citenamefont {Zapata},
  \citenamefont {Albert}, \citenamefont {Parentani},\ and\ \citenamefont
  {Sols}}]{Zapata2011}%
  \BibitemOpen
  \bibfield  {author} {\bibinfo {author} {\bibfnamefont {I}~\bibnamefont
  {Zapata}}, \bibinfo {author} {\bibfnamefont {M}~\bibnamefont {Albert}},
  \bibinfo {author} {\bibfnamefont {R}~\bibnamefont {Parentani}}, \ and\
  \bibinfo {author} {\bibfnamefont {F}~\bibnamefont {Sols}},\ }\bibfield
  {title} {\enquote {\bibinfo {title} {{Resonant Hawking radiation in
  Bose-Einstein condensates}},}\ }\href {\doibase
  10.1088/1367-2630/13/6/063048} {\bibfield  {journal} {\bibinfo  {journal}
  {New J. Phys.}\ }\textbf {\bibinfo {volume} {13}},\ \bibinfo {pages} {063048}
  (\bibinfo {year} {2011})}\BibitemShut {NoStop}%
\bibitem [{\citenamefont {Amico}\ \emph {et~al.}(2021)\citenamefont {Amico},
  \citenamefont {Boshier}, \citenamefont {Birkl}, \citenamefont {Minguzzi},
  \citenamefont {Miniatura}, \citenamefont {Kwek}, \citenamefont {Aghamalyan},
  \citenamefont {Ahufinger}, \citenamefont {Anderson}, \citenamefont {Andrei},
  \citenamefont {Arnold}, \citenamefont {Baker}, \citenamefont {Bell},
  \citenamefont {Bland}, \citenamefont {Brantut}, \citenamefont {Cassettari},
  \citenamefont {Chetcuti}, \citenamefont {Chevy}, \citenamefont {Citro},
  \citenamefont {De~Palo}, \citenamefont {Dumke}, \citenamefont {Edwards},
  \citenamefont {Folman}, \citenamefont {Fortagh}, \citenamefont {Gardiner},
  \citenamefont {Garraway}, \citenamefont {Gauthier}, \citenamefont
  {G\"unther}, \citenamefont {Haug}, \citenamefont {Hufnagel}, \citenamefont
  {Keil}, \citenamefont {Ireland}, \citenamefont {Lebrat}, \citenamefont {Li},
  \citenamefont {Longchambon}, \citenamefont {Mompart}, \citenamefont {Morsch},
  \citenamefont {Naldesi}, \citenamefont {Neely}, \citenamefont {Olshanii},
  \citenamefont {Orignac}, \citenamefont {Pandey}, \citenamefont
  {P\'erez-Obiol}, \citenamefont {Perrin}, \citenamefont {Piroli},
  \citenamefont {Polo}, \citenamefont {Pritchard}, \citenamefont {Proukakis},
  \citenamefont {Rylands}, \citenamefont {Rubinsztein-Dunlop}, \citenamefont
  {Scazza}, \citenamefont {Stringari}, \citenamefont {Tosto}, \citenamefont
  {Trombettoni}, \citenamefont {Victorin}, \citenamefont {Klitzing},
  \citenamefont {Wilkowski}, \citenamefont {Xhani},\ and\ \citenamefont
  {Yakimenko}}]{Amico2021}%
  \BibitemOpen
  \bibfield  {author} {\bibinfo {author} {\bibfnamefont {L.}~\bibnamefont
  {Amico}}, \bibinfo {author} {\bibfnamefont {M.}~\bibnamefont {Boshier}},
  \bibinfo {author} {\bibfnamefont {G.}~\bibnamefont {Birkl}}, \bibinfo
  {author} {\bibfnamefont {A.}~\bibnamefont {Minguzzi}}, \bibinfo {author}
  {\bibfnamefont {C.}~\bibnamefont {Miniatura}}, \bibinfo {author}
  {\bibfnamefont {L.-C.}\ \bibnamefont {Kwek}}, \bibinfo {author}
  {\bibfnamefont {D.}~\bibnamefont {Aghamalyan}}, \bibinfo {author}
  {\bibfnamefont {V.}~\bibnamefont {Ahufinger}}, \bibinfo {author}
  {\bibfnamefont {D.}~\bibnamefont {Anderson}}, \bibinfo {author}
  {\bibfnamefont {N.}~\bibnamefont {Andrei}}, \bibinfo {author} {\bibfnamefont
  {A.~S.}\ \bibnamefont {Arnold}}, \bibinfo {author} {\bibfnamefont
  {M.}~\bibnamefont {Baker}}, \bibinfo {author} {\bibfnamefont {T.~A.}\
  \bibnamefont {Bell}}, \bibinfo {author} {\bibfnamefont {T.}~\bibnamefont
  {Bland}}, \bibinfo {author} {\bibfnamefont {J.~P.}\ \bibnamefont {Brantut}},
  \bibinfo {author} {\bibfnamefont {D.}~\bibnamefont {Cassettari}}, \bibinfo
  {author} {\bibfnamefont {W.~J.}\ \bibnamefont {Chetcuti}}, \bibinfo {author}
  {\bibfnamefont {F.}~\bibnamefont {Chevy}}, \bibinfo {author} {\bibfnamefont
  {R.}~\bibnamefont {Citro}}, \bibinfo {author} {\bibfnamefont
  {S.}~\bibnamefont {De~Palo}}, \bibinfo {author} {\bibfnamefont
  {R.}~\bibnamefont {Dumke}}, \bibinfo {author} {\bibfnamefont
  {M.}~\bibnamefont {Edwards}}, \bibinfo {author} {\bibfnamefont
  {R.}~\bibnamefont {Folman}}, \bibinfo {author} {\bibfnamefont
  {J.}~\bibnamefont {Fortagh}}, \bibinfo {author} {\bibfnamefont {S.~A.}\
  \bibnamefont {Gardiner}}, \bibinfo {author} {\bibfnamefont {B.~M.}\
  \bibnamefont {Garraway}}, \bibinfo {author} {\bibfnamefont {G.}~\bibnamefont
  {Gauthier}}, \bibinfo {author} {\bibfnamefont {A.}~\bibnamefont {G\"unther}},
  \bibinfo {author} {\bibfnamefont {T.}~\bibnamefont {Haug}}, \bibinfo {author}
  {\bibfnamefont {C.}~\bibnamefont {Hufnagel}}, \bibinfo {author}
  {\bibfnamefont {M.}~\bibnamefont {Keil}}, \bibinfo {author} {\bibfnamefont
  {P.}~\bibnamefont {Ireland}}, \bibinfo {author} {\bibfnamefont
  {M.}~\bibnamefont {Lebrat}}, \bibinfo {author} {\bibfnamefont
  {W.}~\bibnamefont {Li}}, \bibinfo {author} {\bibfnamefont {L.}~\bibnamefont
  {Longchambon}}, \bibinfo {author} {\bibfnamefont {J.}~\bibnamefont
  {Mompart}}, \bibinfo {author} {\bibfnamefont {O.}~\bibnamefont {Morsch}},
  \bibinfo {author} {\bibfnamefont {P.}~\bibnamefont {Naldesi}}, \bibinfo
  {author} {\bibfnamefont {T.~W.}\ \bibnamefont {Neely}}, \bibinfo {author}
  {\bibfnamefont {M.}~\bibnamefont {Olshanii}}, \bibinfo {author}
  {\bibfnamefont {E.}~\bibnamefont {Orignac}}, \bibinfo {author} {\bibfnamefont
  {S.}~\bibnamefont {Pandey}}, \bibinfo {author} {\bibfnamefont
  {A.}~\bibnamefont {P\'erez-Obiol}}, \bibinfo {author} {\bibfnamefont
  {H.}~\bibnamefont {Perrin}}, \bibinfo {author} {\bibfnamefont
  {L.}~\bibnamefont {Piroli}}, \bibinfo {author} {\bibfnamefont
  {J.}~\bibnamefont {Polo}}, \bibinfo {author} {\bibfnamefont {A.~L.}\
  \bibnamefont {Pritchard}}, \bibinfo {author} {\bibfnamefont {N.~P.}\
  \bibnamefont {Proukakis}}, \bibinfo {author} {\bibfnamefont {C.}~\bibnamefont
  {Rylands}}, \bibinfo {author} {\bibfnamefont {H.}~\bibnamefont
  {Rubinsztein-Dunlop}}, \bibinfo {author} {\bibfnamefont {F.}~\bibnamefont
  {Scazza}}, \bibinfo {author} {\bibfnamefont {S.}~\bibnamefont {Stringari}},
  \bibinfo {author} {\bibfnamefont {F.}~\bibnamefont {Tosto}}, \bibinfo
  {author} {\bibfnamefont {A.}~\bibnamefont {Trombettoni}}, \bibinfo {author}
  {\bibfnamefont {N.}~\bibnamefont {Victorin}}, \bibinfo {author}
  {\bibfnamefont {W.~von}\ \bibnamefont {Klitzing}}, \bibinfo {author}
  {\bibfnamefont {D.}~\bibnamefont {Wilkowski}}, \bibinfo {author}
  {\bibfnamefont {K.}~\bibnamefont {Xhani}}, \ and\ \bibinfo {author}
  {\bibfnamefont {A.}~\bibnamefont {Yakimenko}},\ }\bibfield  {title} {\enquote
  {\bibinfo {title} {{Roadmap on Atomtronics: State of the art and
  perspective}},}\ }\href {https://doi.org/10.1116/5.0026178} {\bibfield
  {journal} {\bibinfo  {journal} {AVS Quantum Science}\ }\textbf {\bibinfo
  {volume} {3}},\ \bibinfo {pages} {039201} (\bibinfo {year}
  {2021})}\BibitemShut {NoStop}%
\bibitem [{\citenamefont {de~Nova}\ \emph {et~al.}(2016)\citenamefont
  {de~Nova}, \citenamefont {Finazzi},\ and\ \citenamefont
  {Carusotto}}]{deNova2016}%
  \BibitemOpen
  \bibfield  {author} {\bibinfo {author} {\bibfnamefont {J.~R.~M.}\
  \bibnamefont {de~Nova}}, \bibinfo {author} {\bibfnamefont {S.}~\bibnamefont
  {Finazzi}}, \ and\ \bibinfo {author} {\bibfnamefont {I.}~\bibnamefont
  {Carusotto}},\ }\bibfield  {title} {\enquote {\bibinfo {title}
  {{Time-dependent study of a black-hole laser in a flowing atomic
  condensate}},}\ }\href {\doibase 10.1103/PhysRevA.94.043616} {\bibfield
  {journal} {\bibinfo  {journal} {Phys. Rev. A}\ }\textbf {\bibinfo {volume}
  {94}},\ \bibinfo {pages} {043616} (\bibinfo {year} {2016})}\BibitemShut
  {NoStop}%
\bibitem [{\citenamefont {Menotti}\ and\ \citenamefont
  {Stringari}(2002)}]{Menotti2002}%
  \BibitemOpen
  \bibfield  {author} {\bibinfo {author} {\bibfnamefont {Chiara}\ \bibnamefont
  {Menotti}}\ and\ \bibinfo {author} {\bibfnamefont {Sandro}\ \bibnamefont
  {Stringari}},\ }\bibfield  {title} {\enquote {\bibinfo {title} {{Collective
  oscillations of a one-dimensional trapped Bose-Einstein gas}},}\ }\href
  {\doibase 10.1103/PhysRevA.66.043610} {\bibfield  {journal} {\bibinfo
  {journal} {Phys. Rev. A}\ }\textbf {\bibinfo {volume} {66}},\ \bibinfo
  {pages} {043610} (\bibinfo {year} {2002})}\BibitemShut {NoStop}%
\bibitem [{\citenamefont {de~Nova}\ \emph {et~al.}(2017)\citenamefont
  {de~Nova}, \citenamefont {Sols},\ and\ \citenamefont {Zapata}}]{deNova2017b}%
  \BibitemOpen
  \bibfield  {author} {\bibinfo {author} {\bibfnamefont {J.~R.~M.}\
  \bibnamefont {de~Nova}}, \bibinfo {author} {\bibfnamefont {F.}~\bibnamefont
  {Sols}}, \ and\ \bibinfo {author} {\bibfnamefont {I.}~\bibnamefont
  {Zapata}},\ }\bibfield  {title} {\enquote {\bibinfo {title} {{Quantum
  Transport in the Black-Hole Configuration of an Atom Condensate Outcoupled
  Through an Optical Lattice}},}\ }\href {\doibase 10.1002/andp.201600385}
  {\bibfield  {journal} {\bibinfo  {journal} {Annalen der Physik}\ }\textbf
  {\bibinfo {volume} {529}},\ \bibinfo {pages} {1600385} (\bibinfo {year}
  {2017})}\BibitemShut {NoStop}%
\bibitem [{\citenamefont {Wu}\ and\ \citenamefont {Niu}(2003)}]{Wu2003}%
  \BibitemOpen
  \bibfield  {author} {\bibinfo {author} {\bibfnamefont {Biao}\ \bibnamefont
  {Wu}}\ and\ \bibinfo {author} {\bibfnamefont {Qian}\ \bibnamefont {Niu}},\
  }\bibfield  {title} {\enquote {\bibinfo {title} {{Superfluidity of
  Bose-Einstein condensate in an optical lattice: Landau-Zener tunnelling and
  dynamical instability}},}\ }\href
  {http://iopscience.iop.org/1367-2630/5/1/104} {\bibfield  {journal} {\bibinfo
   {journal} {New J. Phys.}\ }\textbf {\bibinfo {volume} {5}},\ \bibinfo
  {pages} {104} (\bibinfo {year} {2003})}\BibitemShut {NoStop}%
\bibitem [{\citenamefont {Michel}\ and\ \citenamefont
  {Parentani}(2015)}]{Michel2015}%
  \BibitemOpen
  \bibfield  {author} {\bibinfo {author} {\bibfnamefont {Florent}\ \bibnamefont
  {Michel}}\ and\ \bibinfo {author} {\bibfnamefont {Renaud}\ \bibnamefont
  {Parentani}},\ }\bibfield  {title} {\enquote {\bibinfo {title} {Nonlinear
  effects in time-dependent transonic flows: An analysis of analog black hole
  stability},}\ }\href {\doibase 10.1103/PhysRevA.91.053603} {\bibfield
  {journal} {\bibinfo  {journal} {Phys. Rev. A}\ }\textbf {\bibinfo {volume}
  {91}},\ \bibinfo {pages} {053603} (\bibinfo {year} {2015})}\BibitemShut
  {NoStop}%
\bibitem [{\citenamefont {Ribeiro}\ \emph {et~al.}(2022)\citenamefont
  {Ribeiro}, \citenamefont {Baak},\ and\ \citenamefont
  {Fischer}}]{Ribeiro2022}%
  \BibitemOpen
  \bibfield  {author} {\bibinfo {author} {\bibfnamefont {Caio C.~Holanda}\
  \bibnamefont {Ribeiro}}, \bibinfo {author} {\bibfnamefont {Sang-Shin}\
  \bibnamefont {Baak}}, \ and\ \bibinfo {author} {\bibfnamefont {Uwe~R.}\
  \bibnamefont {Fischer}},\ }\bibfield  {title} {\enquote {\bibinfo {title}
  {{Existence of steady-state black hole analogs in finite
  quasi-one-dimensional Bose-Einstein condensates}},}\ }\href {\doibase
  10.1103/PhysRevD.105.124066} {\bibfield  {journal} {\bibinfo  {journal}
  {Phys. Rev. D}\ }\textbf {\bibinfo {volume} {105}},\ \bibinfo {pages}
  {124066} (\bibinfo {year} {2022})}\BibitemShut {NoStop}%
\bibitem [{\citenamefont {Shammass}\ \emph {et~al.}(2012)\citenamefont
  {Shammass}, \citenamefont {Rinott}, \citenamefont {Berkovitz}, \citenamefont
  {Schley},\ and\ \citenamefont {Steinhauer}}]{Shammass2012}%
  \BibitemOpen
  \bibfield  {author} {\bibinfo {author} {\bibfnamefont {I.}~\bibnamefont
  {Shammass}}, \bibinfo {author} {\bibfnamefont {S.}~\bibnamefont {Rinott}},
  \bibinfo {author} {\bibfnamefont {A.}~\bibnamefont {Berkovitz}}, \bibinfo
  {author} {\bibfnamefont {R.}~\bibnamefont {Schley}}, \ and\ \bibinfo {author}
  {\bibfnamefont {J.}~\bibnamefont {Steinhauer}},\ }\bibfield  {title}
  {\enquote {\bibinfo {title} {{Phonon Dispersion Relation of an Atomic
  Bose-Einstein Condensate}},}\ }\href {\doibase
  10.1103/PhysRevLett.109.195301} {\bibfield  {journal} {\bibinfo  {journal}
  {Phys. Rev. Lett.}\ }\textbf {\bibinfo {volume} {109}},\ \bibinfo {pages}
  {195301} (\bibinfo {year} {2012})}\BibitemShut {NoStop}%
\bibitem [{\citenamefont {Deuar}\ \emph {et~al.}(2009)\citenamefont {Deuar},
  \citenamefont {Sykes}, \citenamefont {Gangardt}, \citenamefont {Davis},
  \citenamefont {Drummond},\ and\ \citenamefont {Kheruntsyan}}]{Deuar2003}%
  \BibitemOpen
  \bibfield  {author} {\bibinfo {author} {\bibfnamefont {P.}~\bibnamefont
  {Deuar}}, \bibinfo {author} {\bibfnamefont {A.~G.}\ \bibnamefont {Sykes}},
  \bibinfo {author} {\bibfnamefont {D.~M.}\ \bibnamefont {Gangardt}}, \bibinfo
  {author} {\bibfnamefont {M.~J.}\ \bibnamefont {Davis}}, \bibinfo {author}
  {\bibfnamefont {P.~D.}\ \bibnamefont {Drummond}}, \ and\ \bibinfo {author}
  {\bibfnamefont {K.~V.}\ \bibnamefont {Kheruntsyan}},\ }\bibfield  {title}
  {\enquote {\bibinfo {title} {Nonlocal pair correlations in the
  one-dimensional bose gas at finite temperature},}\ }\href {\doibase
  10.1103/PhysRevA.79.043619} {\bibfield  {journal} {\bibinfo  {journal} {Phys.
  Rev. A}\ }\textbf {\bibinfo {volume} {79}},\ \bibinfo {pages} {043619}
  (\bibinfo {year} {2009})}\BibitemShut {NoStop}%
\bibitem [{\citenamefont {Larr\'{e}}\ \emph {et~al.}(2012)\citenamefont
  {Larr\'{e}}, \citenamefont {Recati}, \citenamefont {Carusotto},\ and\
  \citenamefont {Pavloff}}]{Larre2012}%
  \BibitemOpen
  \bibfield  {author} {\bibinfo {author} {\bibfnamefont {P~\'{E}.}\
  \bibnamefont {Larr\'{e}}}, \bibinfo {author} {\bibfnamefont {A.}~\bibnamefont
  {Recati}}, \bibinfo {author} {\bibfnamefont {I}~\bibnamefont {Carusotto}}, \
  and\ \bibinfo {author} {\bibfnamefont {N}~\bibnamefont {Pavloff}},\
  }\bibfield  {title} {\enquote {\bibinfo {title} {{Quantum fluctuations around
  black hole horizons in Bose-Einstein condensates}},}\ }\href {\doibase
  10.1103/PhysRevA.85.013621} {\bibfield  {journal} {\bibinfo  {journal} {Phys.
  Rev. A}\ }\textbf {\bibinfo {volume} {85}},\ \bibinfo {pages} {13621}
  (\bibinfo {year} {2012})}\BibitemShut {NoStop}%
\bibitem [{\citenamefont {Mora}\ and\ \citenamefont {Castin}(2003)}]{Mora2003}%
  \BibitemOpen
  \bibfield  {author} {\bibinfo {author} {\bibfnamefont {Christophe}\
  \bibnamefont {Mora}}\ and\ \bibinfo {author} {\bibfnamefont {Yvan}\
  \bibnamefont {Castin}},\ }\bibfield  {title} {\enquote {\bibinfo {title}
  {{Extension of Bogoliubov theory to quasicondensates}},}\ }\href {\doibase
  10.1103/PhysRevA.67.053615} {\bibfield  {journal} {\bibinfo  {journal} {Phys.
  Rev. A}\ }\textbf {\bibinfo {volume} {67}},\ \bibinfo {pages} {053615}
  (\bibinfo {year} {2003})}\BibitemShut {NoStop}%
\bibitem [{\citenamefont {de~Nova}\ and\ \citenamefont
  {Sols}(2022)}]{deNova2022}%
  \BibitemOpen
  \bibfield  {author} {\bibinfo {author} {\bibfnamefont {J.~R.~M.}\
  \bibnamefont {de~Nova}}\ and\ \bibinfo {author} {\bibfnamefont
  {F.}~\bibnamefont {Sols}},\ }\bibfield  {title} {\enquote {\bibinfo {title}
  {{Continuous-time crystal from a spontaneous many-body Floquet state}},}\
  }\href {\doibase 10.1103/PhysRevA.105.043302} {\bibfield  {journal} {\bibinfo
   {journal} {Phys. Rev. A}\ }\textbf {\bibinfo {volume} {105}},\ \bibinfo
  {pages} {043302} (\bibinfo {year} {2022})}\BibitemShut {NoStop}%
\bibitem [{\citenamefont {Jacquet}\ \emph {et~al.}(2022)\citenamefont
  {Jacquet}, \citenamefont {Joly}, \citenamefont {Claude}, \citenamefont
  {Giacomelli}, \citenamefont {Glorieux}, \citenamefont {Bramati},
  \citenamefont {Carusotto},\ and\ \citenamefont {Giacobino}}]{Jacquet2022}%
  \BibitemOpen
  \bibfield  {author} {\bibinfo {author} {\bibfnamefont {Maxime}\ \bibnamefont
  {Jacquet}}, \bibinfo {author} {\bibfnamefont {Malo}\ \bibnamefont {Joly}},
  \bibinfo {author} {\bibfnamefont {Ferdinand}\ \bibnamefont {Claude}},
  \bibinfo {author} {\bibfnamefont {Luca}\ \bibnamefont {Giacomelli}}, \bibinfo
  {author} {\bibfnamefont {Quentin}\ \bibnamefont {Glorieux}}, \bibinfo
  {author} {\bibfnamefont {Alberto}\ \bibnamefont {Bramati}}, \bibinfo {author}
  {\bibfnamefont {Iacopo}\ \bibnamefont {Carusotto}}, \ and\ \bibinfo {author}
  {\bibfnamefont {Elisabeth}\ \bibnamefont {Giacobino}},\ }\bibfield  {title}
  {\enquote {\bibinfo {title} {Analogue quantum simulation of the hawking
  effect in a polariton superfluid},}\ }\href
  {https://link.springer.com/article/10.1140/epjd/s10053-022-00477-5}
  {\bibfield  {journal} {\bibinfo  {journal} {The European Physical Journal D}\
  }\textbf {\bibinfo {volume} {76}},\ \bibinfo {pages} {152} (\bibinfo {year}
  {2022})}\BibitemShut {NoStop}%
\bibitem [{\citenamefont {Chin}\ \emph {et~al.}(2010)\citenamefont {Chin},
  \citenamefont {Grimm}, \citenamefont {Julienne},\ and\ \citenamefont
  {Tiesinga}}]{Chin2010}%
  \BibitemOpen
  \bibfield  {author} {\bibinfo {author} {\bibfnamefont {Cheng}\ \bibnamefont
  {Chin}}, \bibinfo {author} {\bibfnamefont {Rudolf}\ \bibnamefont {Grimm}},
  \bibinfo {author} {\bibfnamefont {Paul}\ \bibnamefont {Julienne}}, \ and\
  \bibinfo {author} {\bibfnamefont {Eite}\ \bibnamefont {Tiesinga}},\
  }\bibfield  {title} {\enquote {\bibinfo {title} {Feshbach resonances in
  ultracold gases},}\ }\href {\doibase 10.1103/RevModPhys.82.1225} {\bibfield
  {journal} {\bibinfo  {journal} {Rev. Mod. Phys.}\ }\textbf {\bibinfo {volume}
  {82}},\ \bibinfo {pages} {1225--1286} (\bibinfo {year} {2010})}\BibitemShut
  {NoStop}%
\bibitem [{\citenamefont {Zapata}\ and\ \citenamefont
  {Sols}(2009)}]{Zapata2009a}%
  \BibitemOpen
  \bibfield  {author} {\bibinfo {author} {\bibfnamefont {I}~\bibnamefont
  {Zapata}}\ and\ \bibinfo {author} {\bibfnamefont {F}~\bibnamefont {Sols}},\
  }\bibfield  {title} {\enquote {\bibinfo {title} {{Andreev Reflection in
  Bosonic Condensates}},}\ }\href {\doibase 10.1103/PhysRevLett.102.180405}
  {\bibfield  {journal} {\bibinfo  {journal} {Phys. Rev. Lett.}\ }\textbf
  {\bibinfo {volume} {102}},\ \bibinfo {pages} {180405} (\bibinfo {year}
  {2009})}\BibitemShut {NoStop}%
\bibitem [{\citenamefont {Coutant}\ and\ \citenamefont
  {Parentani}(2014)}]{Coutant2014}%
  \BibitemOpen
  \bibfield  {author} {\bibinfo {author} {\bibfnamefont {Antonin}\ \bibnamefont
  {Coutant}}\ and\ \bibinfo {author} {\bibfnamefont {Renaud}\ \bibnamefont
  {Parentani}},\ }\bibfield  {title} {\enquote {\bibinfo {title} {{Undulations
  from amplified low frequency surface waves}},}\ }\href {\doibase
  10.1063/1.4872025} {\bibfield  {journal} {\bibinfo  {journal} {Physics of
  Fluids}\ }\textbf {\bibinfo {volume} {26}},\ \bibinfo {pages} {044106}
  (\bibinfo {year} {2014})}\BibitemShut {NoStop}%
\bibitem [{\citenamefont {Bossard}\ \emph {et~al.}(2023)\citenamefont
  {Bossard}, \citenamefont {James}, \citenamefont {Aucouturier}, \citenamefont
  {Fourdrinoy}, \citenamefont {Robertson},\ and\ \citenamefont
  {Rousseaux}}]{Bossard2023}%
  \BibitemOpen
  \bibfield  {author} {\bibinfo {author} {\bibfnamefont {Alexis}\ \bibnamefont
  {Bossard}}, \bibinfo {author} {\bibfnamefont {Nicolas}\ \bibnamefont
  {James}}, \bibinfo {author} {\bibfnamefont {Camille}\ \bibnamefont
  {Aucouturier}}, \bibinfo {author} {\bibfnamefont {Johan}\ \bibnamefont
  {Fourdrinoy}}, \bibinfo {author} {\bibfnamefont {Scott}\ \bibnamefont
  {Robertson}}, \ and\ \bibinfo {author} {\bibfnamefont {Germain}\ \bibnamefont
  {Rousseaux}},\ }\bibfield  {title} {\enquote {\bibinfo {title} {How to create
  analogue black hole or white fountain horizons and laser cavities in
  experimental free surface hydrodynamics?}}\ }\href
  {https://arxiv.org/abs/2307.11022} {\bibfield  {journal} {\bibinfo  {journal}
  {arXiv preprint arXiv:2307.11022}\ } (\bibinfo {year} {2023})}\BibitemShut
  {NoStop}%
\bibitem [{\citenamefont {Balbinot}\ \emph {et~al.}(2005)\citenamefont
  {Balbinot}, \citenamefont {Fagnocchi},\ and\ \citenamefont
  {Procopio}}]{Balbinot2005a}%
  \BibitemOpen
  \bibfield  {author} {\bibinfo {author} {\bibfnamefont {Roberto}\ \bibnamefont
  {Balbinot}}, \bibinfo {author} {\bibfnamefont {Serena}\ \bibnamefont
  {Fagnocchi}}, \ and\ \bibinfo {author} {\bibfnamefont {Giovanni~P}\
  \bibnamefont {Procopio}},\ }\bibfield  {title} {\enquote {\bibinfo {title}
  {{Backreaction in Acoustic Black Holes}},}\ }\href {\doibase
  10.1103/PhysRevLett.94.161302} {\bibfield  {journal} {\bibinfo  {journal}
  {Phys. Rev. Lett.}\ }\textbf {\bibinfo {volume} {94}},\ \bibinfo {pages}
  {161302} (\bibinfo {year} {2005})}\BibitemShut {NoStop}%
\bibitem [{\citenamefont {Baak}\ \emph {et~al.}(2022)\citenamefont {Baak},
  \citenamefont {Ribeiro},\ and\ \citenamefont {Fischer}}]{Baak2022}%
  \BibitemOpen
  \bibfield  {author} {\bibinfo {author} {\bibfnamefont {Sang-Shin}\
  \bibnamefont {Baak}}, \bibinfo {author} {\bibfnamefont {Caio C.~Holanda}\
  \bibnamefont {Ribeiro}}, \ and\ \bibinfo {author} {\bibfnamefont {Uwe~R.}\
  \bibnamefont {Fischer}},\ }\bibfield  {title} {\enquote {\bibinfo {title}
  {{Number-conserving solution for dynamical quantum backreaction in a
  Bose-Einstein condensate}},}\ }\href {\doibase 10.1103/PhysRevA.106.053319}
  {\bibfield  {journal} {\bibinfo  {journal} {Phys. Rev. A}\ }\textbf {\bibinfo
  {volume} {106}},\ \bibinfo {pages} {053319} (\bibinfo {year}
  {2022})}\BibitemShut {NoStop}%
\bibitem [{\citenamefont {Butera}\ and\ \citenamefont
  {Carusotto}(2023)}]{Butera2023}%
  \BibitemOpen
  \bibfield  {author} {\bibinfo {author} {\bibfnamefont {Salvatore}\
  \bibnamefont {Butera}}\ and\ \bibinfo {author} {\bibfnamefont {Iacopo}\
  \bibnamefont {Carusotto}},\ }\bibfield  {title} {\enquote {\bibinfo {title}
  {Numerical studies of back reaction effects in an analog model of
  cosmological preheating},}\ }\href {\doibase 10.1103/PhysRevLett.130.241501}
  {\bibfield  {journal} {\bibinfo  {journal} {Phys. Rev. Lett.}\ }\textbf
  {\bibinfo {volume} {130}},\ \bibinfo {pages} {241501} (\bibinfo {year}
  {2023})}\BibitemShut {NoStop}%
\bibitem [{\citenamefont {de~Nova}\ \emph
  {et~al.}(2014{\natexlab{b}})\citenamefont {de~Nova}, \citenamefont
  {Gu\'{e}ry-Odelin}, \citenamefont {Sols},\ and\ \citenamefont
  {Zapata}}]{deNova2014a}%
  \BibitemOpen
  \bibfield  {author} {\bibinfo {author} {\bibfnamefont {J~R~M}\ \bibnamefont
  {de~Nova}}, \bibinfo {author} {\bibfnamefont {D}~\bibnamefont
  {Gu\'{e}ry-Odelin}}, \bibinfo {author} {\bibfnamefont {F}~\bibnamefont
  {Sols}}, \ and\ \bibinfo {author} {\bibfnamefont {I}~\bibnamefont {Zapata}},\
  }\bibfield  {title} {\enquote {\bibinfo {title} {{Birth of a quasi-stationary
  black hole in an outcoupled Bose-Einstein condensate}},}\ }\href
  {http://stacks.iop.org/1367-2630/16/i=12/a=123033} {\bibfield  {journal}
  {\bibinfo  {journal} {New J. Phys.}\ }\textbf {\bibinfo {volume} {16}},\
  \bibinfo {pages} {123033} (\bibinfo {year} {2014}{\natexlab{b}})}\BibitemShut
  {NoStop}%
\end{thebibliography}%

\end{document}